\pdfoutput=0

\documentclass[twoside,12pt]{article}
\usepackage{textcomp} 
\usepackage{graphicx}
\usepackage{amsmath}  
\usepackage{amssymb} 
\usepackage{caption}
\usepackage{subcaption}

\def\Journal#1#2#3#4{{#1} {#2} ({#4}) {#3} }

\def\NPA{{\em Nucl. Phys.} A}

\def\PLB{{\em Phys. Lett.} B}

\def\PRL{\em Phys. Rev. Lett.}

\def\PREP{\em Phys. Rep.}

\def\PRD{{\em Phys. Rev.} D}
\def\PRE{{\em Phys. Rev.} E}
\def\PRC{{\em Phys. Rev.} C}

\newcommand{\be}{\begin{equation}}
\newcommand{\ee}{\end{equation}}
\newcommand{\bea}{\begin{eqnarray}}
\newcommand{\eea}{\end{eqnarray}}

\begin{document}
%\title{ \vspace{1cm} Radio detection of Cosmic-Ray Air Showers\\ and High-Energy Neutrinos }
\title{Radio detection of Cosmic-Ray Air Showers\\ and High-Energy Neutrinos \\ {\small published by \emph{ELSEVIER} in \emph{Progress in Particle and Nuclear Physics} 93 (2017) 1-68 http://dx.doi.org/10.1016/j.ppnp.2016.12.002} }
\author{Frank G.\ Schr\"oder$^1$\\
\\
$^1$Institut f\"ur Kernphysik, Karlsruhe Institute of Technology (KIT), Germany}
\maketitle

\begin{abstract}
In the last fifteen years radio detection made it back to the list of promising techniques for extensive air showers, firstly, due to the installation and successful operation of digital radio experiments and, secondly, due to the quantitative understanding of the radio emission from atmospheric particle cascades.
The radio technique has an energy threshold of about $100\,$PeV, which coincides with the energy at which a transition from the highest-energy galactic sources to the even more energetic extragalactic cosmic rays is assumed.
Thus, radio detectors are particularly useful to study the highest-energy galactic particles and ultra-high-energy extragalactic particles of all types.
Recent measurements by various antenna arrays like LOPES, CODALEMA, AERA, LOFAR, Tunka-Rex, and others have shown that radio measurements can compete in precision with other established techniques, in particular for the arrival direction, the energy, and the position of the shower maximum, which is one of the best estimators for the composition of the primary cosmic rays. 
The scientific potential of the radio technique seems to be maximum in combination with particle detectors, because this combination of complementary detectors can significantly increase the total accuracy for air-shower measurements.
This increase in accuracy is crucial for a better separation of different primary particles, like gamma-ray photons, neutrinos, or different types of nuclei, because showers initiated by these particles differ in average depth of the shower maximum and in the ratio between the amplitude of the radio signal and the number of muons.
In addition to air-shower measurements, the radio technique can be used to measure particle cascades in dense media, which is a promising technique for detection of ultra-high-energy neutrinos.
Several pioneering experiments like ARA, ARIANNA, and ANITA are currently searching for the radio emission by neutrino-induced particle cascades in ice.
In the next years these two sub-fields of radio detection of cascades in air and in dense media will likely merge, because several future projects aim at the simultaneous detection of both, high-energy cosmic-rays and neutrinos.
SKA will search for neutrino and cosmic-ray initiated cascades in the lunar regolith and simultaneously provide unprecedented detail for air-shower measurements.
Moreover, detectors with huge exposure like GRAND, SWORD or EVA are being considered to study the highest energy cosmic rays and neutrinos.
This review provides an introduction to the physics of radio emission by particle cascades, an overview on the various experiments and their instrumental properties, and a summary of methods for reconstructing the most important air-shower properties from radio measurements.
Finally, potential applications of the radio technique in high-energy astroparticle physics are discussed.
\end{abstract}

\eject

\clearpage

\tableofcontents

\clearpage

\section{Prologue}
In the last years interest in the radio technique has greatly increased for both cosmic-ray and neutrino detection at high energies starting around $100\,$PeV. 
Current experiments detect cosmic rays by the radio emission of particle cascades in air, and neutrinos are searched for by the radio emission of particle cascades in dense media such as the Antarctic ice or the lunar regolith. 
In both cases the principles of the radio emission by the particle cascades are the same, with some difference due to the length scales of the shower development related to the density of the medium.
The different length scales are the reason that air-shower detection is mostly done at frequencies below $100\,$MHz compared to a few $100\,$MHz for radio detection in dense media.
Moreover, they make the situation more complex for atmospheric showers, because two emission mechanisms contribute significantly to the radio signal, which are the geomagnetic and the Askaryan effects, while in dense media the geomagnetic effect is negligible.
From this point of view air showers constitute the more general case, and the radio emission by showers in dense media can be seen as a simplified case.
This is one of the reasons why the radio signal of air showers is the main focus of this review, where differences to the situation in dense media will be mentioned where important.

The focus on air showers also reflects the progress of current experiments, since several antenna arrays are successfully measuring cosmic-ray air showers, while radio experiments searching for neutrinos are mostly in the prototype phase aiming at a first detection.
In any case, the separation between air-shower detection for cosmic rays and dense media for neutrino detection will become less strict in the future.
Planned projects aim at air showers for neutrino detection and at particle cascades in the lunar regolith for cosmic-ray detection, respectively, but this foreseen merge of the two fields of cosmic-ray and neutrino detection might still take a few year.

For cosmic rays, radio detection is competitive already now.
Due to the availability of digital electronics and the feasibility of computing-intensive analysis techniques current radio arrays achieve similar precision as other technologies for air-shower detection. 
Several air-shower arrays already feature radio antennas in addition to optical and particle detectors, and others likely will follow. 
This combination of radio and other complementary detection techniques can be used to maximize the accuracy for the properties of the primary cosmic-rays, in particular the arrival direction, the energy, and the particle type. 
In this sense the radio technique for extensive air-showers has just crossed the threshold from prototype experiments and proof-of-principle demonstrations to their application for serious cosmic-ray science. 
This has been possible not only because of the technological advances, but also because the radio emission of air showers is finally understood on a quantitative level: current air-shower-simulation tools can predict the absolute value of the radio amplitude in agreement with measurements.

In summary, this review gives an extensive overview on recent developments in the radio-detection technique for extensive cosmic-ray air showers, and also includes related topics, in particular the search for ultra-high-energy neutrinos. 
The article covers the various experimental setups used for detection, the instrumental properties and methods, and the results achieved by air-shower experiments which successfully have detected cosmic rays.

In addition to providing an overview for the community of this research field of astroparticle physics, this review is hopefully of help for anybody who wants to start experimental work or analysis in this field. 
In comparison with other reviews on the radio detection of air showers \cite{HuegeReview2016, FilonenkoRadioReview2015}, this review is more extensive on practical experimental aspects, such as the design of radio experiments, treatment of background, or analysis methods for measurements by antenna arrays. 
Ref.~\cite{BrayReview2016} provides an extensive review on the search for particle cascades initiated in the lunar regolith by high-energy neutrinos using radio telescopes, and a summary on the situation of neutrino detection in ice can be found in Ref.~\cite{ConnollyNeutrinoReview2016}.
Other reviews and papers might as well be appropriate for overviews on related topics, like the general situation in ultra-high-energy cosmic-ray physics \cite{Bluemer2009, Antoine2011}, or other detection techniques for air showers \cite{Haungs2003}. 

\clearpage
 
\section{Introduction}

Cosmic rays are extrasolar particles, mainly atomic nuclei, with energies from less than a few $100\,$MeV up to at least a few $100\,$EeV. 
While at energies below $10^{14}\,$eV their composition and flux is accurately measured as a function of energy by space- and balloon-borne experiments \cite{SeoBalloonReview2012, SparvoliSatelliteReview2013, Maestro_CRdirect_ICRC2015}, at higher energies the flux of cosmic rays is simply too low to obtain sufficient statistics with the size of direct detectors.
Instead the atmosphere acts as a calorimeter and air showers are detected, i.e., cascades of secondary particles initiated by the primary cosmic rays. 
Neutral particles like photons and neutrinos are produced as secondary particles by cosmic-rays at the source or during propagation. 
Unlike charged cosmic rays or photons, most neutrinos do not initiate air showers, but pass the atmosphere without interaction due to their low cross section.
Then some of them initiate particle cascades below ground in the Earth where the higher density makes an interaction more likely than in the atmosphere. 
The radio signal emitted by all of these particle cascades in air or dense media provides one of a few possible detection techniques. 
The following sections briefly summarize the status of cosmic-ray physics in the relevant energy range, of air-shower physics, and of other established detection techniques.

\subsection{Cosmic rays} 
\begin{figure*}[t]
  \centering
  \includegraphics[width=0.99\linewidth]{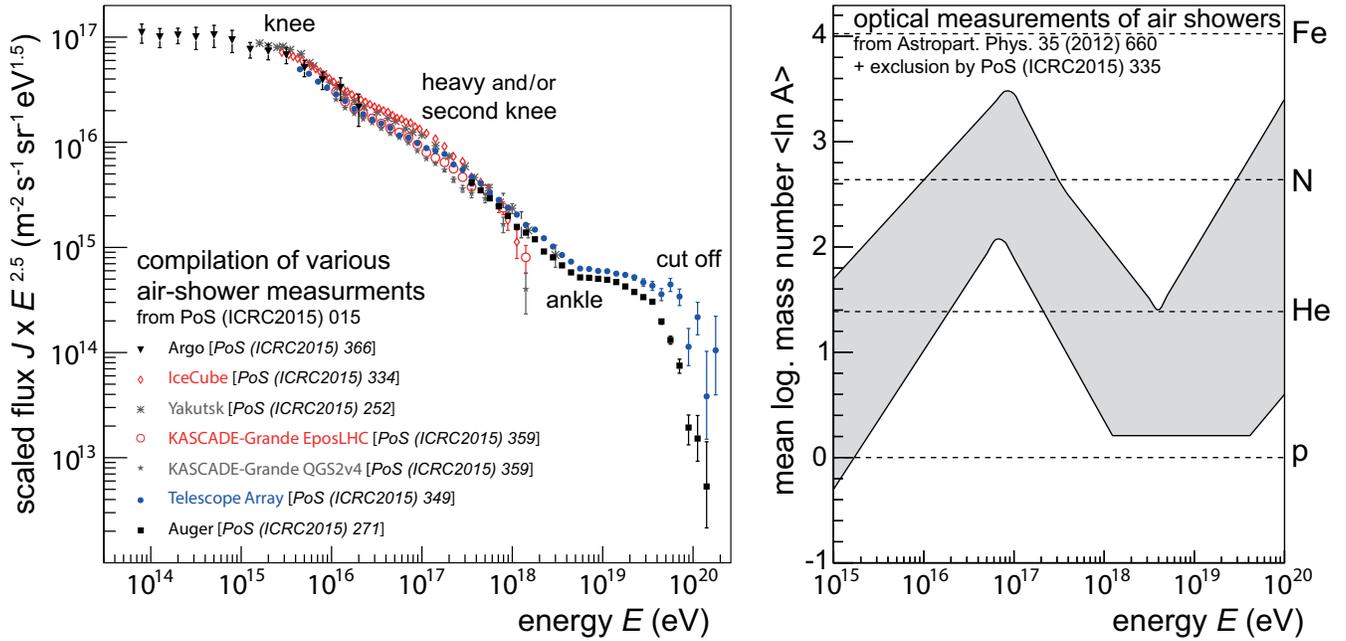}
  \caption{Cosmic-ray energy spectrum reconstructed from various air-shower measurements (left figure, modified from Ref.~\cite{Verzi_CRoverview_ICRC2015}) and average logarithmic mass derived from optical measurements of the atmospheric depth of the shower maximum (right figure, modified from Ref.~\cite{KampertUnger2012}). 
  The gray band includes systematic uncertainties due to the hadronic interaction models used for interpretation and additionally takes into account the exclusion of a pure proton composition by Ref.~\cite{AugerAnkleMixedCompPLB2016}.}
  \label{fig_energySpectrum}
\end{figure*}

The name \lq cosmic rays\rq~often is restricted to charged particles only, i.e., electrons, positrons, protons and atomic nuclei, where electrons and positrons are much less abundant. 
They have been measured only at lower energies up to a few TeV \cite{Maestro_CRdirect_ICRC2015}, and are not considered further in this review which, consequently, restricts the name cosmic rays to all kinds of nuclei including protons.
After many decades of air-shower measurements the energy spectrum of ultra-high-energy cosmic rays has been confined with greater accuracy, but the origin of the highest-energy particles is still unknown. 
While the majority of cosmic rays at least up to the energy of the \lq knee\rq~around $3$ to $5 \cdot 10^{15}\,$eV originate from inside of our galaxy, the origin of cosmic rays at highest energies is probably extra-galactic \cite{Giacinti2012}. 
The transition from galactic to extra-galactic origin is assumed to start at an energy of approximately $10^{17}\,$eV, which roughly corresponds to the lower threshold for the radio technique. 
Splitting the energy spectrum into two components, namely heavy and light nuclei, at this energy a softening\footnote{The terms \lq soft\rq~and \lq hard\rq~refer to the energy spectrum of a radiation. 
The spectrum is softer / harder if the flux decreases faster / slower with increasing energy, i.e., hard radiation has a higher fraction of high-energetic particles than soft radiation.} of the heavy component \cite{KG_heavyKnee}, and a hardening of the light component is observed \cite{KG_lightAnkle}.
Probably this is due to extragalactic cosmic rays whose flux is assumed to decrease slower with energy than for galactic cosmic rays in these energy range.
This means that the knee at $3$ to $5 \cdot 10^{15}\,$eV probably is due to the maximum acceleration energy of the dominant galactic sources for light nuclei \cite{KG_elementSpectra2009}, and the \lq heavy knee\rq~at $10^{17}\,$eV correspondingly is due to the maximum energy for heavy nuclei. 
Moreover, several experiments reported another kink in the energy spectrum around $3$ to $4 \cdot 10^{17}\,$eV called the \lq second knee\rq~\cite{HoerandelErice2006, Tunka133_UHECR2014}, but it is not yet clear whether this is the same or a different feature than the \lq heavy knee\rq~when taking into account all systematic uncertainties of the different experiments and their interpretation. 
The \lq ankle\rq~ in the energy spectrum at approximately $10^{18.7}\,$eV \cite{AugerInclinedEnergySpectrum2015} could mark the end of the transition from galactic to extragalactic cosmic rays. 
Up to the ankle the fraction of light nuclei increases, but a pure proton composition is excluded by recent measurements \cite{AugerXmaxImplications2014, AugerAnkleMixedCompPLB2016}, in contrast to earlier assumptions \cite{DipModel2005}.
At even higher energies the composition becomes heavier again, but current measurements are not very accurate as shown in figure \ref{fig_energySpectrum}. 

This observation is consistent with the assumption that a different type of source is responsible for cosmic rays at higher energies. 
Whatever the acceleration mechanism the maximum energy should be proportional to the charge Z of the nuclei. 
Since protons and light nuclei like alpha particles are more abundant in the universe than heavier nuclei, the natural assumption is that heavy nuclei have a significant abundance among the cosmic rays only above the maximum energy for protons and helium nuclei. 
Since different sources could have a different maximum energy, the situation is more complex, but the general picture remains true. 
Thus, a composition-sensitive measurement is crucial to understand the properties of the cosmic-ray accelerators.
Around $10^{19.6}\,$eV a cutoff in the cosmic-ray spectrum is observed \cite{AugerFluxSuppresion2008}. 
The reason for this cutoff remains unclear. 
It could be that protons and nuclei lose energy due to interaction with the cosmic microwave background by the so-called GZK (Greisen-Zatsepin-Kuzmin) effect \cite{Allard2012}, or the cutoff could mark the maximum energy of the cosmic-ray accelerators \cite{AugerMatteoICRC2015}. 
Again, more accurate measurements of the composition are required to distinguish which of the two different scenarios is the dominating reason for the cutoff. 

While statistical composition measurements are sufficient to make further progress in understanding the origin of cosmic rays, measurements of the mass of individual cosmic-ray particles would provide another advantage.
Due to galactic and intergalactic magnetic fields the cosmic nuclei are deflected on their way to Earth, therefore, the arrival direction usually does not point back to the source. 
However, weak anisotropies in the arrival directions have been observed both at low energies in the GeV to PeV range \cite{ARGOanisotropy2015, HAWCanistropy2014, IceTopAnisotropy2013}, as well as at the highest energies \cite{AugerAnisotropy2015}. 
Moreover, a warm spot \cite{AugerWarmSpot2015}, and a hotspot \cite{TAhotspot2014} potentially indicate sources at the highest energies. 
Since the deflection in the magnetic fields depends on the rigidity (= momentum divided by charge) of the particles, there is hope that sources can be identified more easily once the anisotropy can be studied separately for light and heavy nuclei.
Thus, the potentially increased accuracy for the primary particle type is one of the scientific motivations for adding radio antennas to existing cosmic-ray observatories - especially since other established techniques for estimation of the primary particle type, namely air-fluorescence and air-Cherenkov detection, can be operated only during clear nights.

\subsection{Gamma rays and neutrinos}
In contrast to charged cosmic rays, gamma rays and neutrinos are neutral particles and travel in straight lines without deflection by galactic or extragalactic magnetic fields.
Hence, multi-messenger astronomy with photons and neutrinos could yield a possible breakthrough in finding the sources of ultra-high-energy cosmic rays, since these neutral particles point back directly to their origin \cite{BeckerMultimessengerReview2008, Multimessenger_UHECR2012, AMON2012}.
It is very likely that every source of cosmic rays emits also photons and neutrinos as secondary particles, e.g., as decay products of pions produced in hadronic interactions of the cosmic rays with material in or around the source. 
Moreover, high-energy photons and neutrinos are expected to be produced during the propagation of ultra-high-energy cosmic rays, in particular by the GZK effect. 
For all relevant production mechanisms the energy of the photons and neutrinos is at least an order of magnitude lower than the energy per nucleon of the primary particle, which implies that the photons and neutrino fluxes are the lower the heavier the composition of the primary particles is.
Above a cosmic-ray energy of approximately $10^{19.5}\,$eV the photons of the cosmic microwave background have sufficient energy in the center-of-mass system to excite cosmic-ray protons and nuclei. 
In the subsequent decay photons and neutrinos are produced, whose energy is highest for the decay of excited protons and much lower for the decay of nuclei. 
Therefore, the expected flux of neutrinos and photons depends strongly on the composition of the charged cosmic rays and provides an independent check of composition measurements.

So far the highest energy photons discovered have energies of a few $100\,$TeV, and photons of higher energy might be absorbed during their way to Earth by interaction with background radiation \cite{FunkGammaReview2015}.
Because the mean free path length of PeV photons is of the order of galactic distances, the lack of detection of higher energy photons thus is an additional hint that ultra-high-energy cosmic rays originate from extragalactic sources. 
For neutrinos the highest detected energy is a few PeV \cite{IceCubeNeutrinosScience2013}, which still is many orders of magnitude below the highest energy of cosmic-rays of a few $100\,$EeV.
Unlike photons, neutrinos are not expected to be absorbed and can travel over cosmological distances, but it is not yet sure if their flux at EeV energies will be sufficient for detection in the near future. 
Even once neutrinos at EeV energies will have been detected, it might take long until first sources are statistically revealed, since the missing horizon for neutrino propagation implies that neutrinos arrive from all directions and individual sources are washed out by a large isotropic background.
In summary, the search for ultra-high-energy photons and neutrinos is very worthwhile, but cosmic-ray air showers currently are the only guaranteed window to the Universe at ultra-high energies.

\subsection{Air showers}
\label{sec_airShower}
Air showers are cascades of secondary particles initiated by primary cosmic-ray particles hitting the atmosphere as discovered in the 1930's \cite{Kohlhoerster1938, AugerEhrenfestMaze1939}. 
These cascades are conceptually similar to the ones developing in dense media, like water or ice, or in calorimeters used for particle detection in high-energy physics.
However, air showers are much more extensive with length scales of kilometers while particle cascades in dense media develop within meters. 
This makes a difference for the radio emission by air-showers and neutrino-induced showers in dense media as explained in chapter \ref{sec_radioEmission}.
The focus of this section is on air showers and in particular some general aspects relevant for the radio emission.
More detailed information on air showers can be found elsewhere, e.g., in the didactic introduction by Matthews \cite{Matthews2005}. 

In a simplistic picture an air shower develops as follows: 
in the first interaction of the primary particle in the atmosphere many unstable secondary particles are produced, among them pions and other mesons. 
In particular, the neutral pions will decay almost immediately into photons, which initiate an electromagnetic cascade: 
by pair production electrons and positrons are created which themselves create further photons, e.g., by bremsstrahlung and the inverse Compton effect. 
Moreover, some positrons annihilate with electrons of air atoms producing further photons, and some photons ionize air atoms so that the electrons become part of the air shower. 
Thus, the electromagnetic cascade develops a negative net charge excess during the shower development of $20-30\,\%$ of the total number of electrons and positrons \cite{ScholtenARENA2010}.

The other particles produced in the first interaction partly decay, and partly interact again producing further particles. 
A decay into electrons or photons feeds the electromagnetic cascade, a production of hadrons keeps the shower development going. 
Depending on the primary energy, after a certain number of generations the shower development runs out of fuel: 
the energy of the secondary particles becomes so low that ionization starts to dominate over bremsstrahlung, and less new particles are produced than are absorbed by interactions with the air. 
Thus, the number of particles in the shower reaches a maximum, which is the closer to the ground the higher the energy of the primary particle is.

In the shower development only decays into muons and neutrinos have to be treated separately.
In first order these particles do not contribute to the further shower development, since their interaction probability in the atmosphere is low, and they have a high chance to reach the ground. 
Neutrinos are generally invisible for air-shower detectors and constitute an intrinsic systematic uncertainty when estimating the energy of the primary particle from the measured air-shower energy. 
The muonic component of the shower is invisible to most detection methods, except for dedicated particle detectors or for Cherenkov-light detection in water or ice. 

The type of the primary particle has only little influence on the shower development, but nevertheless this little influence practically is the only way to statistically determine the composition of the primary cosmic rays at high energies. 
Primary photons, electrons and positrons (though not yet detected at ultra-high energies) would directly start an electromagnetic cascade causing muon-poor air showers. 
For primary hadrons, i.e., protons and nuclei, the cross section is roughly proportional to the number of nucleons. 
Thus, heavy nuclei on average interact earlier in the atmosphere and additionally the shower develops faster with fewer generations of secondary particles.
Hence, the shower maximum is higher up in the air, less of the total energy is transferred into the electromagnetic component, and the number of muons is larger than for showers initiated by light nuclei of the same energy.

Consequently, the atmospheric depth of the shower maximum, $X_\mathrm{max}$, and the electron-muon ratio are the two main shower observables available for estimating the type of the primary particle ($X$ refers to the column density of traversed atmosphere measured in g/cm\textsuperscript{2}. 
According to the paradigm of shower universality the whole air shower can be described to good approximation by only a few observables \cite{ShowerUniversalityGiller2004, ShowerUniversalityKG2008}, in particular direction, energy, $X_\mathrm{max}$, and electron-muon ratio. 
Thus, measuring additional observables does add only little information, and further composition-sensitive observables are not independent, but can be derived from the other observables. 
Therefore, the benefit of different detection technologies for air showers, including the radio technique, depends primarily on the achievable accuracies for the energy, for $X_\mathrm{max}$, and for the electron-muon ratio.

Since an air shower typically has millions to billions of particles undergoing stochastic interaction processes, it is practically impossible to calculate probability distributions of air-shower observables. 
Hence, for analyses and interpretations of measurements, air showers are generated by Monte Carlo simulations. 
In the recent years most simulations have been done with the programs CORSIKA \cite{HeckKnappCapdevielle1998} and AIRES \cite{AIRESreference}, which both feature extensions to calculate the radio emission generated by the air showers \cite{LudwigREAS3_2010, HuegeCoREAS_ARENA2012, AlvarezMuniz_ZHAires_air2012}. 
These programs rely on hadronic interaction models to simulate the air shower, such as FLUKA \cite{FLUKA_2007} or URQMD \cite{URQMD} for interactions at lower energy and Sybill \cite{SybillPRD2009,SybillICRC2015}, QGSJET \cite{OstapchenkoQGSjetII2006}, and EPOS \cite{EPOS_LHC, EPOS_orig}, for interactions at the highest energies. 
In principle these models can also be used for particle cascades in dense media, for which additionally GEANT4 has been used in combination with end-point and ZHS formalisms \cite{JamesEndPoint2010, ZHS_1991, ZHS_1992} for calculating the radio emission \cite{ZillesSLAC_Genat4_2015}.

The hadronic-interaction models constitute a major systematic uncertainty for the interpretation of any air-shower measurement, because the center-of-mass energy in the first interactions significantly exceeds the maximum energy studied at accelerators like LHC at CERN. 
Moreover, interactions in the forward direction, i.e., at high pseudorapidities, are very relevant for air showers, but not well studied at accelerator experiments. 
Therefore, the hadronic interaction models use extrapolations and postulated assumptions for ultra-high-energy interactions. 
Until now, no hadronic interaction model has been able to describe all air-shower measurements consistently. 
In particular, the models predict fewer muons on ground than measured \cite{KGmodelsComparison2015, AugerMuonComparison2015}. 
However, the electromagnetic component of air-showers, which is the one relevant for the radio emission, is described consistently. 
Thus, it can be assumed that the choice of a certain hadronic interaction model has little impact on the simulation and interpretation of the radio emission by air showers, but this has not yet been investigated in detail.

\begin{figure*}
  \centering
  \includegraphics[height=9cm]{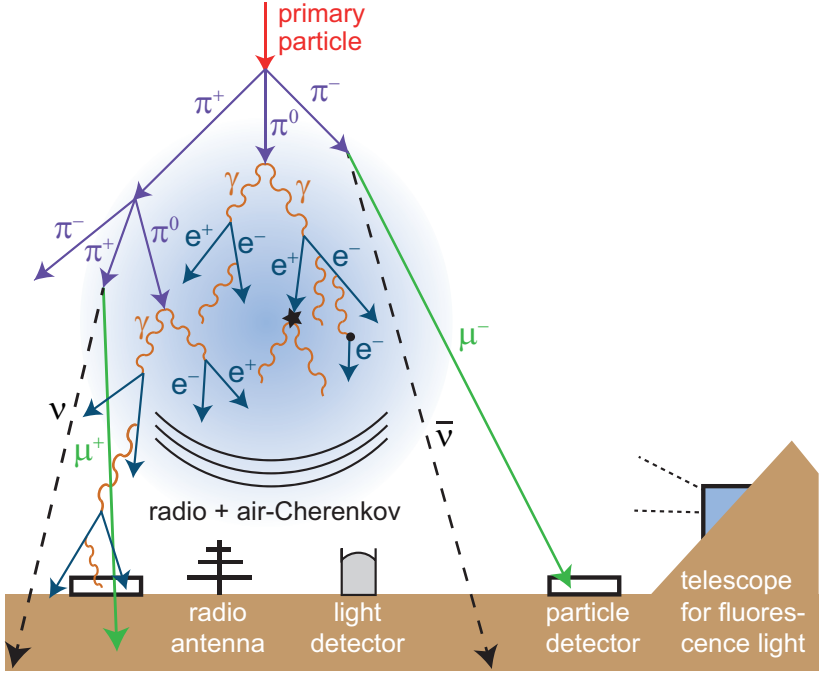}
  \,\,\,\,
  \includegraphics[height=9cm]{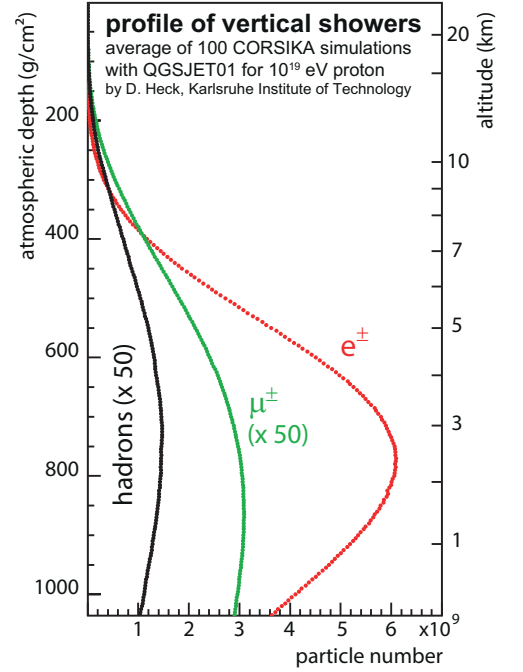}  
  \caption{Left: Simplified sketch of an air shower and possible detection techniques. 
  Real air showers contain more particle types than displayed, but most detection techniques (radio, air-Cherenkov and fluorescence light) are only sensitive to electrons and positrons. 
  Particle detectors also measure muons, which generally reach further out than the electromagnetic component. 
  Right: Longitudinal shower profile for different particle types.}
  \label{fig_showerSketch}
\end{figure*}

\subsection{Detection techniques for air showers}

Since the discovery of air-showers several detection techniques have been developed and are still used (see figure \ref{fig_showerSketch}).
These can be classified into two main categories: 
First of all the direct detection of air-shower particles on ground or underground. 
Secondly the measurement of electromagnetic radiation generated directly or indirectly by the electromagnetic component of the air shower (with only a minor contribution by the muonic component), in particular air-fluorescence and air-Cherenkov light at optical and ultraviolet frequencies, and radio emission. 
Other emission processes of the electromagnetic shower component, like molecular bremsstrahlung, have been proposed \cite{GorhamMolecularBremsstrahlung2008}, but could not yet be experimentally confirmed \cite{MicrowaveExperiments_ICRC2013, CROME_PRL2014}. 
In all cases the shower direction is accessible by measuring arrival times of the signal, and the energy is accessible by integration of the measured signal strengths, i.e., either of the amount of measured particles on ground or the amount of measured electromagnetic radiation. 
The composition-sensitive variables $X_\mathrm{max}$ and electron-muon ratio are usually more difficult to measure, e.g., for radio measurements $X_\mathrm{max}$ is not directly visible, but can be reconstructed from several properties of the measured signal (cf.~chapter \ref{sec_reconstruction}).

Independent of the technique, the low flux of primary cosmic rays at high energies requires large experiments to compensate. 
With the fluorescence technique large atmospheric volumes can be observed with a single telescope sensitive to showers at distances up to a few $10\,$km. 
For all other techniques large ground arrays have to be built with areas up to several $1000\,$km\textsuperscript{2}, where the detector spacing depends on the technique and the targeted energy.
The spacing ranges from a few meters like at KASCADE \cite{Apel2010KASCADEGrande} up to more than a kilometer at the Pierre Auger Observatory \cite{AugerNIM2015}.

In the past all kinds of particle detectors have been used to construct arrays for the detection of air-showers. 
Current experiments mostly use scintillation detectors \cite{AMIGA_JINST2016, TA_NIM2012} and water-Cherenkov detectors \cite{AugerSD2010, IceTopNIM2013} for measuring secondary air-shower particles. 
To obtain information on the electron-muon ratio of the air shower the type of the secondary particles on ground has to be separated in the detector signal. 
For both detector types this is possible using several approaches, but only within limited accuracy when many muons and electrons at different energies arrive at the same time. 
General strategies to distinguish muons from electrons are the use of absorbers, since electrons are absorbed faster than muons, or the combination of two different particle detectors with a different response for electrons and muons.
Moreover, by analyzing the time structure of the signal in individual stations the position of the shower maximum and, thus, its atmospheric depth $X_\mathrm{max}$ can be estimated \cite{AugerSD_Xmax}. 
However, until now it could not be demonstrated that $X_\mathrm{max}$ measurements by particle-detector arrays can achieve the same accuracy as radiation techniques such as air-fluorescence, air-Cherenkov, and radio detection.

The fluorescence technique relies on light emitted by excited nitrogen molecules in the air traversed by the air shower \cite{HiRESfinalResults, AugerFD_NIM}. 
This technique provides direct sensitivity to the position of the shower maximum and a calorimetric measurement of the energy. 
Furthermore, it currently sets the benchmark for the accuracy of the absolute energy scale - with $14\,\%$ total systematic uncertainty for fluorescence detection at the Pierre Auger Observatory \cite{AugerNIM2015} originating mainly from instrumental uncertainties and uncertain atmospheric conditions during an individual measurement. 
While for the interpretation of particle measurements at ground hadronic interaction models constitute a significant uncertainty, measurements of fluorescence light are easier to interpret.
Similar to air-Cherenkov light and to the radio emission by air showers also the fluorescence light is hardly sensitive to the poorly understood muonic component, but mostly to the well-understood electromagnetic component of the air shower.

Many of the electrons and positrons produced in the particle cascades are faster than the speed of light in air and emit Cherenkov light. 
Detection of Cherenkov light is the main technique for showers in dense media, in particular for the detection of high-energy neutrinos: optical modules measure the Cherenkov-light of neutrino-induced particle cascades in ice \cite{IceCubeFirstYear2006} or water \cite{BaikalNeutrino2006, AntaresNim2011}.
Nonetheless, this technique is also used for atmospheric showers: 
for very-high-energy gamma astronomy at energies less than a PeV, imaging telescopes detecting the Cherenkov light of photon-induced air showers are the technique of choice \cite{FunkGammaReview2015}. 
At higher energies ground arrays of photomultipliers like Tunka-133 \cite{Tunka133_NIM2014} detect Cherenkov light emitted by cosmic-ray air showers in the near ultraviolet, since this band provides an optimal signal-to-background ratio. 
At these wavelengths the emission of Cherenkov light is incoherent since the typical spacing between electrons in the air shower is much larger than the wavelength \cite{RisseHeck2003}. 
As consequence the light intensity is roughly proportional to the energy of the electromagnetic component, like for the fluorescence light, i.e., the radiation energy of the Cherenkov light and of the fluorescence light each scale linearly with the shower energy. 
Thus, detection of air-Cherenkov light also yields an accurate measurement of the air-shower energy. 
The position of the shower maximum can be reconstructed from the steepness of the lateral distribution, i.e., how rapidly the intensity decreases with distance from the shower axis, and from the pulse shape when the distance to the shower axis is known.

The main characteristics of the radio emission by cascades in any media are similar to those of air-Cherenkov emission, i.e., the radio emission is beamed in the forward direction of the shower and has similar sensitivity to the shower maximum \cite{TunkaRex_XmaxJCAP2016}. 
However, the mechanisms causing the radio emission are very different, namely geomagnetic deflection of electrons and positrons in air and time variation of the net-charge excess of the shower in all media, as explained in chapter~\ref{sec_radioEmission}. 
Moreover, due to the larger wavelengths of the order of meters radio emission is mostly coherent, which means that the field strength (= amplitude) is roughly proportional to the energy of the shower, i.e., the intensity and radiation energy of the radio signal scale quadratically with the shower energy. 
Compared to the air-fluorescence and air-Cherenkov techniques the main advantage of the radio technique is that it can be operated around the clock, not limited to dark nights with a clear sky. 
Furthermore, in contrast to optical and ultraviolet light, absorption of radio waves in air is negligible for frequencies below $50\,$GHz \cite{ITUatmosphericAttenuation2015}, which covers the complete relevant frequency range (most experiments use $30-80\,$MHz).
This is an additional advantage for the radio technique in particular relevant for inclined showers, because their shower maximum is more distant to the detector and the signal has a larger way through the air.

In summary, of all the different detection techniques air-fluorescence measurements currently provide the highest accuracy for the energy and the shower maximum:
about $14\,\%$ for the absolute energy and better than $20\,$g/cm\textsuperscript{2} for the shower maximum, which corresponds to less than the average difference between showers initiated by protons and helium nuclei, respectively \cite{AugerNIM2015}. 
The air-Cherenkov and radio techniques achieve about the same accuracy \cite{Tunka133_NIM2014, TunkaRex_XmaxJCAP2016, BuitinkLOFAR_Xmax2014}, but systematic uncertainties have not yet been studied as intensively as for air-fluorescence detection.
Of the radiation techniques radio detection is the only one that works during daytime and bad weather. 
Furthermore, radio detection provides the intrinsic advantage of higher sensitivity to inclined showers, but, as we will see, has the disadvantage of a higher detection threshold. 
The advantages are shared by particle detectors for air-shower muons, which provide additional sensitivity to the type of the primary particle when combined with a measurement of the electromagnetic shower component.
Consequently, the highest possible accuracy for air-shower measurements is only achievable with hybrid observatories combining a particle detector array with at least one of the radiation techniques. 
Currently running hybrid arrays are the Telescope Array (fluorescence telescopes + scintillators) \cite{TA_NIM2012}, the Pierre Auger Observatory (fluorescence telescopes + water-Cherenkov detectors + scintillators) \cite{AugerNIM2015}, and the Tunka experiment (air-Cherenkov detectors + scintillators) \cite{TAIGA_2014}, of which the latter two additionally feature radio extensions.

\subsection{Particle cascades in dense media}
\label{sec_DenseMediaCascades}
Particle cascades in dense media are known since long in high-energy physics \cite{AmaldiCalorimeterReview1981}, and explained in various text books on experimental particle physics and detector design.
The development of a shower after the collision of a high-energy particle with matter is the essence of calorimeter detectors.
The main difference to air showers is that the higher density leads to much shorter length scales, i.e., the shower develops over centimeters to meters instead of kilometers. 
This implies that particles are more likely to interact before decaying which has some impact on the relative sizes of the electromagnetic and muonic components. 
Also other parameters depend on the properties of the medium, e.g., the critical energy at which energy loss of electrons by ionization and bremsstrahlung is equal. 
Details are taken into account by Monte Carlo simulation codes for particle cascades \cite{ZillesSLAC_Genat4_2015, HeckKnappCapdevielle1998}, and an extensive discussion is beyond the scope of this review. 
Ref.~\cite{DenseMediaComparison2006} discusses how these details affect the radio signal for the three media considered most important for radio-detection of neutrinos, which are ice, salt, and the lunar regolith.

In principle, particle cascades in dense media can be detected by the same techniques as air-showers, and indeed particle detectors have been deployed in underground laboratories and mines \cite{EMMA2011}. 
These underground particle detectors measure mostly muons from air showers, and a major challenge for neutrino searches is the separation of particles created in the medium by neutrino interactions from air-shower muons.
Only cascades or muons initiated underground are a clear indication for neutrinos, since this implies that the primary particle traversed the atmosphere without interaction. 
The separation of underground interactions requires either a veto for air-shower muons or large detector volumes in order to distinguish whether a cascade or a muon track started inside or outside of the detector medium. 
Considering the low flux of neutrinos above TeV energies requiring cubic-kilometer-size detectors \cite{IceCubeNeutrinosScience2013}, conventional particle detectors are too expensive and other techniques are used for the detection at high energies.

At the moment the state-of-the-art technique for high-energy neutrinos is the detection of Cherenkov light emitted by muon tracks or showers in transparent media, in particular in ice \cite{IceCubeFirstYear2006} or water \cite{BaikalNeutrino2006, AntaresNim2011}. 
For this purpose large networks of photomultipliers have been deployed with precise relative timing. 
The amount of detected light provides a measure for the energy of the neutrino initiating the track or the cascade, and the arrival direction of the neutrino can be estimated from the light-arrival time in the individual photomultipliers. 

As more economic alternatives for detectors even larger than a few cubic kilometers, acoustic \cite{Karg_AcousticOverviewIce_ARENA2012, Graf_AcousticOverviewSea_ARENA2012} and radio detectors are under study \cite{ARA_2015, ARIANNA_2015, GreenlandInIce_2016}.
The rapid heat deposit of a particle cascade in the medium is expected to cause a pulse of sound \cite{AskaryanAcoustic1979, Learned1979}, as has been experimentally observed for particle and laser beams \cite{Lahmann2015}. 
Thus, particle cascades initiated by neutrinos in suitable dense media might be detectable with acoustic sensors.
At the South Pole the acoustic absorption length in ice has been measured to be approximately $300\,$m \cite{SouthPole_AcousticAttenuation2011} compared to about $1\,$km for radio waves \cite{IceAttenuationRICE2005}, and to $100-250\,$m for optical Cherenkov light in the deep ice \cite{IceCubeLightAttenuation2013}.
This would require a relatively dense spacing of acoustic detectors with about ten times more detector strings than for radio arrays, which is important since the deployment of detectors in the ice is a major cost driver.
Hence, in ice the acoustic technique might play a role not as stand-alone method, but in hybrid detectors together with the radio technique in order to observe a fraction of the neutrinos simultaneously with both techniques. 
In water the situation is better for the acoustic technique, since the attenuation length is a few km in sea water and even larger than $10\,$km in fresh water in the relevant frequency range around $10\,$kHz \cite{NahnhauerRICAP2011, LahmannARENA2016}.
Therefore, research on prototype experiments is going on in particular in the Mediterranean Sea and at Lake Baikal.
While Lake Baikal has the advantage of less sound attenuation, its disadvantage compared to the Mediterranean Sea is its lower temperature which is close to the temperature of maximum water density.
This decreases the amplitude of the sound pulse emitted by particle cascades, because the amplitude is correlated with the density change of the water due to heat deposit. 
For these reasons it is not yet clear what would be the optimal site on Earth for an acoustic neutrino detector. 
In any case, the thresholds of discussed acoustic arrays are currently at neutrino energies around an EeV, which is an order of magnitude higher than the estimated threshold of radio arrays in ice.

The radio signal is emitted by cascades in dense media due to the Askaryan effect \cite{Askaryan1962, Askaryan1965} (as explained in later sections) and is expected to be detectable with sparse radio networks for neutrino energies around $100\,$PeV.
In contrast to the acoustic technique, the radio technique has been proven not only at accelerator experiments \cite{Saltzberg_SLAC_Askaryan2001}, but also in nature, namely in air showers \cite{AugerAERApolarization2014}.
Thus, it is not a principle question if this method works for neutrino detection, but the main issue is that the detection volume has to be sufficiently large for the yet unknown flux of neutrinos at energies above $10\,$PeV. 
While in salt the attenuation length of radio waves is only $30-300\,$m \cite{SalsaARENA2010}, which probably is too short for large detector volumes, in ice larger attenuation lengths of up to $1.5\,$km at the South Pole and about $1\,$km in Greenland have been measured \cite{IceAttenuationRICE2005, IceAttenuationARIANNA2011, IceAttenuationGreenland2015}.
This makes the radio technique promising for the search of ultra-high-energy neutrinos.
Several prototype experiments in Antarctica and Greenland are currently aiming at the first detection of EeV neutrinos and in later stages will have thresholds of $10-100\,$PeV.

Finally, the Moon can be used as detector for particles at even higher energies.
The radio emission of showers initiated by ZeV neutrinos or cosmic rays in the lunar regolith would be observable on Earth with current radio telescopes \cite{BrayReview2016, WesterborkCRlimit2010}, but the existence of particles with such high energy is speculative. 
Next-generation radio telescopes, such as the SKA, are expected to be sensitive enough for lunar particle cascades in the energy range around $100\,$EeV \cite{SKAlunar_ICRC2015}.
This will make the same technique of lunar observation applicable for non-speculative cosmic-ray detection at the highest energies.

\clearpage
 
\section{Historic overview}
Radio emission by particle cascades in the atmosphere and in dense media was already predicted more than 50 years ago \cite{Askaryan1962, KahnLerche1966}, and soon after discovered for air showers \cite{Jelly1965}.
While theoretical calculations for radio emission by air showers and cascades in dense media have been performed at all times since then, the experimental history of the field is divided into two main periods:
An 'analog' epoch of air-shower measurements in the 1960's and the beginning of the 1970's, and a 'digital' epoch starting at the end of the 1990's with prototype experiments for neutrino search and the revival of the radio technique for air-shower measurements in the 2000's.
The results of the air-shower measurements in the analog epoch have been reviewed by Allan in 1971 \cite{AllanReview1971}, and only little progress had been made in the remaining years of the last century.
Since this review concentrates on the digital epoch, the summary on the analog epoch is restricted to a few sentences, despite the immense credit the researchers of this period deserve for their pioneering work. 

In the analog epoch experiments have been performed at several locations, e.g., Jodrell bank near Manchester \cite{Jelly1965}, Moscow \cite{Vernov1967}, Haverah Park in England \cite{Allan1964}, Medicina in Italy \cite{Medicina1977}, Penticton in Canada \cite{Hough1970}, Chacaltaya mountain in Bolivia \cite{Hazen1969}, and other places.
These first experiments basically relied on taking photos of oscilloscopes triggered by simple air-shower detectors, like arrays of Geiger counters. 
In this pioneering phase a wide range of frequency bands was explored:
radio emission by air-showers was not only discovered in the range of $30 - 80\,$MHz mostly used today, but also at higher frequencies up to several $100\,$MHz \cite{Fegan1968, Spencer1969, AllanFreqSpecAndNoise1970, Charman1970}.
Moreover, detection at lower frequencies of a few MHz \cite{AllanFreqSpecAndNoise1970, Stubbs1971}, and even around $100\,$kHz had been reported \cite{ClayICRC1973}, but the origin of these radio signals remains questionable, and it is not clear in which way the detected signals were correlated with the coincident air showers \cite{ClayICRC1975}.
This issue of the radio signal at low frequencies remains unsolved until today, even though low-frequency measurements were one of the very few radio activities still ongoing in the 1980's in Japan \cite{AGASAlowFreq1985, AGASAlowFreq1993}, and in the USSR \cite{YakutskLowFreq}.

The main results of the analog epoch can be very briefly summarized as follows: 
Most of the qualitative features of the radio signal have been discovered, in particular its dependence on the shower energy and the geomagnetic angle, i.e., the angle between the shower axis and the Earth's magnetic field. 
Also various more subtle effects on the lateral distribution of the radio signal had been studied, like its dependence on the position of the shower maximum \cite{AllanICRC1971, Hough1973}. 
However, no general consensus had been achieved, and different experiments showed a significant disagreement in the absolute strength of the radio signal. 
Finally, the measurements were insufficient to reconstruct air-shower parameters like energy and $X_\mathrm{max}$ with an accuracy competitive to other techniques. 

In the 1970's the cosmic-ray community almost stopped research on the radio technique and instead focused on other techniques, like observation of air-Cherenkov and air-fluorescence light. 
It seems that several reasons played a role for favoring other techniques at this time, and it is not clear to me which reason had been the most important one.
Maybe the main reason was the unsatisfactory accuracy of air-shower parameters achieved at this time, or maybe that radio emission by distant thunderstorms disturbed the measurements. 
Luckily, both shortcomings have mostly been solved by recent digital experiments as will be outlined later in this review.

Another reason for focusing on other techniques at that time might have been the small detectable radio footprint which has a diameter of only $200-300\,$m for vertical showers, compared to a few km for the footprint of the air-shower particles. 
This means that huge arrays of several $100$ or $1000\,$km\textsuperscript{2} would have to be equipped much denser with radio antennas than with particle detectors. 
Only at the start of the digital epoch it was realized that for very inclined showers the radio footprint becomes of comparable size as the particle footprint \cite{Gousset2004}, which could make arrays with large antenna spacings technically and economically feasible.  

Nevertheless, interest in the radio technique had never totally deceased.
In the 1980's concepts for large-scale radio arrays for neutrinos and cosmic rays were developed \cite{Markov1986}, and radio antennas were operated at the air-shower array in Yakutsk \cite{YakutskOldMeas2013}. 
First digital radio measurements were performed at low frequencies up to a few MHz by the AGASA and the EAS-TOP experiments \cite{AGASAdigital1991, EASRADIO1993}.
New theoretical calculations in the 1990's for the radio emission in dense media \cite{ZHS_1991} were followed by the RICE experiment consisting of radio antennas deployed in the Antarctic ice at the AMANDA neutrino observatory \cite{RICE_1998}. 
In the early 2000's the Askaryan effect was experimentally confirmed for dense media at the SLAC accelerator \cite{Saltzberg_SLAC_Askaryan2001}.
This provided additional confidence for the feasibility of neutrino detection with the radio technique, and was one of the motivations for the ANITA experiment searching for radio emission from showers induced by neutrinos in the Antarctic ice \cite{ANITAlite_PRL_2006}.

Moreover, large digital antenna arrays have been planned for radio astronomy, and new theoretical calculations for air showers based on the geosynchrotron model have been presented thinking about how to additionally use such antenna arrays for air-shower detection \cite{FalckeGorham2003, HuegeFalcke2003, Suprun2003}. 
Although it is not clear whether the geosynchrotron model is theoretically accurate, it helped to revive the field. 
LOPES \cite{FalckeNature2005} was built at the Karlsruhe Institute of Technology (KIT) as a prototype station for the astronomical antenna array LOFAR \cite{SchellartLOFAR2013}, and provided the proof-of-principle for the radio detection of cosmic rays with digital interferometry. 
Approximately at the same time another antenna array, CODALEMA \cite{CODALEMA_2005}, confirmed the potential of the digital radio technique, and other prototype experiments followed soon afterwards.

Nowadays a second generation of digital antenna arrays is operating and successfully detecting air showers. 
The focus of research has turned from the pure proof of feasibility to the optimization of the technique and of analysis methods. 
The goal is to compete in the precision for the reconstructed air-shower parameters with the established detection techniques.
Recent results summarized in this review show that this goal has mostly been achieved by now.
The next step will be to study whether the combination of radio with other detection techniques, e.g., simultaneous muon measurements of the same shower, indeed increases the total accuracy for the properties of the primary cosmic-ray particle. 
In this sense the radio technique has recently reached maturity and might bring a real advance in cosmic-ray physics in the next years. 
Consequently, with GRAND \cite{GRAND_ICRC2015} and SKA \cite{SKA_ICRC2015} two third-generation radio arrays are planned: 
GRAND will measure inclined showers initiated by neutrinos and cosmic rays with world-leading total exposure, and SKA will provide ultimate measurement precision for individual air-showers.
At the same time research is going on for radio detection in dense media, in particular with the construction of ARA and ARIANNA, which both search for neutrino-initiated particle cascades in the Antarctic ice.

\clearpage 
\section{Radio emission by particle cascades}
\label{sec_radioEmission}

The radio emission of particle cascades is coherent radiation generated by relativistic electrons and positrons in the electromagnetic component of the shower, with negligible contribution of other shower particles. 
As described later in this section several mechanisms contribute to the total emission, in particular the geomagnetic deflection of charged particles and the Askaryan effect, i.e., radiation due to the time variation of the net charge excess, which both have some general features in common.
The geomagnetic effect generally dominates in air showers and is negligible in dense media where the Askaryan effect dominates.
However, the Askaryan effect is not negligible for air showers, such that both the Askaryan and the geomagnetic effect are important in air.
Thus, this chapter focuses on the more general case of radio emission by air showers and mentions the difference to radio emission in dense media at the end.

\subsection{General features}

Independent of the emission mechanisms, the emission is coherent if the wavelength is larger than the emission region. 
As a consequence of coherence the amplitude of the radio emission scales linearly and the power quadratically with the number of electrons in the shower, similar to the emission of a free-electron laser. 
For this reason the total power in the radio signal scales quadratically with the energy of the primary particle \cite{AERAenergyPRL}, since the number of electrons in the shower is approximately proportional to the primary energy.
Most of the radiation is produced where the electron density is highest. 
In terms of shower development this is around the time and location when the number of relativistic electrons in the shower reaches its maximum, which often is simply called the shower maximum. 
Although air showers can have lateral extensions over kilometers, the electron density quickly decreases with lateral distance \cite{Lafebre2009}, and the emission is mostly created within the first meter from the shower axis. 
This means that the total lateral extension of the shower given by the Moliere radius, which is in the order of $100\,$m for air showers, is \emph{not} the relevant length scale, since the electron density at a few meter distance from the shower axis is too low for significantly contributing to the radio emission.
In this respect the main difference between cascades in air and in dense media relevant for the radio emission is not their lateral, but their longitudinal extension.

Since the typical thickness of the air-shower front is of the order of a meter, the emission is coherent and strongly amplified at wavelengths of a meter or larger for air showers, and correspondingly at smaller wavelengths larger than about $10\,$cm for the more compact showers in dense media. 
Thus, radio emission of air showers is typically observed at frequencies below $100\,$MHz, and radio emission in dense media at frequencies of several $100\,$MHz, where higher frequencies generally bring the advantage of lower background (cf.~section \ref{sec_background}). 
However, the situation is more complex than this simple argument indicates:
the relevant effective thickness of the shower front depends on the observer angle and even for air showers the speed of light in the medium (= atmosphere) plays a role, since coherence requires that the radio wave is in phase with the shower front during the emission. 
Generally, at larger distances to the shower axis full coherence is only achieved for larger wavelengths, which implies that measurements at lower frequencies allow for larger observation angles and larger detector spacings.

\begin{figure}[t]
  \centering
  \includegraphics[width=0.52\linewidth]{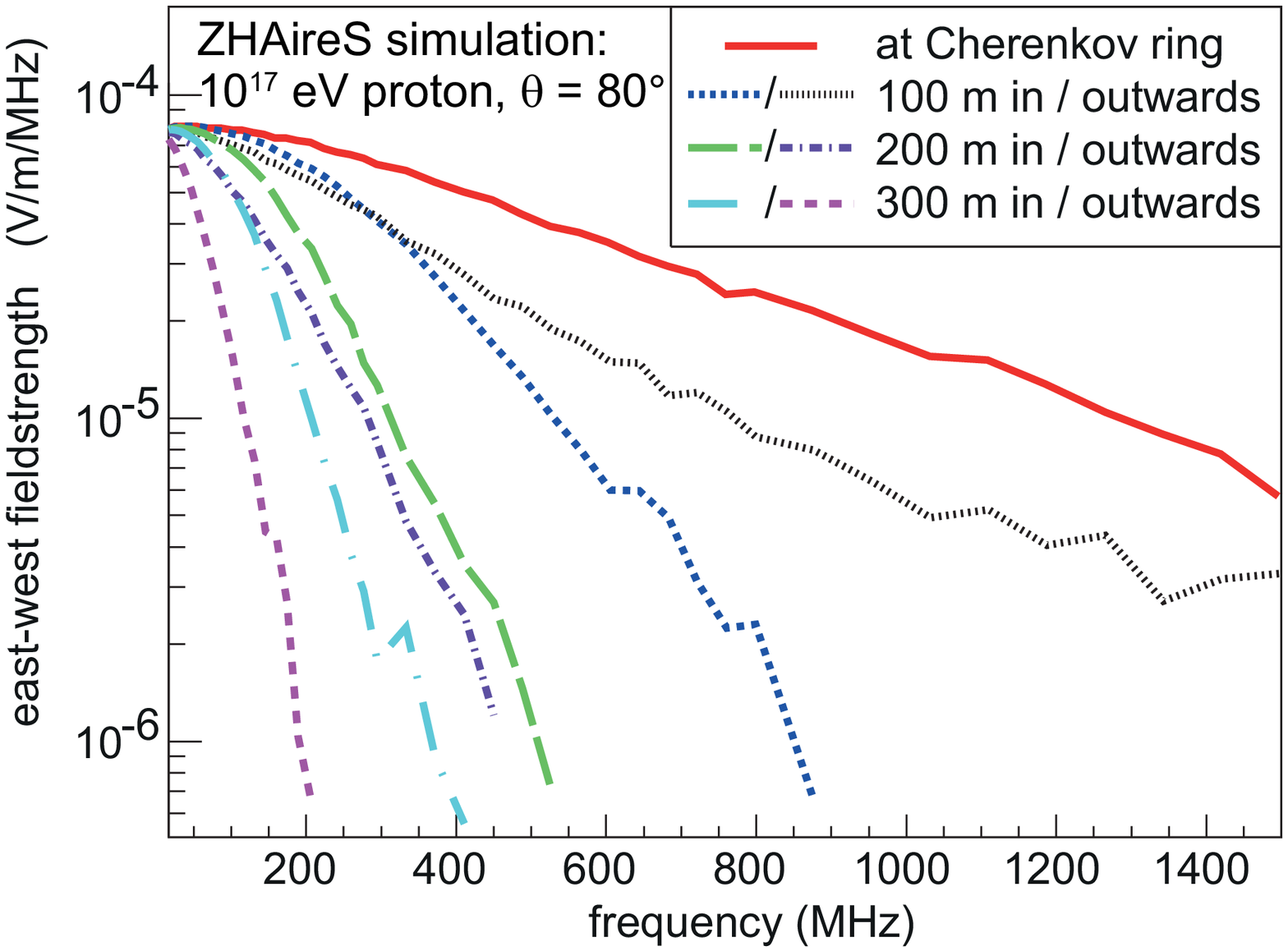}
  \hfill
  \includegraphics[width=0.46\linewidth]{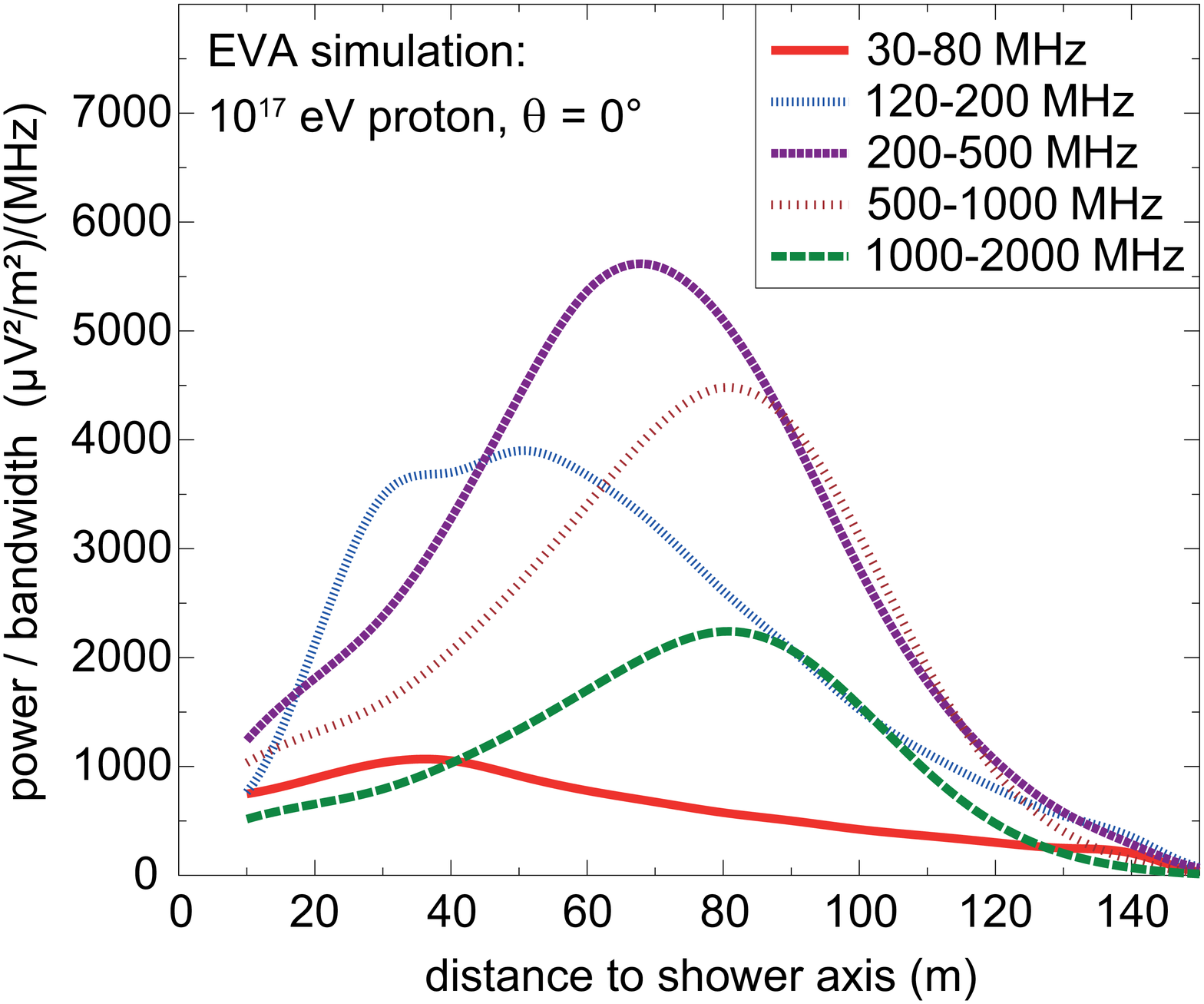}
  \caption{Left: Frequency spectrum of the radio emission simulated for a very inclined air shower. 
  Due to the inclination there is a difference between the spectrum inwards (= against the arrival direction of the shower projected on ground) and outwards (= following the arrival direction on ground). 
  Right: Signal strengths over distance to the showers axis in different bandwidths for a vertical shower. As visible in both figures, the emission extends up to GHz frequencies when measured at the Cherenkov angle, which corresponds to a distance of about $80\,$m in the left figure. 
  (Left figure slightly modified from Ref.~\cite{AlvarezMuniz_ANITAsims_2012}, right figure slightly modified from Ref.~\cite{deVries2013}).}
  \label{fig_frequencySpectra}
\end{figure}

\begin{figure}
  \centering
  \includegraphics[width=0.50\linewidth]{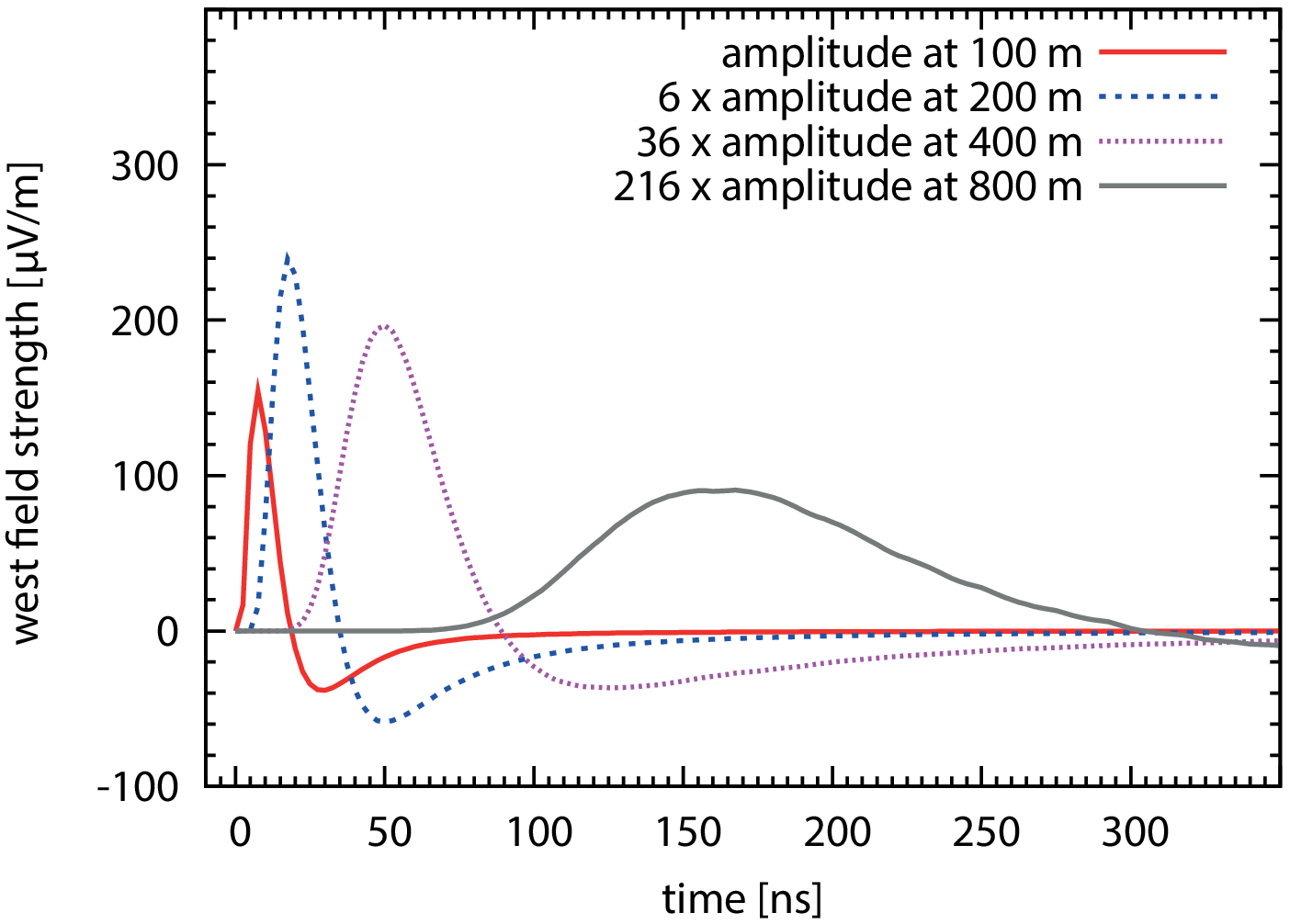}
  \hfill
  \includegraphics[width=0.47\linewidth]{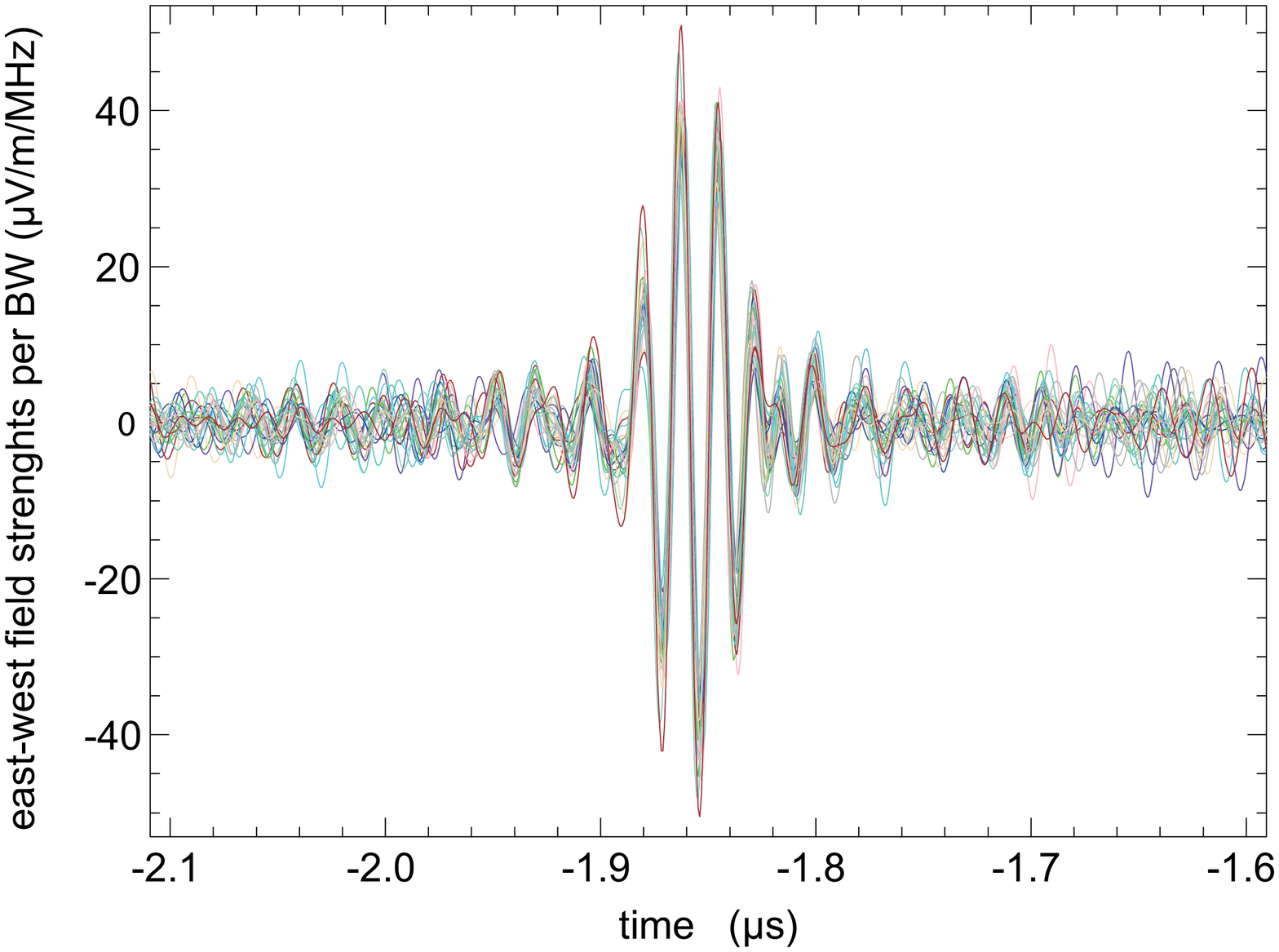}
  \caption{Left: MGMR pulses for a $0.1\,$EeV vertical shower at different distances to the shower axis. 
  Right: one air-shower event measured by different LOPES antennas. 
  The oscillating structure of both the radio pulse in the middle and the background is determined by the instrumental response of LOPES (left figure slightly modified from Ref.~\cite{REAS_MGMR_comparison_arxiv2012}, right figure slightly modified from Ref.~\cite{SchroederTimeCalibration2010}).}
  \label{fig_examplePulses}
\end{figure}

Moreover Cherenkov-like effects are important for both, air showers and cascades in dense media.
For air showers they come into play close to the shower axis ($d \lesssim 150\,$m for air showers on ground, and larger for very inclined showers):
The typical electron in the shower is only slightly slower than the speed of light in vacuum, and not necessarily slower than the speed of radio waves in air or whatever other medium. 
The propagation speed of the radio waves is defined by the refractive index $n$, which in air depends on the density and slightly on the humidity with $n \approx 1.0003$ at ground \cite{ITUrefractiveIndex2015}. 
At a certain angle, namely the Cherenkov angle of $\theta_c \approx \arccos 1/n \approx 1^\circ$ in air, radio waves and ultra relativistic particles propagate roughly at the same speed. 
Thus, at this angle radiation is coherent up to much smaller wavelengths corresponding to several GHz \cite{AlvarezMuniz_ANITAsims_2012, deVries2013}. 
Therefore, a Cherenkov ring with a typical diameter of around $200\,$m (depending on observation level and shower inclination) is seen in the radio footprint of air showers on ground, in particular at higher frequencies (see figure \ref{fig_frequencySpectra}) \cite{LOFARcherenkovRing2014}. 

\begin{figure}[t]
  \centering
  \includegraphics[width=0.6\linewidth]{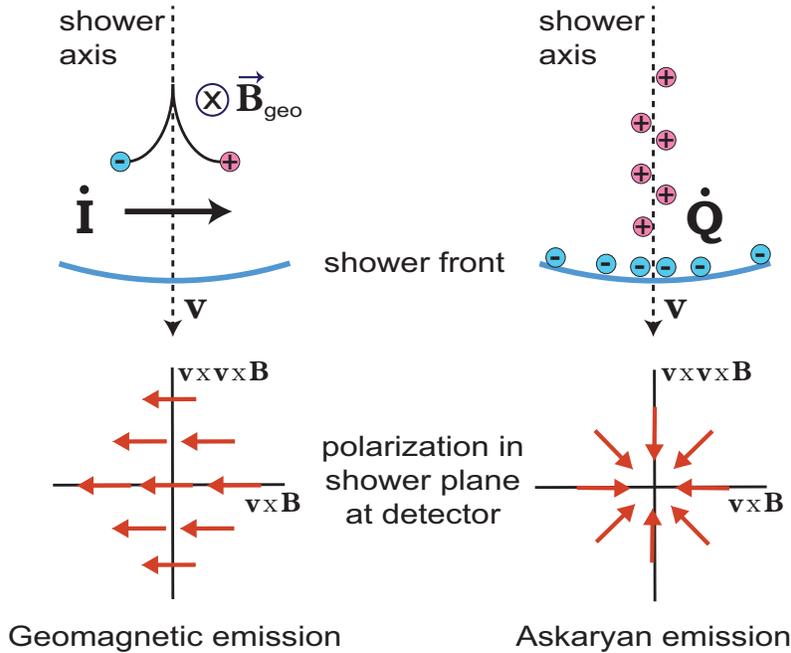}
  \caption{Emission mechanisms proven relevant at typical observation frequencies (few MHz to few GHz). 
  The geomagnetic emission due to the induction of a transverse current is polarized in the direction of the geomagnetic Lorentz force. 
  In air it is typically stronger than the radially polarized Askaryan emission due to the time variation of the net charge excess in the shower front, which is the main mechanism in dense media. 
  For air showers the radio emission is the coherent sum of both mechanisms, where depending on the local orientation of the electric-field vectors the interference of both mechanisms can be constructive or destructive. 
  The direction of the electric field can be determined according to Lenz's law, which says that the induced field counteracts its cause of origin. 
  Thus, for the geomagnetic effect the orientation is constant, but for the Askaryan effect it changes after shower maximum when the charge excess starts to decrease again (figure from Ref.~\cite{SchroederCRIS2015}).}
  \label{fig_emissionMechanisms}
\end{figure}

For showers in dense media the refractive index is much larger and a significant emission strength is only observed close to the Cherenkov angle \cite{AlvarezMuniz_ZHAires_ice2012}, which is about $56^\circ$ in ice.
The emission is also strongest and extends to highest frequencies of several GHz exactly at the Cherenkov angle, but at lower frequencies below $100\,$MHz remains visible at a much wider angular range of about $20^\circ$ around the Cherenkov angle \cite{AlvarezMuniz_ZHAires_ice2012}.
Obviously, the diameter of the Cherenkov ring depends strongly on the distance of the observer.
Whatever the medium, these Cherenkov-like features do not depend on the actual emission mechanism: 
the Cherenkov ring is not only expected for Cherenkov light emitted by particles faster than the speed of light in the medium, but for any kind of coherent electromagnetic emission. 
To say it clearly: radio emission by particle showers is \emph{not} Cherenkov light at MHz and GHz frequencies, but caused by other emission mechanisms.

Corresponding to the broad frequency spectrum radio pulses are short in time with typical pulse widths from a few ns inside the Cherenkov ring up to a few $100\,$ns at distances far away from the shower axis.
This means that the radio pulse contains only a few oscillations at each frequency, which makes air-shower pulses very different to radio signals used for technical purposes like communication. 
Thus, one has to be careful when trying to apply general theorems of radio engineering on the radio signal emitted by air-showers. 
Due to the short nature of the radio pulse its measured shape does significantly depend on the bandwidth of the measurement device: 
Figure \ref{fig_examplePulses} shows how the pulse shape depends on distance for radio emission by air showers simulated with unlimited bandwidth.
It also shows that the pulses of a real air-shower measured by LOPES in a typical band of $43-74\,$ MHz have a completely different structure, which is almost equal for all antennas although measured at different distances to the shower axis.
Consequently, these very general considerations mean that the main information contained in a measured radio pulse is only its amplitude and arrival time. 
At least at the typical frequency bands below $100\,$MHz, more detailed information like the exact pulse shape is hidden behind the instrumental response of the experiment and difficult to extract.

\begin{figure*}[t]
  \centering
  \includegraphics[width=0.45\linewidth]{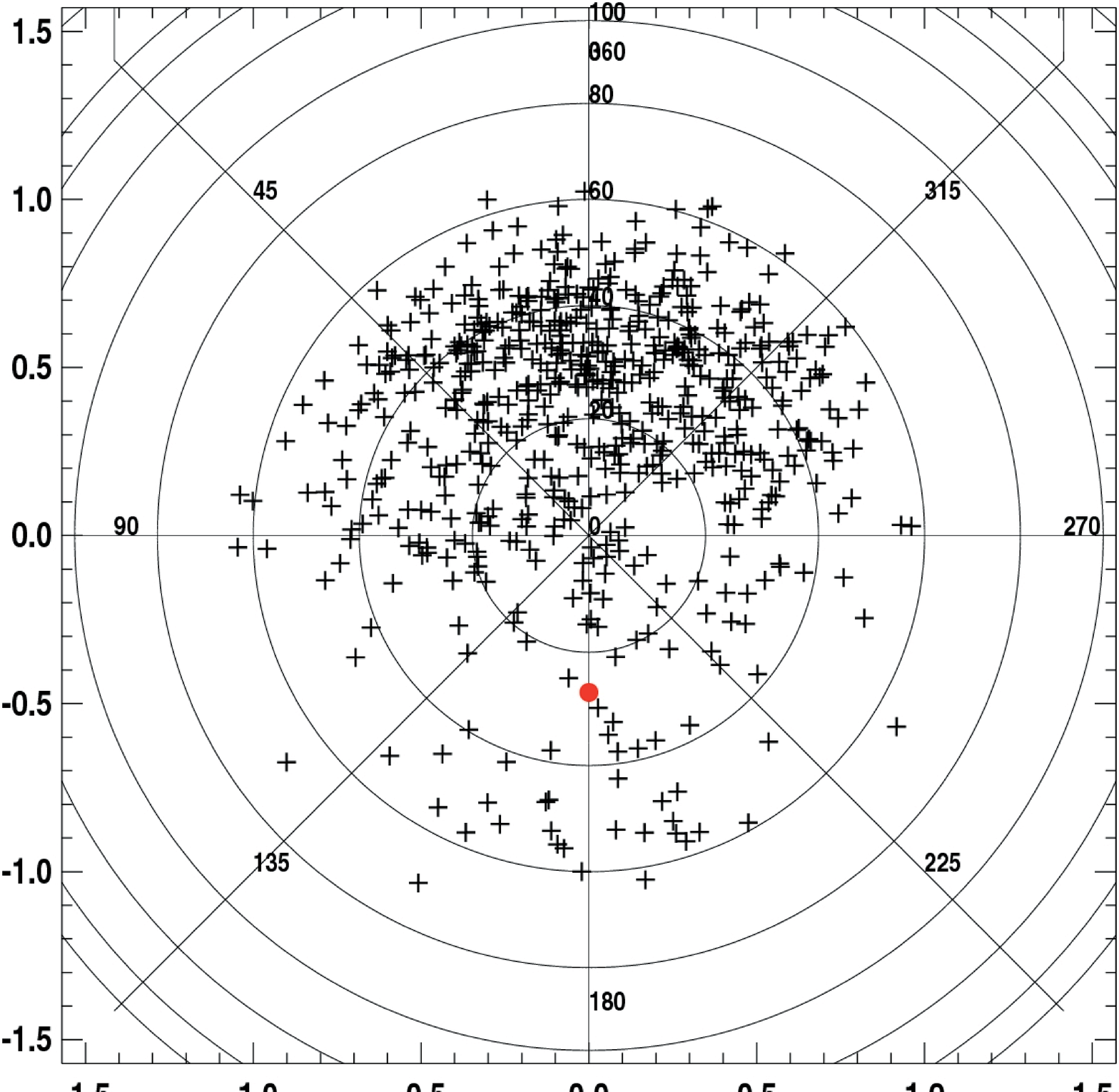}
  \hfill
  \includegraphics[width=0.49\linewidth]{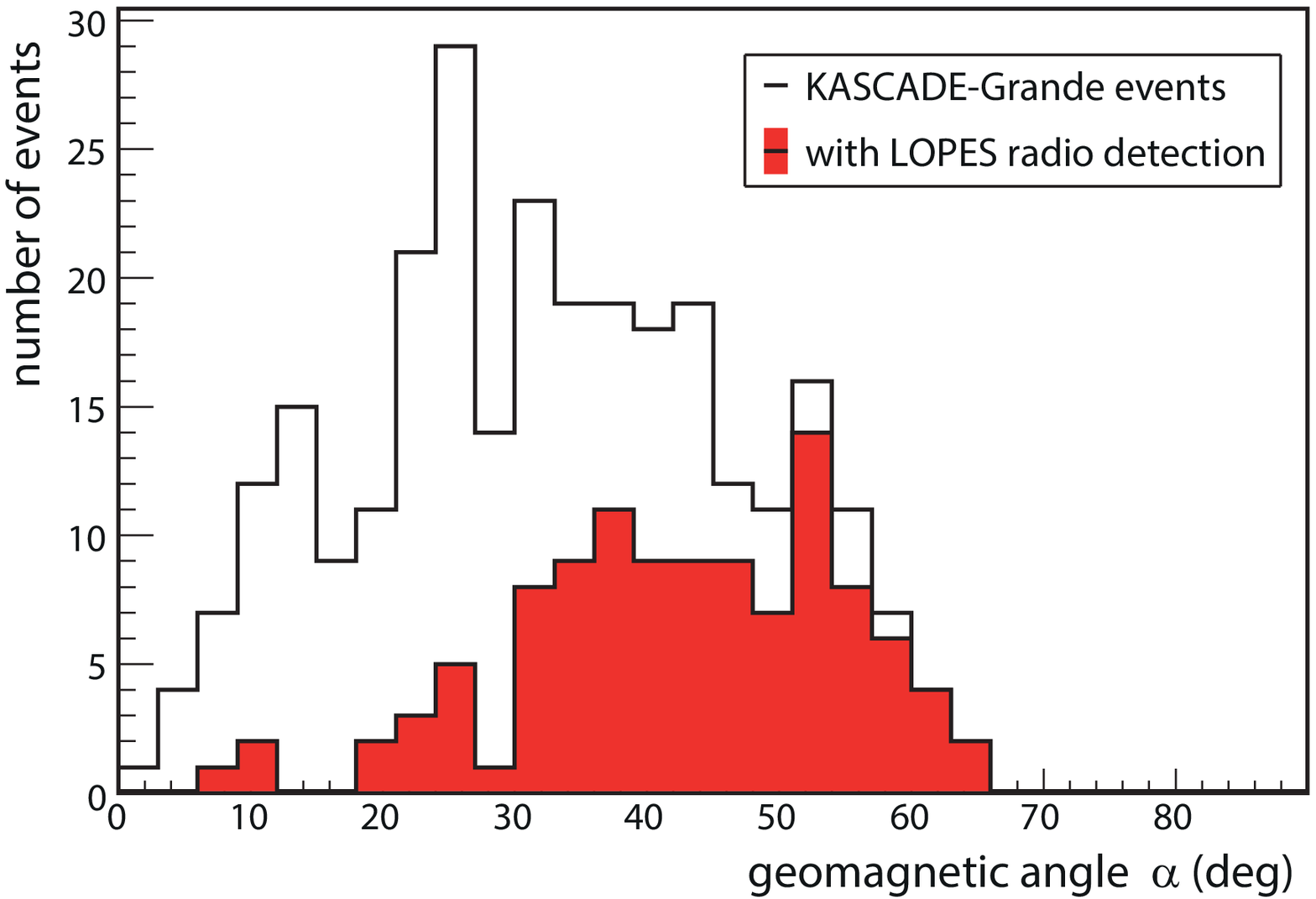}
  \caption{Left: Sky map of events detected with CODALEMA; crosses denote the arrival direction of detected air-showers, the dot marks the direction of the geomagnetic field (from Ref.~\cite{CODALEMA_Geomagnetic}). 
  Right: fraction of KASCADE-Grande events seen with the LOPES radio detector over the angle $\alpha$ between the axis of the air shower and the geomagnetic field (slightly modified from Ref.~\cite{2010ApelLOPESlateral}).}
  \label{fig_geomagneticSkymap}
\end{figure*}

\subsection{Emission mechanisms}
\label{sec_emissionMechanisms}

The most important emission mechanisms are the geomagnetic deflection of the electrons and positrons in the shower and the Askaryan effect, i.e., emission due to the time variation of the net charge in the shower front (see figure \ref{fig_emissionMechanisms}). 
As said, in air the Askaryan effect typically is weaker than the geomagnetic effect, but for showers in dense media which are important for neutrino detection, the situation is reversed - not because the Askaryan effect would be much stronger than in air, but simply because geomagnetic emission is negligible for compact particle showers in dense media.
During thunderstorms acceleration by atmospheric electric fields is relevant and can even dominate over the other two effects in emission strengths. 

Other emission mechanisms should exist, but have not yet been proven experimentally to be of relevant strengths at the usual observation frequencies of a few MHz up to a few GHz, in particular: 
ordinary Cherenkov-light emission, emission or reflection by the plasma disk left behind by the shower, molecular bremsstrahlung \cite{TAelectronBeam2016, Samarai2016}, or transition radiation and bremsstrahlung when the shower enters the ground \cite{deVries2015, RevenuSuddenDeath_ICRC2013}. 
Recently it has been suggested that air showers should emit incoherent radiation of significant power at frequencies above $100\,$GHz \cite{Filonenko2016}, but this topic requires more extensive studies and no experiments have searched for this emission, yet. 
Even if this emission exists, detection efforts have to consider that atmospheric absorption is not negligible at frequencies above $50\,$GHz \cite{ITUatmosphericAttenuation2015}.
Finally, transition radiation could be relevant when a particle cascades initiated in a dense medium passes the boarder to another medium, e.g., the surface between ice and air or the border of the lunar regolith \cite{Filonenko2012, MotlochTransitionRaditation2016}.
Only the first three effects, i.e., geomagnetic deflection, the Askaryan effect, and electric fields in the atmosphere are the mechanisms experimentally proven in air showers, and are discussed in more detail here.

\subsubsection{Geomagnetic effect}
The Lorentz force of the geomagnetic field deflects electrons and positrons in opposite directions which induces a transverse drift current in the air varying in time since the number of electrons and positrons changes during shower development \cite{KahnLerche1966, Colgate1967}. 
This is conceptually the same as what happens in a Hertz dipole. 
Therefore, the geomagnetic emission is linearly polarized orthogonally to the direction of the geomagnetic field.\footnote{
There has been some dispute whether the geomagnetic emission can be described equally well or even better as geo-synchrotron radiation, since electrons and positrons are deflected on curves (with very large radii). 
A geo-synchrotron model predicted the correct order of magnitude for the radio amplitude \cite{HuegeFalcke2003}, but no typical synchrotron features have been experimentally observed, and are predicted by CoREAS simulations only for high frequencies of several GHz \cite{HuegeCoREAS_ARENA2012}. 
Thus, the question is still open to which extent the geo-synchrotron and the transverse current models are just different pictures of the same process. 
For frequently used microscopic simulations codes, like CoREAS or ZHAires, this question of the macroscopic picture is unimportant because they calculate the emission of individual particles.}

The amplitude of the emission is proportional to the Lorentz force, i.e., proportional to the local strength of the geomagnetic field and proportional to $\sin \alpha$, where $\alpha$ is the angle between the geomagnetic field and the shower axis. 
However, slight deviations from an exact, linear proportionality are discussed for high values of the geomagnetic Lorentz force, which needs deeper investigation \cite{GlaserShortAuthor2016}.
The radiated energy increases with the duration of the emission process, corresponding to the longitudinal length of the shower maximum. 
Thus, radio emission should be slightly stronger for inclined air showers, since inclined showers developed higher up in the atmosphere.
Due to the lower atmospheric density at higher altitude inclined air showers extend over longer distances than vertical showers developing closer to ground, and the duration of the geomagnetic radio emission lasts slightly longer.
For all zenith angles the shower maximum has a much larger longitudinal extension for air showers (scale of kilometers) than for particle cascades in dense media (scale of meters), which is the principle reason why geomagnetic emission is negligible in dense media. 
The mostly geomagnetic origin of the radio signal emitted by air showers has been confirmed by many experiments including the historic ones \cite{AllanReview1971}, and causes a north-south asymmetry in the detection efficiency: 
the amplitude, and thus the detection threshold, depend on the arrival direction of the air shower \cite{CODALEMA_Geomagnetic} (see figure \ref{fig_geomagneticSkymap}).
Nowadays this dependence of the detection rate on the geomagnetic angle already has been converted into a cross-check if observed radio signals really originate from air showers or from some other source, which usually is one of the first analyses made by new radio experiments.

\begin{figure*}
  \centering
  \includegraphics[width=0.6\linewidth]{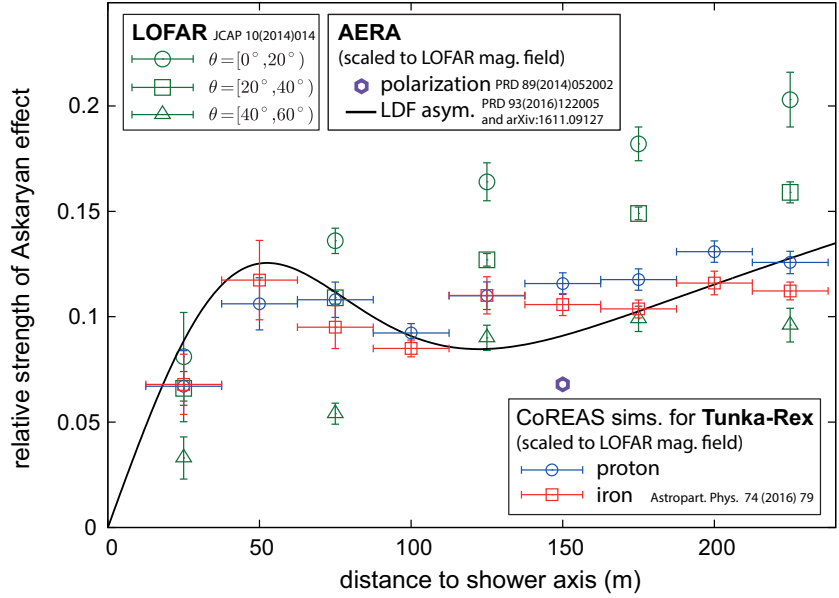}
  \caption{Relative strength of the Askaryan emission compared to the geomagnetic emission derived either from polarization measurements or from the asymmetry of the radio footprint. 
  LOFAR measurements \cite{CorstanjePolarizationICRC2015} in bins of zenith angle and axis distance are compared to CoREAS simulations made for the situation of Tunka-Rex \cite{KostuninTheory2015}, to a value derived from AERA polarization measurements \cite{AugerAERApolarization2014}, which is valid for a typical distance of $150\,$m and a zenith angle of roughly $40^\circ$, and to the value implied in the asymmetry of the two-dimensional lateral-distribution function used by AERA (solid line: calculated from LDF of Ref.~\cite{AERAenergyPRD} for the zenith angle range of $40^\circ - 50^\circ$, a geomagnetic angle of $45^\circ$, and a footprint width of $\sigma = 152\,$m \cite{KostuninARENA2016}). 
  The Tunka-Rex and AERA values have been scaled to the LOFAR situation according to the local strengths of the geomagnetic field (figure from Refs.~\cite{KostuninPhDThesis2015, KostuninARENA2016}).}
  \label{fig_LOFARpolarizationDependence}
\end{figure*}

\begin{figure*}
  \centering
  \includegraphics[width=0.48\linewidth]{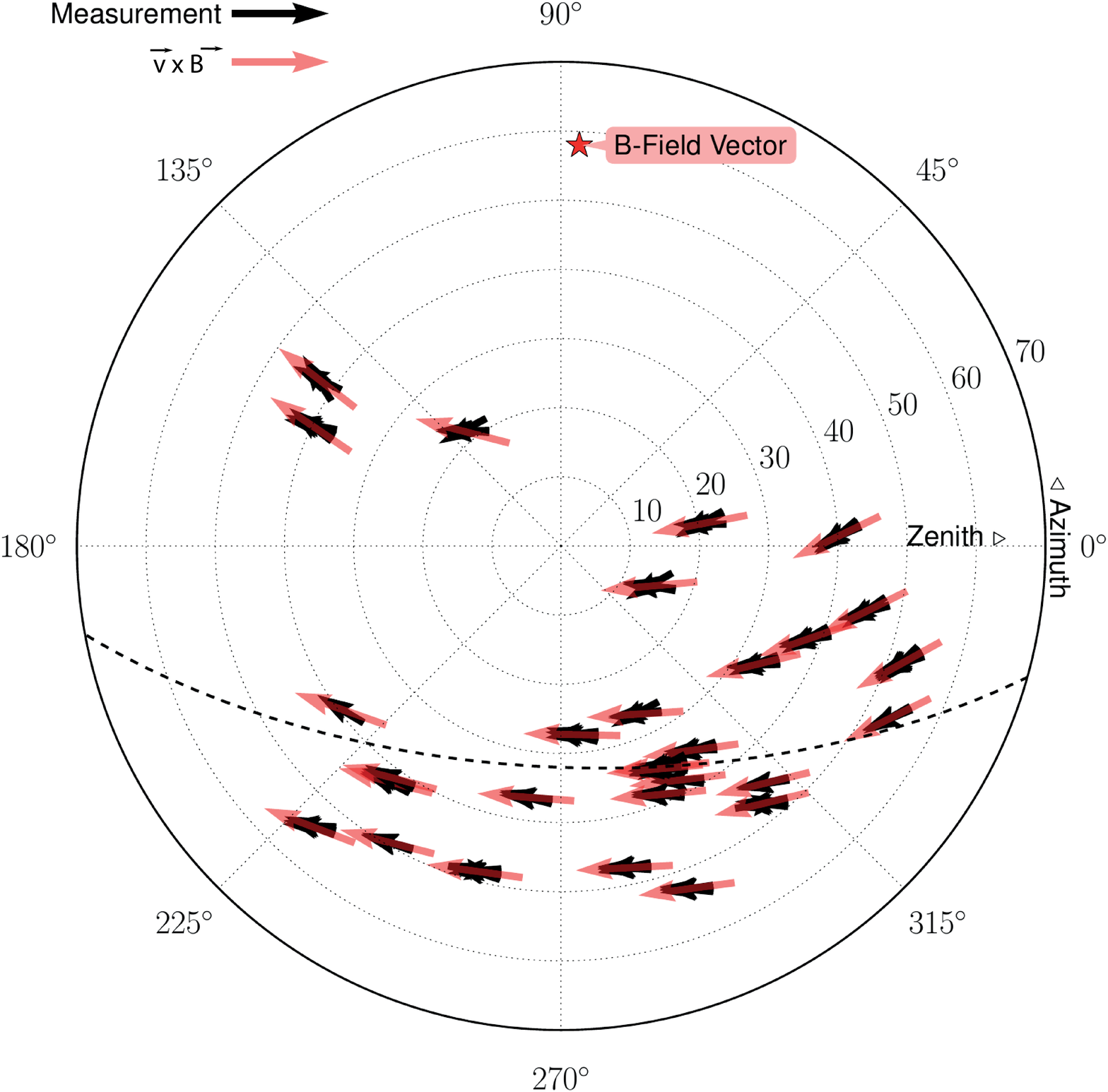}
  \hfill
  \includegraphics[width=0.35\linewidth]{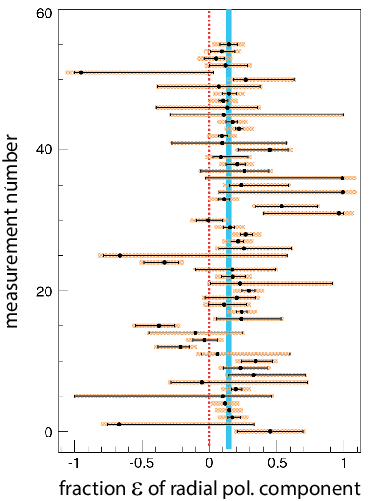}
  \caption{Polarization measurements of AERA (from Ref.~\cite{QaderECRS2014}). 
  The measured polarization agrees approximately with the polarization expected from the dominant geomagnetic emission mechanism (left figure), and the deviations indicate a radial polarization component with average relative strengths of $14\,\%$ (right figure).}
  \label{fig_askaryanAERApol}
\end{figure*}

\subsubsection{Askaryan effect}
Askaryan emission is radio emission caused by a time-variation of the net charge-excess in the shower front \cite{Askaryan1962}: 
on first view, time variation of the total charge seems to violate charge conservation. 
However, as explained in section \ref{sec_airShower}, the shower front accumulates a negative net charge excess relevant for the radio emission and charge is totally conserved due to a positively charged plasma created behind the shower. 
Because of the high electron density at the shower axis the net charge excess is highest there and the excess of electrons can be as large as $20-30\,\%$ of the total number of electrons and positrons \cite{ScholtenARENA2010}. 
In a simplistic view the shower is a point charge whose strength changes during the shower development. 
Since electric-field lines of a point charge are radial, Askaryan emission is radially polarized and zero in the center at the shower axis \cite{AugerAERApolarization2014}.

For dense media Askaryan emission is the only one of relevance at radio frequencies, and for air showers the radio signal observed on ground is the combination of the Askaryan and the geomagnetic effect (see figure \ref{fig_emissionMechanisms}).
This interference of both effects causes the maximum of the radio amplitude to be slightly displaced from the shower core, which is defined as the point where the shower axis (= prolongation of the primary-particle track) enters the ground \cite{CODALEMA_Askaryan}, i.e., the radio footprint is asymmetric for air showers\footnote{For showers in dense media, the Askaryan effect is the only one relevant. Thus, the radio signal is radially symmetric around the shower axis, which implies that the radio amplitude must be zero at the axis. This is why radio emission has a ring-like structure around the shower axis in dense media, where the diameter corresponds to the Cherenkov angle in the medium.}. 
The relative strength of the Askaryan effect to the geomagnetic effect ($a$ or $\epsilon$ in different literature) depends mostly on the Lorentz force for the particular shower, which itself depends on the shower direction and the local magnetic field strengths. 
Also the distance to the shower axis, the shower inclination (see figure \ref{fig_LOFARpolarizationDependence}, \cite{SchellartLOFARpolarization2014}), and possibly the observation altitude and the depth of the shower maximum play a role \cite{KostuninTheory2015}. 
Recently different experiments published values for the relative strength of the Askaryan effect in air based either on the polarization of the radio signal or on the asymmetry for the footprint on ground. 
Here the values of the relative strength of the Askaryan effect are given together with the local strength of the geomagnetic field $B$: $(14\pm2)\,\%$ for $B = 24\,$\textmu T at AERA \cite{AugerAERApolarization2014} (see figure \ref{fig_askaryanAERApol}), $8.5\,\%$ for $B = 60\,$\textmu T \cite{KostuninTheory2015} at Tunka-Rex (value based on CoREAS simulations), and $(3.3\pm1.0)\,\%$ for inclined air showers at $25\,$m to  $(20.3\pm1.3)\,\%$ for near-vertical showers at $225\,$m for  $B = 49\,$\textmu T at LOFAR \cite{SchellartLOFARpolarization2014}. 
Moreover, the lateral-distribution function used to fit the asymmetric radio footprints measured by AERA implicitly contains a value consistent with the CoREAS simulations made for Tunka-Rex \cite{KostuninARENA2016}.

In first order, the relative strength of the Askaryan effect should be anti-proportional to the local strength of the geomagnetic field. 
Taking into account the typical distance to the shower axis of $150\,$m and zenith angle of roughly $40^\circ$ for the published AERA events, the AERA value of $(14\pm2)\,\%$ is expected to be twice as large as the LOFAR values at these distances and zenith angles ranging from $(9.0\pm0.6)\,\%$ to $(14.9\pm0.3)\,\%$. 
The weaker value derived from CoREAS simulations for Tunka-Rex would correspond to roughly $10.5\,\%$ at the LOFAR magnetic field, and $21.3\,\%$ at the AERA magnetic field.
Consequently, there is some tension between the polarization measurement of AERA on the one hand, and LOFAR, Tunka-Rex, and the AERA value based on the footprint asymmetry on the other hand, but generally measurements based on the polarization and on the footprint asymmetry seem to be consistent. 
Since the AERA polarization result is based on limited statistics of the first events recorded by AERA, it could be that some systematic uncertainties and selection biases have been underestimated. 
Future follow-up analyses at AERA with larger statistics, analyses based on Tunka-Rex measurements instead of simulations, and possible measurements by other experiments like CODALEMA can finally clarify this issue.
Moreover, recently a slight phase delay of up to $1\,$ns between geomagnetic and Askaryan emission has been measured by LOFAR in the band of $30-80\,$MHz \cite{ScholtenLOFAR_elipticity2016}, which was not taken into account in the reported measurements of $a$: without a thorough check it is difficult to tell which values of $a$ refer to the total strength of the Askaryan effect and which ones refer only to the part of the Askaryan emission which is in phase with the geomagnetic emission.
Given that a phase delay of $1\,$ns corresponds to about $20\,\%$ of the delay required for maximum ellipticity, the relative effect on the value of $a$ likely will be smaller than $20\,\%$, which still might be sufficient to explain the difference between both AERA results.

\begin{figure*}
  \centering
  \includegraphics[width=0.9\linewidth]{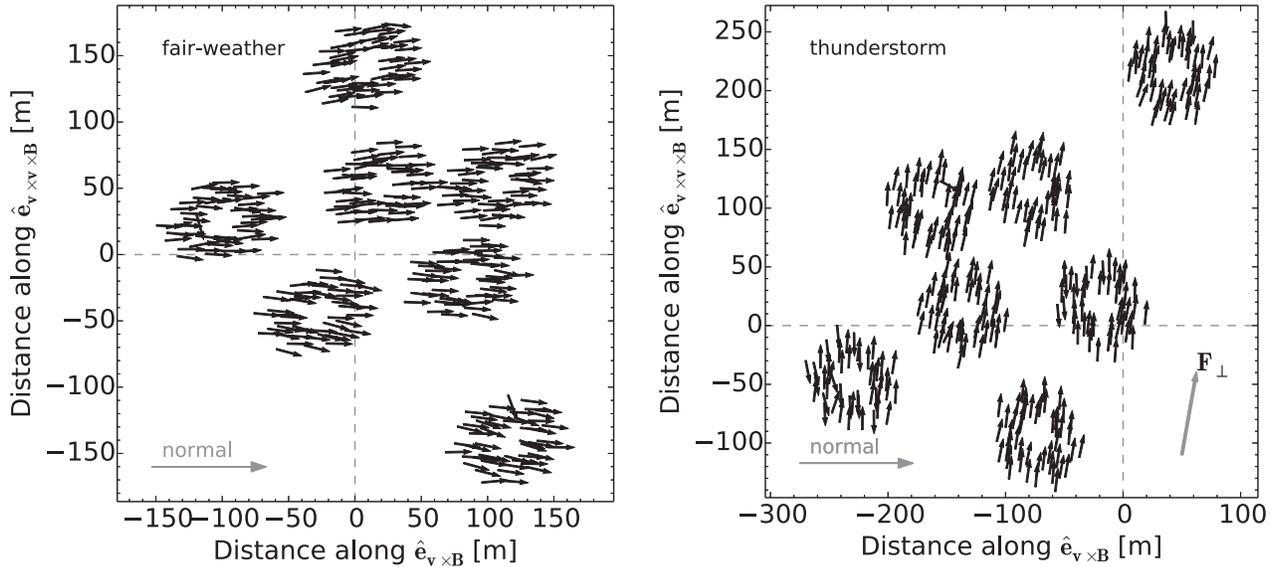}
  \caption{Directions of the radio electric-field vector of the air-shower emission measured by LOFAR during normal weather conditions (left) and during a thunderstorm (right). 
  The deviation in the thunderstorm case from the \lq normal\rq~direction expected for the geomagnetic emission can be explained by the atmospheric electric field of the thundercloud accelerating the air-shower electrons and positrons (from Ref.~\cite{LOFARthunderstormICRC2015}).}
  \label{fig_thunderStormPol}
\end{figure*}

\subsubsection{Atmospheric electric field}
Like the geomagnetic field, the atmospheric electric field also accelerates electrons and positrons in opposite directions and induces time-variable transverse currents. 
During normal weather conditions the atmospheric field is a few $100\,$V/m and its force is negligible against the geomagnetic Lorentz force \cite{BuitinkThunderstromSims2010}. 
However, during rain the field can be significantly stronger, and during thunderstorms it can exceed several $10\,$kV/m. 
Then the acceleration by electric fields can be the dominant source of radio emission, and the total radio amplitude as well as the polarization of the radio emission can be significantly enhanced compared to fair weather conditions (see figure \ref{fig_thunderStormPol}) \cite{LOPESthunderstorm2007, LOPESthunderstorm2011, SchellartLOFARthunderstorm2015}. 

Since the direction of the atmospheric electric field varies with height, and can even be opposite for different layers in thunderstorm clouds, radio measurements during thunderstorm are hardly usable for cosmic-ray physics. 
In addition to the strong effects during thunderstorms, LOPES measurements have shown a small effect on the radio emission by air showers starting at atmospheric electric-fields of about $3\,$kV/m, which occurred during less than $5\,\%$ of the total time, in particular when heavy rain clouds were over the array.
Therefore, radio experiments are either switched off during thunderstorms and heavy rain, or they have to monitor the strengths of the atmospheric electric field and exclude from analysis the small fraction of their data measured during high-field conditions. 
Finally, the effect of atmospheric electric fields can be simulated \cite{BuitinkThunderstromSims2010, Trinh2015}, which enables using radio measurements of air showers for probing thunderstorm clouds, as has recently been demonstrated by LOFAR measurements~\cite{SchellartLOFARthunderstorm2015}. 

\begin{figure}[t]
  \centering
  \includegraphics[width=0.99\linewidth]{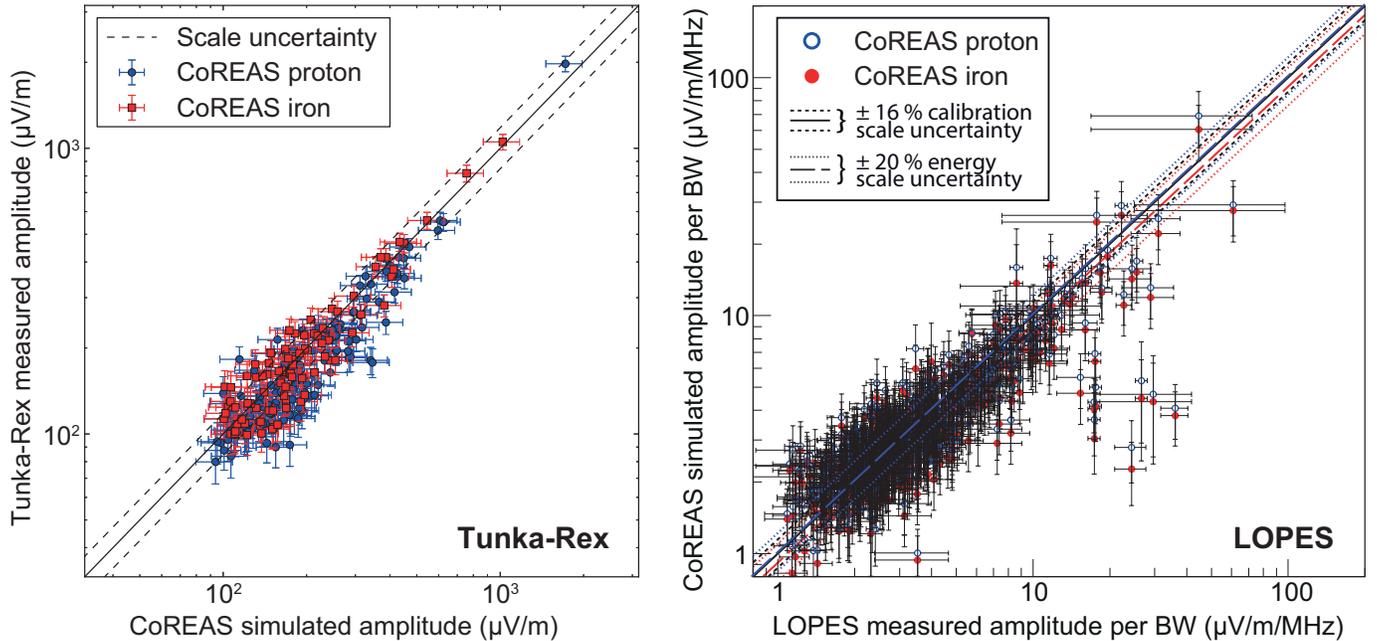}
  \caption{Comparison of CoREAS simulations of the radio amplitude with measured data. 
  Within the scale uncertainties the simulated amplitudes are compatible with the measurements when assuming protons or iron nuclei as primary particles.
  This covers the extreme cases assuming that nuclei heavier than iron have negligible abundance among the primary cosmic rays.
  Left: Comparison with Tunka-Rex measurements (amplitude = total electric-field strength) recorded at different axis distances (slightly modified from Ref.~\cite{TunkaRex_NIM2015}). 
  Right: Comparison with LOPES measurements (amplitude = east-west field strength) interpolated for each event to an axis distance of $100\,$m and normalized to the effective bandwidth of $43-74\,$MHz (slightly modified from Refs.~\cite{LOPESimprovedCalibration2016, LinkICRC2015}).}
  \label{fig_comparisonCoREAS}
\end{figure}

\subsection{Simulation codes}
\label{sec_simulations}
Numerous calculations have been performed of the radio signal emitted by showers in air and in dense media, and several simulation programs have been developed \cite{BelovTheoryOverview_ARENA2012}. 
While the situation for dense media is simpler, for air showers only recently general agreement was achieved that the geomagnetic effect, the Askaryan effect, and the Cherenkov-like effects due to the refractive index of the atmosphere are all of relevance. 
These effects now are included explicitly or implicitly in all recent simulation programs like EVA \cite{Werner_EVA2012}, ZHAireS \cite{AlvarezMuniz_ZHAires_air2012}, SELFAS \cite{Marin_SELFAS2_2012}, and CoREAS \cite{HuegeCoREAS_ARENA2012}. 
The different simulation codes differ in complexity, level of detail and required computing resources, e.g., ZHAireS and CoREAS are full Monte Carlo simulations of the radio signal emitted by the individual electrons and positrons in air showers, which makes them accurate but also computing intensive. 
A review on the differences and a comparison on the results of different simulation codes is available in reference \cite{HuegeTheoryOverview_ARENA2012}.

Experimental tests of simulated properties like the amplitude and its dependence on various shower parameters are limited by systematic uncertainties of the experiments, in particular the uncertainty of the energy scale (ranging from $14\,\%$ \cite{AugerNIM2015} to $50\,\%$ \cite{ThoudamLORAspectrum2015}) and the uncertainty of the absolute amplitude calibration (ranging from $14\,\%$ \cite{AERAenergyPRL, AERAantennaJINST2012} to $18\,\%$ \cite{LOPEScalibrationNehls2008, LOPESimprovedCalibration2016, TunkaRex_NIM2015, LOFARcalibration2015}). 
Within these uncertainties most simulation codes seem to reproduce radio measurements of air-showers, in particular CoREAS has been tested extensively (see figure \ref{fig_comparisonCoREAS}) \cite{LOPESimprovedCalibration2016, TunkaRex_NIM2015, LOFARcalibration2015, AERA_ICRC2013}.
Moreover, at the Stanford Linear Accelerator recently Askaryan and magnetic emission of particle showers induced by accelerated electron bunches have been measured under controlled laboratory conditions \cite{SLAC_T510_PRL2016}. 
These measurements agree within a systematic uncertainty of about $40\,\%$ in amplitude with simulations which are based on the same principles also used for CoREAS and ZHAires simulations for air showers as well as for ZHS simulations for particle cascades in dense media.
This provides independent proof that the emission mechanisms are understood to at least this level of $40\,\%$ accuracy - in addition to the air-shower measurements confirming the CoREAS simulations with higher accuracy of about $20\,\%$, but with different systematic uncertainties.
However, a systematic comparison which would check several simulation codes against a larger statistics of measured air-showers is still missing. 
This will be necessary to experimentally test more subtle effects on the radio emission, e.g., the effect of the hadronic interaction models used for the air-shower simulation, or the effect of the atmospheric conditions like density or humidity, which influence not only the air-shower development, but also the refractive index \cite{ITUrefractiveIndex2015}.

\begin{figure*}[p]
  \centering
  \includegraphics[width=0.9\linewidth]{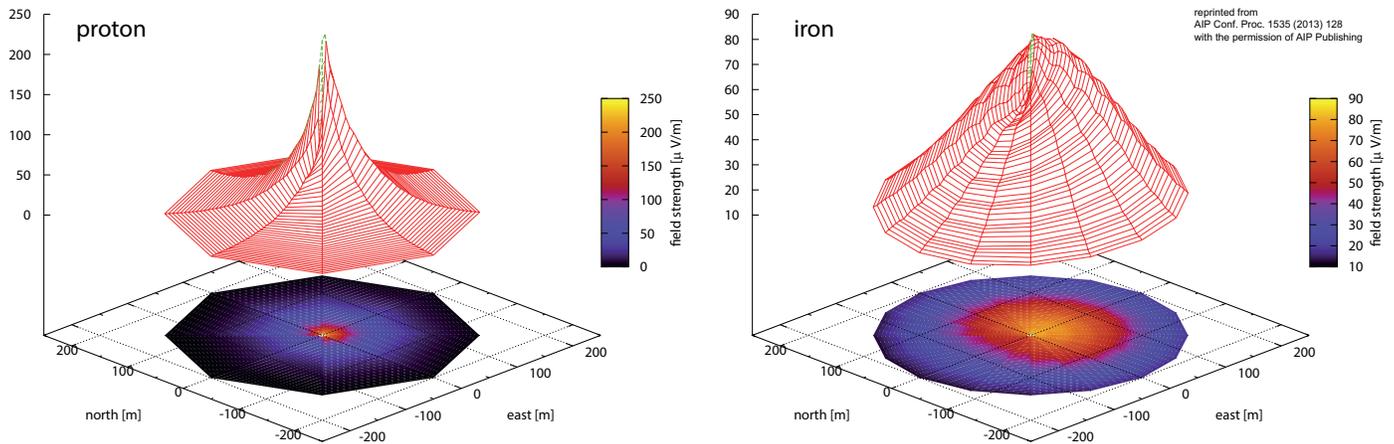}
  \caption{Time evolution of the electric-field vector at different positions relative to the shower axis of an air shower simulated by CoREAS for the conditions of the LOPES experiment. 
  The middle plot does not indicate the polarization, but the lateral distribution of the total radio amplitude (from Ref.~\cite{HuegeCoREAS_ARENA2012}).}
  \label{fig_CoREASpolarization}
\end{figure*}

\begin{figure*}[p]
  \centering
  \includegraphics[width=0.99\linewidth]{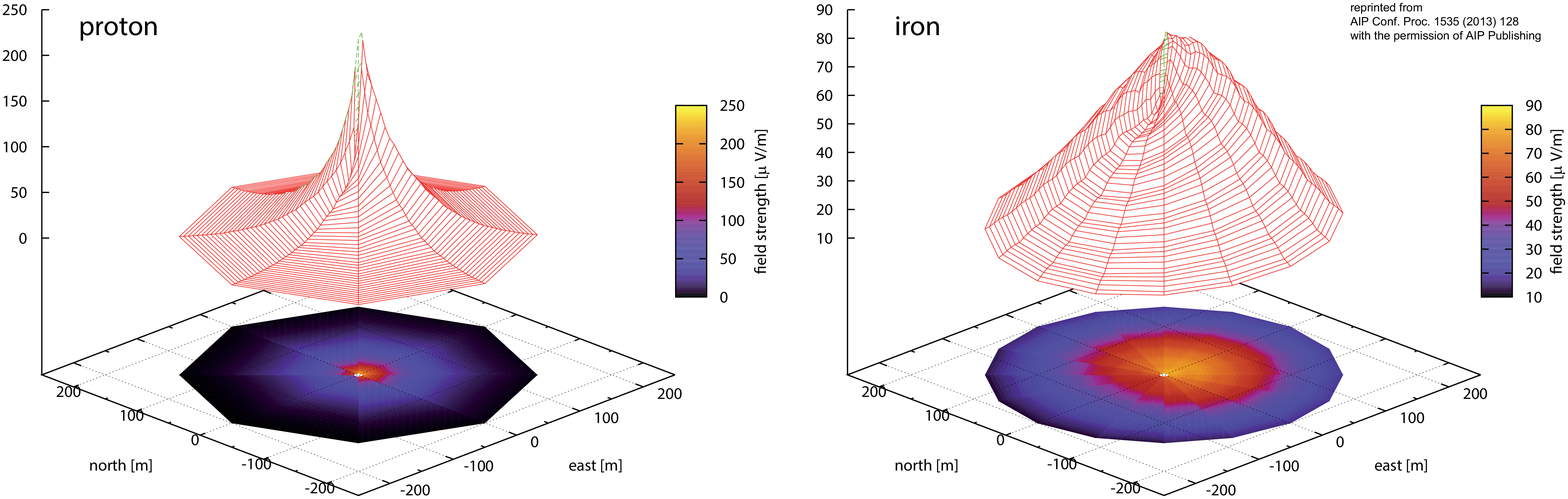}
  \caption{Footprints of two air showers simulated by CoREAS for the conditions of the LOPES experiment. 
  Both, the color-code and the height of the cone are the total electric-field strength of the radio signal. 
  The steepness of the footprint and its shape depend mainly on the distance from the observation plane to the shower maximum, which statistically correlates with the mass of the primary particle (from Ref.~\cite{HuegeCoREAS_ARENA2012}).}
  \label{fig_CoREASfootprints}
\end{figure*}

\subsection{Polarization of the radio signal}
\label{sec_polarization}
The polarization\footnote{Although the term polarization is not well defined for a short radio pulse whose duration is not longer than a few oscillations of the relevant frequencies, often the direction of the electric field and its time evolution is referred to as polarization \cite{SchellartLOFARpolarization2014, AugerAERApolarization2014}.} of the radio signal emitted by air showers is not uniform, but depends on the position relative to the shower axis \cite{HuegeCoREAS_ARENA2012}, since it is determined by the interference of the geomagnetic and Askaryan emissions. 
The electric-field vector of the geomagnetic emission always points in the direction of the geomagnetic Lorentz force, but the electric-field vector of the Askaryan emission points towards the shower axis, so that the resulting direction depends on the relative strength of both emission mechanisms and on the azimuth angle relative to the shower axis.

Although the polarization of each individual emission mechanism is assumed to be linear, CoREAS simulations predict that the net polarization is slightly elliptical, except for the particular locations where Askaryan and geomagnetic emission are completely aligned (see figure \ref{fig_CoREASpolarization}). 
Recently, LOFAR published a first measurement of the ellipticity in the frequency band of $30 - 80\,$MHz, which indicates that the emission by the Askaryan and geomagnetic effects are not in phase \cite{ScholtenLOFAR_elipticity2016}.
Consequently, both types of emissions must originate from different parts of the shower, or at least their strengths relative to each other are not constant, but vary with the shower development as indicated by simulation studies \cite{deVries2013, GlaserShortAuthor2016}.
The size of the delay between both effects depends on the position relative to the shower axis and is maximum at about $100\,$m distance from the axis, i.e., close to the Cherenkov angle. 
There the relative delay between geomagnetic and Askaryan emission has been measured to approximately $1\,$ns, which is about $20\,\%$ of the delay required for maximum ellipticity, where circular polarization would be achieved only for showers arriving under the particular geomagnetic angle at which the Askaryan and geomagnetic emissions have equal strength.
Previously, many analyses implicitly or explicitly made the approximation of exactly linear polarization, which now should be rechecked, in particular, when aiming for high accuracy.

In summary, the polarization is approximately linear in the direction of the geomagnetic Lorentz force with a slightly varying direction and ellipticity due to the weaker Askaryan effect and with a second-order sensitivity to the longitudinal shower development.

\begin{figure*}[p]
  \centering
  \includegraphics[width=0.54\linewidth]{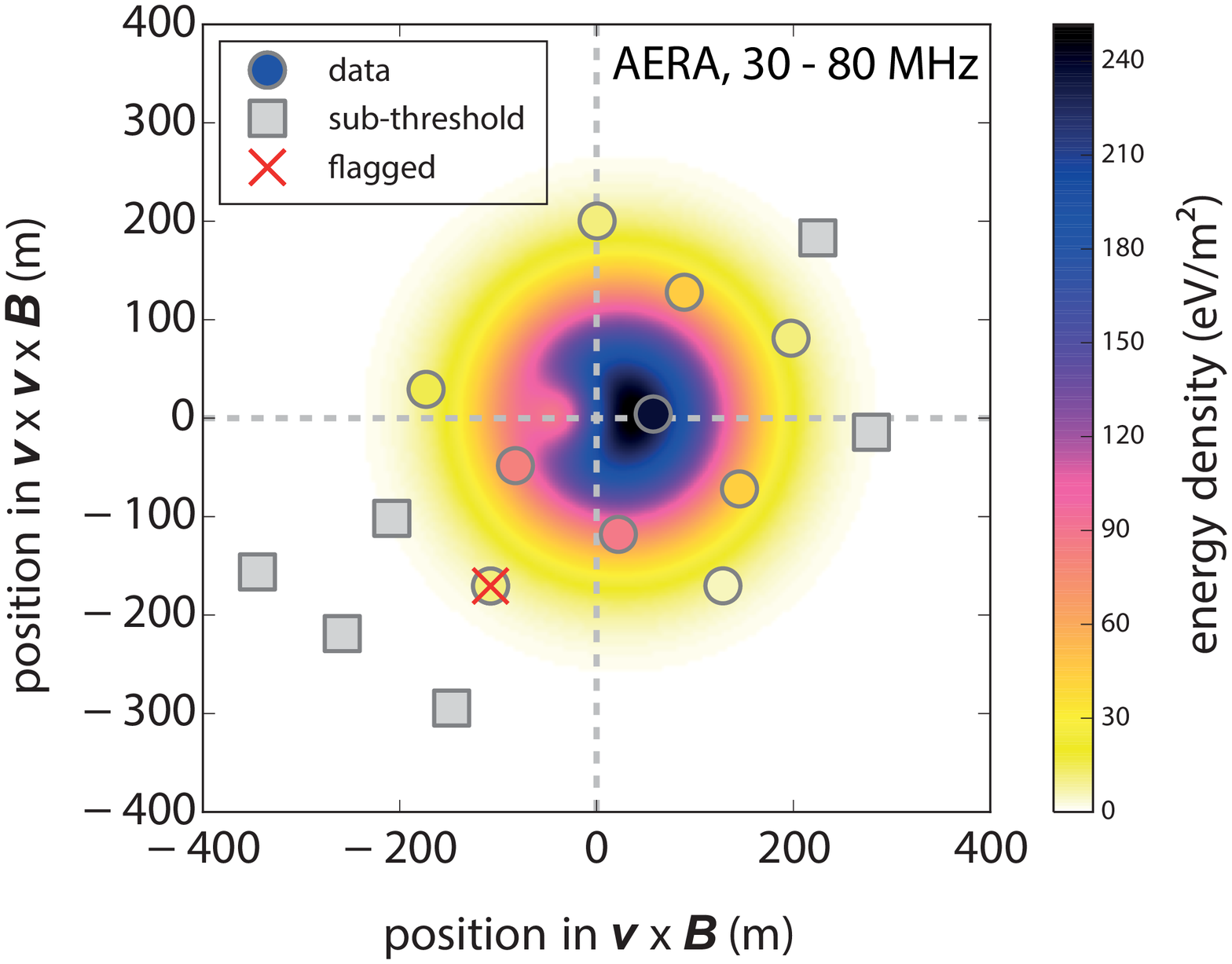}
  \hfill
  \includegraphics[width=0.44\linewidth]{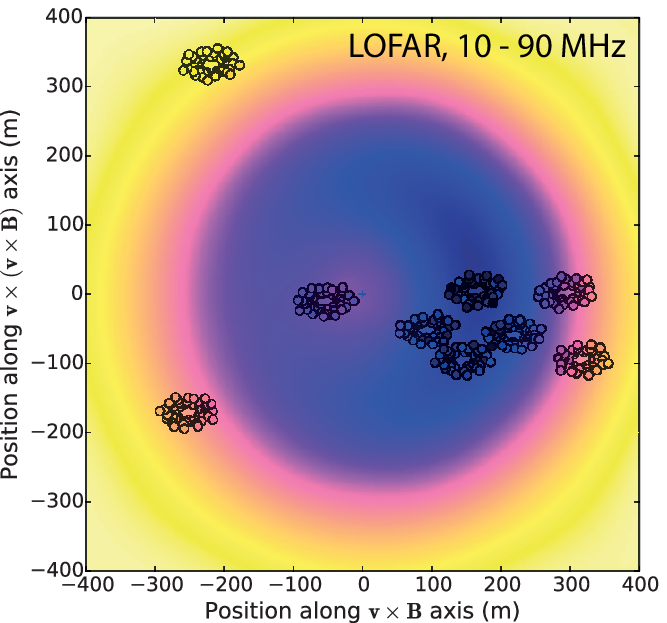}
  \caption{Footprint of an AERA event (left, energy density in absolute units) and a LOFAR air-shower event (right, power in arbitrary units). 
  The circles are the measurements by antenna stations, for AERA overlaid on a parametrization of the footprint, and for LOFAR overlaid on the best fitting out of 75 CoREAS simulations made for this specific event, i.e., the color inside a circle is equal to the surrounding color exactly when the measured signal is equal to the simulated signal (slightly modified from Refs.~\cite{GlaserEnergyICRC2015} and \cite{LOFAR_Xmax_ICRC2015}, respectively).}
  \label{fig_footprints}
\end{figure*}

\begin{figure*}[p]
  \centering
  \includegraphics[width=0.54\linewidth]{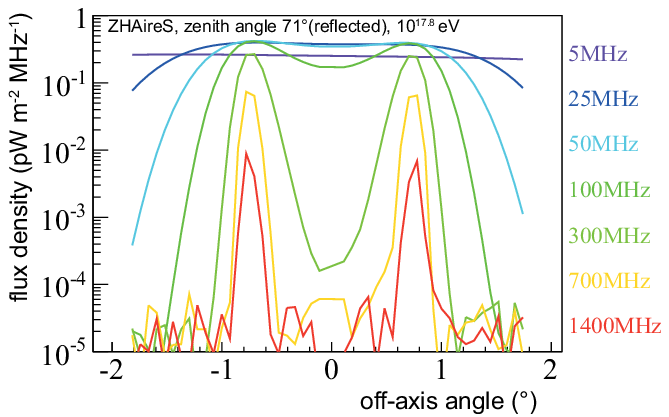}
  \hfill
  \includegraphics[width=0.44\linewidth]{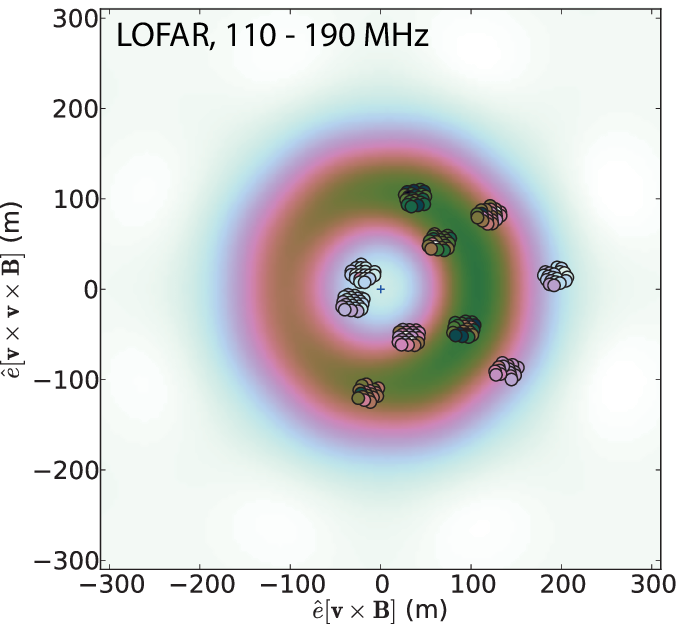}
  \caption{Left: ZHAireS simulation of the reflected radio emission of an air shower; the higher the observation frequency the stronger pronounced is the enhancement at the Cherenkov angle (slightly modified from Ref.~\cite{ReflectedZHAiresS_ICRC2015}).
  Right: This is qualitatively confirmed by a LOFAR measurement of the direct radio signal at higher frequencies (slightly modified from Ref.~\cite{NellesLOFAR_ECRS2014})}
  \label{fig_highFreqLDF}
\end{figure*}

\subsection{Footprint - the lateral distribution of the radio signal}
\label{sec_footprint}

Traditionally the lateral distribution of an air-shower observable describes its dependence on the distance to the shower axis, which means that a classical lateral-distribution function (LDF) is a one-dimensional function in the shower plane, i.e., the plane perpendicular to the shower axis.
For dense media this is a valid description since the radio signal is a radially symmetric around the shower axis with maximum amplitude at the Cherenkov angle. 
However for air showers, even in the shower plane, the radio signal is generally asymmetric around the shower axis due to the interference of the geomagnetic and Askaryan effects. 
Thus, the radio footprint has to be described by a two-dimensional LDF depending not only on the distance, but also on the azimuthal angle around the shower axis (see figure \ref{fig_CoREASfootprints} for a simulation example). 
The coordinate system in the shower plane can be chosen arbitrarily, but a natural choice is to align the $x$-axis with the geomagnetic Lorentz force pointing in $\vec{v} \times \vec{B}$ direction (with $\vec{v}$ the direction of the shower axis, and $\vec{B}$ the direction of the geomagnetic field, cf.~section \ref{sec_signalProcessing}).

The shape of the footprint depends strongly on the distance to the shower maximum and on the frequency range.
Typically it features a kind of bean shape because the signal is enhanced at the Cherenkov ring and at the same time features an asymmetry in the direction of the geomagnetic Lorentz force (see figure \ref{fig_footprints}). 
The size of the asymmetry depends on the relative strength of the Askaryan effect to the geomagnetic effect.
Since recently the radio emission of air showers has been observed not only directly, but also after reflection on ice by ANITA \cite{ANITA_CR_PRL_2010}, the features of reflected radio signals have been studied with simulations as well \cite{ReflectedZHAiresS_ICRC2015}.
No major differences have been found compared to direct emission except for a phase flip by $180^\circ$ observed by ANITA. 
Whether the radio emission is observed directly or after reflection, the Cherenkov ring is washed out at lower frequencies, but is very sharp at high frequencies of several $100\,$MHz or more (see figure \ref{fig_highFreqLDF}).

Several functions have been suggested to describe the footprint, in particular a pragmatic phenomenological solution now used by LOFAR and AERA, which is accurate to a few percent and has been used to fit the AERA event in figure \ref{fig_footprints}.
This LDF is a two-dimensional Gaussian with amplitude $A_+$, width $\sigma_+$, and center $(X_c, Y_c)$ different from the shower core, from which a Gaussian of smaller amplitude $A_-$ displaced by $x_-$ in the direction of the geomagnetic Lorentz force is subtracted \cite{NellesLOFAR_LDF2014}. 
The subtraction describes the enhancement at the Cherenkov ring (i.e., the total amplitude in the center can be smaller than at the ring) and the displacement describes the asymmetry:

\begin{equation}
P(x^{\prime},y^{\prime}) = A_+ \cdot \exp\left(\frac{-[(x^{\prime}-X_c)^2+(y^{\prime}-Y_c)^2]}{\sigma_+^2}\right) - A_-\cdot  \exp\left(\frac{-[(x^{\prime}-(X_c+x_-))^2+(y^{\prime}-Y_c)^2]}{(C_1\cdot e^{C_2\cdot \sigma_+})^2}\right)
\end{equation}
with $P(x^{\prime},y^{\prime})$ the time-integrated power of the radio signal at position $(x^{\prime},y^{\prime})$ in the shower plane, and $C_i$ constants (see Ref.~\cite{NellesLOFAR_LDF2014} for details).

\begin{figure}[t]
  \centering
  \includegraphics[width=0.99\linewidth]{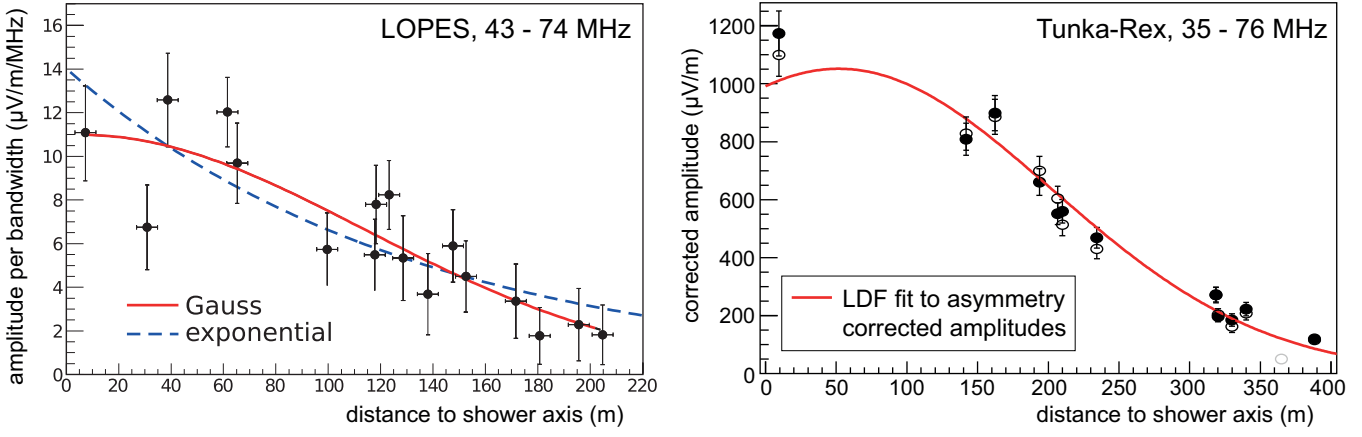}
  \caption{Examples of one-dimensional lateral-distribution functions fit to air-shower events measured by LOPES (left) and by Tunka-Rex after asymmetry correction (right, open circles are before, filled circles after asymmetry correction). 
  For typical events the size of the asymmetry as well as the difference between a Gaussian and an exponential LDF is smaller than the individual measurement uncertainties (modified from Refs.~\cite{2014ApelLOPES_MassComposition} and \cite{TunkaRex_XmaxJCAP2016}).}
  \label{fig_LDFexamples}
\end{figure}

For many practical purposes the radio footprint can still be approximated by a one-dimensional function describing the signal strengths over distance to the shower axis and neglecting the azimuthal asymmetry (see figure \ref{fig_LDFexamples}). 
This is in particular valid in three cases: 
First, if other measurement uncertainties dominate over the asymmetry. 
Second, if there is a sufficient number of antennas at different azimuthal angles around the shower axis.
Then a one-dimensional LDF implicitly averages over the asymmetry and can be used to determine the average amplitude or slope of the footprint at a certain distance. 
Third, after correction of the asymmetry:
since the average size of the asymmetry is known for certain experiments (cf.~figure \ref{fig_LOFARpolarizationDependence} in section \ref{sec_emissionMechanisms}) it is possible to approximately restore the azimuthal symmetry by correcting the signal strength measured at individual positions \cite{KostuninTheory2015, TunkaRex_XmaxJCAP2016}:

\begin{equation}
P_\mathrm{cor}(r_\mathrm{axis}) = \frac{P(r_\mathrm{axis}, \phi_\mathrm{axis})}{a_\mathrm{rel}^2 + 2a_\mathrm{rel} +1}
\end{equation}
with $P_\mathrm{(cor)}(r_\mathrm{axis})$ the (asymmetry-corrected) power at distance $r_\mathrm{axis}$ from the shower axis, $\phi_\mathrm{axis}$ the azimuth angle relative to the shower axis with $\phi_\mathrm{axis} = 0$ in the direction of the geomagnetic Lorentz force, and $a_\mathrm{rel} = \epsilon /\sin \alpha$ the relative strengths of the Askaryan effect, where $\epsilon$ is the strength of the Askaryan effect (cf.~figure \ref{fig_LOFARpolarizationDependence}) and $\alpha$ is the geomagnetic angle (= angle between shower axis and geomagnetic field).
This means that for $\alpha = 0$ no correction is applied since the geomagnetic effect vanishes and the radio signal is solely due to the Askaryan effect which already is radially symmetric around the shower axis.

The asymmetry-corrected footprint can be described by a one-dimensional Gaussian LDF with only three free parameters \cite{2014ApelLOPES_MassComposition, KostuninTheory2015}:
\begin{equation}
P(r_\mathrm{axis}) = P_0 \exp (- \eta_1 r_\mathrm{axis} + \eta_2 r_\mathrm{axis}^2)
\end{equation}
with $P_0$ the power at the shower axis, which depends on the energy and the distance to the shower maximum, and $\eta_1$ and $\eta_2$ free parameters describing the slope of the exponential tail of the LDF and the flattening towards the shower axis, respectively. 
Often not the amplitude at the shower axis, but at a reference distance of about $100\,$m is chosen, because there the signal strength depends least on the position of the shower maximum and in good approximation only on the shower energy. 
This distance is roughly at the position of the Cherenkov ring, where the emission of all parts of the shower arrives approximately simultaneously.
However, the Cherenkov ring and the reference distance might coincide only by chance since such a reference distance with minimal influence of the shower maximum on the amplitude exists also in simulations without refractive index \cite{HuegeUlrichEngel2008}.
Instead of the Gaussian LDF sometimes an exponential LDF is used (i.e., $\eta_2 = 0$) \cite{AllanReview1971, 2010ApelLOPESlateral}, which ignores the flattening towards the shower axis, but for many purposes is a sufficient approximation, e.g., when interested in the slope of the exponential tail sensitive to the longitudinal shower development \cite{2012ApelLOPES_MTD}.
Further details on the reconstruction of the energy and the shower maximum using lateral-distribution functions are explained in chapter \ref{sec_reconstruction}.

\begin{figure}[t]
  \centering
  \includegraphics[width=0.65\linewidth]{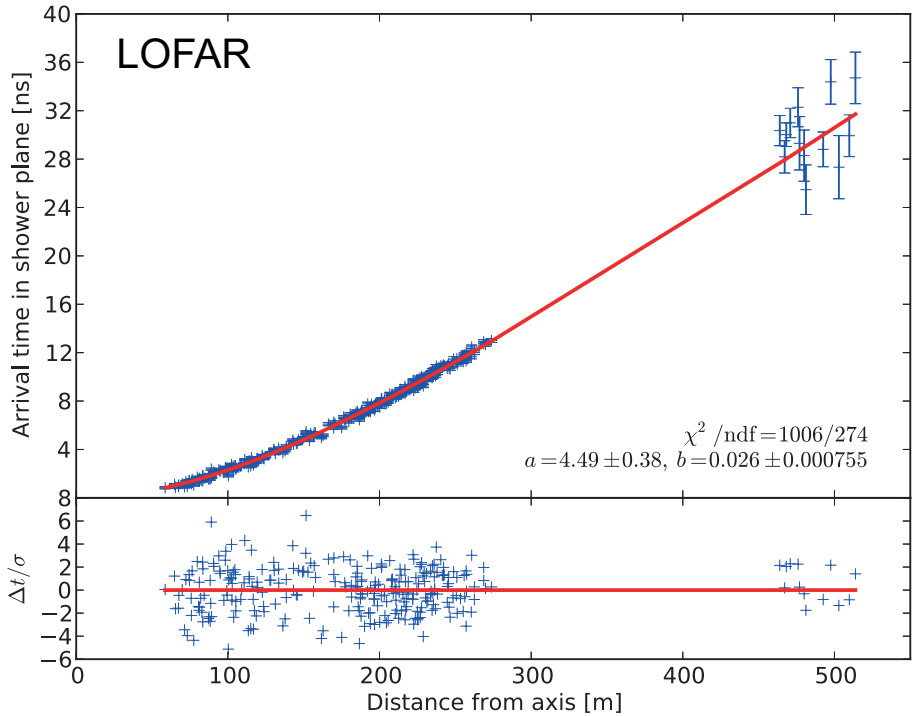}
  \caption{Radio wavefront of an air shower measured by LOFAR determined by fitting the arrival time of different antennas projected to the shower plane (from Ref.~\cite{CorstanjePolarizationICRC2015}). 
  The parameter $a$ is the offset of the hyperboloid at the center to the asymptotic cone, and $b = \sin \rho$ defines the angle $\rho$ of the asymptotic cone to the shower plane.}
  \label{fig_LOFARwavefront}
\end{figure}

\subsection{Wavefront}
\label{sec_wavefront}
By definition the radio wavefront is the surface perpendicular to the propagation direction of the radio signal. 
This theoretical definition, however, lacks practical application since the propagation cannot be measured directly. 
Instead, the arrival time of the radio signal can be measured in the individual antennas, and the distribution of the arrival times is used to reconstruct the shape of the wavefront. 
This makes the reconstructed wavefront depend on the properties of the measurement and the analysis procedure, like the used bandpass filters, or the methods how to determine the pulse time \cite{SchroederICRC2015}. 
Nevertheless, some general, qualitative features of the radio wavefront of air showers, in particular its hyperbolic shape, can be assumed to be independent of the measurement details. 
While the absolute value of the steepness of the hyperboloid depends on the measurement procedure, for each given procedure the steepness should also depend on the zenith angle and the depth of shower maximum: the more distant the shower maximum the flatter the wavefront \cite{LOPESwavefront2014}.

\begin{figure}[t]
  \centering
  \includegraphics[width=0.49\linewidth]{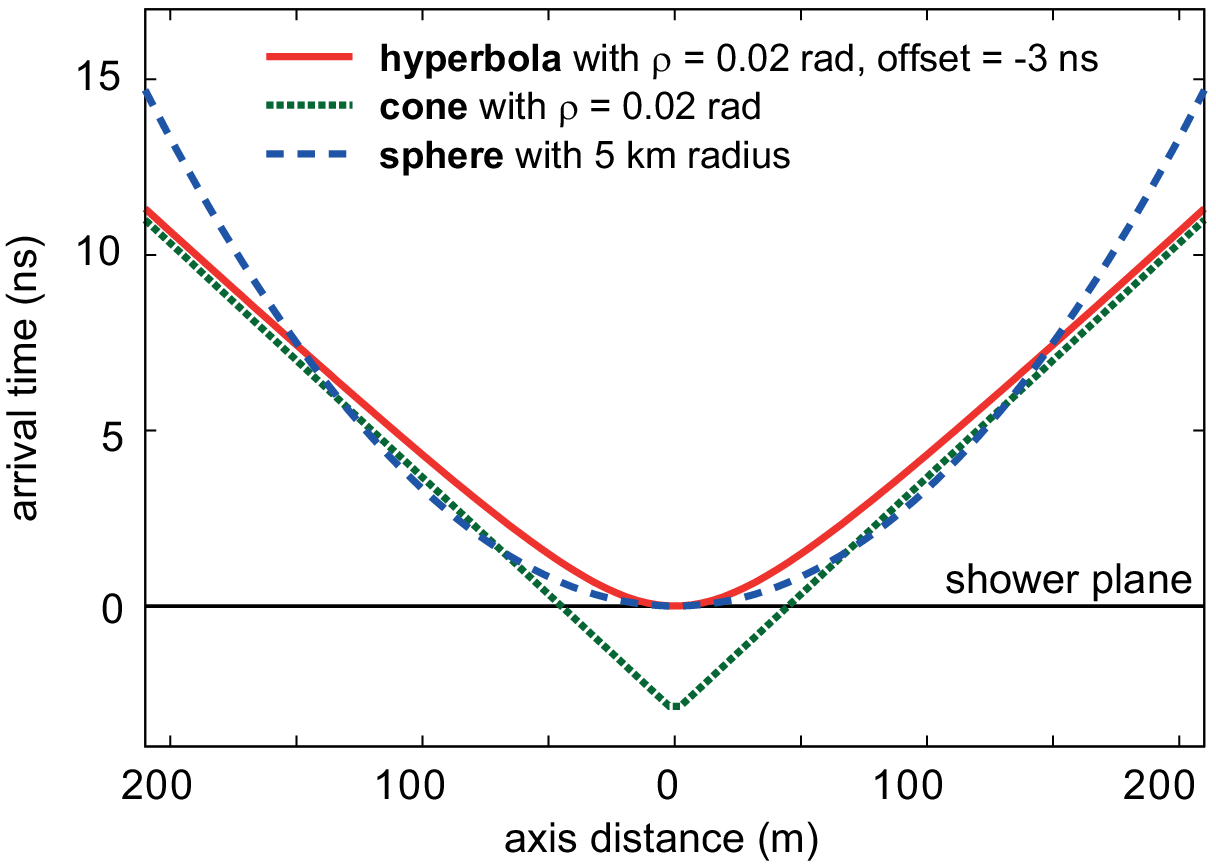}
  \hfill
  \includegraphics[width=0.49\linewidth]{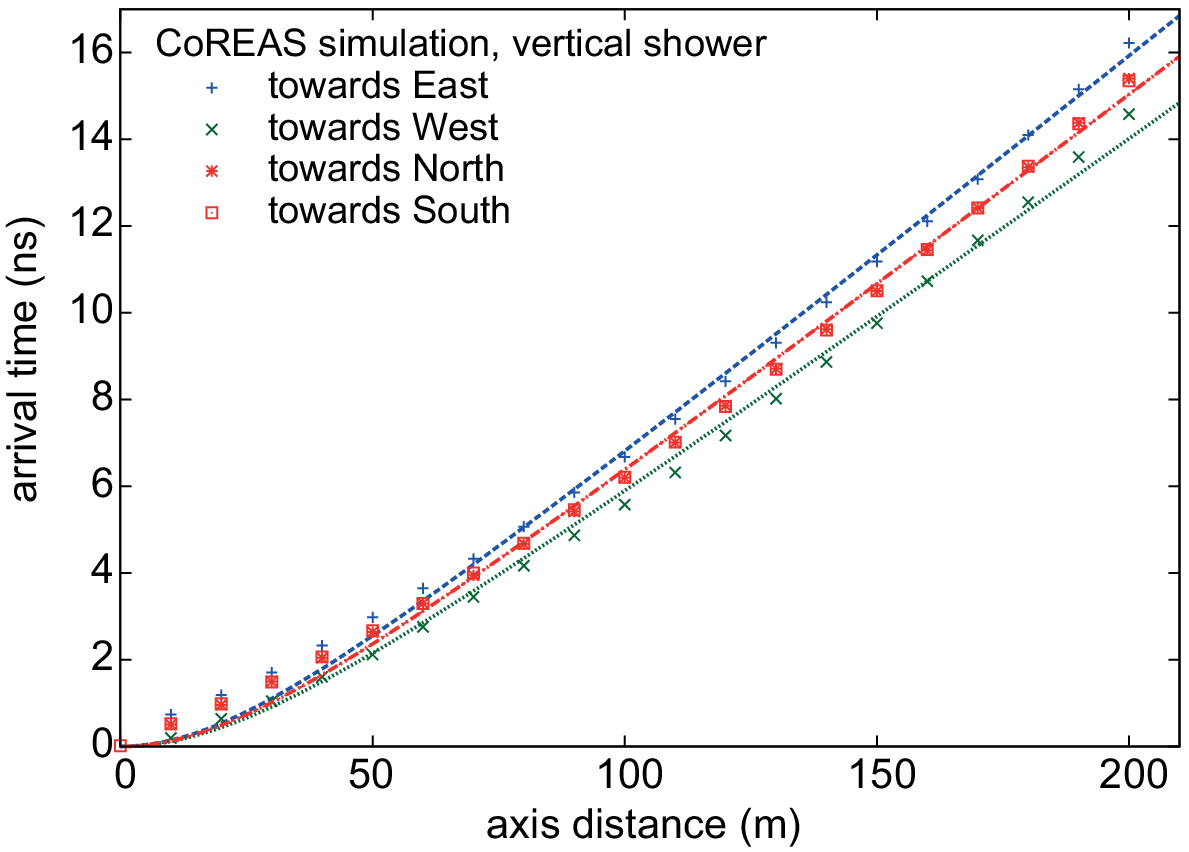}
  \caption{Left: Different wavefront models for comparison. 
  The hyperbolic wavefront is approximately spherical close to the shower axis, and approaches a cone at larger distances. 
  Right: According to CoREAS simulations of a vertical air shower, the radio wavefront features a small east-west asymmetry perpendicular to the geomagnetic field, but is identical towards North and South which, however, has not yet been confirmed by measurements (both figures from Ref.~\cite{LOPESwavefront2014})}
  \label{fig_wavefront}
\end{figure}

\begin{figure}[t]
  \centering
  \includegraphics[width=0.49\linewidth]{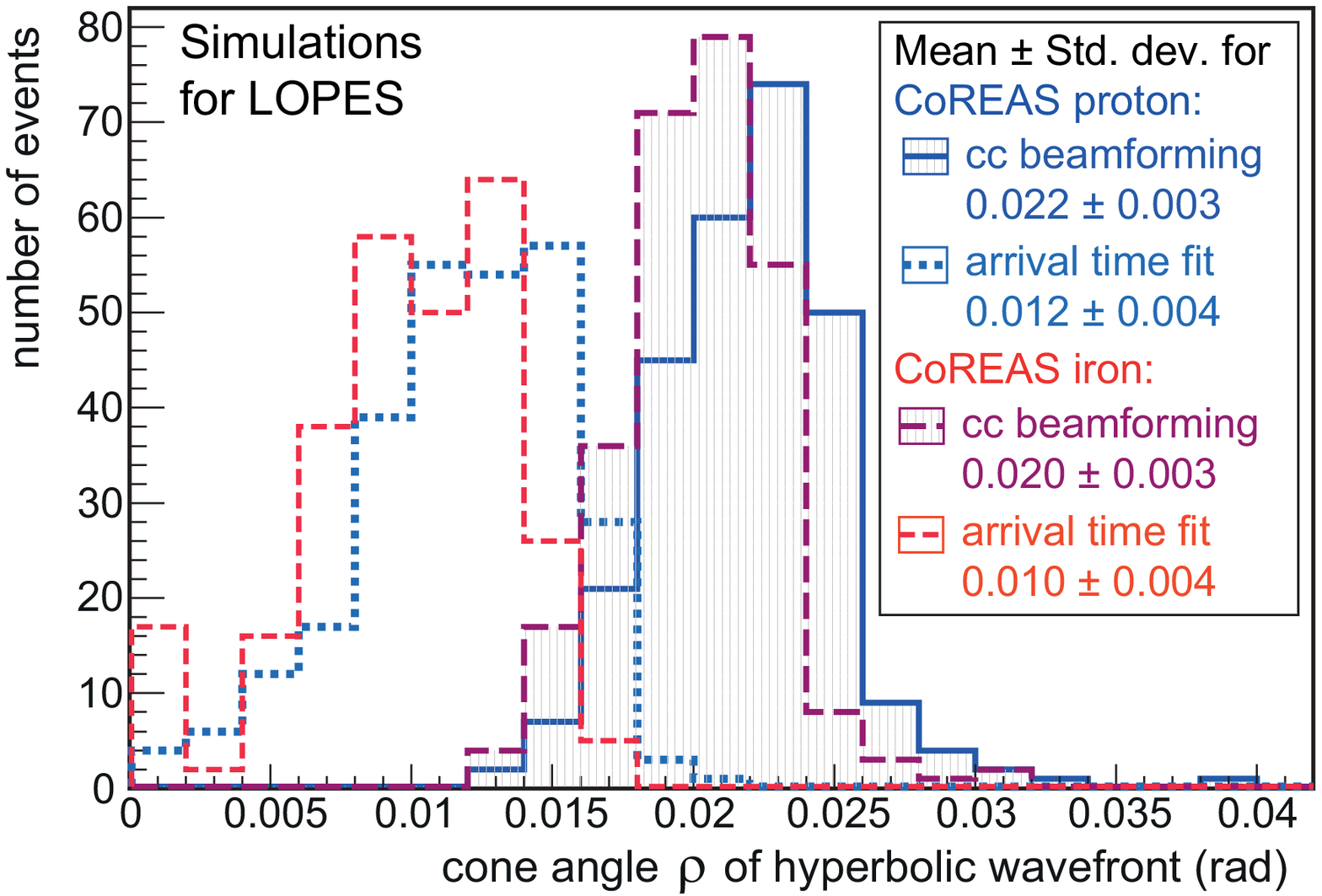}
  \hfill
  \includegraphics[width=0.49\linewidth]{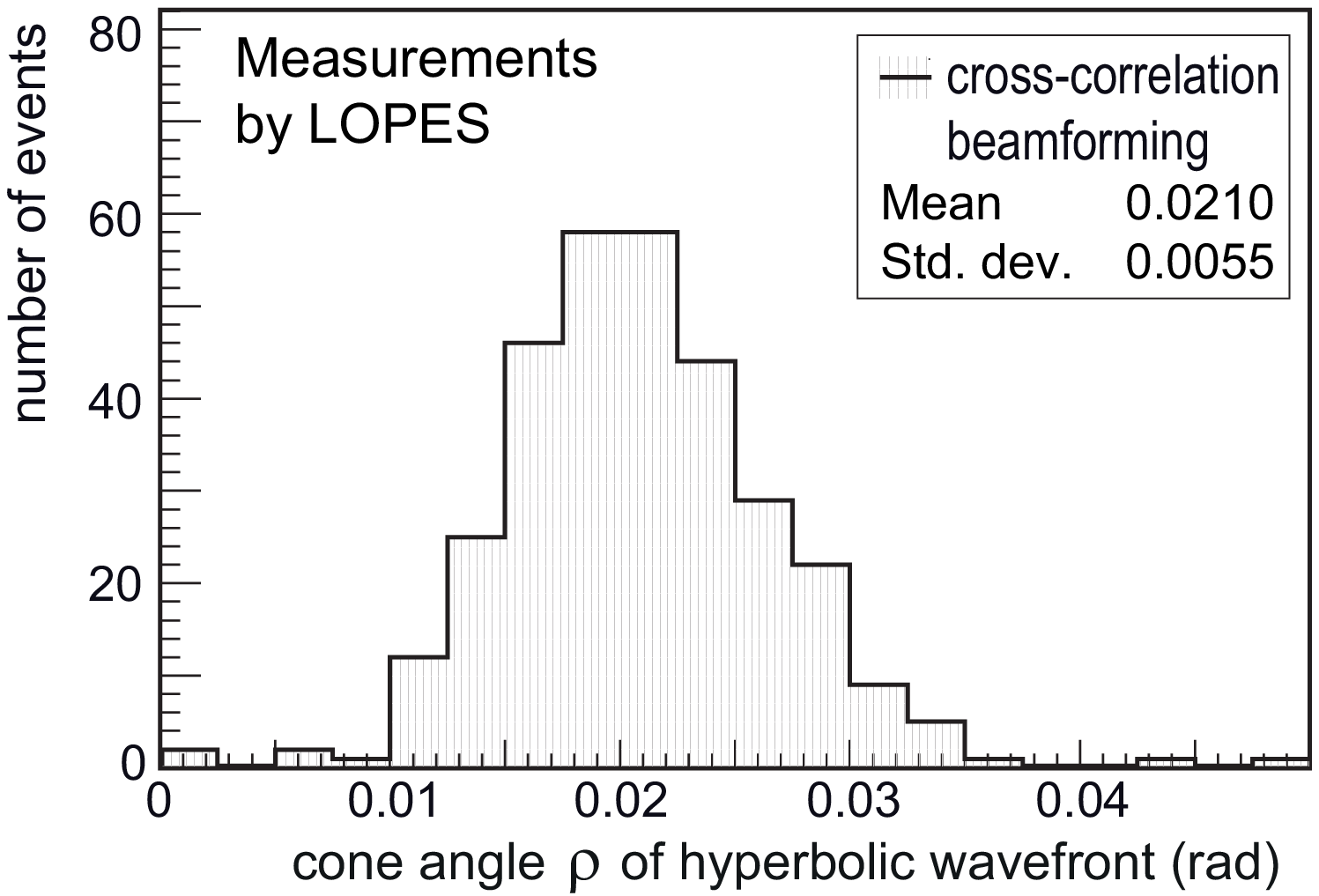}
  \caption{Left: CoREAS simulations of the wavefront including the LOPES detector simulation for protons and iron nuclei as primary particles. 
  The cone angle of the hyperbolic wavefront depends strongly on the method how the wavefront is determined (maximum time in individual antennas or cross-correlation beamforming, cf.~section \ref{sec_beamforming}).
  Right: LOPES measurements of the cone angle determined with cross-correlation beamforming (left figure modified from Ref.~\cite{SchroederICRC2015}, right figure from Ref.~\cite{LOPESwavefront2014}).}
  \label{fig_LOPESwavefront}
\end{figure}

Studies based on air-shower measurements of the LOPES and LOFAR experiments as well as CoREAS simulations agree that the radio wavefront is of approximately hyperbolic shape \cite{CorstanjeLOFAR_wavefront2014, LOPESwavefront2014} (see figures \ref{fig_LOFARwavefront} and \ref{fig_wavefront}).
The wavefront is neither a plane, nor a sphere, nor a cone, although all these shapes can be valid approximations of the hyperboloid under certain conditions: 
in the center the hyperboloid is approximately spherical.
With increasing distance from the shower axis the hyperboloid approaches a cone, with an angle to the shower plane of at most a few degrees (see figure \ref{fig_LOPESwavefront}).
Consequently, a sphere could be a good approximation for distant showers measured by compact detectors, a cone can be a valid approximation for sparse arrays, and a plane is a sufficient approximation when a direction uncertainty of a few degrees is unimportant, or simply when the number of antennas is insufficient to determine the wavefront shape more accurately. 

The nature of the hyperbolic shape can be understood in a simple way: 
Compared to the typical extension of radio arrays of larger than $100\,$m, the radio emission of the shower comes from a point source moving along the shower axis roughly with the speed of light. 
This point source radiates only for a certain time around the shower maximum, so it has a finite track length.
This track length is not negligible against the array dimensions and, thus, the shower cannot be approximated by a static point source (which would cause a spherical wavefront). 
While an infinite track length at light speed would lead to a conical wavefront, a finite length leads to a hyperboloid which has a conical shape distant from the axis, and a spherical shape close to the axis. 
Two conclusions can be made from this simple picture: 
First, the exact shape depends on the distance from the end of the track to the array, and by this to the distance of the shower maximum. 
Second, since the radio signal is created at the shower axis, there is no reason why the radio wavefront should be of similar shape as the particle front of the shower!

\begin{figure}[t]
  \centering
  \includegraphics[width=0.6\linewidth]{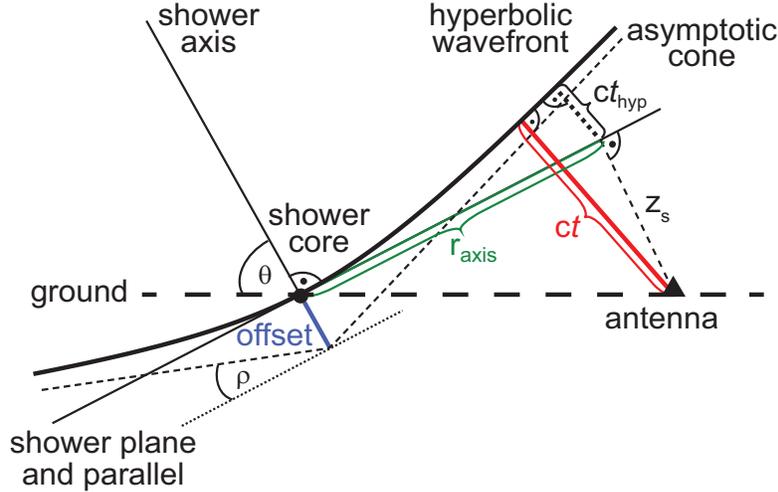}
  \caption{Sketch of the hyperbolic wavefront.
   LOPES used the arrival time $t$ at ground to describe the wavefront as function of the distance to the shower axis $r_\mathrm{axis}$ and the distance to the shower plane $z_s$ with the cone angle $\rho$ and an offset parameter called $b$ in LOPES notation. 
   LOFAR uses the arrival time $t_\mathrm{hyp}$ in the shower plane as coordinate, names the offset parameter $a$ and uses $b = \sin \rho$ as shape parameter. 
   The mistake of the LOFAR method is $ct - (ct_\mathrm{hyp} + z_s)$ which in many practical cases is negligible against measurement uncertainties (figure modified from Ref.~\cite{LOPESwavefront2014}).}
  \label{fig_wavefrontQuantities}
\end{figure}

The same hyperbolic wavefront can be described by different formulas and no consensus has been achieved on the notation.
LOFAR used the following formula to describe the arrival time $t$ in the shower plane as a function of the distance to the shower axis $r_\mathrm{axis}$:

\begin{equation}
c \, t_\mathrm{hyp}(r_\mathrm{axis}) = -a_\mathrm{offset} + \sqrt{a_\mathrm{offset}^2+b_\mathrm{shape}^2r_\mathrm{axis}^2}
\label{eq_LOFARhyp}
\end{equation}
with $c$ the speed of light, $a_\mathrm{offset}$ the offset of the asymptotic cone of the hyperboloid to the shower plane at the core; $b_\mathrm{shape}$ is related to the steepness of the hyperboloid, since $\rho = \arcsin b_\mathrm{shape}$ is the angle between the shower plane and the asymptotic cone. 
Confusingly, the parameter $a_\mathrm{offset}$ in this LOFAR notation is $c \cdot b_\mathrm{offset}$ in the LOPES notation, and the sign of this parameter depends on whether the positive direction along the shower axis is with or against the propagation direction of the shower.
To apply equation (\ref{eq_LOFARhyp}) to measured or simulated events the arrival time in each antenna first has to be corrected for the distance $z_s$ of the antenna to the shower plane.

This description of the wavefront by the arrival time in shower plane neglects that the radio signal propagates not exactly in the direction of the shower axis, i.e., not perpendicular to the shower plane, but instead perpendicular to the wavefront. 
Thus, a mistake is made roughly proportional to the distance $z_s$ from the individual antennas to the shower plane.
Luckily the error is small in many cases since the angle between the wavefront and the shower plane is small (cf.~figure \ref{fig_wavefrontQuantities}). 
Nevertheless, a more accurate approach followed by LOPES is to describe the wavefront shape based on the arrival times at the real antenna positions defined by their distance to the shower plane $z_s$, and their distance to the shower axis $r_\mathrm{axis}$:

\begin{equation}
c \, t(r_\mathrm{axis},z_s) = \sqrt{(r_\mathrm{axis}\sin\rho)^2+(c\cdot b_\mathrm{offset})^2} + z_s\cos\rho + c\cdot b_\mathrm{offset}
\label{eq_hyperboloid}
\end{equation}
with $c$ the speed of light, $b_\mathrm{offset}$ the offset of the asymptotic cone of the hyperboloid to the shower plane at the core, and $\rho$ the angle between the shower plane and the asymptotic cone. 
The distance of an antenna to the shower plane depends on the azimuthal position relative to the shower axis and on the zenith angle $\theta$.
Under the approximation $\cot \theta \cdot \cos \rho \approx 1$, equation (\ref{eq_hyperboloid}) is equivalent to the simpler equation (\ref{eq_LOFARhyp}), which allows for a more intuitive way of one-dimensional plotting, since the arrival time $t$ depends only on the coordinate $r_\mathrm{axis}$, instead of the two coordinates $r_\mathrm{axis}$ and $z_s$. 
To check for a specific analysis if this simplification is valid or not, the size of the error has to be compared to the arrival-time precision of a particular experiment.

Going into more details the radio wavefront slightly differs from a hyperboloid (see right-hand side of figure \ref{fig_LOPESwavefront}), but the difference is so small that it has been ignored so far. 
CoREAS simulations show that the wavefront is slightly asymmetric, similar to the footprint, though on a smaller scale of about $1\,$ns only, which would currently make LOFAR the only experiment precise enough for an experimental test. 
The reason for this asymmetry has not yet been investigated, but there is a reasonable explanation. 
Since the Askaryan and geomagnetic emissions are not completely in phase, their interference ought to affect not only the pulse amplitude, but also the pulse shape. 
For a given method to measure the pulse time, e.g., the time of the maximum, the arrival time then depends on the pulse shape and can be different for antennas with constructive or destructive interference of both emission mechanisms. 
This again shows that the exact wavefront shape is very sensitive to the details of how the arrival time is determined, which has to be taken into account in any interpretation of the wavefront shape, e.g., for $X_\mathrm{max}$ reconstruction (cf.~chapter \ref{sec_reconstruction}) and in any comparison of measured or simulated wavefronts.

\begin{figure}[t]
  \centering
  \includegraphics[width=0.49\linewidth]{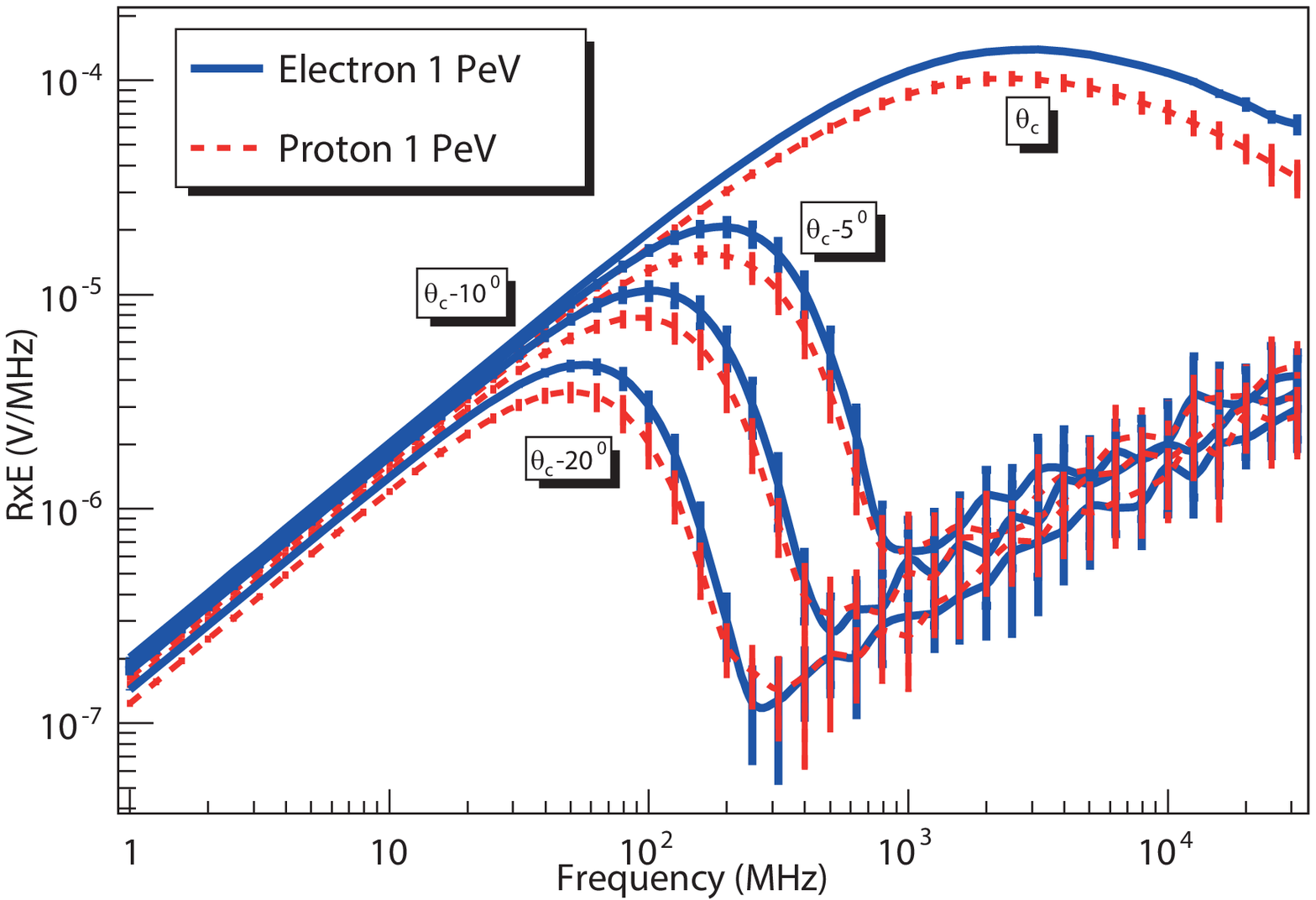}
  \hfill
  \includegraphics[width=0.49\linewidth]{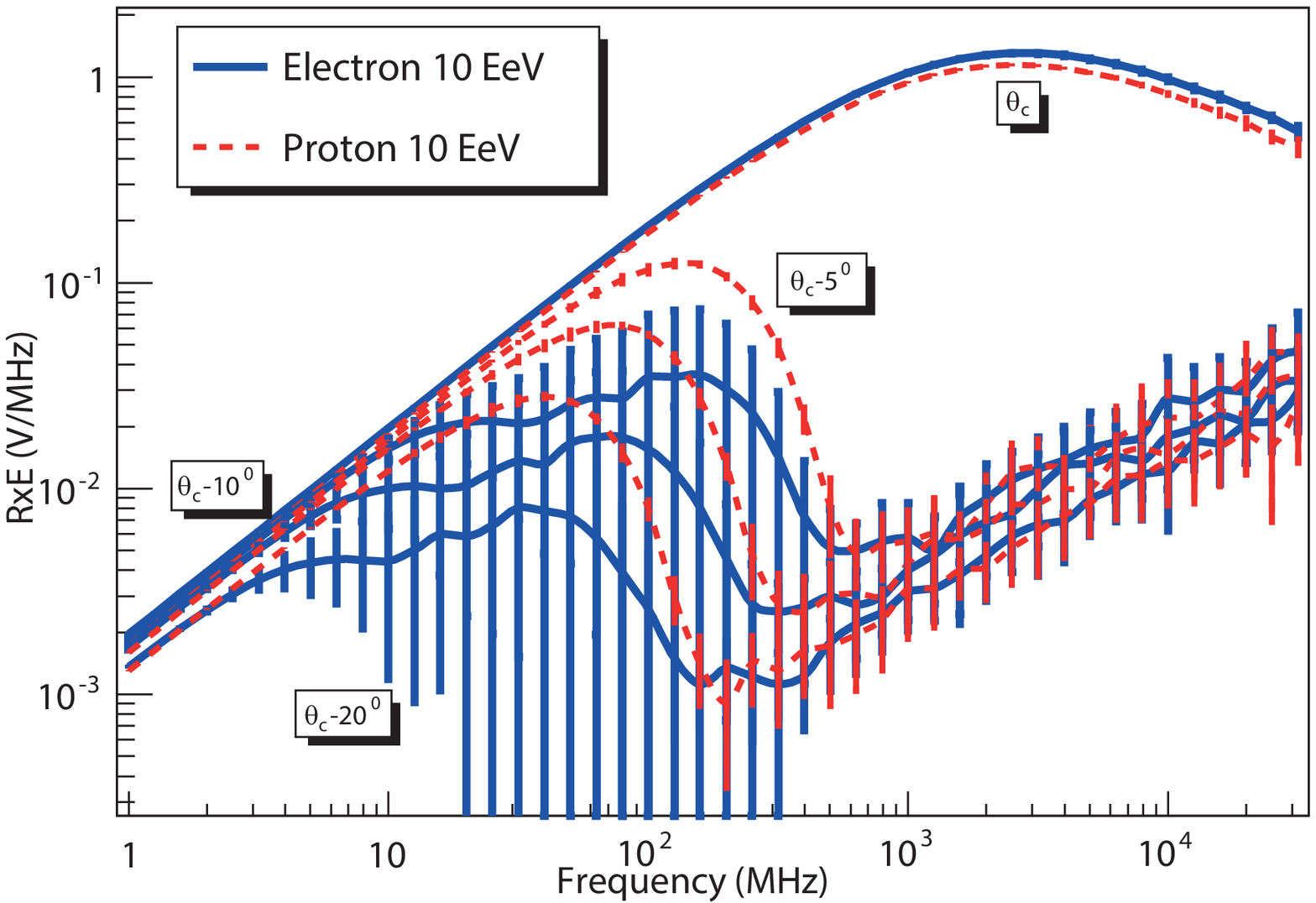}
  \caption{Simulated radio signal of showers in ice initiated by electrons and protons of different energy, where electron-initiated cascades can be a consequence of charged-current neutrino interactions, and proton-initiated cascades are very similar to hadronic cascades originating from all types of neutrino interactions. 
  Vertical bars mark shower-to-shower fluctuations.
  The frequency spectrum depends strongly on the observation angle relative to the Cherenkov angle $\theta_c$, and additionally on the type of the cascade, i.e., whether purely electromagnetic or hadronic (from Ref.~\cite{AlvarezMuniz_ZHAires_ice2012}).}
  \label{fig_frequencySpectrumDenseMedia}
\end{figure}

\subsection{Features of the radio signal in dense media}
\label{sec_radioFeaturesDenseMedia}
Since natural showers have not yet been measured in dense media, the knowledge about the features of the radio emission in dense media is based on calculations, simulations and accelerator measurements \cite{Saltzberg_SLAC_Askaryan2001, ANITA_SLAC_PRL_2007, SLAC_T510_PRL2016}. 
Nonetheless, it can be assumed that recent simulation codes describe the properties approximately correctly and predict at least the correct order of magnitude of the absolute amplitude, because simulations relying on the same principles have been experimentally confirmed for air showers and accelerator experiments \cite{LOPESimprovedCalibration2016, TunkaRex_NIM2015, SLAC_T510_PRL2016}.
Although the emission in dense media is by the same Askaryan effect relevant also for air showers, there are some differences in the resulting radio signal, mainly due to the much higher refractive index in dense media.
 
Independent of the primary particle, showers in dense media have a scale of meters, which is very compact compared to the typical distance from the shower to the detector scaling from $100\,$m for in-ice arrays up to $400,000\,$km for lunar observations. 
In particular for experiments using ice as medium the distance can vary drastically from shower to shower due to the random location of the neutrino interaction initiating the shower.
This is a difference to air showers for which the typical shower-to-shower fluctuations make up for only a few $10\,\%$ in distance between the detector and the shower maximum. 
Thus, for dense media the properties of the radio emission are typically not given as functions of distance to the shower axis, but instead as functions of the observation angle, i.e., either the angle relative to the shower axis or relative to the Cherenkov angle of the medium.

The radio signal in dense media has a ring-like structure similar to the radio footprint of air showers above $100\,$MHz, but the Cherenkov angle is much larger in dense media ($56^\circ$ in ice compared to about $1^\circ$ in air). 
Therefore, the radio signal in dense media has a significant strength only at the Cherenkov ring whose width depends on the observation frequency (see figure \ref{fig_frequencySpectrumDenseMedia}). 
For all frequency bands the radio signal has a hole around the shower axis, i.e., the signal is zero at the axis - in contrast to the radio signal of air showers.
For this reason the concept of a radio wavefront relative to the shower plane makes no sense for showers in dense media. 
Instead it might be useful to study the radio wavefront relative to the surface perpendicular to the Cherenkov cone, where the wavefront should be flatter the more distant the vertex of the Cherenkov cone is. 
The polarization of showers in dense media is simply that of the Askaryan effect: the electric-field vector is radially oriented with the shower axis in the center. 
So both the amplitude pattern and the polarization are radially symmetric with respect to the shower axis.

There are some differences between hadronic and electromagnetic showers in dense media, but the type of the primary particle initiating hadronic showers has only little impact on the shower.
Hadronic showers initiated by neutrino interactions are hardly distinguishable from those initiated by protons when the shower energy is equal \cite{ACoRNE2007, AlvarezMuniz_ZHAires_ice2012}.
Hence, even though the primary particle is assumed to be a neutrino, showers initiated by protons and electrons are investigated in simulations in order to study the radio emission in dense media. 
Electron initiated showers are rarer for neutrino interactions, since except for the Glashow resonance of anti-electron neutrinos at about $6\,$PeV, neutrinos have much larger cross sections for interactions with  nuclei than for interactions with electrons.
Still, electromagnetic showers can occur together with a hadronic cascade due to charged-current interactions of electron neutrinos or by tau decays, where the taus originate from charged-current interactions of tau neutrinos.

For a given shower energy, the amplitude of the radio signal of different showers varies only slightly depending on the fraction of the primary energy transferred in the electromagnetic shower component.
For hadronic showers this fraction increases with energy, and the difference in total amplitude between hadronic and electromagnetic showers shrinks with increasing energy \cite{AlvarezMuniz_ZHAires_ice2012, AlvarezMuniz_ice2010}. 
However, other differences between hadronic and electromagnetic showers become more important with increasing primary energy, especially differences in the frequency spectrum and the angular distribution of the radio signal around the Cherenkov cone (see figure \ref{fig_frequencySpectrumDenseMedia}). 
Moreover, shower-to-shower fluctuations become smaller for hadronic showers with increasing energy, but due to the LPM effect larger for electromagnetic showers, i.e, in some cases electromagnetic showers can extend over several $10\,$m even in dense media such as ice \cite{LPMice1998}. 
It is not yet clear how accurate these features can be measured by real detectors, but at least in principle the differences between hadronic and electromagnetic cascades could be converted into a tool for statistically determining the fraction of electron neutrinos in the total neutrino flux.

%\clearpage

\section{Detectors and Measurements}
%\textbf{Figure: Maps of experiments.}

This chapter reviews modern digital techniques for the radio measurement of air showers and cascades in ice initiated by cosmic rays and neutrinos.
Related experiments, such as radar detection, radio observation of the Moon, accelerator experiments, and acoustic detection are additionally mentioned.
In the later sections of this chapter there is again a focus on air-shower experiments shortly summarizing concepts for dense media at the end: features of different antenna arrays are compared with each other and, finally, techniques for calibration and for the reconstruction of signal properties are discussed. 
Reconstruction methods for air-shower parameters based on the measurements of the radio signal are reviewed in chapter \ref{sec_reconstruction}.

\begin{table}[p]
\centering
\caption{Modern radio experiments for high-energy cosmic rays and neutrinos. 
Current radio arrays for neutrino detection are still under construction and the starting year refers to first prototype setups.} \label{tab_experiments}
\vspace{0.3cm}
\small
\begin{tabular}{lcccc}
\hline
Name of&Operation  & \multicolumn{2}{c}{aiming at} & medium of \\
experiment& period & cosmic rays & neutrinos & radio emission \\
\hline
Yakutsk & since 1972 & x & & air \\
RICE& 1999 - 2010 &  & x & ice \\
LOPES & 2003 - 2013 & x & & air \\
CODALEMA& since 2003 &  x & & air \\
ANITA(-lite)& first flight 2004 & x & x & air + ice \\
AURA& 2006 -2009 &  & x & ice \\
TREND & 2009 - 2014 & x & & air \\
AERA & since 2010 & x & & air \\
ARA& since 2010 &  & x & ice \\
LOFAR & since 2011 & x & x & air + moon \\
Tunka-Rex & since 2012 & x & & air \\ 
ARIANNA& since 2012 & x & x & air + ice \\
TAROGE& since 2014 & x & & air \\ 
GNO& tests since 2015 &  & x & ice \\
SKA-low & planned & x & x & air + moon \\
GRAND & planned & x & x & air + mountain \\
\hline
\end{tabular}
\end{table}

\begin{figure}[p]
  \centering
  \includegraphics[width=0.99\linewidth]{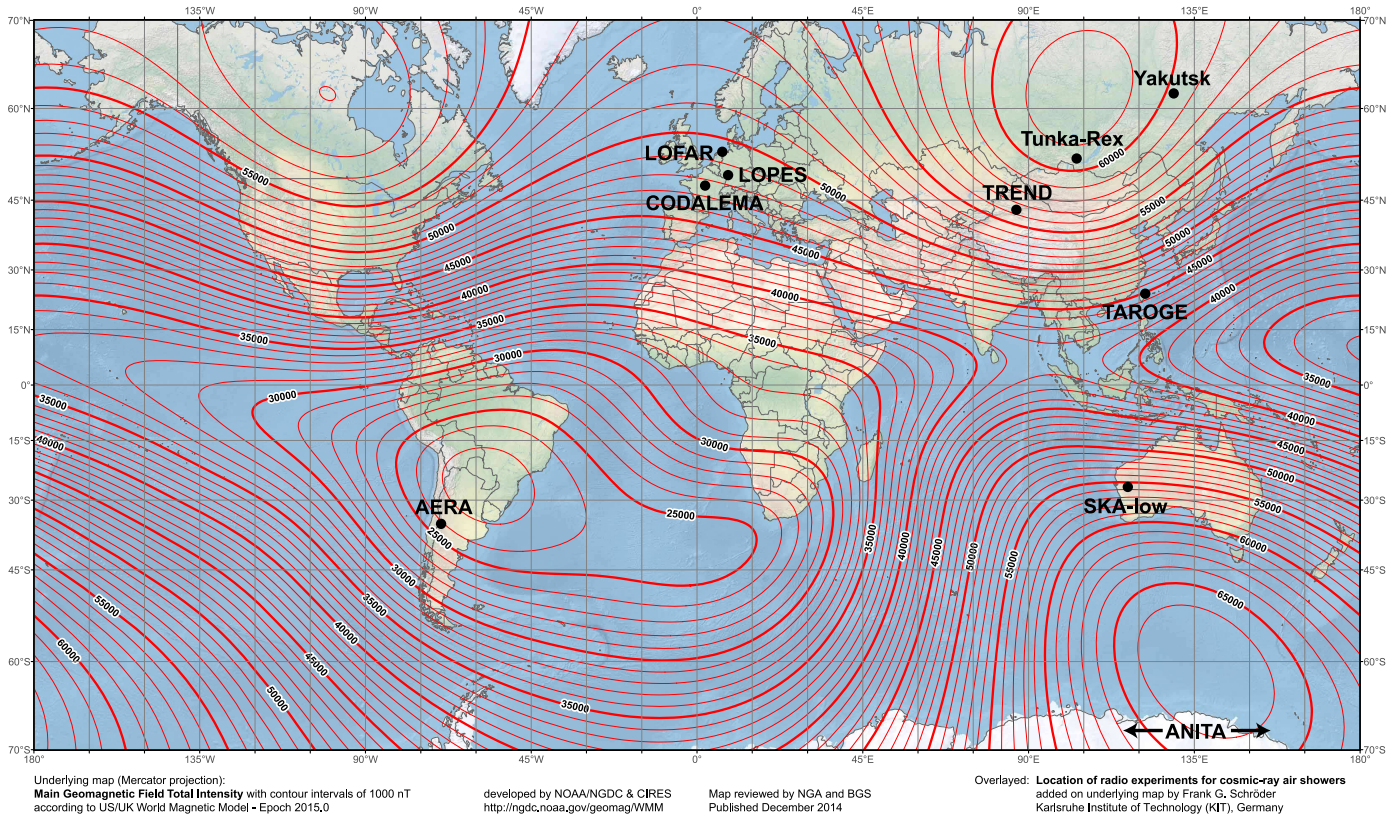}
  \caption{Map of the total geomagnetic field strengths (world magnetic model \cite{WorldMagneticModel2015}) and the location of various radio experiments detecting cosmic-ray air showers (figure from Ref.~\cite{SchroederCRIS2015}).}
  \label{fig_mapExperiments}
\end{figure}

Table \ref{tab_experiments} gives an overview on major modern radio experiment for cosmic-ray and neutrino detection, and table \ref{tab_experimentsSites} provides additional information on the experimental sites of antenna arrays for air showers. 
In particular, the local geomagnetic field and to a certain extent also the altitude determine the features of the radio signal emitted by air showers. 
The amplitude of the radio signal is roughly proportional to the strength of the geomagnetic field. 
This slightly influences the detection threshold which is around $100\,$PeV for all running air-shower arrays. 
The larger the zenith angle of the magnetic field the larger the asymmetry of the detection efficiency with respect to the azimuth of the arrival direction. 
The locations of different experiments are marked on a map of the geomagnetic field strength in figure \ref{fig_mapExperiments}. 
From pure geomagnetic field considerations the ideal site would be in the middle of Siberia. 
However, experimental sites for radio arrays are usually chosen by other criteria, like other air-shower arrays already in place, available infrastructure and political reasons.
For experiments searching for neutrino-induced showers in dense media, especially in ice, the geomagnetic field is unimportant, and they are not indicated on the map.
Current prototype experiments aiming at neutrino detection are located at the South Pole, on the Ross Ice Shelf, and in Greenland.

\begin{table}[t]
\centering
\caption{Selected digital antenna arrays used for air-shower detection, and the values of the international geomagnetic reference model {IGRF} for the experimental site \cite{PotsdamBfield2016}: 
the zenith angle of the geomagnetic field, $\theta_\mathrm{geo} = 90^\circ - |\mathrm{inclination}|$, and the magnetic field strength $B_\mathrm{geo}$. 
Also the numbers of antennas, the approximate areas covered with antennas, and the nominal frequency bands are given. 
However, many experiments were used also in configurations different from the one stated in the table, and many analyses are based on subsets of antennas and smaller sub-bands.} \label{tab_experimentsSites}
\vspace{0.3cm}
\small
\begin{tabular}{lccrrccccr}
\hline
Name of& Latitude & Longitude &Altitude& $\theta_\mathrm{geo}$ &$B_\mathrm{geo}$ &Number of& Area &Band\\
experiment& &  & in m & in $^\circ$  &in \textmu T&antennas& in km\textsuperscript{2} & in MHz \\
\hline
LOPES & $49^\circ06$' N & $8^\circ26$' E & $110$ & $25.2^\circ$ & $48.4$ & $30$ & $0.04$ & $40-80$\\
Yakutsk & $61^\circ 42$' N & $129^\circ 24$' E & $100$ & $13.8^\circ$ & $59.7$ & $6$ & $0.1$ & $32$\\
CODALEMA & $47^\circ 23$' N & $2^\circ 12$' E & $130$ & $27.4^\circ$ & $47.6$ & $60$ & $1$ & $2-200$\\
TREND & $42^\circ 56$' N & $86^\circ 41$' E & $2650$ & $26.7^\circ$ & $56.3$ & $50$ & $1.2$ & $50-100$\\
AERA & $35^\circ06$' S & $69^\circ30$' W & $1550$ & $54.0^\circ$ & $24.0$ & $153$ & $17$ & $30-80$\\
LOFAR & $52^\circ55$' N & $6^\circ52$' E & $5$ & $22.1^\circ$ & $49.3$ & $\approx 300$ & $0.2$ & $10-240$\\
Tunka-Rex & $51^\circ49$' N & $103^\circ04$' E & $675$ & $18.1^\circ$ & $60.4$ & $44$ & $1$ & $30-80$\\ 
South Pole & $90^\circ$ S & - & $2834$ & $17.9^\circ$ & $54.8$ & few & - & various\\ 
ARIANNA & $78^\circ45$' S & $165^\circ00$' E & $400$ & $4.7^\circ$ & $58.1$ & $8 \times 4$ & $5$ & $50-1000$\\ 
SKA-low & $26^\circ 41$' S & $116^\circ 38$' E & $370$ & $29.8^\circ$ & $55.5$ & $60,000$ & $1$ & $50-350$\\
\hline
\end{tabular}
\end{table}

\subsection{Modern radio experiments}

\subsubsection{LOPES}
LOPES (\underline{LO}FAR \underline{p}rototyp\underline{e} \underline{s}tation) was a digital radio interferometer at the site of the KASCADE-Grande experiment \cite{FalckeNature2005} in Karlsruhe, Germany. 
Triggered by the KASCADE and Grande particle detector arrays LOPES measured the radio signal in the nominal band of $40-80\,$MHz (effective band $43-74\,$MHz). 
A major advantage of LOPES was the possibility to study correlations between features of the radio signal and parameters of the particle cascade accurately reconstructed by KASCADE-Grande. 
A disadvantage was the radio-loud environment: many LOPES results were sufficient for proof-of-principle demonstrations, but limited in precision by the high background. 
Starting 2003 with 8 east-west-aligned prototype antennas of the LOFAR radio observatory, LOPES demonstrated that air showers can be detected via digital radio interferometry, despite the high radio background of the site. 
Several other proof-of-principle detections for the digital technique followed, e.g., the detection of distant and inclined events \cite{LOPESdistant2006, LOPESinclined2007}.
Also the influence of atmospheric electric fields on the radio signal was confirmed \cite{LOPESthunderstorm2007, LOPESthunderstorm2011}.
Then in 2005, LOPES was extended to 30 antennas, and half of them were rotated from east-west to north-south orientation at the end of 2006 in order to study the polarization of the radio signal \cite{LOPESpolarization_ARENA2008}. 

Most physics results have been obtained with this LOPES-30 setup operated until 2009, which provided statistics of about 500 high-quality events \cite{HuegeARENA_LOPESSummary2010, SchroederLOPESsummaryARENA2012}. 
In particular, LOPES achieved an angular resolution of better than $0.7^\circ$ for air showers \cite{NiglDirection2008, LOPESwavefront2014}, an energy precision and accuracy of at least $20\,\%$ \cite{2014ApelLOPES_MassComposition, LOPES_ECRS2012}, and experimentally demonstrated the sensitivity of the radio signal to the longitudinal shower development \cite{2012ApelLOPES_MTD}.
However, due to the high measurement uncertainties of LOPES, the achieved $X_\mathrm{max}$ precision was not competitive to other experiments \cite{LOPESwavefront2014, 2014ApelLOPES_MassComposition}.

Moreover, additional antennas have been deployed to test self-triggering on the radio signal \cite{Gemmeke2007}, and detection at low frequencies below $1\,$MHz \cite{LinkDiplomaThesis2009}. 
Although both had been demonstrated at sites with lower background earlier \cite{Jelly1965, RAugerSelfTrigger2012, HoughLowFreq1971, AllanLowFreq1972}, the LOPES efforts have been unsuccessful, probably because of the large background in the urban area of the research center. 
In the last stage of LOPES starting in 2010, antennas have been exchanged to so-called tripole antennas consisting of three crossed dipoles for better sensitivity to inclined showers \cite{LOPES_3D_2012}. 
However, a drastic increase of background was discovered in this period, making this technique useless at the LOPES site.
Finally, in 2013 LOPES was dismantled together with its KASCADE host experiment, but the data are planned to be made public and open access within KCDC, the \underline{K}ASCADE \underline{c}osmic-ray \underline{d}ata \underline{c}enter \cite{KCDC_ECRS2014}. 
The latest and probably final results of LOPES are available in reference \cite{LinkPhDThesis2016}.

\subsubsection{CODALEMA, EXTASIS, and NenuFAR}
CODALEMA (\underline{co}smic-ray \underline{d}etection \underline{a}rray with \underline{l}ogarithmic \underline{e}lectro-\underline{m}agnetic \underline{a}ntennas) is located at the Nan\c{c}ay Radio Observatory in France, and started air-shower detection in 2003, roughly at the same time as LOPES. 
The situation of CODALEMA is complementary to LOPES: CODALEMA is located in a radio-quiet site enabling accurate radio measurements, but the co-located particle detector array is simple and limited in accuracy. 
In its first stage, CODALEMA made use of the decametric array already existing at the site and mainly used for radio astronomy by adding a small particle-detector array triggering on air showers \cite{CODALEMA_2005, CODALEMA_FrequencySpectrum}. 
In later stages dedicated antenna stations with self-triggering capability were installed, and the triggered events were cross-checked with the coincident measurements of an array of 13 scintillators detecting air-shower particles. 
With these measurements CODALEMA provided evidence for the geomagnetic and Askaryan emission mechanisms \cite{CODALEMA_Geomagnetic, CODALEMA_Askaryan}.

Currently CODALEMA consists of a $1\,$km\textsuperscript{2}-large, sparse array of 57 autonomous antenna stations operating in the frequency band of $20-200\,$MHz, and a compact array of cabled antennas triggered by the scintillator array.
Moreover, a prototype low-frequency setup named EXTASIS (\underline{ext}inction of \underline{a}ir-\underline{s}hower \underline{i}nduced \underline{s}ignal) \cite{EXTASIS_ICRC2015} has been installed to search for radiation emitted by the termination of the air shower when entering the ground \cite{RevenuSuddenDeath_ICRC2013}.
In the future, NenuFAR (\underline{n}ew \underline{e}xtension in \underline{N}an\c{c}ay \underline{u}pgrading LO\underline{FAR}) will be installed at the CODALEMA site consisting of 1824 dual-polarized antennas operating in the band of $10-85\,$MHz \cite{NenuFAR2015}. 
This will increase the performance of LOFAR for astronomical observations and also for cosmic-ray detection, since a small fraction of the NenuFAR antennas will be equipped with buffers enabling air-shower measurements.

\subsubsection{LOFAR}
LOFAR (\underline{lo}w \underline{f}requency \underline{ar}ray) is a large-scale digital radio interferometer consisting of more than 40 stations spread over several European countries \cite{LOFARinstrument2013}. 
The central stations stand on the so-called superterp and are used for air-shower detection in parallel to astronomical observations \cite{SchellartLOFAR2013}. 
Triggered by the dedicated small particle-detector array LORA (\underline{LO}FAR \underline{R}adboud \underline{a}ir-shower array) with about $300\,$m diameter \cite{ThoudamLORA2014}, the radio data of those antennas used at this moment for astronomical observations are stored on disk for later air-shower analysis. 
This means that the available radio data vary from event to event, and for the typical air-shower event, data of several hundred radio antennas are available, which is more than for any other running antenna array. 
While the accuracy of the particle-detector array cannot compete with leading cosmic-ray experiments, the strong point of LOFAR is the very accurate and detailed radio measurement of air showers.
These measurements have been exploited to gain deeper insight in the radio emission, e.g., the dependence of the Askaryan emission on the zenith angle and on the distance to the shower axis \cite{SchellartLOFARpolarization2014}, and the change of the radio signal during thunderstorms \cite{SchellartLOFARthunderstorm2015}.
Moreover, LOFAR so far yields the most precise radio measurements of the depth of the shower maximum, $X_\mathrm{max}$, of about $20\,$g/cm\textsuperscript{2} \cite{BuitinkLOFAR_Xmax2014} (see section \ref{sec_XmaxReconstruction}), which has already been exploited to estimate the mass composition of cosmic rays in the energy range around $10^{17}\,$eV \cite{LOFAR_Nature2016}.

\begin{figure}[t]
  \centering
  \includegraphics[width=0.59\linewidth]{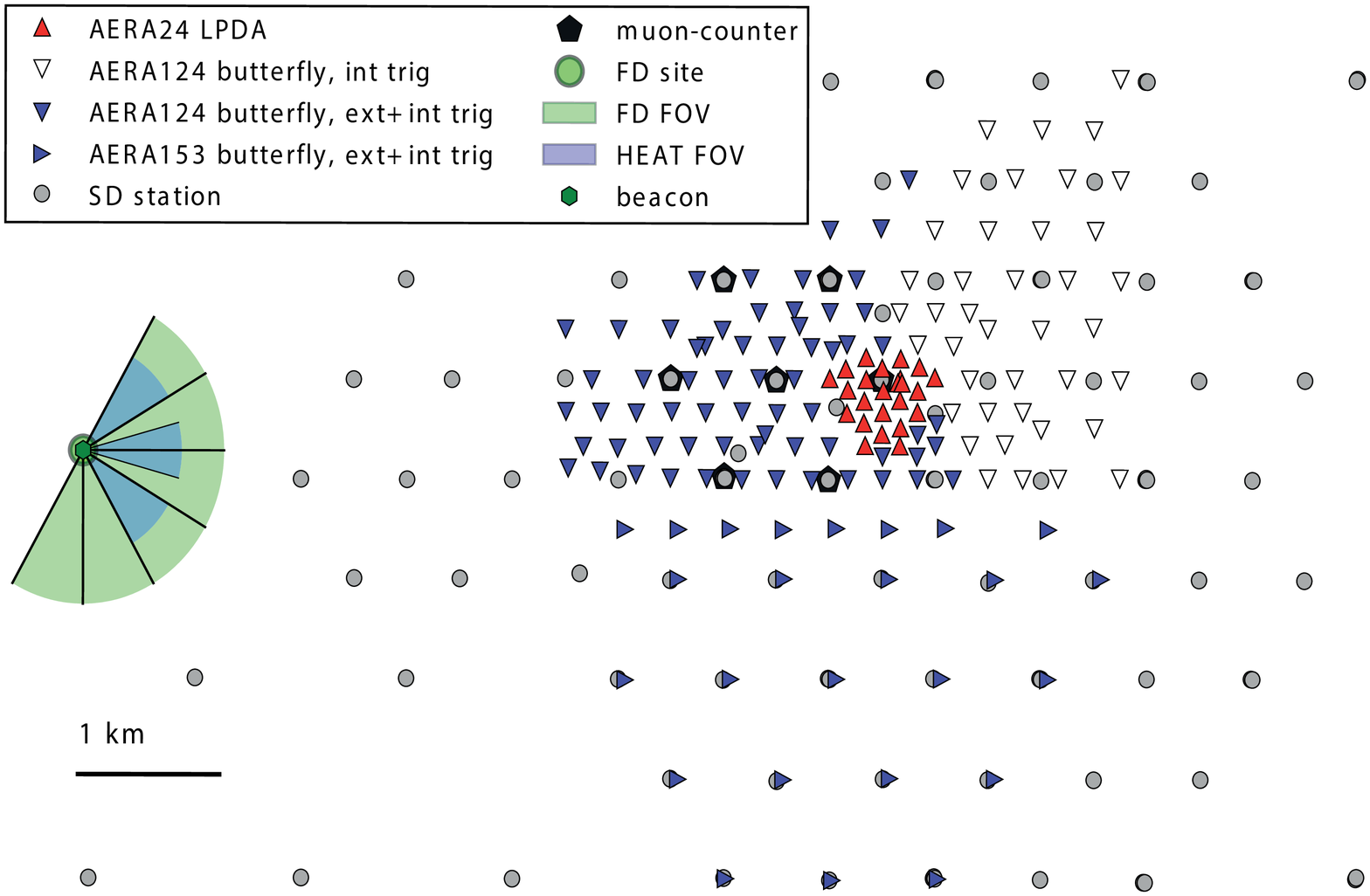}
  \hfill
  \includegraphics[width=0.4\linewidth]{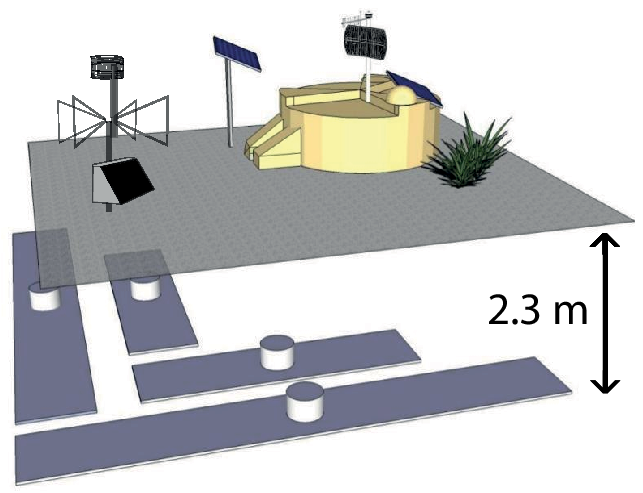}
  \caption{Map of AERA and sketch of the various detectors operated jointly in this area of the Pierre Auger Observatory (figures from Refs.~\cite{AERAoverviewICRC2015} and \cite{Holt_TAUP}, abbreviations: SD = surface detector, FD = fluorescence detector, HEAT = high-elevation Auger telescopes).}
  \label{fig_mapAERA}
\end{figure}

\subsubsection{AERA}
AERA (\underline{A}uger \underline{e}ngineering \underline{r}adio \underline{a}rray) is one of the enhancements of the Pierre Auger Observatory in Argentina \cite{AugerNIM2015}.
With more than 150 autonomous antenna stations operating in the frequency band of $30-80\,$MHz, AERA is the largest antenna array for air-shower detection covering an area of about $17\,$km\textsuperscript{2}.  
AERA is at the same location inside of the Pierre Auger Observatory as other enhancements, in particular AMIGA (\underline{A}uger \underline{m}uon and \underline{i}nfill \underline{g}round \underline{a}rray) which features water-Cherenkov detectors on the surface as well as underground scintillators for muon detection \cite{AMIGA_JINST2016}. 
Moreover, AERA currently  is the only radio experiment measuring in coincidence with fluorescence telescopes, which enables cross-calibration of energy and $X_\mathrm{max}$ measurements in the EeV energy range (see figure \ref{fig_mapAERA}). 

AERA aims at a variety of different technical and scientific goals, which makes it a pathfinder for future applications of the radio technique, but sometimes limits its performance for a specific goal. 
Technical goals of AERA are to demonstrate that the radio technique can be applied to large-scale arrays, and the evaluation of various engineering aspects, e.g., different types of antennas and designs of autonomous stations \cite{AERAantennaJINST2012}. 
Scientifically, the first goal of AERA was to better understand the physics of the radio emission, e.g., by measuring the polarization of the radio signal \cite{AugerAERApolarization2014}, which confirmed the picture presented in chapter \ref{sec_radioEmission}. 
Now AERA mainly aims at the development and improvement of reconstruction methods for shower parameters (direction, energy, $X_\mathrm{max}$) and their scientific application, in particular for the measurement of the absolute flux and mass composition of the primary cosmic rays as a function of energy. 
A related topic is the combined analysis of radio and muon measurements, which could improve the total accuracy for energy and mass composition \cite{Holt_TAUP}. 
For all these analyses the Pierre Auger Collaboration developed a proprietary software framework called \lq Offline\rq~\cite{Offline2007}, whose radio extension \cite{RadioOffline2011} is available to other collaborations, too, and is already used by Tunka-Rex.

\subsubsection{Tunka-Rex}
Tunka-Rex (Tunka \underline{r}adio \underline{ex}tension) \cite{TunkaRex_NIM2015} is the radio extension of the Tunka-133 \cite{Tunka133_NIM2014}, and Tunka-Grande \cite{TAIGA_2014} arrays at the same location in Siberia close to Lake Baikal, all dedicated to cosmic-ray research up to energies of a few EeV. 
The main goals of Tunka-Rex have been a cross-calibration of radio and air-Cherenkov measurements and the demonstration that an economic design of the antenna stations does not hamper the performance as cosmic-ray detector.
After accomplishing these goals, the experiment now aims at accurate measurements of the cosmic-ray mass composition in the energy range above $10^{17}$ by combining the information of Tunka-Rex, Tunka-133 and Tunka-Grande.
Tunka-133 is an array of photomultipliers for non-imaging air-Cherenkov detection which triggers Tunka-Rex during clear nights; Tunka-Grande consists of surface and underground scintillator stations triggering Tunka-Rex during daytime and bad weather. 
In the next years all arrays at the site will be integrated in TAIGA (\underline{T}unka \underline{a}dvanced \underline{i}nstrument for cosmic rays and \underline{g}amma \underline{a}stronomy), which features additional imaging and non-imaging air-Cherenkov detectors \cite{TAIGA_2014}.
Tunka-Rex consists of 38 dual-polarized antenna stations on $1\,$km\textsuperscript{2} with a crown of 6 outer stations extending the total area to $3\,$km\textsuperscript{2}, and is planned to be extended by another 19 antenna stations to densify the inner area.
Likely it is the most cost-effective of all radio arrays, since it operates completely in slave mode attached to an already existing data acquisition, and uses economic SALLA antennas \cite{AERAantennaJINST2012} on wooden poles. 
Simulation studies as well as the comparison of Tunka-Rex with Tunka-133 measurements revealed that this cheap design does not affect the accuracy for the shower energy.
Nevertheless, the use of SALLA antennas slightly increases the detection threshold by about $40\,\%$ compared to other antenna types, but this is partly compensated by the high geomagnetic field at the Tunka site.
After observing a direct correlation of the $X_\mathrm{max}$ values reconstructed from radio and air-Cherenkov measurements \cite{TunkaRex_XmaxJCAP2016}, which is one of the classical optical methods, the next goal of Tunka-Rex is to study the cosmic-ray mass composition in the energy range above $10^{17}\,$eV.

\subsubsection{Yakutsk}
The longest-running air-shower detector is located close to Yakutsk \cite{YakutskWebPage2016, YakutskReview2016}. 
In addition to particle and air-Cherenkov detectors, the Yakutsk experiment performed low-frequency radio measurements at $1.9\,$MHz in the years 1972 - 1973, and operated a few antennas in the frequency band around $32\,$MHz in the years 1986 - 1989 and again since 2008 \cite{YakutskICRC2013}. 
Generally the results of Yakutsk confirm the results of other experiment, with one exception: Only at Yakutsk a flattening of the average lateral distribution of the radio signal was seen at distances of several $100\,$m from the shower axis \cite{YakutskICRC2015}. 
Whether this is due to an instrumental or measurement effect, due to background, or due to an additional emission mechanism is not yet known, but so far this flattening of the radio footprint at large distances has not been reported by other large-scale experiments.

\begin{figure}[t]
  \centering
  \includegraphics[width=0.605\linewidth]{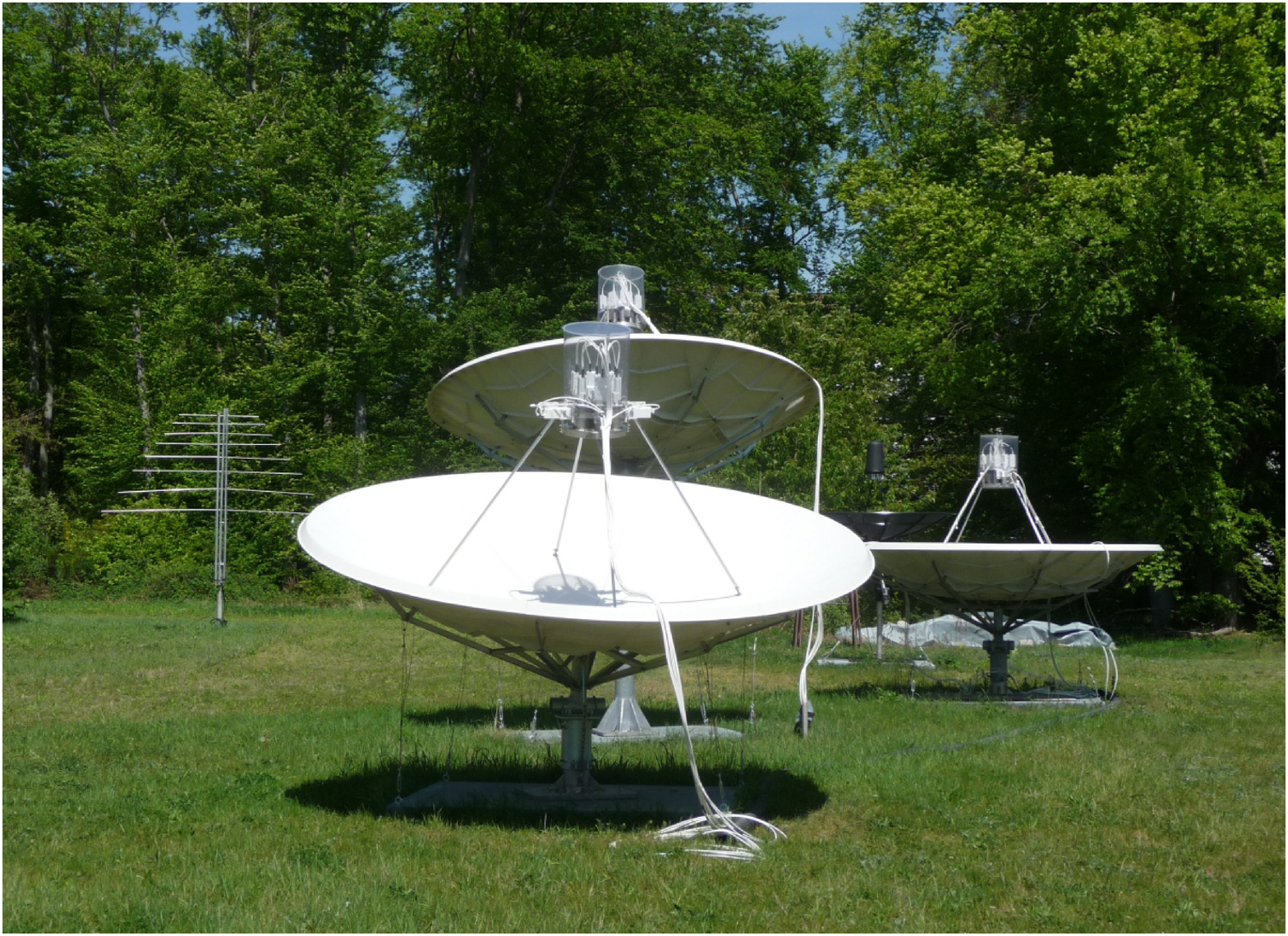}
  \hfill
  \includegraphics[width=0.385\linewidth]{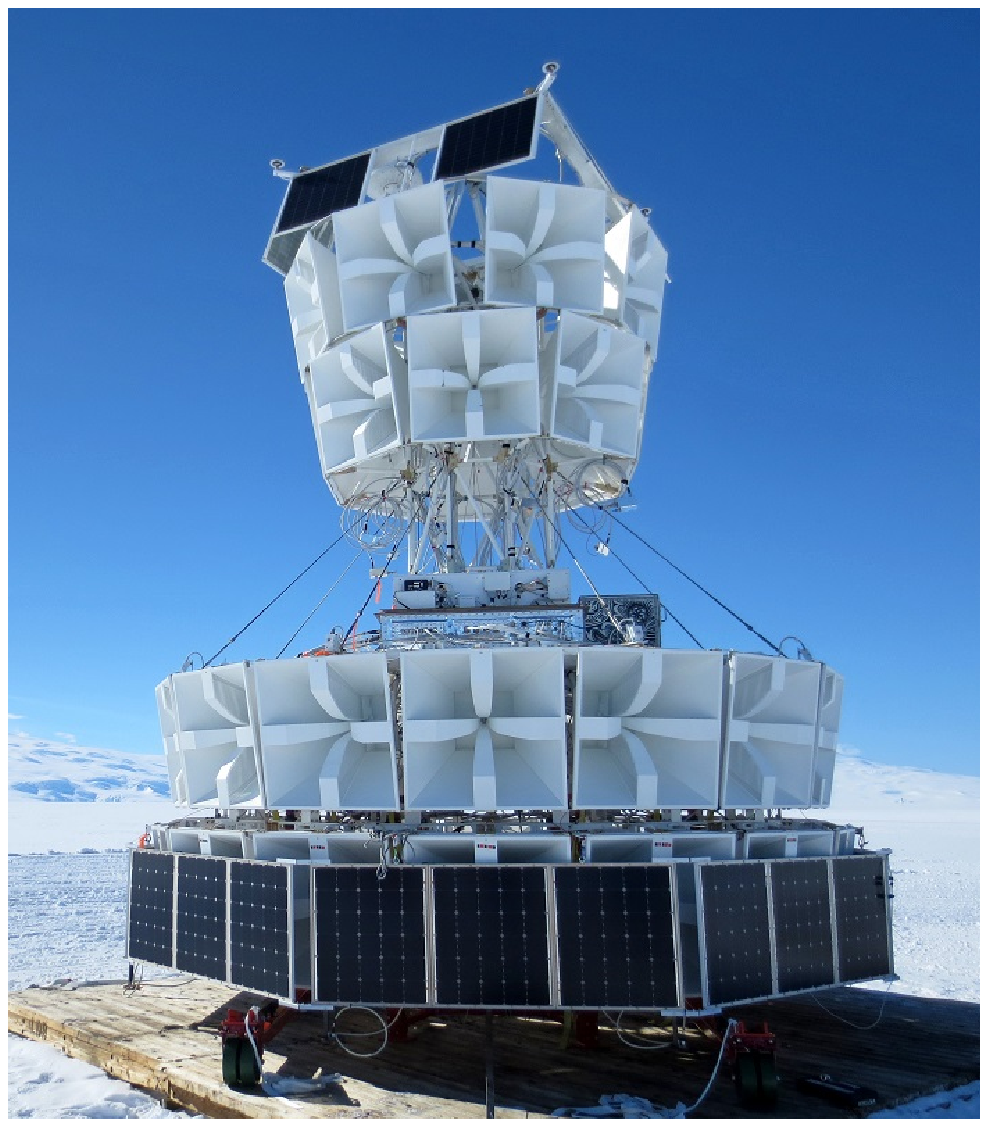}
  \caption{Left: The CROME experiment at the Karlsruhe Institute of Technology (photograph courtesy of R. Smida). Right: ANITA payload before its third flight (photograph courtesy of C. Miki).}
  \label{fig_CROMEandANITA}
\end{figure}

\subsubsection{CROME, EASIER, and other GHz experiments}
CROME (\underline{c}osmic-\underline{r}ay \underline{o}bservation via \underline{m}icrowave \underline{e}mission) was located in the center of the KASCADE-Grande particle detector array, just a few $100\,$m away from LOPES. 
It featured a variety of GHz antennas (see figure \ref{fig_CROMEandANITA}) and detected several air showers in the C band of $3.4-4.2\,$ GHz \cite{CROME_PRL2014}. 
The original goal of CROME was the search of isotropic molecular bremsstrahlung, but no evidence was found.
Instead, the measured GHz signal can be explained by the radio-emission mechanisms described above, i.e., geomagnetic and Askaryan emission extending to higher frequencies close to the Cherenkov angle.

Also EASIER (\underline{e}xtensive \underline{a}ir-\underline{s}hower \underline{i}dentification using \underline{e}lectron \underline{r}adiometer), an array consisting of 61 horn antennas each attached to one surface detector of the Pierre Auger Observatory, detected a few events in the C band \cite{AugerGHz2015}. 
Two other experiments at the Pierre Auger Observatory tried to observe air showers from the side in the same way that fluorescence telescopes do: AMBER (\underline{a}ir-shower \underline{m}icrowave \underline{b}remsstrahlung \underline{e}xperimental \underline{r}adiometer) and MIDAS (\underline{mi}crowave \underline{d}etection of \underline{a}ir \underline{s}howers).
However, in contrast to CROME and EASIER neither AMBER nor MIDAS did detect any air showers \cite{AugerGHz2015}.
This indicates that the GHz emission is mostly beamed in the forward direction and not isotropic, as expected for geomagnetic and Askaryan emission.
Thus, the strength of isotropic molecular bremsstrahlung likely is much weaker than originally predicted in Ref.~\cite{GorhamMolecularBremsstrahlung2008}, which would be in agreement with recent calculations predicting that the signal in air-showers would be too low for serving as an easy and economic detection technique \cite{Samarai2016}. 

\subsubsection{TREND and GRAND}
TREND (\underline{T}ianshan \underline{r}adio \underline{e}xperiment for \underline{n}eutrino \underline{d}etection) is located in a radio-quiet Chinese valley in the XinJiang province, at the same location as the 21CMA radio observatory. 
Starting in 2009 with a 6-antenna prototype, it successfully demonstrated that self-triggering on cosmic-ray air showers is possible in these conditions \cite{TREND_Astropp2011}, and was extended to 50 antennas soon afterwards \cite{TREND_ICGAC2011}. 
TREND is a pathfinder for GRAND (\underline{g}iant \underline{r}adio \underline{a}rray for \underline{n}eutrino \underline{d}etection) which is also planned in Tianshan \cite{GRAND_ICRC2015, GRANDproto_ICRC2015}. 
GRAND will be the largest cosmic-ray detector on Earth consisting of about $100,000$ antennas distributed over a huge area of $200,000\,$km\textsuperscript{2}.
Its main scientific goal will be the detection of neutrinos interacting in the surrounding mountains and initiating air showers emitting radio pulses. 
Many design decisions still have to be made. 
One of the key technological questions is how to distinguish background pulses and cosmic-ray air showers from neutrino-initiated showers. 
In principle this should be possible from signal properties, but the efficiency and purity of neutrino discrimination has to be demonstrated in practice. 
Currently GRANDproto, a small array of 35 antennas, is under construction in order to determine quantitatively the efficiency of autonomous radio detection for air showers, and the proof-of-concept for background discrimination of the technique.
As a next step an engineering array of about $1,000\,$km\textsuperscript{2} is considered, which already would be the world's largest antenna array.

\begin{figure}[t]
  \centering
  \includegraphics[width=0.99\linewidth]{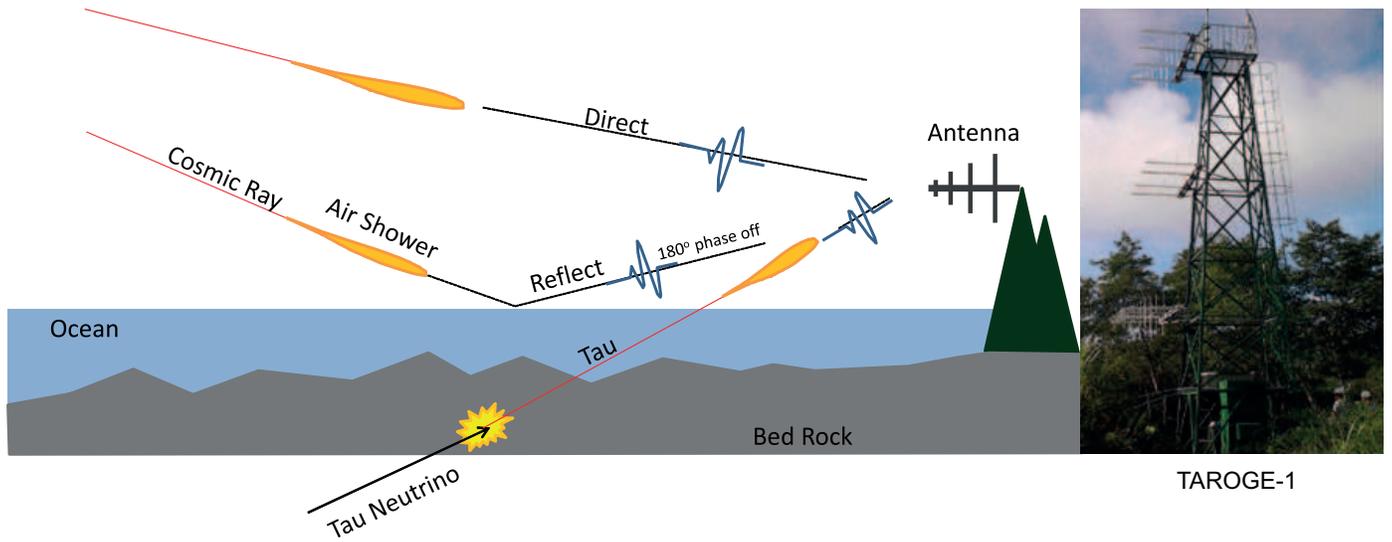}
  %\hfill
  \caption{Concept and photo of the TAROGE experiment, which is similar to the concept of ANITA with the difference that the TAROGE antennas are on a summit and the ANITA antennas are hanging under a balloon (figure from Ref.~\cite{TAROGE_ICRC2015}).}
  \label{fig_TAROGE}
\end{figure}

\subsubsection{ANITA and planned successors}
ANITA (\underline{An}tarctic \underline{i}mpulsive \underline{t}ransient \underline{a}ntenna) is a balloon-borne experiment flying over Antarctica. 
After an initial flight by its prototype ANITA-lite in 2004 \cite{ANITAlite_PRL_2006}, ANITA had three flights since 2006 with a fourth flight planned for the end of 2016. 
The payload of ANITA (see figure \ref{fig_CROMEandANITA}) featured between 32 and 48 (depending on the flight) horn antennas measuring radio emission at high frequencies of $200 - 1200\,$MHz. 
At this frequency range detection is expected only for showers with favorable geometry: The Cherenkov cone of the shower must hit the antenna, which limits the detection to near-horizontal air showers and neutrino-induced cascades in the overflown ice. 
The main science goal of ANITA is the search for ultra-high-energy neutrinos, and the non-detection led to competitive upper limits at energies $E_\nu > 10^{19}\,$eV \cite{ANITA_neutrino_PRL_2009, ANITA_neutrino_PRD_2012}. 
Nevertheless, ANITA successfully detected several dozens of cosmic-ray induced air showers at energies around $10^{19}\,$eV \cite{ANITA_CR_PRL_2010, ANITA_CR_2016}.
For most of them ANITA did not measure the radio signal directly, but the signal after reflection on the ice \cite{MunizReflectedSim2015}.

ANITA pioneered the balloon-borne technique applying several techniques also used in ground-based arrays, e.g., individual antennas are combined by interferometric beamforming (as for LOPES) and polarization characteristics are used to discriminate air showers from background (as for AERA).
Future projects like EVA (\underline{e}xa\underline{v}olt \underline{a}ntenna) \cite{GorhamEVA2011}, or SWORD (\underline{s}ynoptic \underline{w}ideband \underline{o}rbiting \underline{r}adio \underline{d}etector) \cite{SWORDarxiv2013} plan to further develop this concept:
EVA will be a large antenna array integrated directly in a high-altitude long-duration balloon, i.e., the antennas are part of the balloon itself and not hanging on a balloon as usual balloon-borne experiments do.
SWORD will be a satellite mission observing cosmic-ray and neutrino-induced showers from space. 
This technique was pioneered in the early 2000's by FORTE (\underline{f}ast \underline{o}n-orbit \underline{r}ecording of \underline{t}ransient \underline{e}vents), a satellite built for other purposes, but also used for the search of neutrino-induced showers in ice \cite{FORTE_2004}. 

Satellite missions for cosmic rays bring the potential of large statistics at the highest energies, since even small antenna arrays in space could have yearly exposures for air showers larger than that of the Pierre Auger Observatory \cite{MotlochSatelliteExposure2014}.
While at least a rough energy reconstruction of showers has already been demonstrated by ANITA for the balloon concept \cite{ANITA_CR_2016}, the situation might be more difficult for satellites, since the radio signal from the air showers will be distorted by the ionosphere, which could further decrease the accuracy for the energy and arrival direction.  
Moreover, methods for reconstruction of the primary particle type still have to be developed. 
Nevertheless, in case that more accurate ground based experiments would consistently show that the cosmic-ray composition at the highest energies is pure (e.g., only iron nuclei as expected by some theoretical models \cite{AugerMatteoICRC2015}), then composition sensitivity would be unnecessary and only energy and direction information count. 
Even when optimizing for cosmic-ray detection, neutrino searches can be continued in parallel, where neutrinos would be distinguishable from cosmic rays by the arrival direction and polarization characteristics.

\subsubsection{TAROGE}
TAROGE (\underline{T}aiwan \underline{a}stroparticle \underline{r}adiowave \underline{o}bservatory for \underline{g}eo-synchrotron \underline{e}missions) is a new experiment consisting of two sites in Taiwan \cite{TAROGE_ICRC2015}.
TAROGE aims at the detection of near horizontal showers, similar to the ANITA approach, except that the receiver is on top of a mountain instead of hanging on a balloon (see figure \ref{fig_TAROGE}).  
Depending on the observation angle, the radio signal can be measured directly or after reflection on the ocean.
In principle also neutrino-initiated showers should be detectable provided sufficient discrimination against background pulses, e.g., caused by ships. 
The first site TAROGE-1 is on a $1000\,$m high mountain close to the coast and features twelve log-periodic dipole antennas (cf.~section~\ref{sec_antennaTypes} for a comparison of antenna types) operating in the frequency band of $110-300\,$MHz.
The second site TAROGE-2 is planned on an even higher mountain. 
The clear advantage of TAROGE is its large area overlooked with only few antennas. 
A potential disadvantage is the missing knowledge on how exactly the radio signal is affected by the reflection on the ocean, since water waves could have structures of similar size as the radio wavelengths, which may cause interference effects limiting the measurement accuracy. 
At least for ANITA the surface roughness is one of the major systematic uncertainties when calculating the exposure for air showers detected by their radio emission reflected on the Antarctic ice \cite{ANITA_CR_2016}.
If TAROGE successfully provides a proof-of-principle, its technique could be an interesting approach for the highest energies, because it offers large exposure for relatively low cost, though several mountain stations would have to be combined to compete with the exposure of possible satellite experiments \cite{MotlochSatelliteExposure2014}.

\subsubsection{Air-shower arrays at the South Pole}
For several years the possibility of a radio extension of IceTop, the surface detector of the IceCube neutrino observatory has been under discussion, e.g., in the frame of the RASTA (\underline{r}adio \underline{a}ir-\underline{s}hower \underline{t}est \underline{a}rray) project \cite{RASTA_Boeser2012}. 
Prototype setups at the South Pole proved that there is almost no human-made background, i.e., only thermal and Galactic noise limit the measurements.
Currently ARA (\underline{A}skaryan \underline{r}adio \underline{a}rray) features a few antennas at the surface near the IceTop array \cite{ARA_2012}, in addition to three strings of antennas in the ice whose main focus is the search of neutrino-induced showers. 
Furthermore, the concept of a large-scale radio array is discussed again in the context of future surface extensions for the IceCube observatory \cite{IceTopNIM2013}.

\subsubsection{The SKA}
The SKA (\underline{s}quare \underline{k}ilometer \underline{a}rray) will be a new radio observatory built for a variety of astrophysical science goals.
The low-frequency core of the SKA in Australia will be the densest antenna array in the world featuring about $60,000$ antennas on less than $1\,$km\textsuperscript{2} \cite{SKA_ICRC2015}. 
Provided some technical design modifications in the data-acquisition system and the deployment of a simple particle-detector array for triggering, SKA can be used for the measurement of air showers in addition to astronomical observations.
This enables extremely detailed measurements of the radio signal, and could beat the fluorescence and air-Cherenkov techniques in total accuracy for the cosmic-ray mass composition and energy.
Moreover, the SKA will make a major step in sensitivity for the radio signal of showers initiated in the lunar regolith expecting the first detection of ultra-high-energy cosmic rays by this method, and at the same time significantly improving the limits on ultra-high-energy neutrinos \cite{SKAlunar_ICRC2015}.

\subsubsection{Radio detection in ice: ARA, ARIANNA, and others}
The radio technique in dense media is most advanced in ice, although other media have been investigated as well, in particular salt in the frame of the SalSA (\underline{sal}t \underline{s}ensor \underline{a}rray) project \cite{GorhamSalt2002, SALSA_2009}. 
However, because of the larger attenuation length of radio waves in ice, and because of the large available ice volumes at Antarctica and on Greenland, ice seems to be the medium of choice for future large-scale detectors of several $10$ or $100\,$km\textsuperscript{3} \cite{RadioInIceNEUTEL2015}.
RICE (\underline{r}adio \underline{i}ce \underline{C}herenkov \underline{e}xperiment) \cite{RICE_2006}, and AURA (\underline{A}ntarctic \underline{u}nder-ice \underline{r}adio \underline{a}rray) \cite{Landsman09} were first prototype experiments in the ice at the South Pole, but have been too small for successful neutrino detection.

Nowadays, a next generation of prototype arrays approaches a sensitivity range where first detection of neutrinos can be expected:
ARA (\underline{A}skaryan \underline{r}adio \underline{a}rray) \cite{ARA_2012, ARA_2015} close to IceCube at South Pole, ARIANNA (\underline{A}ntarctic \underline{R}oss \underline{I}ce \underline{S}helf \underline{an}tenna \underline{n}eutrino \underline{a}rray) \cite{ARIANNA_2015} on the Ross Ice Shelf at the Antarctic coast, and GNO (\underline{G}reenland \underline{n}eutrino \underline{o}bservatory) in Greenland \cite{GreenlandInIce_2016}.
ARA and ARIANNA are already under construction aiming at a proof-of-principle that the radio technique in ice indeed can be applied for high-energy neutrino astronomy.

ARA is planned to have 37 stations in its final configuration with a frequency band of $150-850\,$MHz, and with $1\,$km spacing between stations, since the attenuation length of radio waves in ice is of the order of $1\,$km \cite{IceAttenuationRICE2005}. 
At the moment 3 stations are running, and two more will be deployed next year. 
Each station features 20 antennas of which 16 are in the ice and 4 at the surface for air-shower and background measurements.
The vicinity to IceCube might provide a chance to cross-calibrate the radio and the optical techniques for neutrinos, when both the ARA and the IceCube arrays will be extended such that they have a significant overlap.
 
ARIANNA currently consists of 8 stations with 4 antennas each searching for in-ice showers in the frequency band of $50-1000\,$MHz.
Additionally, it has already measured a few cosmic-ray air showers in the band of $50-500\,$MHz \cite{NellesARENA2016}. 
In its final stage ARIANNA is proposed to consists of $36 \times 36$ stations also with $1\,$km spacing between stations.
Then ARIANNA will be sensitive enough to see whether the astrophysical neutrino flux measured by IceCube until a few PeV extends until EeV energies, or whether there is a cut-off at any energy in between.

\subsubsection{Radio observation of the Moon}
Several projects have used radio telescopes built for astronomy to search for showers in the lunar regolith like GLUE (\underline{G}oldstone \underline{l}unar \underline{u}ltra-high-energy neutrino \underline{e}xperiment) \cite{GLUE_2004}, RESUN (\underline{r}adio \underline{E}VLA [\underline{e}xpanded \underline{v}ery \underline{l}arge \underline{a}rray] \underline{s}earch for \underline{u}ltra-high-energy \underline{n}eutrinos) \cite{RESUN_2010}, NuMoon \cite{NuMoon2010}, or LUNASKA (\underline{l}unar \underline{u}ltra-high energy \underline{n}eutrino \underline{a}strophysics with the \underline{SKA}) \cite{LUNASKA_2014}.
So far no neutrino or cosmic-ray events were detected by any of these experiments. 
Given their sensitivity this is consistent with expectations from traditional air-shower arrays and has been used to set limits for the neutrino and cosmic-ray flux at ZeV energies \cite{BrayReview2016}.

Nevertheless, the search for radio pulses from the Moon will be continued with more sensitive observatories.
The observation of the lunar regolith is one of the key science projects of LOFAR \cite{LOFAR_NuMoon_ARENA2012}, but even for LOFAR a detection of neutrinos is expected only in exotic scenarios \cite{BrayReview2016}.
In the future the SKA (see above) and LORD (\underline{l}unar \underline{o}rbital \underline{r}adio \underline{d}etector) \cite{GusevLORD2014}, a radio satellite orbiting the Moon, will have a reasonable chance for detection of ultra-high-energy cosmic rays in addition to neutrinos.

\subsection{Related experimental activities}

\subsubsection{Acoustic detection}
Acoustic detection techniques are similar to radio techniques.
They make use of sound pulses created by the showers when their energy deposit is converted into heat, which leads to a rapid, local expansion of the medium. 
This technique is expected to work not only in solid media like salt and ice, but also in liquids like water. 
Prototype experiments have been built at several locations in sea water \cite{Graf_AcousticOverviewSea_ARENA2012}, e.g., AMADEUS (\underline{A}NTARES [\underline{a}stronomy with a \underline{n}eutrino \underline{t}elescope and \underline{a}byss environmental \underline{res}earch] \underline{m}odules for the \underline{a}coustic \underline{de}tection \underline{u}nder the \underline{s}ea) \cite{AMADEUS_2011}, and ONDE (\underline{o}cean \underline{n}oise \underline{d}etection \underline{e}xperiment) \cite{ONDE_2009} in the Mediterranean Sea, as well as SAUND (\underline{s}tudy of \underline{a}coustic \underline{u}ltra-high-energy \underline{n}eutrino \underline{d}etection) \cite{SAUND_ARENA2010}, and ACoRNE (\underline{a}coustic \underline{co}smic \underline{r}ay \underline{n}eutrino \underline{e}xperiment) \cite{ACoRNE_2008} in the Atlantic Ocean, in fresh water in Lake Baikal \cite{AcousticLakeBaikal_ARENA2012}, and in ice \cite{Karg_AcousticOverviewIce_ARENA2012}, e.g., SPATS (\underline{S}outh \underline{P}ole \underline{a}coustic \underline{t}est \underline{s}etup) as an extension of the IceCube neutrino observatory \cite{SPATS_2012}. 
At the moment it seems that relatively small resources are invested in the further development of the acoustic technique.
Nevertheless, research is going on at Lake Baikal and in the Mediterranean Sea, and acoustic sensors are additionally used for position calibration of optical detectors \cite{AcousticPositioning_2013}, and for acoustic observation in marine biology \cite{AcousticMarineBiology_2012}.
As shortly discussed in section \ref{sec_DenseMediaCascades} and in chapter \ref{sec_neutrinoReco}, the future of the acoustic technique might lie either in under-water detectors, since there is no alternative technique for the EeV energy range in water, or in combined radio-acoustic arrays in ice, since the combination of both techniques could increase the measurement accuracy for the properties of neutrinos initiating cascades in the ice.

\subsubsection{Radar detection}

Even before the first successful detection of air showers at Jodrell bank in 1965 \cite{Jelly1965}, it was predicted \cite{FirstRadarPerdiction1941} and tested experimentally if air showers can be detected by their radar reflection.
However, not air showers were detected, but instead the radar reflection of meteor trails had been discovered \cite{RadarMeasurementOfMeteors1947}. 
Nevertheless, the idea was followed over the years \cite{GrohamRadarCaluclations2001}, but until today the strength of the radio signal reflected by air showers is still not clear. 
Recent calculations predict much weaker signals than thought earlier \cite{StasielakRadarCalculations2016}, which is consistent with the non-detection of any air shower so far.
Thus, even if the radar technique works at all for air showers, it will require expensive high-power radar systems.
Still, experimental search for radar reflection by air showers goes on, for example at the Telescope Array in Utah \cite{TARA_2015}.
Moreover, it is under investigation whether the radar reflection might be stronger from the more compact showers in dense media, like salt \cite{RADARinSalt2012} or ice \cite{RADARinIce2015}, which could provide a suitable technique for neutrinos in the energy range above $10\,$PeV.

\subsubsection{Accelerator experiments}
Several accelerator experiments have been preformed to study the radio emission by particle cascades under laboratory conditions. 
While the advantage is that the conditions can be controlled, the disadvantage is that it is not always clear how to transfer the results to the situation of natural showers. 
Furthermore, additional sources of radiation present only in the accelerator experiments have to be understood, e.g., transition radiation when the accelerator beam enters the target.
For these reasons accelerator experiments have large systematic uncertainties in their quantitative interpretation for the situation of showers in air or in extended dense media, such as the Antarctic ice. 
Nevertheless, these experiments provide an independent check that the emission mechanisms are qualitatively understood.
Several experiments at SLAC (\underline{S}tanford \underline{l}inear \underline{a}ccelerator \underline{c}enter) have measured the Askaryan effect in a variety of different media \cite{Saltzberg_SLAC_Askaryan2001, ANITA_SLAC_PRL_2007, SLAC_T510_PRL2016}, and the first laboratory measurement of the magnetic emission mechanism was performed there \cite{SLAC_T510_PRL2016}.
Moreover, Askaryan emission in ice has recently been measured at the electron accelerator of the Telescope Array \cite{TA_ELS_ICE_ICRC2015}.

Accelerator experiments are also used to search for additional emission mechanisms, and several experiments have been performed to study isotropic molecular bremsstrahlung.
First experiments at SLAC and AWA (\underline{A}rgonne \underline{w}akefield \underline{a}ccelerator) had measured a strong microwave signal \cite{GorhamMolecularBremsstrahlung2008}.
However, these signal might not have been pure isotropic molecular bremsstrahlung.
A later experiment called AMY (\underline{a}ir \underline{m}icrowave \underline{y}ield) \cite{AMY_UHECR2012} is consistent with the existence of weak molecular bremsstrahlung, but final results have not yet been published.
Results by MAYBE (\underline{m}icrowave \underline{a}ir \underline{y}ield \underline{b}eam \underline{e}xperiment) \cite{MAYBE_UHECR2012} indicate that isotropic molecular bremsstrahlung does exist at GHz frequencies, but is much weaker than derived from the original experiments.
Furthermore,its intensity only scales linearly with the beam energy as expected for incoherent emission mechanisms, and not quadratically as expected for coherent emission. 
This weaker signal is consistent with the non-observation of isotropic radiation by various air-shower experiments. 
Hence, independent of the exact strength and nature of molecular bremsstrahlung, its detection is not as promising as alternative technique for air showers as initially thought.
Consequently, current and planned radio experiments rely on the proven Askaryan and geomagnetic emission mechanisms for the detection of neutrinos and cosmic rays.

\subsection{Comparison of detector concepts for air showers}
In order to learn from the experiences of different experiments, it is worthwhile to evaluate the common and different features and concepts of the experiments mentioned above. 
While some design decisions turned out to be generally (un)recommendable, others depend on the physics goal of the experiment.

\subsubsection{General detector layout}
Most radio experiments have not been designed freely in order to optimize the array layout for specific scientific goals, but were either built as extensions to existing cosmic-ray experiments or, like LOFAR and the SKA in the future, simply are special use cases of radio observatories designed for astronomy. 
Thus, only limited experience is available on how to build an optimal antenna array for air-showers, but still some lessons can be learned from differences in existing experiments.

The first lesson is that the cost of a radio array is typically not dominated by the antennas and their electronics, but by the infrastructure of the data acquisition and the experimental site. 
So far, AERA is the most autonomous radio array. 
Although exploiting the general infrastructure of the Pierre Auger Observatory in Argentina, it features its own data acquisition and communication systems, and radio stations operating remotely on solar power \cite{SchroederAERA_PISA2015}. 
The clear advantage is that within little limitations, the layout of the array could be chosen independently of the co-located surface-detector array. 
Nevertheless, this freedom has not been exploited to optimize AERA for a single goal, but instead a graded design with different station spacings from $175\,$m to $750\,$m has been selected, in order to test possible different configurations. 
The larger spacing might be fully sufficient for inclined showers \cite{AERAinclinded_ARENA2014}, and would enable covering large areas at reasonable costs. 
A definite conclusion based on experimental results is expected in the next years.

Tunka-Rex features a similar detector spacing of $200\,$m, but in contrast to AERA with an order of magnitude lower cost per station, since it is completely attached to the existing data acquisition of the Tunka-133 and Tunka-Grande arrays, and operates in slave mode only \cite{TunkaRex_NIM2015}. 
Signal and power transfer is via simple coaxial cables. 
At least for the measurement of the shower energy, this economic design seems to perform equally well as the AERA design \cite{TunkaRex_XmaxJCAP2016, AERAenergyPRD}.
LOPES and CODALEMA are in between these extremes: LOPES was attached to the KASCADE infrastructure, but featured its own data acquisition, and CODALEMA has a cabled setup of autonomous stations.
The disadvantage of any cabled design is the scalability: 
Already at Tunka-Rex with distances of about $20\,$m between antennas and data-acquisition electronics the cost for the signal cable is of the same order as for the antennas themselves, and the manpower for burying cables is significant. 
Consequently, cabled radio extensions seem to be the optimal solution for arrays in the order of a square kilometer or less, but huge arrays of many $100$ or even $1000\,$km\textsuperscript{2}, required to study the highest-energy cosmic rays, will be too expensive when equipped with cables.

The array layout also influences the station design: 
Antenna stations of radio extensions, like Tunka-Rex or LOPES, basically consist only of the antennas themselves and the outgoing cables, while for AERA the station is a complex unit consisting of a physics antenna, a communication antenna, a solar panel with battery, local electronics, and in some cases even a small scintillator to provide an additional trigger on air-shower particles \cite{SchroederAERA_PISA2015}. 
Prototypes for complex, but flexible stations including several different detector systems are also studied in the TAXI project (\underline{t}ransportable \underline{a}rray for e\underline{x}tremely large area \underline{i}nstrumentation studies) \cite{Karg_TAXI_ARENA2014}.
This might be a solution for next-generation arrays, since maximum accuracy for the primary particle type requires the combination of several detector systems anyway \cite{AugerUpgrade_CRIS2015, AugerUpgradePRD2016}, and the combination of particle and radio detectors in the same station enables triggering the whole station including the antenna on the particle signal.

Given a fixed area and number of antenna stations, an open question is the optimal distribution of the antennas within the area, i.e., if a regular or irregular design should be preferred.
No systematic study of advantages or disadvantages exists, yet, only rough arguments: 
An irregular design as featured by LOFAR and the SKA is optimal for interferometric analyses since regularities would enhance instrumental artifacts like side lobes. 
LOPES, but not yet LOFAR, has demonstrated that interferometric beamforming can lower the detection threshold for air-showers in noisy environments \cite{FalckeNature2005}.
A regular design should facilitate the calculation of the exposure and efficiency, and thus reduce systematic uncertainties due to any kind of selection biases.
Consequently, my personal assessment is that an irregular layout will lower the detection threshold, and a regular one will reduce systematic uncertainties in the analyses of high-energy events well above the threshold. 
Whenever important for the design of any new array, this assumption should be tested with simulation studies.

\begin{figure}
  \centering
  \begin{subfigure}[b]{0.552\textwidth}
        \includegraphics[width=\textwidth]{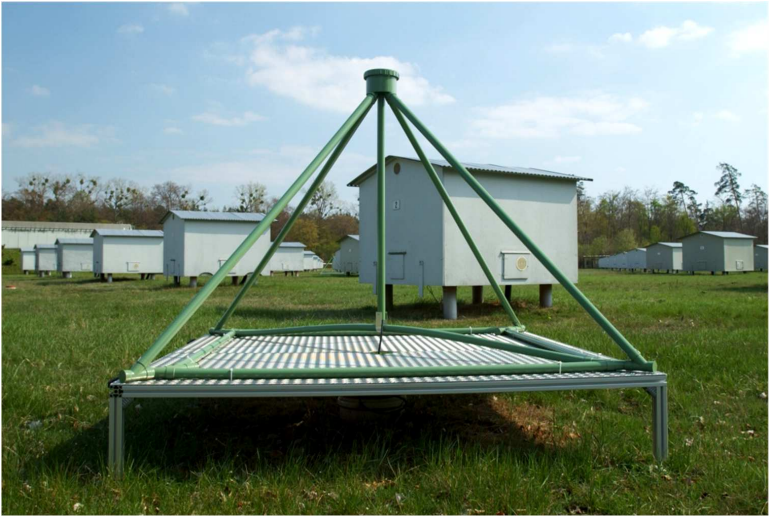}
        \caption{Inverted v-shape dipole at LOPES}
  \end{subfigure}
  \hfill
  \begin{subfigure}[b]{0.438\textwidth}
       \includegraphics[width=\textwidth]{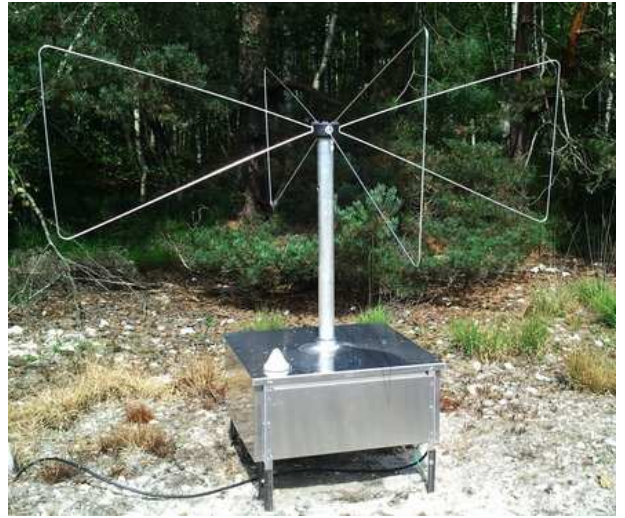}
        \caption{Butterfly at CODALEMA}
  \end{subfigure}
  \par\medskip
  \begin{subfigure}[b]{0.4465\textwidth}
        \includegraphics[width=\textwidth]{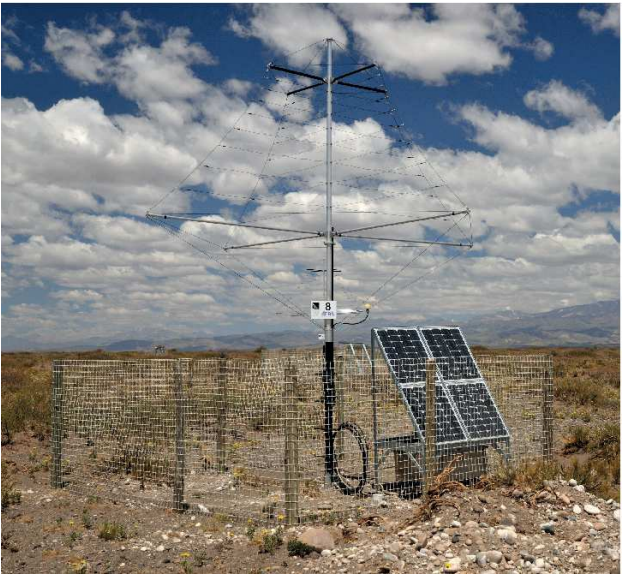}
        \caption{LPDA at AERA}
  \end{subfigure}
  \hfill
  \begin{subfigure}[b]{0.545\textwidth}
       \includegraphics[width=\textwidth]{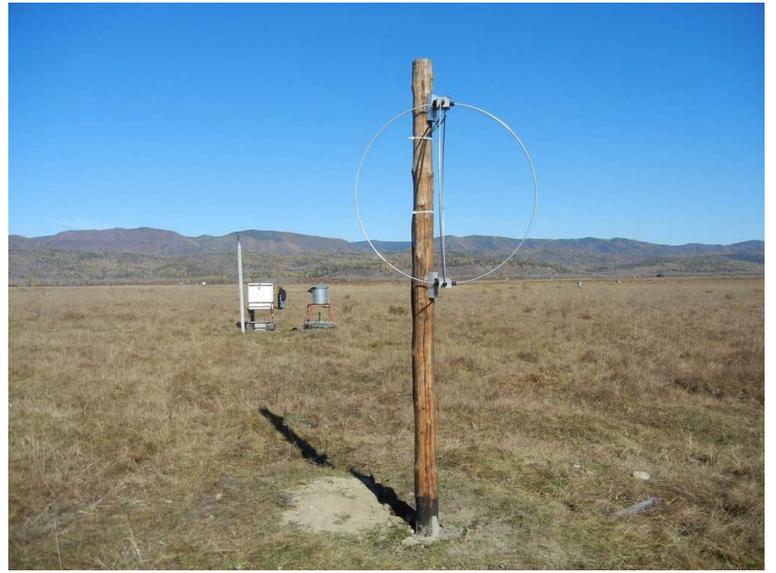}
        \caption{SALLA at Tunka-Rex}
  \end{subfigure}
  \caption{Various antenna types used by different radio arrays for cosmic-ray air showers (photographs (a) to (d) from Refs.~\cite{SchroederLOPES_PISA2009, AERAantennaJINST2012, SchroederAERA_PISA2015, KostuninTunkaRex_RICAP2013}).}
  \label{fig_antennaTypes}
\end{figure}

\subsubsection{Antenna types}
\label{sec_antennaTypes}
Various antenna types have been used by different experiments (see figure \ref{fig_antennaTypes}), and there is not a single best choice. 
Instead different antenna types feature different advantages and disadvantages in terms of sensitivity, measurement uncertainty and cost, where the cost often is not dominated by the antenna structure itself, but by the required resources for assembly and deployment.
This means that depending on the accessibility and infrastructure of the experimental site, on the available funding, and on the physics goals of the experiment a different antenna type might by best, although this decision is not crucial. 
At least in the widely used frequency range below $100\,$MHz the antenna type has little importance for the overall performance of the experiment, because measurements are affected by galactic radio background typically exceeding the internal system noise.
Moreover, some apparent disadvantages, e.g., pulse distortion by non-uniform group delays \cite{AERAantennaJINST2012, SchroederTimeCalibration2010}, can nowadays be corrected during offline digital data processing.

This means that most important are those features which irreversibly degrade the original measurements: 
first, this would be a very low gain for the arrival directions of interest.
As a rule of thumb only antenna gains below $-10\,$dBi, i.e., ten times lower than the gain of an ideal isotropic antenna, significantly increase the detection threshold, since for any higher gains the Galactic noise and not the gain is limiting. 
Second, the systematic uncertainty of the absolute value of the electric-field strength, in particular because of unknown variations of the gain with changing environmental or ground conditions. 
Hence, the antenna characteristics should ideally have negligible influence from temperature, humidity, and ground conditions.

Here is a short personal summary of the main features of different antenna types successfully used by at least one radio array for air-showers:
\begin{itemize}
 \item \textbf{Dipole antennas:} Dipole-like antennas have been used in LOPES \cite{FalckeNature2005}, LOFAR \cite{LOFARinstrument2013}, and CODALEMA \cite{CODALEMA_Geomagnetic}. 
 They are a simple and economic choice, and in principle have easy-to-calculate characteristics. 
 However, experience has shown that the antenna response is more difficult to understand than initially thought - at least for the folded dipoles on a reflective ground plane used by LOPES and LOFAR \cite{SchroederICRC2015}.
 Dipole antennas only have two small disadvantages: first, they require protection from cattle or animals when deployed in the wild, second, their gain slightly depends on the ground. 
 The latter reason is why LOPES and LOFAR deployed their antennas above a metal ground plate or mesh, respectively.
 \item \textbf{SALLA:} An equally economic antenna type is the SALLA (\underline{s}hort / \underline{s}mall \underline{a}periodic \underline{l}oaded \underline{l}oop \underline{a}ntenna) \cite{AERAantennaJINST2012} used by Tunka-Rex \cite{TunkaRex_NIM2015}, which features a resistive load in order to make the antenna gain broad-band and almost independent of ground conditions. 
 The downside is a lower gain, which increases the detection threshold in terms of cosmic-ray energy by about $30\,\%$. 
 Consequently, the SALLA seems to be inadequate when maximum statistics is the goal, but at the same time might be the best choice when aiming at maximum accuracy, i.e., when low systematic uncertainties are important. 
 The simple robust design, and the possibility to deploy the SALLA on poles in a height sufficient for protection from animals without the need of a fence also make them a very cost-effective solution.
 \item \textbf{Butterfly:} Butterfly antennas are in use at CODALEMA and AERA \cite{AERAantennaJINST2012}. 
 Their design makes use of the reflection of the radio signal by the ground, which enhances the total gain. 
 Due to the limitation by Galactic noise this improves the detection threshold only slightly, not yet quantified. 
 The downside is a large dependency on the exact conditions of the environment, which has to be corrected for when aiming at accurate measurements of the radio amplitude and phase.
 \item \textbf{LPDA:} LPDAs (\underline{l}ogarithmic \underline{p}eriodic \underline{d}ipole \underline{a}ntennas) have been used by many experiments, e.g., AERA and LOPES, and will be used by the SKA. 
 Compared to the other antenna types the LPDA is relatively expensive and complicated to deploy. 
 This is paid off by a variety of advantages: LPDAs can cover large bandwidth with high gain and are relatively independent of the ground. 
 Provided sufficient resources are available, the LPDA is probably the best selection for any air-shower array. 
 However, if resources are limited, then all the slight advantages of the LPDA might be outweighed by their costs, simply because for a fixed budget more antenna stations can be constructed when using any of the other antenna types mentioned above.
 \item \textbf{Directional Antennas:} In order to maximize the exposure of air-shower arrays, directional antennas are usually avoided with two main exceptions. 
 First, at GHz frequencies the CROME experiment used highly directional dish antennas.
 They provide a high gain and low internal system noise which is important at GHz frequencies, since the external Galactic noise is much lower compared to MHz frequencies. 
 Second, experiments like ANITA and TAROGE target very inclined showers only, i.e., a very limited zenith angle range. 
 Consequently, using frequency bands above $100\,$MHz the use of directional antennas can lower the detection threshold for inclined showers, e.g., ANITA uses directional horn antennas at several $100\,$MHz \cite{ANITA_neutrino_PRL_2009}.
\end{itemize}

In summary, a large variety of radio antennas has been successfully operated for air-shower detection below $100\,$MHz. 
At higher frequencies there is less experience, but at the same time the choice of antenna type is more critical, because measurements are generally limited by the noise of the receiver system, and not anymore by external, galactic background as for the lower frequencies.

\begin{figure}[t]
  \centering
  \includegraphics[width=0.9\linewidth]{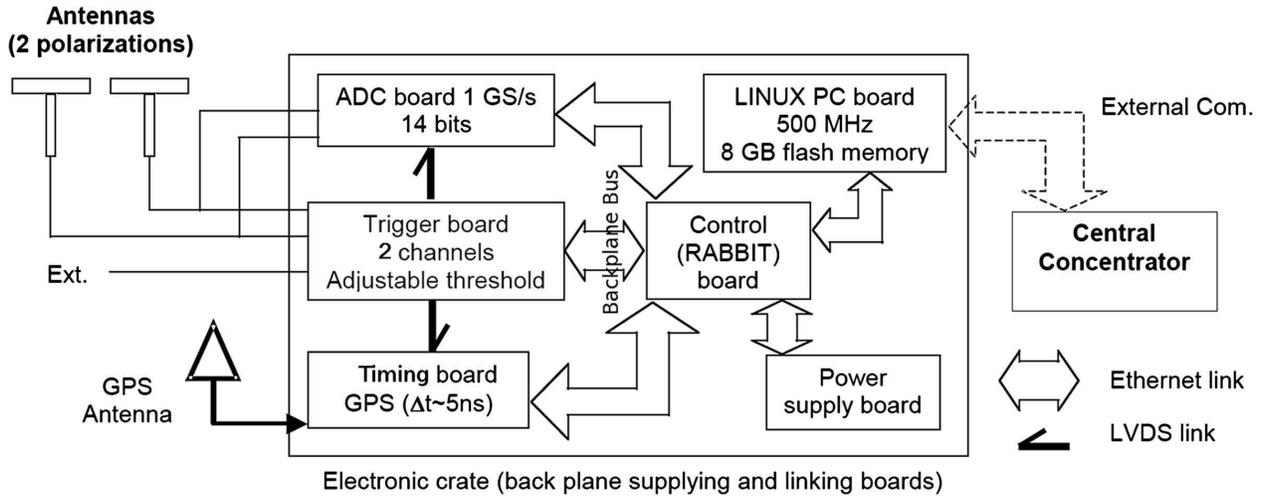}
  \caption{Sketch of the CODALEMA local-station electronics (from Ref.~\cite{CODALEMA_RICAP2012}).}
  \label{fig_CODALEMAelectronics}
\end{figure}

\subsubsection{Electronics for signal processing and data acquisition}
The performance of a radio detector does not only depend on the antenna, but also on the analog and digital electronics of the subsequent signal chain. 
The total response of this signal chain determines the response of the detector to an incoming signal, i.e., the amplification, the delay and also the distortion of the signal shape. 
Usually the signal chain begins with a low-noise amplifier (LNA) either directly integrated into the antenna or connected to it with short cables (short compared to the measured wavelengths). 
Depending on the station design longer or shorter signal cables, often coaxial cables of type RG 213, lead to a main amplifier which is connected to an analog-digital converter (ADC) digitizing the radio signal as a function of time. 
In case of autonomous stations the digitization happens inside of the local electronics and the digitized signal is communicated to a central station for data acquisition; 
in case of cabled array layouts the digitization can take place in central hubs instead of at the antenna, since in the widely used frequency band below $100\,$MHz, the pre-amplified signal can easily be transferred for several $10\,$m. 

In addition to amplifiers, filters are also commonly used in radio detectors, e.g., integrated into the main amplifier, fulfilling several purposes:
first, the sensitive band is restricted at least to the design band of the antenna or to a smaller sub-band, in order to improve the signal-to-noise ratio.
Second, sometimes the band is restricted even further to suppress disturbances of short-wave and FM radio, and potentially other human-made narrow-band communication. 
For this reason, a usual frequency band chosen by several experiments is $30-80\,$MHz. 
Otherwise, the short-wave and FM radio disturbances had to be filtered digitally during offline data processing which would require the whole detector system to handle a much higher dynamic range, e.g., by using more expensive ADCs. 
Third, according to the Nyquist sampling theorem \cite{Nyquist1928}, the sampling frequency must be at least twice the upper band limit to avoid information loss. 
Then the full waveform can be recovered by upsampling (see section \ref{sec_signalProcessing}). 
In principle the sampling rate can be decreased when the lower limit of the band is at least half of the upper limit: 
LOPES operated in the so-called second Nyquist domain digitizing the signal recorded within $40-80\,$MHz with a sampling rate of $80\,$MHz \cite{FalckeNature2005, HuegeLOPES_ARENA2010}. 
However, with state-of-the-art electronics higher sampling rates have become inexpensive, such that the short-wave radio disturbances nowadays are the main reason for using high-pass filters in addition to low-pass filters.

Other design features depend on the station design, e.g., a self-trigger on the radio signal requires local computing power and, thus, a CPU or an FPGA \cite{RAugerSelfTrigger2012, CODALEMA_selftriggerECRS2012, AERA_selftriggerIEEE2012}, while an external trigger can be implemented in a simpler way (see figure \ref{fig_CODALEMAelectronics} for the design of an externally triggered CODALEMA station for example). 
Depending on the design of the array trigger a certain local buffer depth is also necessary. 
Ideally data are not only stored after a local trigger, but continuously and read out in case of a trigger somewhere else in the array. 
Such externally-triggered readout is particularly useful for interferometric analysis techniques exploiting the information contained in sub-threshold stations. 

Some experiments also took care to minimize variations of the group delay over frequency, since these lead to pulse distortion and can lower the signal-to-noise ratio. 
However, this is only a problem for local self-triggering on the radio signal, not in case of external triggering, because the response of any linear component in the signal chain can be corrected for during data analysis. 
This means that in good approximation the effect of pulse distortion can be reverted \cite{SchroederTimeCalibration2010, RadioOffline2011}, and does not harm the reconstruction of the measured electric field originating from the air shower. 
Thus, the specific properties of the signal response are less important for the performance of a radio detector as long as they a measurable and, consequently, correctable during data analysis.

Finally, the clock design of the array is important. 
With cabled setups a central clock can easily be distributed with nanosecond precision.
Even sub-nanosecond precision is available with state-of-the art electronics, e.g., with WhiteRabbit via Ethernet \cite{WhiteRabbitICRC2015}. 
In setups with autonomous stations, each station needs its own clock. 
Current experiments decided for economic GPS clocks providing a timing precision of only a few nanoseconds, which can however be improved by the use of a reference beacon (cf.~section \ref{sec_timing}) \cite{SchroederTimeCalibration2010, AERAtimingJINST2016}.

\subsection{Calibration techniques}
\label{sec_calibration}
The accuracy of the reconstructed air-shower observables of interest like direction, energy, and position of the shower maximum, depends on the accuracy of the underlying radio measurements.
In particular the accuracy of the arrival time and amplitude are important, and for some reconstruction methods also the accuracy of the frequency spectrum of the air-shower radio signal. 
The measurements of all these observables are affected by the properties of the individual antenna stations, and in the case of the arrival time also by the synchronization of the full array. 
Thus, the response of the detector stations has to be known sufficiently well for all relevant arrival directions of the radio signal as well as for all relevant environmental conditions in order to interpret the measurements correctly. 
Sufficiently well means that the final measurement uncertainty should ideally be limited only by irreducible background, but not by the knowledge of the detector response - 
a goal which  has not yet been achieved by present radio arrays, although lately significant progress has been made towards more accurate amplitude \cite{AERAantennaJINST2012,TunkaRex_NIM2015, LOPESimprovedCalibration2016, LOFARcalibration2015} and time calibration \cite{SchroederTimeCalibration2010, AERAtimingJINST2016, CorstanjeLOFARtimeCalibration2016}.

Amplitude and time measurements can be calibrated absolutely or relatively:
Absolute amplitude calibration refers to the amplitude scale of the array. 
This is particularly important for the comparison of different experiments with each other and with theoretical predictions. 
Moreover, an absolute measurement of the radio amplitude is the prerequisite for an independent measurement of the energy scale of cosmic-ray air showers, without the need to rely on the cross-calibration with another cosmic-ray detector. 
Absolute time calibration is important for the correlation of cosmic-ray events with the detection of other astroparticle messengers like neutrinos, photons, or gravitational waves. 
However, even cheap GPS clocks are sufficiently accurate for this purpose, in contrast to the relative synchronization of antenna stations with each other which requires much higher precision.

Relative calibration has the goal of ensuring that the response of the whole array is sufficiently homogeneous. 
Fluctuations between stations should not contribute significantly to the final uncertainty for the reconstruction of the air-shower properties, like the amplitude for energy reconstruction, or the arrival time for direction reconstruction. 
Due to production variations as well as different ground conditions at different antenna stations, this usually requires calibration measurements of the individual parts and/or end-to-end measurements in the finally deployed array.
Whatever the method, the calibration finally consists of the determination of the response of each individual antenna station, or at least of the average response and its uncertainty or spread over the array.

\begin{figure}[t]
  \centering
  \includegraphics[width=0.32\linewidth]{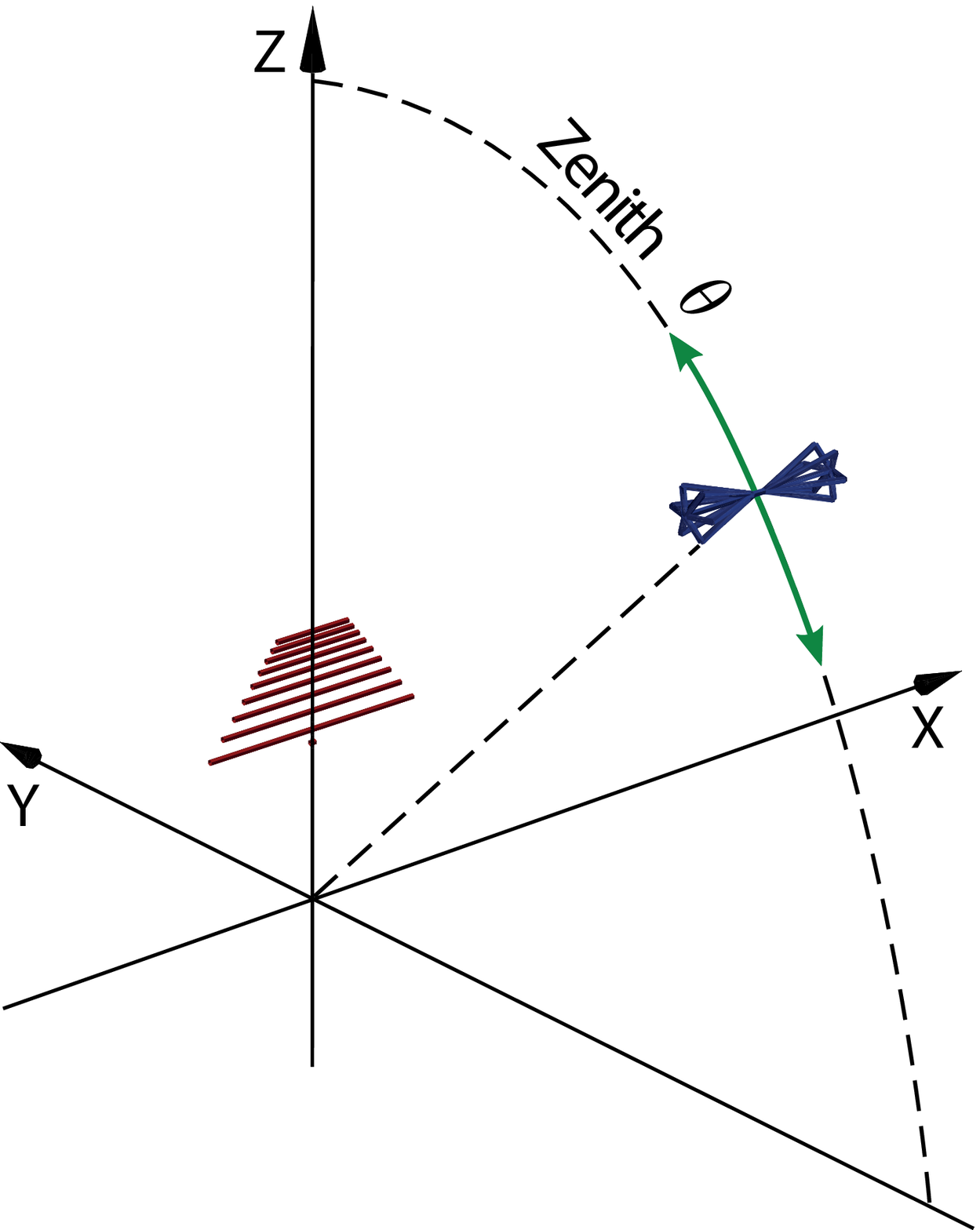}
  \hfill
  \includegraphics[width=0.275\linewidth]{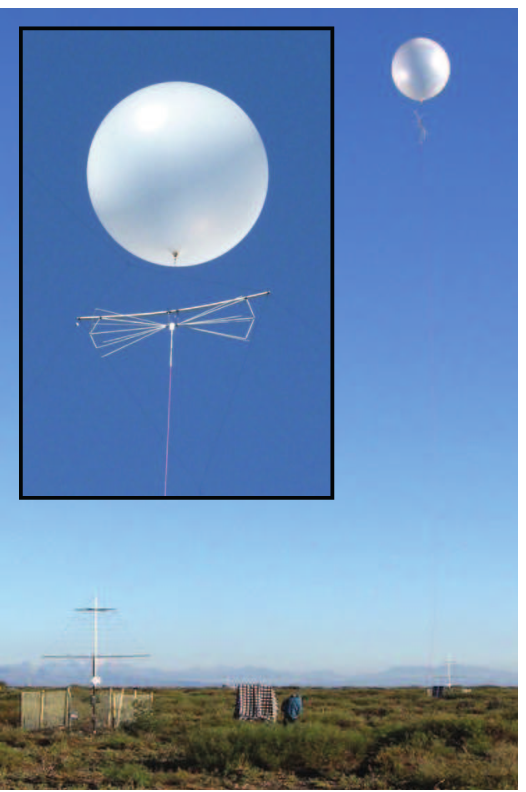}
  \hfill
  \includegraphics[width=0.37\linewidth]{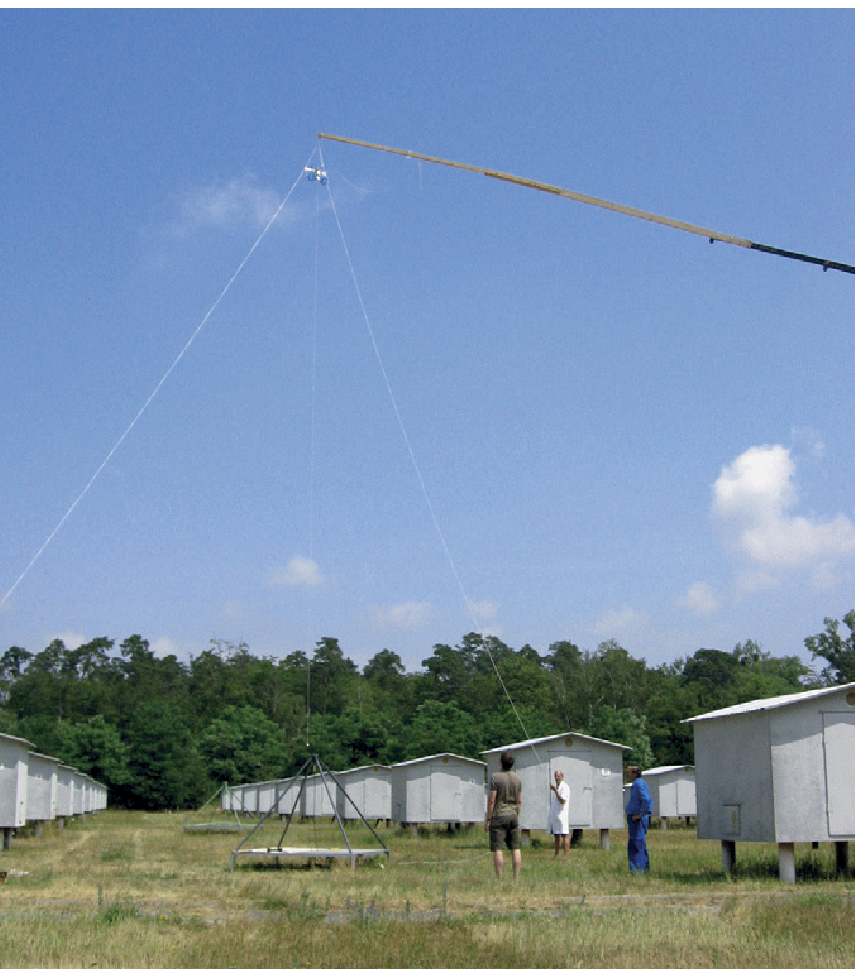}
  \caption{Calibration of the antenna gain with reference sources in the field. Left: measurement principle; middle: calibration of AERA; right: calibration of LOPES (left and middle picture from Ref.~\cite{AERAantennaJINST2012}, right picture from Ref.~\cite{LOPEScalibrationNehls2008}).}
  \label{fig_calibrationMeasurements}
\end{figure}

\subsubsection{Response of antenna stations}
The principle physics signal measurable at each antenna station is the electric-field strength vector as a function of time. 
However, a single antenna cannot measure the vector, but only one projection. 
Moreover, the time resolution and the frequency bandwidth are limited:
even broadband antenna types have a certain minimum and maximum frequency they are sensitive to, as have other components in the signal chain like amplifiers, cables, and filters. 
Usually bandpass filters are used to further limit the frequency band, which means that the calibration can be restricted to this well-defined band. 

The purpose of the calibration at station level is to determine the response of the complete signal chain from the antenna to the ADC within the frequency range measured, i.e., 
how the electric-field vector is converted by the antenna to a voltage signal, and how the following signal chain changes amplitude and phase. 
If a station has several differently aligned antennas (channels) measuring different projections of the the electric-field vector, the response of each channel has to be known. 
Of all components in the signal chain only the antenna has a response depending on the arrival direction in addition to the frequency. 

Determining the response of the antenna is the most difficult part of the whole calibration. 
Antennas for higher frequencies of several $100\,$MHz are usually small enough for accurate measurements of their free-space response in anechoic chambers by the two-antenna method, as has been done for the ANITA horn antennas \cite{ANITA2PhdThesis}. 
Antennas for the frequency band of $30-80\,$MHz are larger, which would require large anechoic chambers.
As additional difficulty the free-space response would have to be converted to the response of the antenna for realistic ground conditions. 
To my knowledge none of the currently running air-shower arrays has performed such calibration measurements with the two-antenna method in anechoic chambers, but instead calibration measurements of single antennas directly in the field. 
Furthermore, all modern air-shower arrays have used simulations to account for the direction dependence of the antenna gain pattern \cite{AERAantennaJINST2012, LOPEScalibrationNehls2008, TunkaRex_NIM2015, LOFARcalibration2015}. 
Codes like NEC2 \cite{NEC2_IEEE_2004} calculate the effect of the antenna on the amplitude and phase of the recorded radio signal as a function of frequency, arrival direction, and orientation of the electric-field vector. 
Different grounds can be included in the calculation, which depending on the antenna type can significantly affect the response function.

Most experiments cross-checked the simulated antenna response against calibration measurements, particularly the frequency-dependent amplitude response.
Sometimes the phase response is also determined using an external reference antenna which is placed at known positions in the air using a crane, a flying drone, or a balloon (see figure \ref{fig_calibrationMeasurements} for examples). 
These measurements are used to test the simulated antenna response for a limited number of arrival directions. 
Moreover, the absolute amplitude scale can be determined when the reference source is calibrated absolutely, as it has been done for LOPES, LOFAR, Tunka-Rex, and AERA.

\begin{figure}[t]
  \centering
  \includegraphics[width=0.45\linewidth]{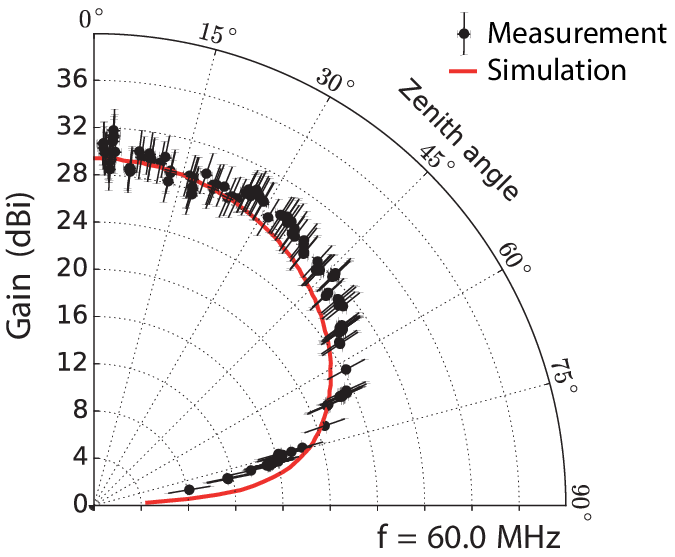}
  \hfill
  \includegraphics[width=0.43\linewidth]{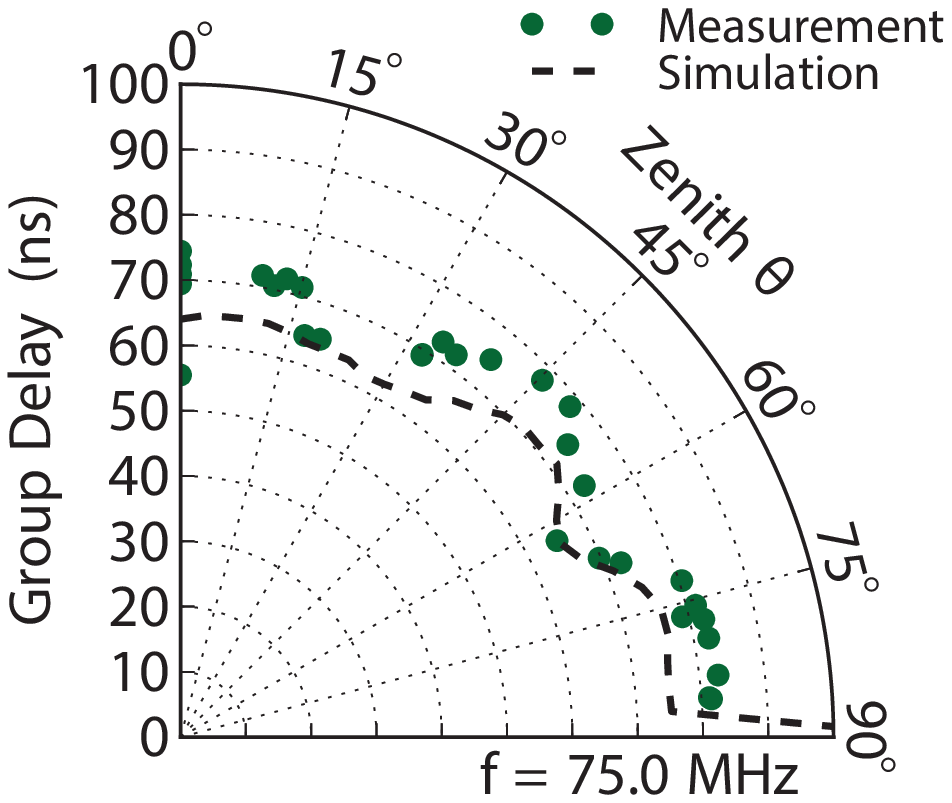}
  \caption{Examples for calibration measurements compared to simulated antenna characteristics. 
  Left: gain of a LOFAR antenna at $60\,$MHz over zenith angle.
  Right: group delay of an LPDA antenna at AERA at $75\,$MHz (slightly modified from Refs.~\cite{HoerandelLOFARcalibICRC2015} and \cite{AERAantennaJINST2012}, respectively).}
  \label{fig_calibrationResults}
\end{figure}

Measurements at AERA and LOFAR confirm that to first order the direction dependence is described correctly by the simulations, but deviations are not negligible, in particular for large zenith angles (see figure \ref{fig_calibrationResults} for examples). 
The deviations currently are one of the largest systematic uncertainties in radio measurements of air showers.
For example, LOPES observed a discrepancy in the zenith dependence of the amplitude between CoREAS simulations and measurement, but it was not possible to decide whether this indicates a shortcoming of the antenna model or of the CoREAS simulations \cite{LinkICRC2015}.
Furthermore, for several experiments the observed deviations between calibration measurements and simulations depend on the frequency, which translates into a systematic uncertainty when measuring the frequency spectrum of the radio signal.
Finally, calibration uncertainties in the group delay cause a systematic uncertainty for wavefront measurements. 
Consequently, further improvements in the antenna calibration are one of the most urgent issues in order to further improve the measurement accuracy of air-shower observables.

In comparison to the antenna calibration, the response of the remaining components in the signal chain can be measured relatively easily with network analyzers in laboratories. 
If all components are linear and have the same impedance, the combined response function is easy to calculate by multiplication of the individual response functions in Fourier space (= in the frequency domain). 
However, the situation of matched impedances is not automatically given, in particular not for the connection of the antenna to the remaining signal chain:
the impedance of antennas typically changes with frequency and does not match the usual $50\,\Omega$ impedance of other analog components. 
A balun (balanced unbalanced transformer) can reduce, but not totally eliminate this problem, such that any remaining mismatch has to be taken into account when calculating or simulating the total response of the whole signal chain including the antenna \cite{HillerThesis2016}. 

A conceptually simpler way of calibration avoiding the risk of errors, e.g., due to impedance mismatches, is provided by end-to-end calibrations:
the full signal chain including the antenna is calibrated at once by using an external reference source, or by any other known external signal, like galactic radio noise. 
However, the intensity of the Galactic noise depends on the observed part of the sky, which itself depends on the zenith characteristics of the antenna, on the latitude of the experiment, and changes with time due to the rotation of the Earth. 
At least for LOFAR using Galactic noise led to less accurate calibration results than using an artificial reference source \cite{LOFARcalibration2015}. 
Consequently, using an absolutely calibrated reference source is the recommended method, and Galactic noise can be used as an independent cross-check.  

Most experiments take a mixed approach between end-to-end calibrations, laboratory measurements, and simulations of the instrumental response: 
the absolute amplitude scale is calibrated (or at least cross-checked) with an end-to-end calibration, and the directional dependence of the antenna response is determined with simulations.
Laboratory measurements of the individual parts are then used to correct for small deviations from the average behavior, e.g., due to production fluctuations.

Overall, calibration accuracies in the order of $15-20\,\%$ have been achieved for the absolute amplitude by current antenna arrays, where large contributions to this uncertainty come from the reference source and the simulated antenna models. 
Smaller contributions to the uncertainty of the order of $5-10\,\%$ concern the homogeneity over the whole array and the stability over time, in particular due to varying environmental and ground conditions.
The latter is close to the minimum uncertainty achieved for the energy reconstruction for air-shower simulations under ideal conditions \cite{2014ApelLOPES_MassComposition, GlaserShortAuthor2016}.
This means that apart from the absolute amplitude scale and the antenna models, other calibration uncertainties are already at a tolerable level.

\subsubsection{Time Synchronization}
\label{sec_timing}
For accurate arrival time measurements it is necessary, but not sufficient to determine the group delay of each antenna and the attached signal chain.
In addition, all measurements in the array must be synchronous, but before sophisticated calibration efforts have been performed significant asynchronicities in the order of $10\,$ns have been observed in AERA \cite{AERAtimingJINST2016}, Tunka-Rex \cite{SchroederTunkaRex_PISA2015}, and even more in LOPES \cite{SchroederTimeCalibration2010}.
Only LOFAR seems to feature a sub-nanosecond precise time synchronization due to a sophisticated clock system \cite{CorstanjeLOFARtimeCalibration2016}.
For the purpose of direction reconstruction a timing uncertainty in the order of $10\,$ns might be tolerable: 
depending on the array size, this uncertainty leads to resolutions of at least $1^\circ$.
This is more than enough for physics analysis of charged cosmic rays because they are deflected on their way to Earth by magnetic fields.
However, interferometric analysis methods or wavefront measurements require a synchronization accuracy of about $1\,$ns, which is why methods have been developed to correct for clock drifts.

Pulsers and pingers are used to periodically check the time synchronization not only for air-shower arrays, but also for balloons and experiments in dense media \cite{Silvestri05, Hoffman06}.
Nevertheless, for continuous monitoring of the time synchronization such calibration pulses bring the disadvantage that they might look similar to the signal and cause dead time.
Instead continuous-wave signals can be used for permanent monitoring and time calibration \cite{Landsman09, SchroederTimeCalibration2010}.
In particular narrow-band transmitters inside the measured frequency band are used by several antenna arrays to check the relative timing between antennas and to correct for offsets, drifts and jumps. 
These narrow-band signals can easily be filtered during offline data analysis and only marginally degrade the measurement quality of air-shower radio pulses. 
In contrast to broad-band noise, narrow-band transmitters feature a stable phase relation at the receiving antennas over time if the transmitter and the receiving antennas remain at constant positions. 
Variations in the timing (e.g., clock drifts) of one antenna then cause corresponding variations in the measured phasing. 
Depending on the measurement accuracy of the phasing, and depending on the size and nature of the observed timing variations, different strategies and algorithms have been developed all based on this principle.

LOPES \cite{SchroederTimeCalibration2010} and AERA \cite{AERAtimingJINST2016} deployed a dedicated artificial transmitter, a beacon, transmitting continuous sine-wave signals at known frequencies whose phasing is used to correct for time drifts during offline data processing. 
Tunka-Rex uses such a beacon only for occasional cross-checks \cite{SchroederTunkaRex_PISA2015}.
LOFAR uses narrow-band transmitters already present in their band (e.g., FM radio channels) for cross-check \cite{CorstanjeLOFARtimeCalibration2016}, an idea pioneered by LOPES, which used the phasing of a TV transmitter before deploying the dedicated beacon \cite{FalckeNature2005}.

The time calibration by these phasing-methods has been successfully confirmed by independent calibration measurements based on pulses emitted by commercial airplanes \cite{AERAtimingJINST2016}, and by dedicated calibration drones \cite{CorstanjeLOFARtimeCalibration2016}. 
Thus, the method can be considered state-of-the-art, and the only open question is the achievable accuracy for the correction of observed timing drifts. 
LOFAR has shown that precisions in the order of $0.1\,$ns are achievable for cabled arrays of a few $100\,$m extension.
AERA demonstrated better than $2\,$ns precision for an array of autonomous stations with an extension of a few km.
It is not yet clear whether this can be easily improved to below $1\,$ns as desired for interferometry, or whether yet unknown systematic uncertainties or propagation effects limit the accuracy of the beacon method for large extensions.
With the steady improvement of satellite-based timing better clocks might solve this problem for future arrays and make later timing corrections obsolete.

\begin{figure}[t]
  \centering
  \includegraphics[width=0.5\linewidth]{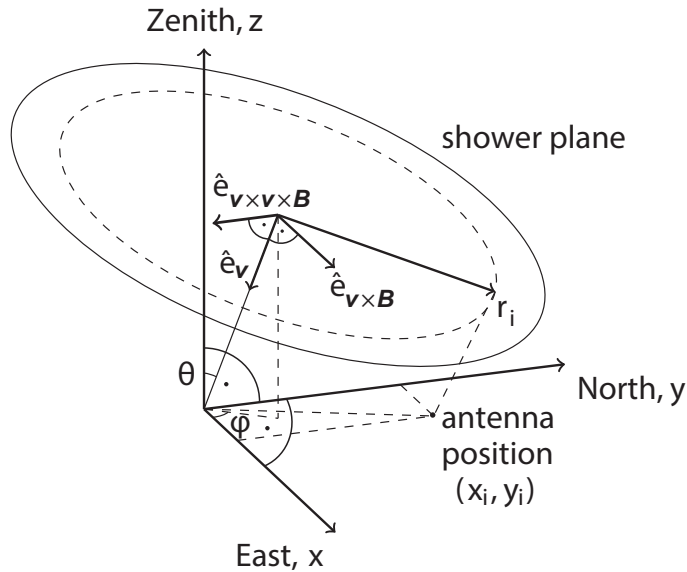}
  \caption{Typical coordinate systems used for the reconstruction of the electric-field vector of air-shower radio signals: $\theta$ and $\phi$ are the zenith and azimuth angle of the shower, respectively, $(x_i, y_i)$ is the antenna position on ground, $r_i$ the antenna position projected into the shower plane, and the absolute value of $r_i$ is the distance to the shower axis (modified from Ref.~\cite{CorstanjePolarizationICRC2015}).}
  \label{fig_coordinateSystems}
\end{figure}

\begin{figure}[t]
  \centering
  \includegraphics[width=0.65\linewidth]{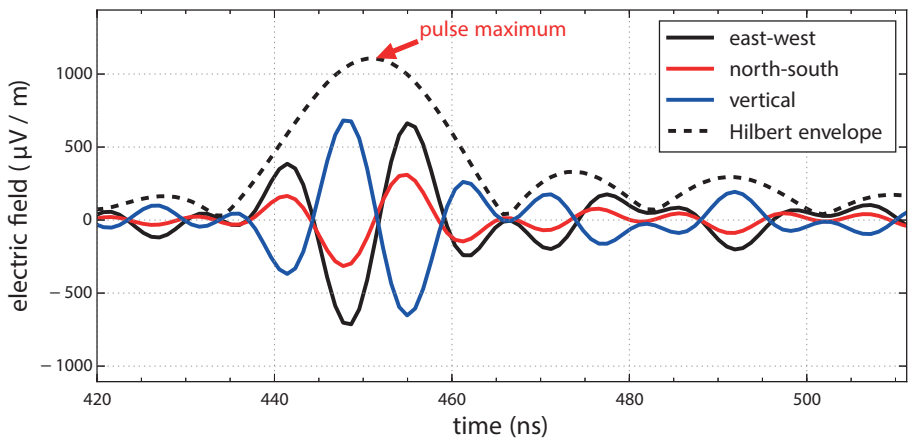}
  \caption{AERA example event: reconstructed components of the electric-field vector measured by one antenna and the corresponding Hilbert envelope (from Ref.~\cite{GlaserEnergyICRC2015}).}
  \label{fig_AERApulse}
\end{figure}

\subsection{Digital signal processing and beamforming}
\label{sec_signalProcessing}
\label{sec_beamforming}

The goal of signal processing is the reconstruction of physically meaningful observables from the measured raw data. 
This means the conversion of traces of ADC counts, which correspond to voltages recorded by the different polarization channels of antenna stations, into the properties of the electric field arriving at the antenna stations, which are the amplitude, arrival time, polarization, and more complex quantities like the signal shape or the frequency spectrum.
In a subsequent step described later in chapter \ref{sec_reconstruction}, these quantities can be used to reconstruct the properties of the air shower and the primary particle, like its direction, energy and mass. 

Already the first step is non-trivial since the complete detector response to an incoming signal has to be known and unfolded. 
In particular, the response of the antenna depends on the arrival direction of the signal, which for random air showers is not known in advance and has to be determined from the measurement itself. 
This is complicated since the direction assumed for unfolding the antenna response usually also influences the reconstructed arrival time of the signal and thus the reconstructed arrival direction. 
Luckily the effect of the direction on the reconstructed arrival time is in the order of only a few $10\,$ns for commonly used antenna types, which is small compared to the propagation time of the radio signal over the typical extension of antenna arrays. 
Although mathematically it is not guaranteed that there is a unique solution to this hen-egg problem of direction reconstruction, iterative algorithms usually converge to a solution for the incoming direction of real measured air showers.
Once the direction is known, the response of the antenna and the subsequent electronics can be unfolded under the realistic assumption that all electronic components are linear, i.e., their response can be inverted in a unique way.

For practical reasons any unfolding of detector responses is done in Fourier space: 
the response of any components in the signal chain can be understood as a frequency-dependent gain and phase shift of the signal, which means that in Fourier space the inverse response can simply be multiplied to the raw data. 
This is at least true for components like filters and amplifiers acting on a single polarization channel. 
Only for the antenna a more complex model is required, which describes for each polarization channel of the antenna the response to an incoming electric-field vector. 
In principle the electric-field vector consists of three components as functions of time, but only two components contain information on the air-shower signal, since for transverse radio waves the component orthogonal to the radio wavefront must be zero, i.e., the orthogonal component contains only background. 
Since the radio wavefront has an angle of typically less than $2^\circ$ to the shower plane (cf. section \ref{sec_wavefront}), often the approximation is used that the electric-field vector is perpendicular to the shower plane and zero in direction $\vec{v}$ of the shower axis. 

There are two widely used coordinate systems to display the reconstructed electric-field vector as shown in figure \ref{fig_coordinateSystems}.
One simply takes the east-west, north-south and vertical height axes of the antenna (e.g., the example event shown in figure \ref{fig_AERApulse}), the other one is motived by the physics of the two emission mechanisms and oriented by the shower direction $\vec{v}$ and the direction of the geomagnetic field $\vec{B}$:
The reconstructed electric-field vector has two non-zero components, one in $\vec{v} \times \vec{B}$ direction dominated by the geomagnetic radio emission, and one in $\vec{v} \times (\vec{v} \times \vec{B})$ direction dominated by Askaryan emission.

Through unfolding of the detector response the components of the electric-field vector are reconstructed as a functions of time in bins of the original sampling rate. 
In case that the sampling rate is not an order of magnitude higher than the upper limit of the measurement band, this can impact the reconstruction accuracy, although according to the Nyquist sampling theorem \cite{Nyquist1928} all information is in principle available. 
This problem can be solved by \textbf{upsampling} the signal during the offline analysis, e.g., by zero-padding: 
the frequency spectrum obtained by Fourier transformation is extended by adding zeros after the highest frequency, such that it has an integer number $n$ times more samples. 
The back-transformed time series then also has $n$ times finer sampling where the new samples correspond to the correct interpolation of the electric field, since the zero-padding method implicitly takes into account that the signal is composed only of frequency components inside of the measurement band. 

Finally, more sophisticated parameters can be determined based on the upsampled electric-field vector.
Stokes parameters can be used as a measure for the polarization of the signal \cite{AugerAERApolarization2014, SchellartLOFARpolarization2014}, although it is not clearly defined what \lq polarization\rq~means for a signal which only lasts for few osculations (cf.~section \ref{sec_polarization}).
Many experiments use the Hilbert envelope of the signal, which gives the maximum instantaneous amplitude.
However, one has to keep in mind that these quantities might be calculated wrongly in case of noise and background, because they arrive from random directions and not only from the direction assumed for the unfolding of the antenna response. 
This means that only the signal, but not the noise amplitude is a physical quantity with a meaningful unit and absolute scale, and great care has to be taken to determine the influence of noise on a specific analysis.

\begin{figure}[t]
  \centering
  \includegraphics[width=0.7\linewidth]{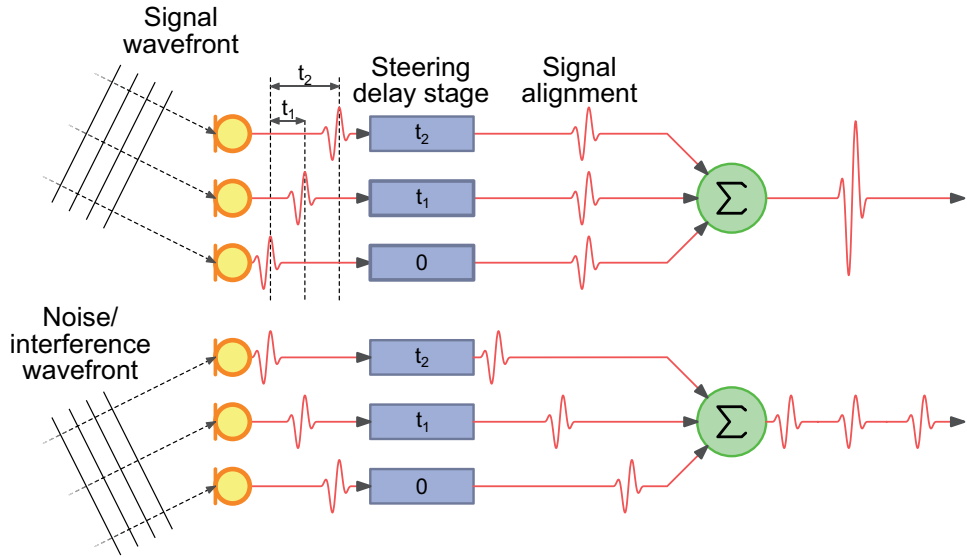}
  \caption{Principle of beamforming for a plane-wave signal (from Ref.~\cite{labbookpages}).}
  \label{fig_beamforming}
\end{figure}

\begin{figure}[t]
  \centering
  \includegraphics[width=0.48\linewidth]{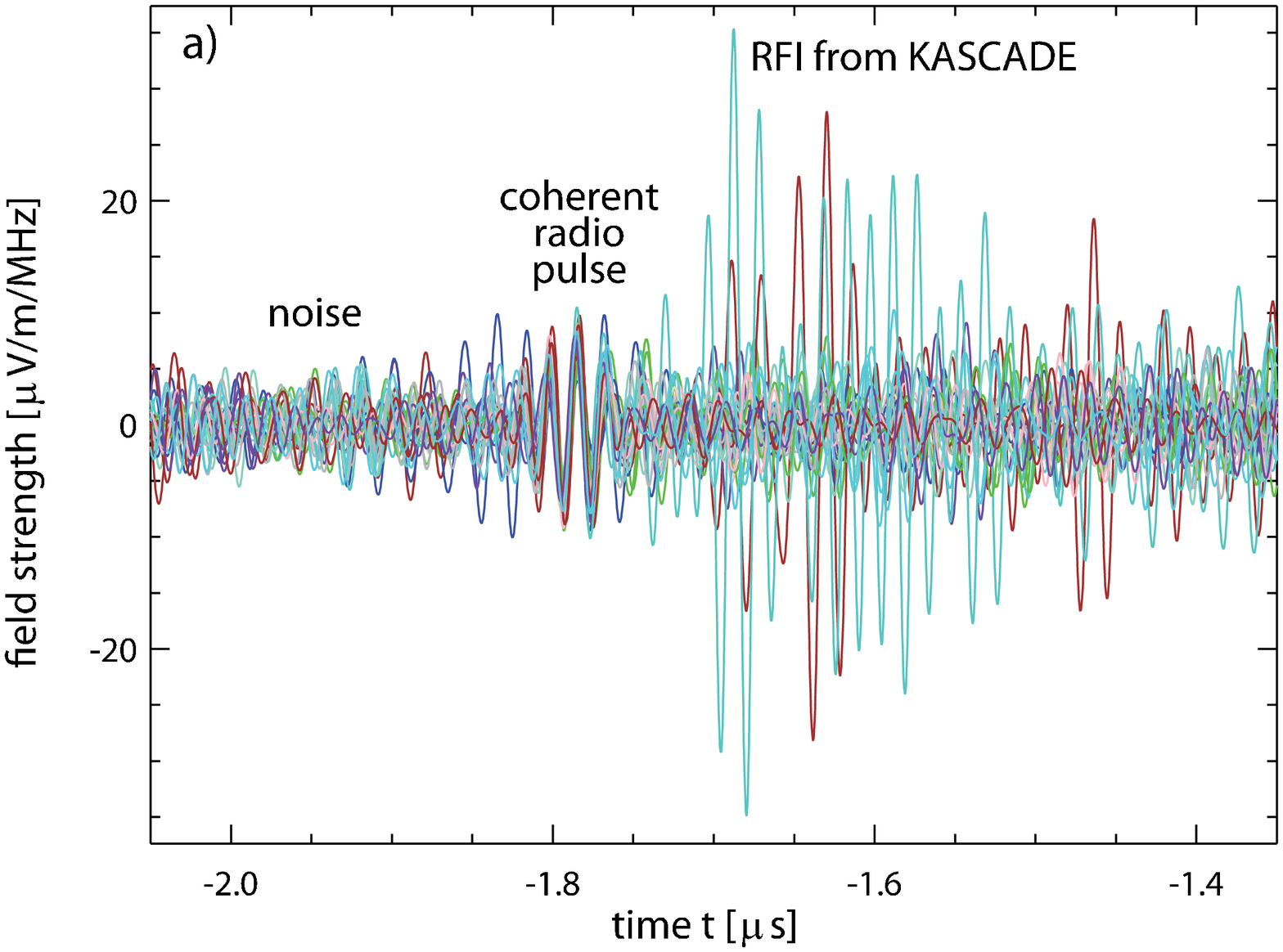}
  \hfill
  \includegraphics[width=0.48\linewidth]{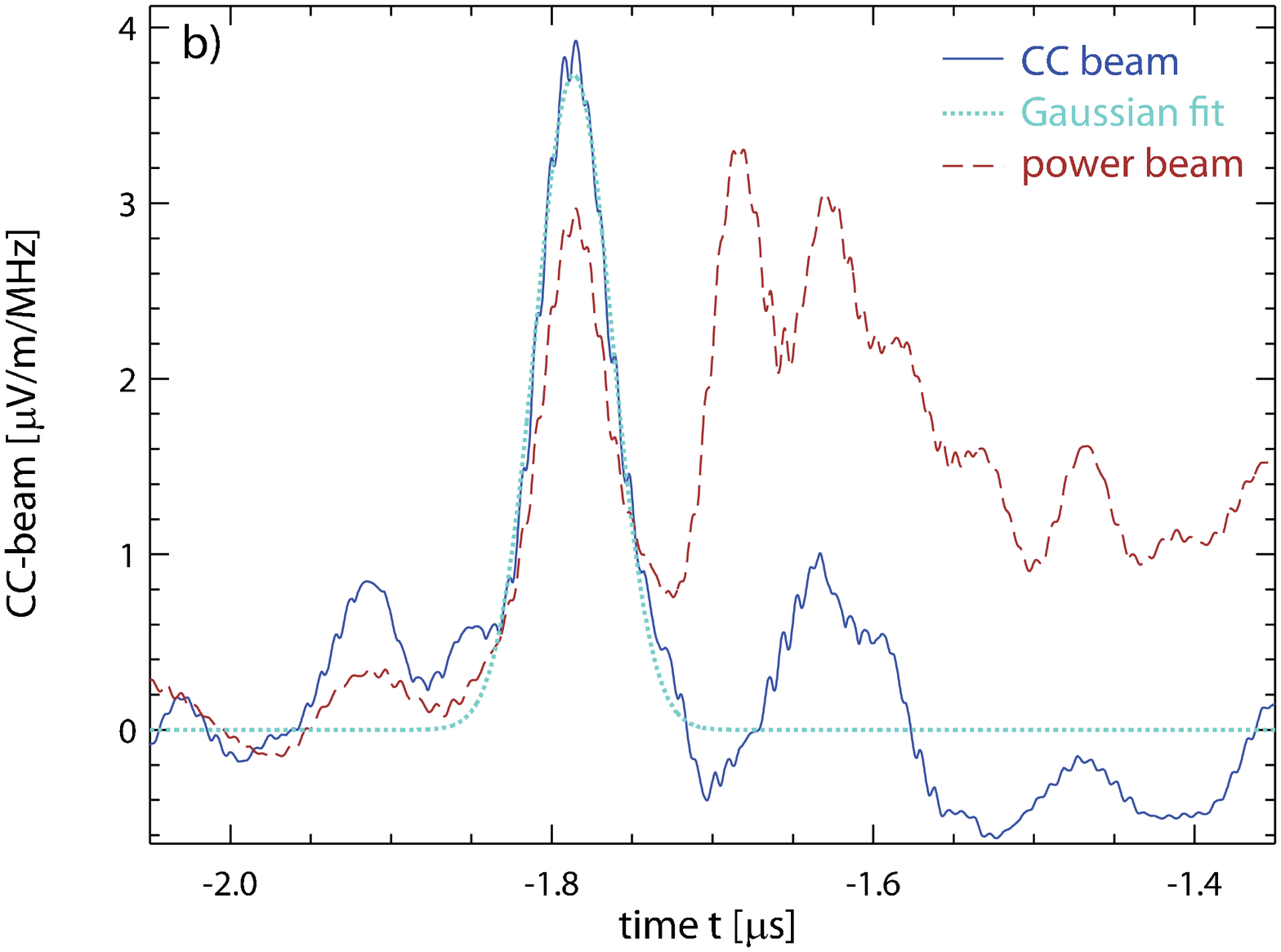}
  \caption{LOPES example event. a) Traces of individual antennas after upsampling, and after the shift according to the arrival time of the signal in each antenna. 
  b) Cross-correlation and power beam (from Ref.~\cite{2013ApelLOPESlateralComparison}).}
  \label{fig_beamformingExampleEvent}
\end{figure}

\textbf{Beamforming} is a method which reduces the impact of noise by combining the measurement of all antennas into one single signal. 
For each antenna the trace of a channel, its Hilbert envelope, or the electric-field vector is digitally shifted in time according to the expected arrival time of the radio signal in that antenna. 
Thereby the expected arrival time depends on the shower direction as well as on the shape of the radio wavefront.
Then all traces are summed up, and if the shift corresponds to the real arrival time the signal is much stronger in the sum than in the original antennas (see figure \ref{fig_beamforming}). 
For the practical application of beamforming a computing-intensive multi-dimensional fit or a fine-graded grid of the possible arrival directions and wavefront shapes is required, in order to search for the parameters yielding the maximum signal.
The benefit of this method is that the signal-to-noise ratio is enhanced, since the measured external noise is a mixture from many different arrival directions, with only a small part coming from approximately the same direction as the signal. 
If the timing accuracy of the array permits, even the phase information in the signal can be exploited by calculating a cross-correlation instead of the simple sum. 
This is an interferometric method applied by LOPES \cite{FalckeNature2005}, which further enhances the signal-to-noise ratio (see figure \ref{fig_beamformingExampleEvent}) and enables air-shower detection in radio-loud environments. 
Furthermore, ANITA successfully applied cross-correlation beamforming for self-triggered detection of air-showers at higher frequencies \cite{ANITA_Interferometery_2015}, and the technique is also supposed to significantly lower the detection threshold for in-ice detection of neutrinos \cite{PhasedArraysForNeutrinos_ICRC2015}.

Several software tools have been developed by different experiments for the purpose of signal processing and for the reconstruction of air-shower properties:
the most sophisticated probably is the radio extension of the Auger Offline software \cite{RadioOffline2011} written in C++, which upon request is available to other experiments and is already used by Tunka-Rex. 
LOPES and LOFAR have their own software written in C++ and Python, respectively, which both are available as open source, but are not yet re-used by other experiments. 
At least in the LOPES software only a simplified reconstruction of the electric-field vector is implemented, because not all LOPES stations were equipped with two polarization channels. 
In summary, my personal suggestion is to ask the Pierre Auger Collaboration, if dedicated software will be needed for any new air-shower array.

\begin{figure}[t]
  \centering
  \includegraphics[width=0.98\linewidth]{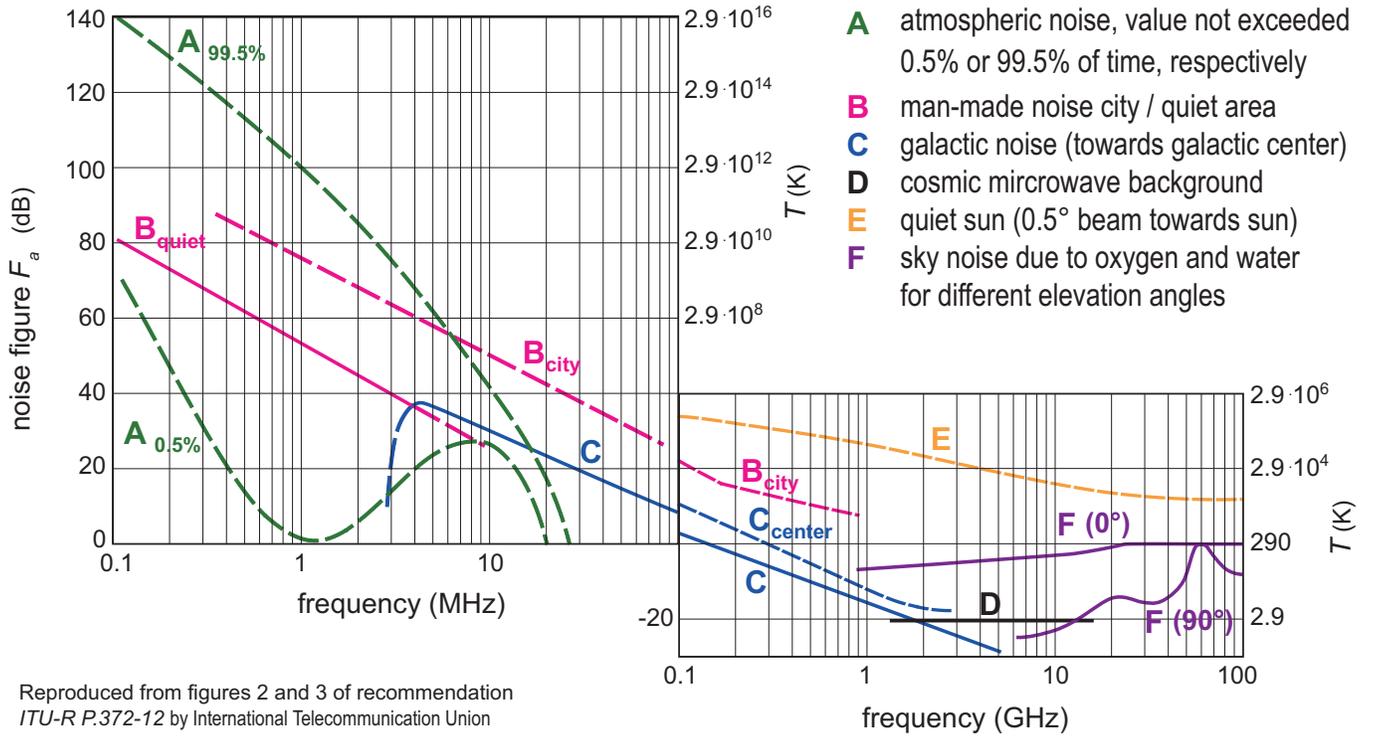}
  \caption{Typical average contributions of external noise to the total radio noise, reproduced from Ref.~\cite{ITUnoise2015}. 
  The actual contribution depends on time, location, and properties of the receiving antenna (e.g., directionality). 
  The atmospheric noise contribution $A$ is mainly due to distant lightning.}
  \label{fig_ITUnoise}
\end{figure}

\subsection{Treatment of background and noise}
\label{sec_background}
Depending on the frequency range used for detection, the location of the experiment and the used hardware, different types of background and noise are relevant. 
Typically there is no strict distinction between the terms 'noise' and 'background' in the literature of this field, although different origins of background could be distinguished. 
First, there is external background from unresolved sources of human and natural origin, where in radio-quiet regions on Earth usually the Galactic noise dominates. 
Second, there is internal noise of the instrument, which becomes the dominant background at higher frequencies of several $100\,$MHz or GHz, because the external background is minimal in this frequency range (see figure \ref{fig_ITUnoise}). 
Since air-shower measurements are usually performed at frequencies below $100\,$MHz, Galactic noise is the main background here.
For the search of neutrino-induced showers in dense media higher frequencies are used typically above $100\,$MHz. 
Hence, not Galactic but instead thermal noise limits radio arrays in dense media, which therefore generally have a lower noise level than air-shower arrays.  

The strength of background often is characterized by its noise temperature $T$, which is the temperature of a black body radiating equal amount of power at the considered frequency. 
Since many background sources are non-thermal, they have non-thermal spectra and their noise temperature changes with frequency. 
The noise figure $F_a$ is a measure for the noise power $p$ at a finite bandwidth $BW$, and is defined relative to the radiating power of a black body at $T_\mathrm{ref} = 290\,$K \cite{ITUnoise2015}:
\begin{equation}
F_a = 10 \lg \frac{p}{kT_\mathrm{ref} BW} \, \mathrm{dB}
\end{equation}
with the Boltzmann constant $k = 1.38 \cdot 10^{-23}\,$J/K.

However, the picture of an average noise level is overly simple.
The contribution of different noise sources depends on time, location and characteristics of the receiving antenna, e.g., its directionality, since many human-made noise sources are close to the horizon. 
Moreover, many noise sources, e.g., lightning and anthropogenic radio frequency interferences (RFI), have a distinct time structure, and can drastically exceed the average level for certain time periods. 
For the historic, analog radio measurements of air showers especially radio background from distant thunderstorms was a severe issue.
For recent digital experiments this problem is practically solved because most of them operate in coincidence with other air-shower detectors providing information of the arrival time and direction of the air showers, which is used to distinguish cosmic-ray initiated pulses from thunderstorm-initiated pulses.
For self-triggered arrays thunderstorm pulses could still be an issue and have to be identified by clever algorithms or separate monitoring of thunderstorm activities, e.g., with lightning-mapping arrays \cite{KrehbielLightningMapping2004}.

Closeby thunderstorms additionally affect the measurement of air showers by their high atmospheric electric field, which can be monitored, e.g., by e-field mills.
Additionally, radio pulses from thunderstorms can be distinguished during later data analysis because they are much broader than air-shower pulses and have a different time structure \cite{LOPESthunderstorm2011}.
More problematic are human-made machines, e.g., cars or transformers connected to power lines, because they often generate narrow pulses which can hardly be distinguished from air-shower pulses. 
This means that not only the average noise level, but also the height and rate of background pulses are relevant and have to be determined by each experiment individually for the relevant time and location.

Adding another complication, real radio background cannot be described by a Gaussian distribution, but has significantly enhanced tails \cite{DoctersThesis2015}. 
As a consequence, also uncertainties of any radio observables might have non-Gaussian tails, which should be taken into account when interpreting results of radio experiments. 
In particular this means that significances given in \lq sigmas\rq~cannot be simply converted to probabilities in the usual way assuming a Gaussian distribution, but the translation has to be done more carefully considering the influence of noise on the specific measurement.

There are at least two common ways to study the influence of noise on a specific measurement, in both cases using real noise samples measured under the same conditions as the experimental air-shower data: 
first, the probability for pure noise passing the quality cuts applied during data analysis can be used to estimate the impact of false-positive detection. 
Second, the measured noise can be digitally added to previously recorded or simulated pulses, to determine measurement uncertainties and biases caused by the realistic background of the experiment. 
As first-order description the average impact on signal observables like the amplitude and arrival time of a pulse can be parameterized as functions of the signal-to-noise ratio, but care has to be taken that additional higher-order effects might exist on derived observables like the shape of the lateral distribution. 

Unfortunately, the community has not yet agreed on a common definition of the signal-to-noise ratio for air-shower measurements. 
Thus, before comparing signal-to-noise ratios of different measurements or analyses, the definitions have to be compared. 
Important differences are whether signal and noise are taken as amplitude (field-strength) or power (field-strength squared) measurements, and how the noise is measured, e.g., as average power, or in the same way as the signal measurements by the average height of peaks. 
The resulting difference for the value of the signal-to-noise ratio can be huge.
I have suggested to use only definitions fulfilling a consistency criterion:
the mean signal-to-noise ratio should be $1$ for the case that pure noise is measured and interpreted as signal \cite{SchroederNoise2010}. 
However, this criterion seldom is applied, and usual definitions of signal-to-noise ratios yield much larger values for measurements of pure noise, i.e., confusingly noise then has an average signal-to-noise ratio significantly larger than $1$.
After all, it has been shown that the ratio between the values of typically used definitions is approximately constant \cite{GlaserMasterThesis2012, SchroederNoise2010}, which at least enables conversion between different definitions.

\begin{figure}[t]
  \centering
  \includegraphics[width=0.9\linewidth]{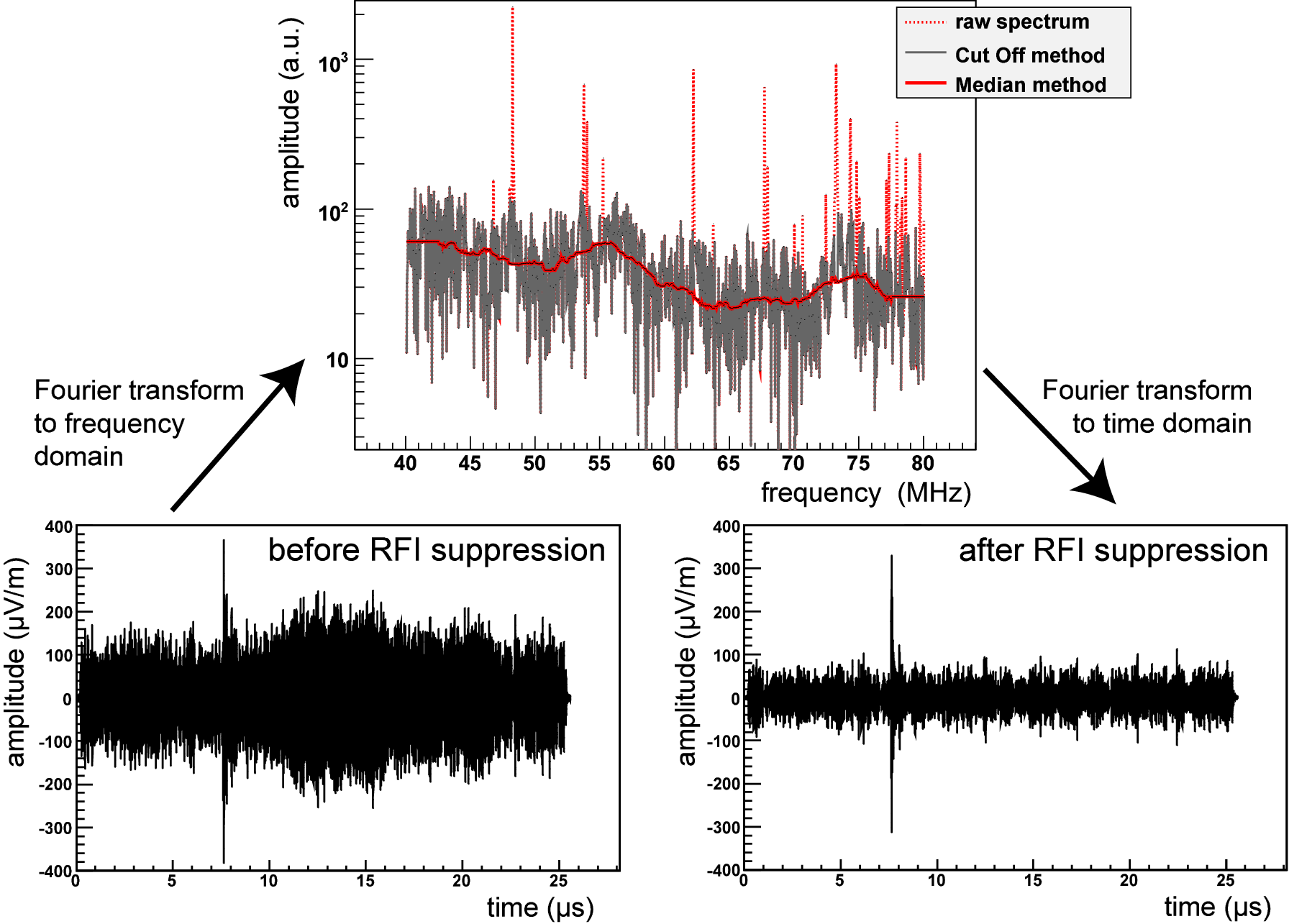}
  \caption{Event measured by LOPES-STAR contaminated by narrow-band radio frequency interferences (RFI), which are filtered in the frequency domain using two different methods:
  the cut off method is the one used by the LOPES standard analysis, the median method now is known for some difficult non-linear features and not recommendable for quantitative data analyses. 
  Similar RFI filters are applied by many experiments during digital data analysis, all of them increasing the signal-to-noise ratio (figure modified from Ref.~\cite{AschPhDThesis2009}).}
  \label{fig_RFIsuppression}
\end{figure}

\subsubsection{Techniques to reduce the influence of background}
Several techniques have been used to reduce the impact of background on air-shower pulses.
In particular anthropogenic radio interferences originating from any kind of radio communication can easily be filtered, since they typically are within narrow frequency bands.
They have a structure different from air-shower pulses, which like any short pulses correspond to a broad frequency spectrum.
Hence, any narrow-band background can easily be distinguished and suppressed by filters, e.g., simple band-stop or notch filters, median filters \cite{KelleyVLVνT2011}, or more sophisticated filter algorithms \cite{RFIfilters_ICRC2013, JansenThesis2016, CorstanjeLOFARtimeCalibration2016}. 
At least for the median filter it has been shown that it does not only improve the signal-to-noise ratio, but also changes the impact of noise on the signal (it invalidates equation (\ref{eq_noiseInfluence}) presented in the next section), which requires some care in its application.
Another easy way to suppress narrow-band interferences is to simply digitally cut them in the frequency spectrum during data analysis (see figure \ref{fig_RFIsuppression}).
However, for all filters it has to be checked in which way they might bias a particular analysis, e.g., by systematically decreasing the amplitude of an air-shower pulse.

Even better noise suppression should be possible with templates, matched filters or any other technique making explicit use of the expected signal shape for air-shower pulses \cite{MallerRICAP2013}. 
This can be useful, in particular for detection of pulses with self-trigger algorithms, but should be applied with care, since the filter could introduce a systematic bias, because the pulse shape depends on the distances to the shower axis and to the shower maximum. 

Finally, interferometric combinations of antennas have been used to reduce the detection threshold at LOPES \cite{FalckeNature2005} and ANITA \cite{ANITA_Interferometery_2015}, in particular cross-correlation beamforming (cf. section \ref{sec_beamforming}). 
This technique exploits that noise is approximately random and uncorrelated in all antennas, while the signal is similar and correlated in all antennas. 
All traces of the individual antennas are shifted in time such that the radio pulses overlap and then are correlated. 
In this way the random noise cancels on average which increases the signal-to-noise ratio.  
The only disadvantages of this technique are that it is computationally expensive, and that it requires very accurate timing of about $1\,$ns relative accuracy for the typical frequency range up to $80\,$MHz \cite{SchroederTimeCalibration2010}. 
Still, as LOPES has shown, the benefit for small-scale arrays featuring at least 10 antennas with signal per event is worth this effort.
For large-scale arrays the situation is not clear, yet: 
first, the improvement of interferometric techniques depends on the number of antennas with signal. 
Second, the shape of the air-shower radio pulse changes with distance to the shower axis, which decreases the degree of signal-correlation between different antennas in case of large antenna spacings. 
Nevertheless, some improvement can be expected by correlating the information contained in several antennas, and this technique is currently under study for larger antenna spacings at AERA.

\begin{figure}[t]
  \centering
  \includegraphics[width=0.45\linewidth]{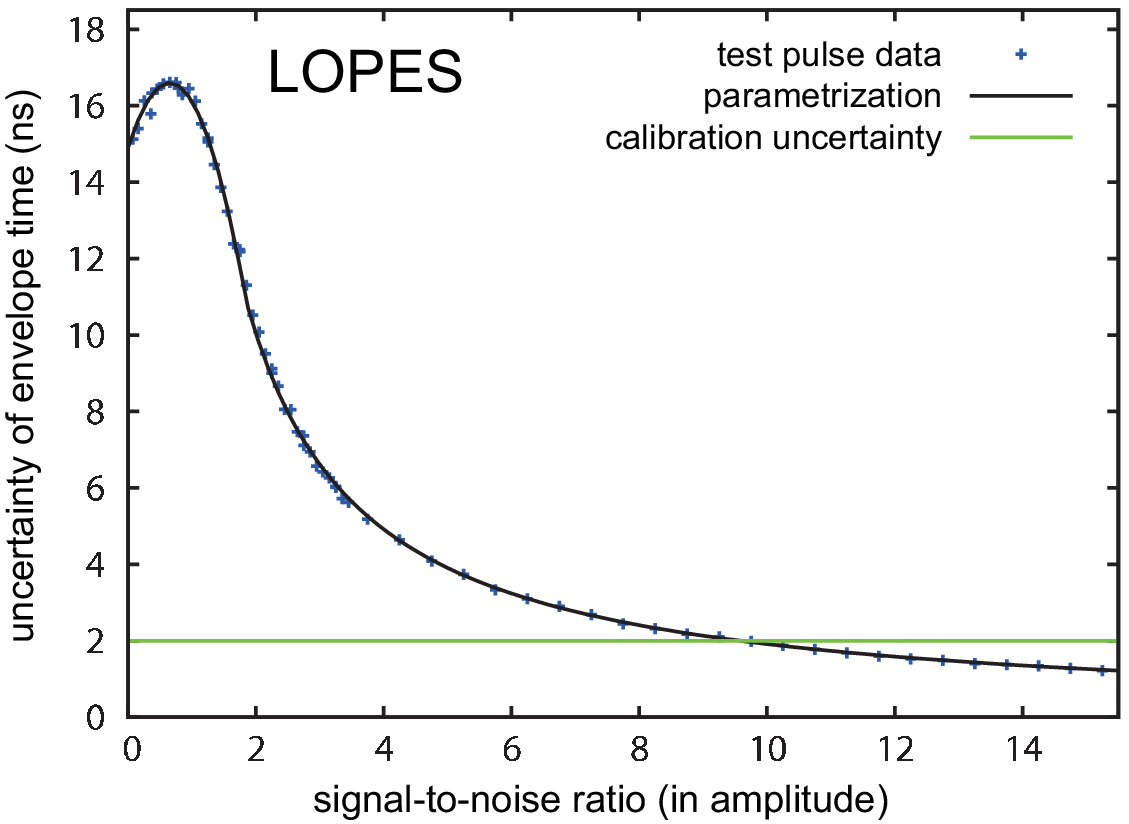}
  \hfill
  \includegraphics[width=0.53\linewidth]{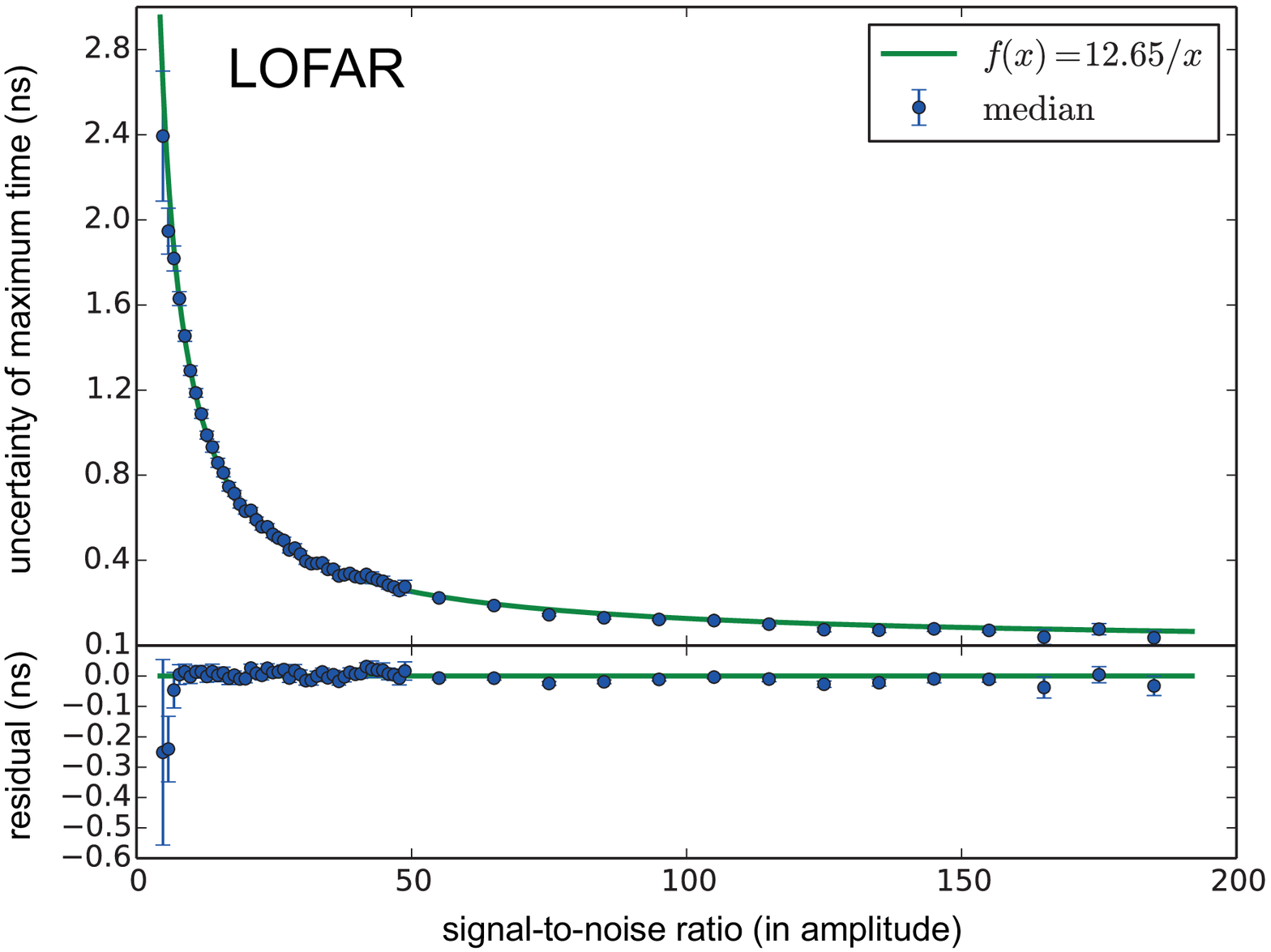}
  \caption{Uncertainty of the maximum time of a radio pulse due to background as a function of the signal-to-noise ratio (left: LOPES \cite{SchroederNoise2010}; right LOFAR \cite{CorstanjeLOFAR_wavefront2014}).}
  \label{fig_timeUncertaintyNoise}
\end{figure}

\begin{figure}[t]
  \centering
  \includegraphics[width=0.45\linewidth]{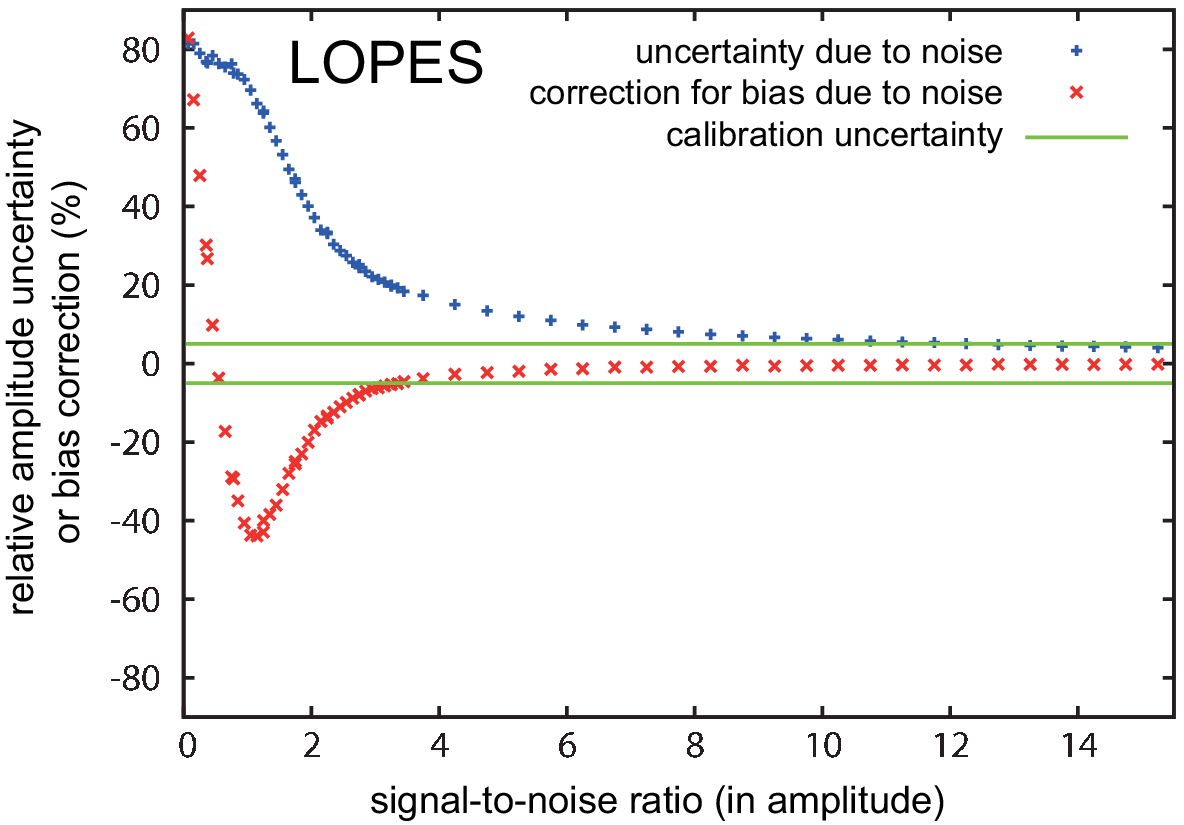}
  \hfill
  \includegraphics[width=0.48\linewidth]{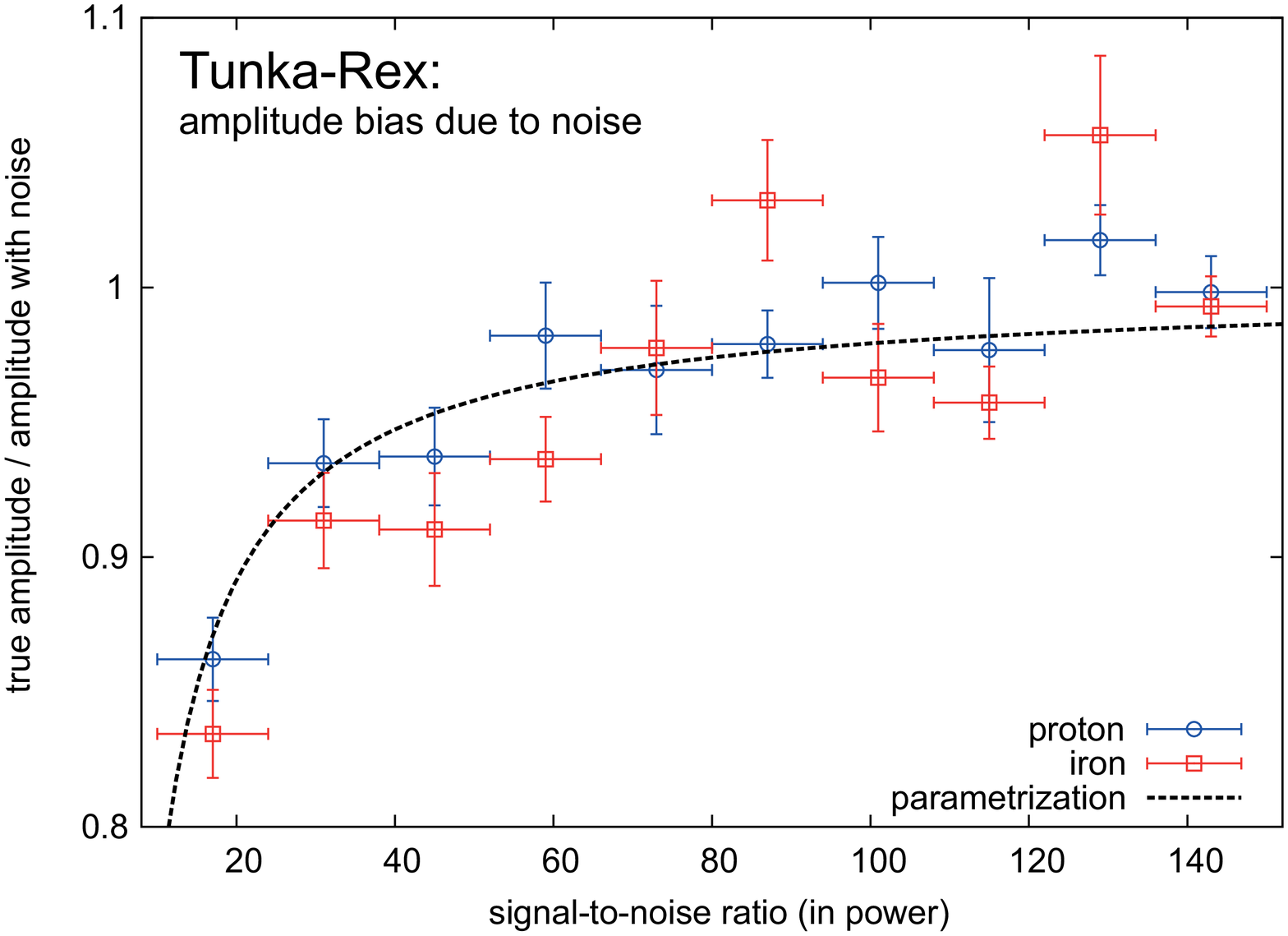}
  \caption{Uncertainty and bias of the pulse height (amplitude) as a function of the signal-to-noise ratio (left: LOPES \cite{SchroederNoise2010}; right Tunka-Rex \cite{TunkaRex_XmaxJCAP2016}).}
  \label{fig_amplitudeUncertaintyNoise}
\end{figure}

\subsubsection{Impact of noise on pulse properties}
\label{sec_noiseInfluence}

Even after the application of reduction techniques, the remaining noise affects the properties of the measured signal. 
Noise causes an uncertainty on the time and amplitude of a radio pulse, and additionally a systematic bias on the amplitude (see figure \ref{fig_timeUncertaintyNoise} and \ref{fig_amplitudeUncertaintyNoise}).
The strengths of these effects depend on the signal-to-noise ratio as well as on the measurement and analysis procedures.

The measurement uncertainty caused by noise has been studied by adding measured noise samples to simulated or generated pulses with pulse shapes similar to real air-shower pulses. 
As expected the uncertainty of both amplitude and time generally decreases with increasing signal-to-noise ratio \cite{SchroederNoise2010, CorstanjeLOFAR_wavefront2014}, except for a potential threshold effect at very low signal-to-noise ratios seen by LOPES, which might be related to the search strategy for pulses hardly distinguishable from noise.
For the uncertainty of the maximum time of the pulse $\sigma_t$, the following relation has been confirmed by LOPES and LOFAR:

\begin{equation}
\sigma_t (SNR) = \mathrm{const} / SNR
\end{equation}
with $SNR$ the signal-to-noise ratio in amplitude.
The constant has been determined to $20.5\,$ns for LOPES, and $12.65\,$ns for LOFAR. 
However, LOFAR and LOPES calculated the $SNR$ in a different way, which leads to $1.67$ times higher values for the case of LOFAR \cite{SchroederPhdThesis}.
Consequently, both results agree within a few percent.
However, before using this formula for estimating the time uncertainty at other experiments, it should first be checked whether it is valid for a particular experiment, since the background and potentially also the method for measuring the pulse time might be different.

For the uncertainty on the amplitude the situation is even less clear, since different experiments used different parameterizations to describe the decrease of the uncertainty with increasing signal-to-noise ratio. 
LOPES used an exponential parametrization.
AERA used the reasonable assumption that the uncertainty of the amplitude (power) is directly proportional to the signal-to-noise ratio in amplitude (power) \cite{GlaserMasterThesis2012}. 
Both assumptions fit well to the data of the experiments.
The latter assumption corresponds to the relation expected when the pulse height is evaluated always at its true position, which means that the relation does not account for the fact that also the pulse maximum is shifted in time by $\sigma_t$.
However, it has not yet been studied how relevant this effect is for the pulse height. 
In any case, for high signal-to-noise ratios the uncertainty due to noise becomes negligible against systematic uncertainties, e.g., due to a small misalignment of the antennas, environmental effects, or calibration uncertainties.

In addition to the uncertainty caused by noise there is a systematic bias: \emph{on average} (and only on average) noise increases the measured pulse amplitude, in contrast to the situation in radio astronomy, where the measured power is \emph{always} increased by noise. 
The reason for this subtle difference is that in usual applications of radio astronomy, the integration period of the measurement is long against the reciprocal of the measurement frequencies. 
Then noise can simply be subtracted by determining the average noise power in a time or space region without signal. 
This is a consequence of energy conservation: the total detected energy is the sum of the signal and noise energies. 
For air-shower detection the typical pulse width is of the same order as the reciprocal of the measurement frequencies, and by chance noise can interfere constructively or destructively in an individual measurement, i.e., increase or decrease the pulse amplitude. 
However, it has been shown for several air-shower experiments that the average influence of noise on the measured signal still corresponds to the expectation of energy conservation \cite{AllanFreqSpecAndNoise1970, SchroederNoise2010, GlaserMasterThesis2012, TunkaRex_XmaxJCAP2016}, i.e., that on average noise increases the pulse amplitude:

\begin{equation}
S^2 + \kappa N^2 = M^2 \pm scatter \, ,
\label{eq_noiseInfluence}
\end{equation}
where $S$ is the true signal amplitude (= field strength), $N$ the noise amplitude, $\kappa$ a normalization factor depending on the way noise is measured (with $\kappa = 1$ when the consistency criterion introduced above is fulfilled), $M$ the measured amplitude, and the scatter reflects the statistical measurement uncertainty due to the possibility of destructive or constructive interference in the individual measurement.

Of course not only the basic pulse properties time and amplitude, but also higher-order observables have uncertainties and potential biases depending on noise.
Furthermore, the exact noise influence depends on the used pipeline (software) for data analysis, in particular the used filters for suppression of narrow-band interferences \cite{TunkaRex_XmaxJCAP2016}, 
and an evaluation of all possible scenarios is beyond the scope of this review. 
Consequently, the noise influence on each target observable has to be studied again for each specific analysis, because it depends on three aspects: the instrument, the external background, and the methods of data processing.

\begin{figure}[t]
  \centering
  \includegraphics[width=0.475\linewidth]{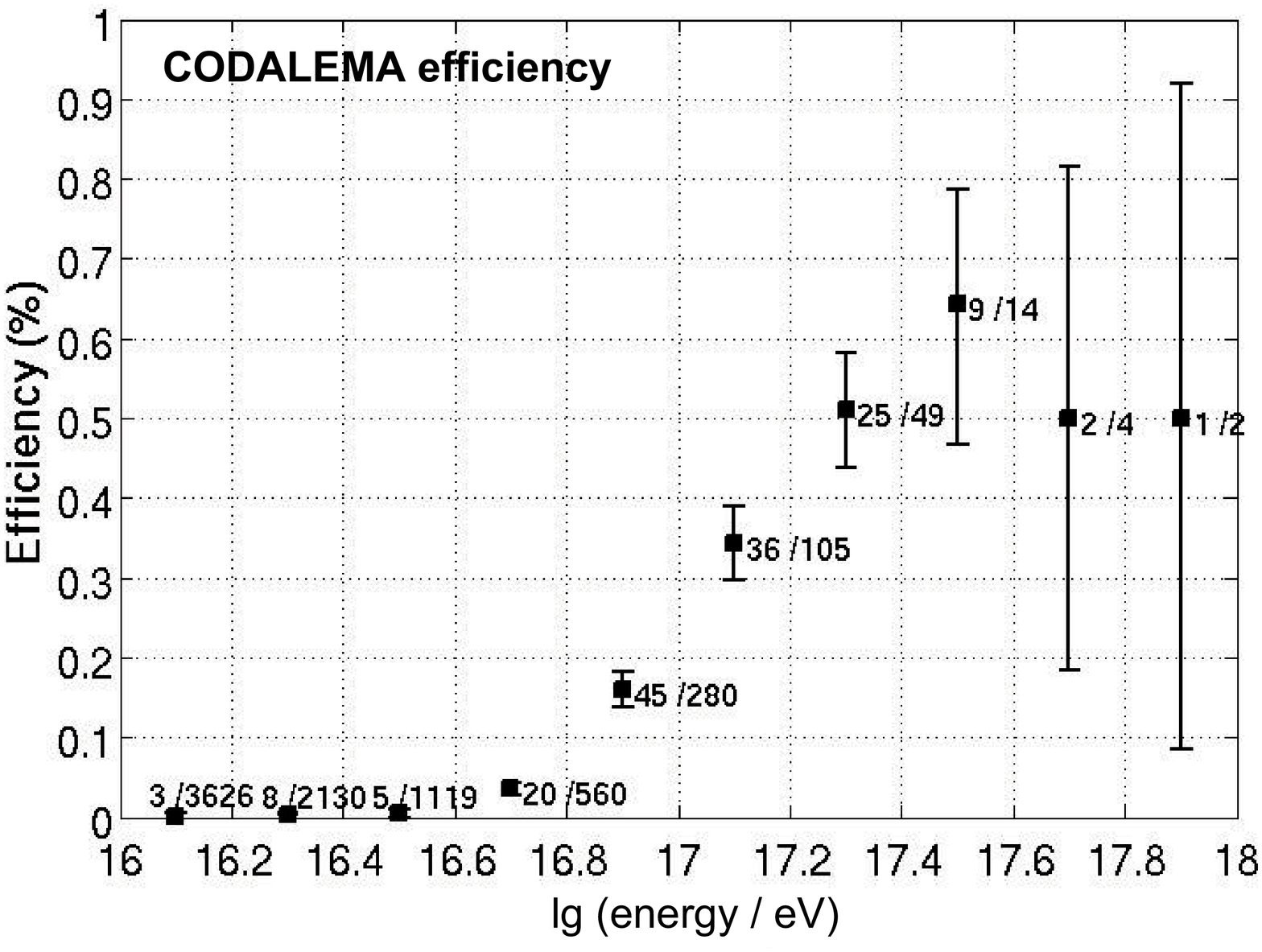}
  \hfill
  \includegraphics[width=0.515\linewidth]{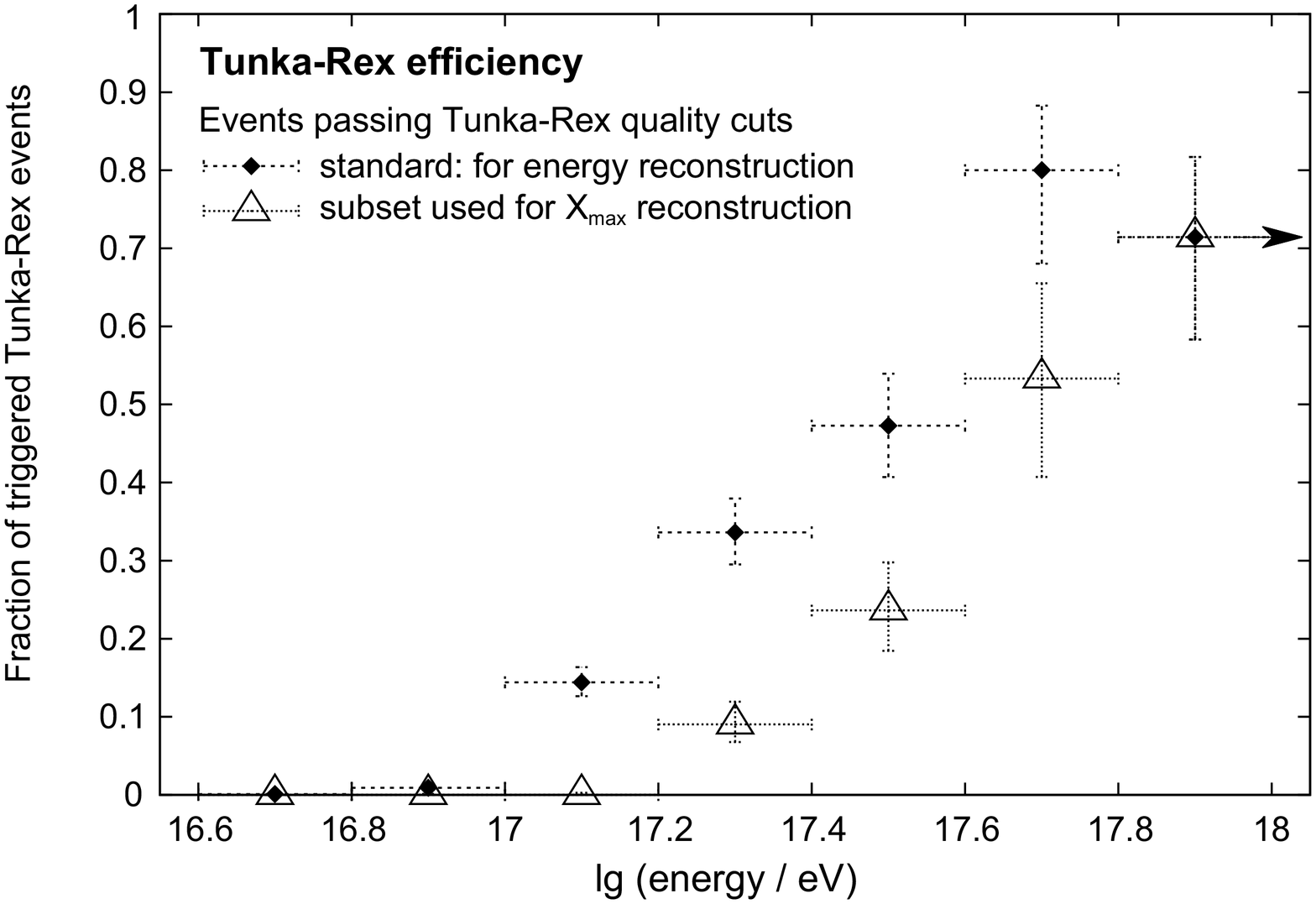} 
  \caption{Efficiency increase with energy of CODALEMA (left, slightly modified from Ref.~\cite{CODALEMA_Geomagnetic}), and Tunka-Rex (right, slightly modified from Ref.~\cite{TunkaRex_XmaxJCAP2016}): 
  for both experiments the fraction of externally triggered radio events passing certain quality cuts is shown; in the case of Tunka-Rex once for weaker quality cuts qualifying a detection of the radio signal and sufficient for energy reconstruction, and additionally for stronger quality cuts used for reconstruction of $X_\mathrm{max}$.}
  \label{fig_efficiency}
\end{figure}

\subsection{Efficiency of air-shower arrays}
The efficiency is the fraction of events fulfilling certain criteria, e.g., sufficient quality for a specific physics analysis.
The basis to determine this fraction are all events in a defined area and range of arrival directions, e.g., the fiducial area of the experiment and the region of the sky which the experiment is sensitive to. 
In our context, the \emph{detection efficiency} is the fraction of all air-shower events in this area and sky region for which a radio signal has been detected. 
\emph{Full efficiency} strictly means that all events are detected, though often a fraction larger than $90\,\%$ or $95\,\%$ is taken as the threshold for full efficiency. 

For \emph{externally triggered} radio arrays, by definition all events are real air showers, but not all recorded events contain a radio pulse of sufficient strength and distinguishable from noise. 
Consequently, there is some freedom in how to define a \lq detection\rq, but most experiments use similar definitions.
Common detection criteria require a minimum signal-to-noise ratio in at least three antennas at the time given by the external air-shower trigger, and additionally the reconstructed arrival direction has to be in agreement with the direction reconstruction of the triggering detector \cite{CODALEMA_Geomagnetic, TunkaRex_NIM2015, AERAenergyPRD}. 
If the triggering detector itself is not fully efficient, then the calculation of the radio-detection efficiency is more complicated and has to take into account that some events, which in principle would fulfill the quality criteria of the radio detector, might not have been triggered.
Nonetheless, the detection threshold of other techniques is typically lower than for the radio technique, such that for most externally-triggered radio arrays the efficiency is simply the fraction of triggered events with detected radio signal (see figure \ref{fig_efficiency} for examples).

For \emph{self-triggering} radio detectors the situation is different, because the detection efficiency is determined by the efficiency of the trigger. 
Thus, for a high detection efficiency first of all a high trigger efficiency is required, which however comes with two types of challenges.
First, high trigger rates require powerful data-acquisition electronics, high communication bandwidths, and large buffer memories, which all increase the cost of an experiment. 
Second, a high efficiency typically goes along with a poor purity because of false positive detections due to noise fluctuations or background pulses. 
All these problems can be avoided by an external trigger, which in my opinion is the preferred option for any radio array operated together with other detection techniques. 

Nevertheless, significant research has been done on the development of self-triggers searching for an optimum compromise between efficiency and purity \cite{MallerRICAP2013, AERA_selftriggerIEEE2012}. 
Self-trigger strategies involve background suppression and filters matched to the expected properties of air-shower radio pulses. 
Moreover, false-positive detections can be recognized to a large extent in later offline analysis, e.g., by clustering analyses, because background pulses often are correlated in time and location \cite{ANITA_CR_PRL_2010}.
Results by TREND \cite{TREND_Astropp2011}, by an AERA prototype setup \cite{RAugerSelfTrigger2012}, by CODALEMA \cite{CODALEMA_selftriggerECRS2012}, and by ANITA \cite{ANITA_CR_PRL_2010} show that the problem can be solved in principle.
Still it is not clear whether self-triggered detection can provide a similar energy threshold as external triggering, and if yes, what the effort in terms of cost, power consumption, communication, and computing resources will be.

Independent of if externally or self-triggered, the lowest detectable energy and the threshold for full efficiency depend on the instrumental properties like antenna spacing, frequency band and the used analysis technique. 
For example cross-correlation beamforming allows for significantly lower thresholds when the radio signal is recorded by several antennas \cite{FalckeNature2005, HuegeFalcke2003, ANITA_Interferometery_2015, PhasedArraysForNeutrinos_ICRC2015}. 
Furthermore, the threshold also depends on zenith angle because the footprint gets larger and fainter for inclined showers, and on the geomagnetic angle because for showers parallel to the geomagnetic field only the weaker Askaryan emission is detectable. 
So far no experiment has sufficient statistics to experimentally check at which energy full efficiency is reached for this case, but this has been studied with simulations. 
Reference \cite{HillerThesis2016} describes a simple model developed for Tunka-Rex and transferable to other antenna arrays for estimation of the efficiency as a function of zenith angle and energy: 
for experiments having the geomagnetic field direction in the field of view, the efficiency-versus-energy curve rises relatively slowly compared to other detection techniques, like particle or air-Cherenkov detection, and full efficiency is achieved at an energy several ten times higher than the threshold energy for first detection. 
Consequently, air-shower arrays like LOPES, LOFAR, CODALEMA, AERA, or Tunka-Rex have detected few events at energies even below $10^{17}\,$eV, and full efficiency for all arrival directions is expected only at energies above $10^{18}\,$eV. 

\begin{figure}[t]
  \centering
  \includegraphics[width=0.99\linewidth]{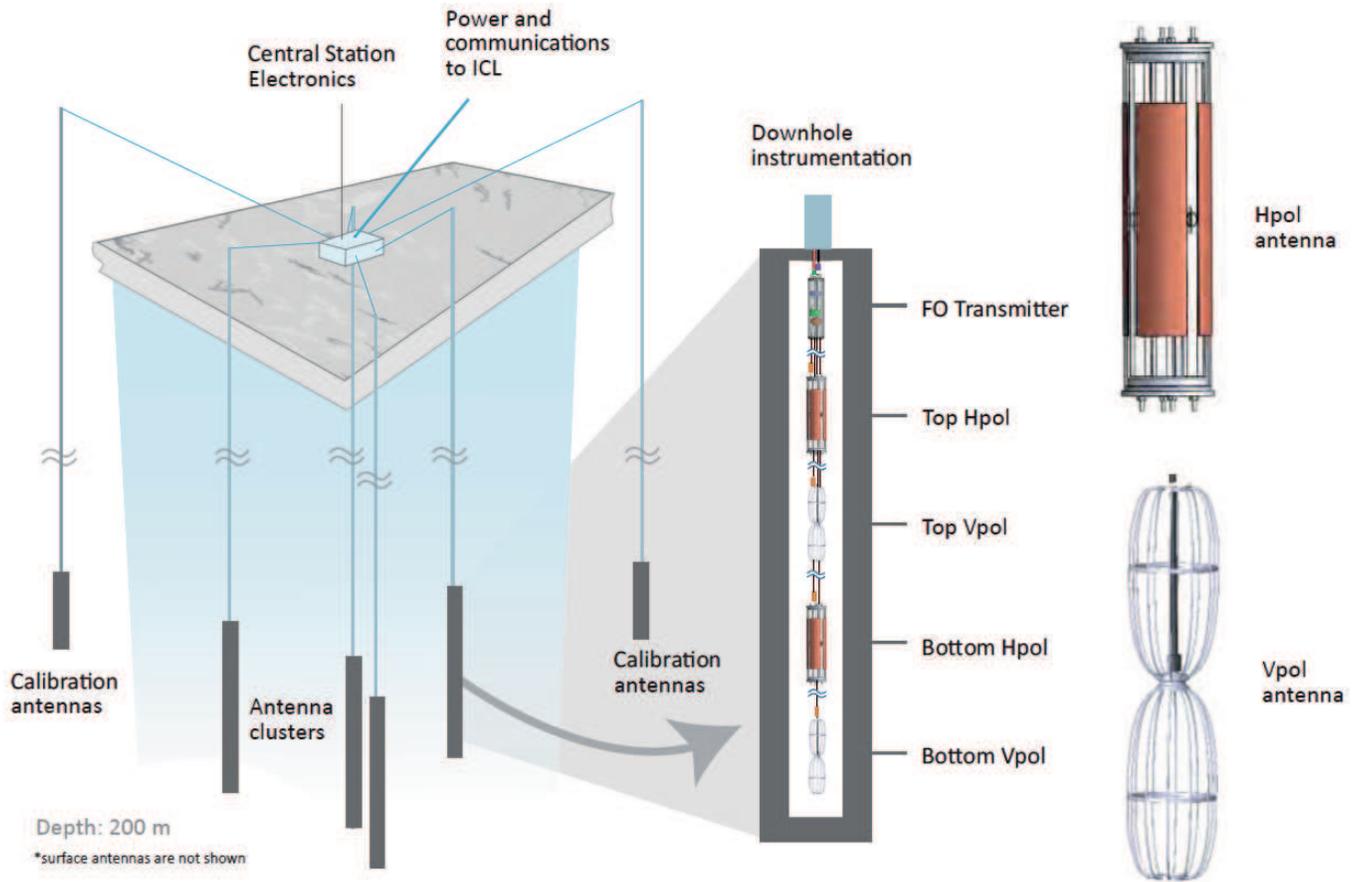}
  \caption{Sketch of one ARA station consisting of four antenna strings each with two vertically and two horizontally polarized antennas in the ice at the South Pole (from Ref.~\cite{ARAsketchICRC2015}).}
  \label{fig_ARAsketch}
\end{figure}

\subsection{Detector concepts in dense media}
There are two main strategies for the detection of radio pulses emitted by showers in dense media. 
Both have in common that they aim at detecting the radio emission close to the Cherenkov angle where the frequency spectrum has a peak at GHz frequencies. 
Therefore detector concepts for dense media typically choose a frequency band between a few $100\,$MHz and a few GHz, in which Galactic noise is negligible against thermal noise.
This enables higher signal-to-noise ratios and slightly lower detection thresholds for dense media compared to air-shower observations typically done below $100\,$MHz. 

The first concept is the observation of the medium from a distant point.
This is the concept of lunar observations with radio telescopes \cite{BrayReview2016, SKAlunar_ICRC2015}, of balloon-borne experiments like ANITA when searching for neutrinos \cite{ANITAupwardEvent2016}, and of future space missions orbiting the Earth \cite{SWORDarxiv2013} or the Moon \cite{Stal2007, GusevLORD2014}.
The main difficulty is to distinguishing background pulses of natural and anthropogenic origin from real cosmic-ray or neutrino events.
ANITA has successfully developed two techniques for this purpose \cite{ANITA_neutrino_PRD_2012, ANITA_CR_2016}. 
First, in contrast to high-energy particles arriving more or less randomly at the Earth, most background events are clustered, i.e., several pulses are received from the same location and can be excluded from analysis. 
Second, the distinct radial polarization of the Askaryan effect can be used as consistency check by comparing the arrival direction of the pulse with the orientation of the measured electric-field vector. 
For lunar observations the main method of background rejecting is the comparison of different beams pointing towards and near the Moon at the same time \cite{BrayReview2016}. 
Any background is assumed to have similar characteristics in all beams pointing at approximately the same direction, but the pulse of a real cosmic-ray or neutrino event should only be visible in the one beam pointing exactly at its origin in the lunar regolith.
Last doubts whether this method works reliable should be ruled out by lunar observations with the planned SKA \cite{SKAlunar_ICRC2015}. 
As consistency check the cosmic-ray flux per energy measured with the SKA can be compared to the one measured by established air-shower arrays such as the Telescope Array or the Pierre Auger Observatory.
  
\begin{figure}[p]
  \centering
  \includegraphics[width=0.7\linewidth]{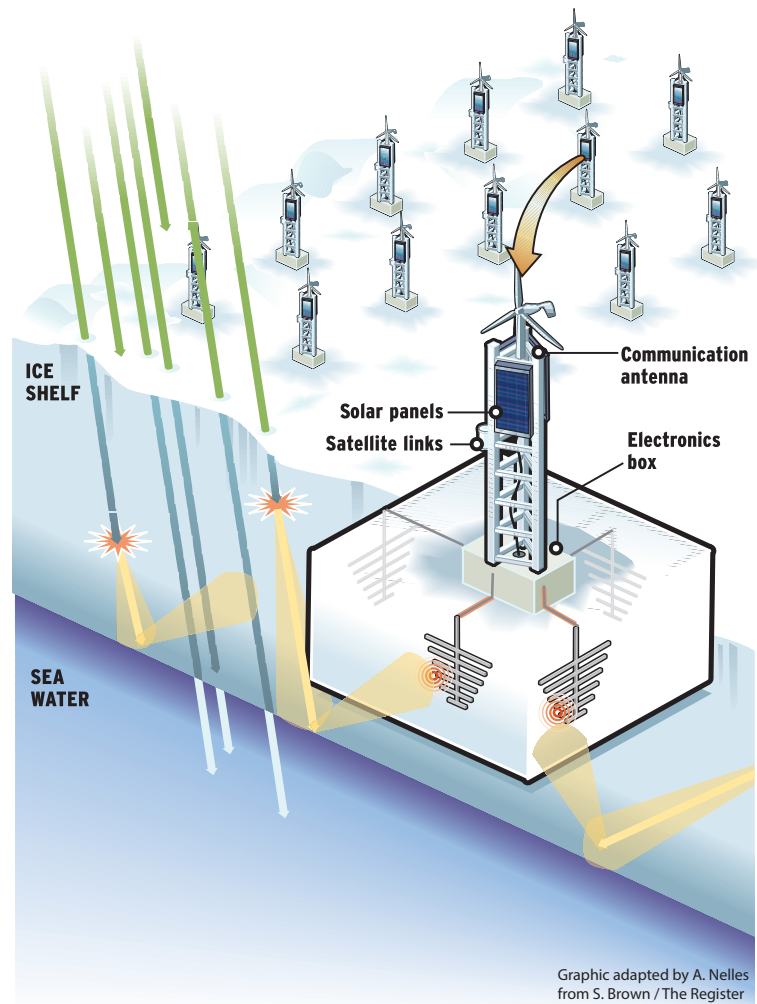}
  \caption{Sketch of the ARIANNA experiment consisting of autonomous stations on the Ross Ice Shelf which detected the radio signal of air showers both directly and after reflection on the ice-water border and search for the radio signal emitted by neutrino-induced showers in the ice (figure provided by the ARIANNA Collaboration).}
  \label{fig_ARIANNAsketch}
\end{figure}

The second concept is the deployment of dedicated antenna arrays in the medium where the radio signal is emitted or on its surface (see figures \ref{fig_ARAsketch} and \ref{fig_ARIANNAsketch}).
This strategy is followed by the in-ice arrays ARA \cite{ARA_2016} and GNO \cite{GreenlandInIce_2016}, and by the on-ice array ARIANNA \cite{ARIANNA_2015}.
Provided sufficiently dense antenna spacing, three-dimensional arrays consisting of antenna strings in the ice can determine whether a signal was generated inside of the instrumented volume or whether it comes from outside. 
By this, neutrino induced showers in the ice can be distinguished from air-showers or background pulses emitted by human activities at the surface.  
The in-ice array ARA features 4 nearby strings per station which enables the use of interferometric beamforming techniques for lowering the detection threshold and for better discrimination of in-ice showers against background from the surface \cite{ARA_2012}.

The advantage of the on-ice array ARIANNA is the cheaper deployment, because no deep holes have to be drilled in the ice, but stations consisting of several antennas each are deployed on the surface burying the antennas only a few meters deep in the snow. 
Since ARIANNA is being built on the floating Ross Ice Shelf it will detect the direct upward emission by neutrino-induced showers as well as the radio signal reflected at the ice-water surface.
This surface reflects also the radio emission by air showers back to the ARIANNA detector stations, what has to be discriminated against neutrino signals.
Upward pointing antennas at the surface can be used for this purpose, because only for atmospheric showers there is radio emission coming from the sky in addition to the reflected signal from below.

In summary, since all radio arrays using ice as medium are still under construction and none of them has detected neutrinos yet, it is too early to draw conclusions on whether the on-ice or the in-ice technique will be more successful.

\clearpage
 
\section{Reconstruction of air-shower properties}
\label{sec_reconstruction}
The application of radio measurements for cosmic-ray science requires the reconstruction of air-shower parameters from the measured radio signal. 
In particular, the radio signal depends on the position and direction of the shower axis, the energy of the electromagnetic shower component, and the distance to the shower maximum, which can be exploited to determine these shower parameters. 
This section explains the status and achievements made by several experiments with a particular focus on the available methods and their performance for the various air-shower parameters\footnote{The further improvement of the introduced methods, and the development of new methods is work in progress for several experiments, such that the latest status is sometimes only available in internal notes, PhD and master theses, but not yet in journal papers.}. 

In a subsequent step these air-shower parameters have to be interpreted to derive the properties of the primary cosmic-ray particles: 
While the shower axis is just the continuation of the arrival direction of the primary particle, the energy of the primary cosmic-ray particle has to be estimated from the energy of the electromagnetic shower component. 
At least statistically the mass composition can be derived from the measured positions of the shower maximum, but shower-to-shower fluctuations and measurement uncertainties are too large to reconstruct the type of the primary particle for an individual air-shower measurement.
These are general problems of all air-shower experiments, not only of radio detectors. 
However, other detection techniques operated jointly with radio arrays can provide complementary information increasing the total accuracy for the properties of the primary particle, especially the combination of radio and muon measurements is under study for this purpose \cite{Holt_TAUP}.
Since a complete elaboration of this transition from air-shower observables to the properties of the primary particle would be beyond the scope of this review, this section focuses on the reconstruction of the air-shower parameters themselves and only sketches some radio-related issues relevant for the determination of the primary cosmic-ray energy and mass composition.

\subsection{Geometry: shower axis}
Due to momentum conservation, the shower develops in the direction of the primary particle. 
Thus, the direction of the shower axis is approximately equal to the arrival direction of the primary particle, and there is no need to distinguish. 
While the direction of the shower axis is a parameter of physical interest, e.g., for cosmic-ray anisotropy studies \cite{AugerAnisotropyChemComp2011}, the point of intersection with a defined ground plane (= shower core) is a technical parameter mostly important for an accurate reconstruction of other observables. 

\begin{figure}[t]
  \centering
  \includegraphics[width=0.5\linewidth]{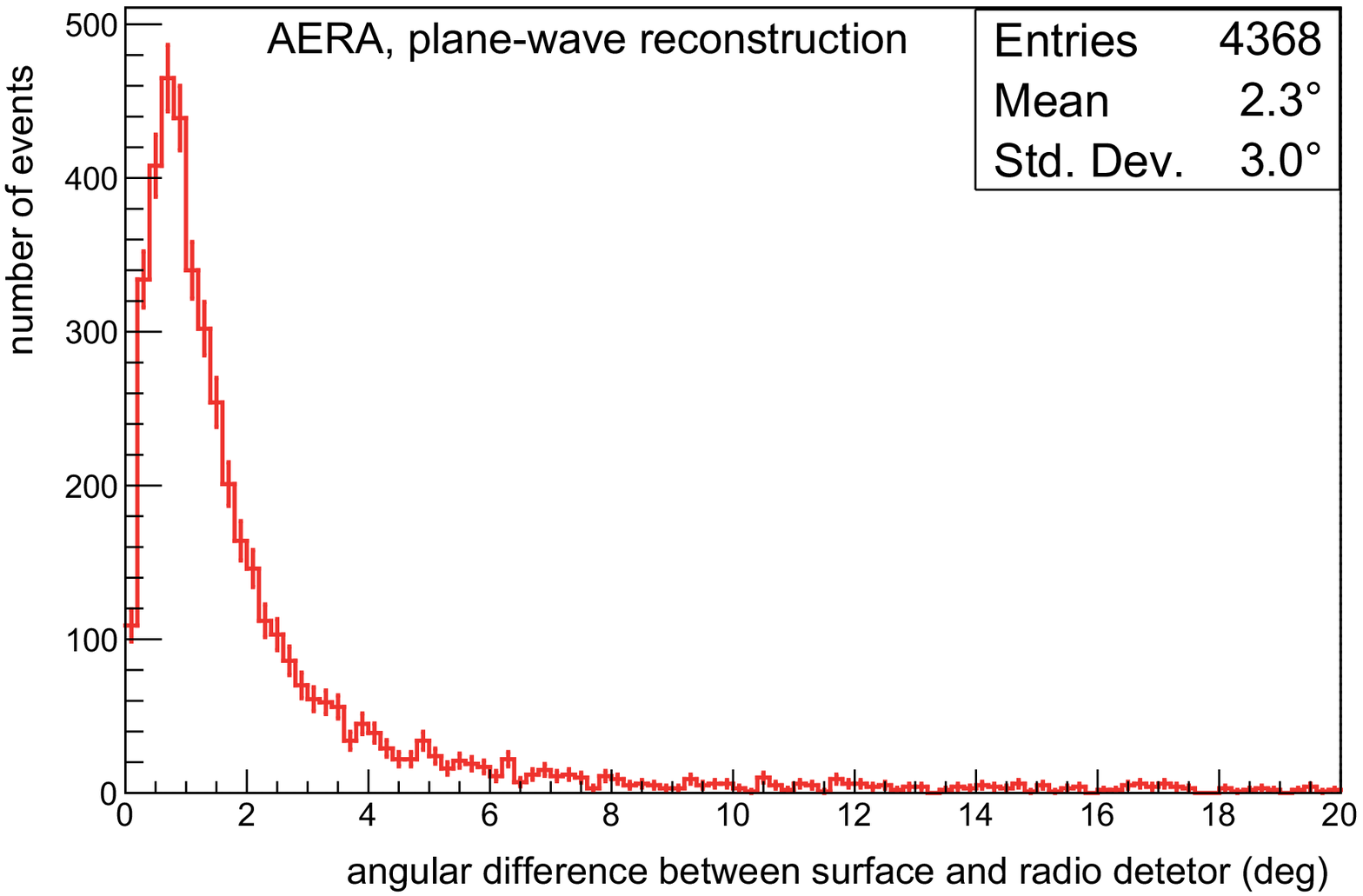}
  \hfill
  \includegraphics[width=0.48\linewidth]{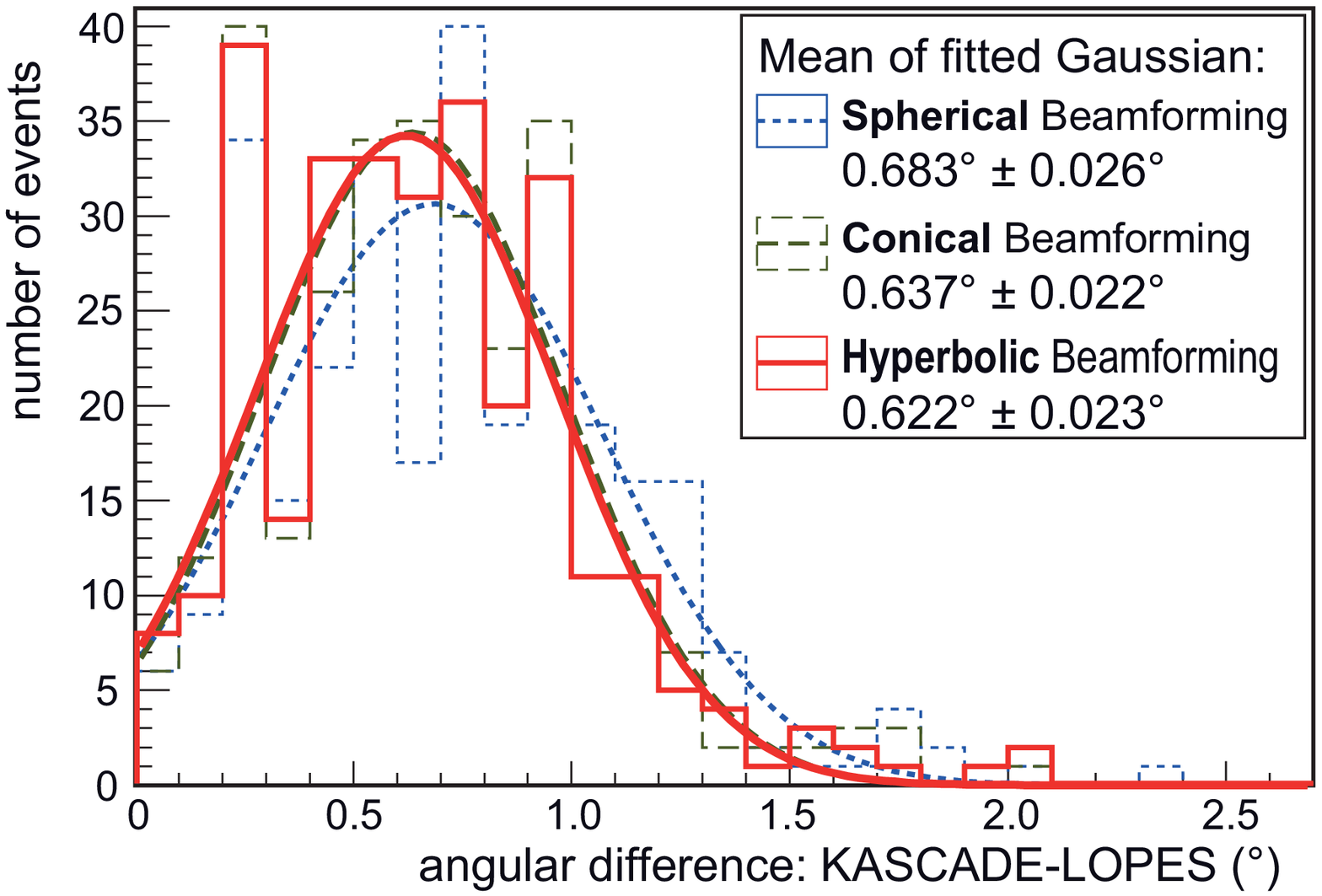} 
  \caption{Difference between the directions reconstructed by a radio detector and the co-located particle detector arrays using different models for the radio wavefront (left AERA, slightly modified from Ref.~\cite{QaderECRS2014}, right LOPES, slightly modified from Ref.~\cite{LOPESwavefront2014}).}
  \label{fig_angularResolution}
\end{figure}

\subsubsection{Direction}
The shower direction can be reconstructed from the arrival time of the radio signal in the individual detector stations by fitting a wavefront model (cf.~section \ref{sec_wavefront}), where simple triangulation corresponds to a plane-wave model. 
Since all wavefront models used so far are azimuthally symmetric around the shower axis, the direction of the shower axis is reconstructed implicitly when determining the shape of the radio wavefront. 

The accuracy achievable with a plane wave model is about $2^\circ$ \cite{TunkaRex_NIM2015, QaderECRS2014}, which already might be sufficient for many applications in cosmic-ray physics. 
The best accuracy has been achieved using a hyperbolic wavefront model, which is significantly better than with a plane-wave model, and slightly better than for conical or spherical models (see figure \ref{fig_angularResolution}).
This holds at least for the small LOPES array \cite{LOPESwavefront2014}, which has been compared to the KASCADE particle detector array featuring an angular resolution of $0.1^\circ$ at high energies \cite{KASCADE_2003}. 
The deviation between the LOPES and the KASCADE directions provides an upper limit for the absolute accuracy: the radio reconstruction of the shower direction is better than $0.7^\circ$, which currently is the best radio accuracy for air showers independently cross-checked by another detector. 
It is difficult to conclude to which extent LOPES is better because of the better wavefront model, and to which extent because of the cross-correlation beamforming technique instead of a simple arrival-time fit. 
My personal assumption is that the wavefront model is crucial, since the angle between the wavefront and the shower plane typically is larger than a degree.
Consequently, sub-degree resolution should be impossible with simple triangulation or a plane-wave reconstruction.
Moreover, for a given antenna array the direction accuracy for air showers can be significantly worse than the pointing resolution for plane-wave radio signals from distant astronomical sources. 
Then in the sub-degree range it seems that other uncertainties, e.g., due to noise and time calibration, become important and can dominate the achievable direction accuracy even when a hyperbolic wavefront is used (cf.~section \ref{sec_noiseInfluence}).

LOFAR, the radio array leading in timing accuracy features a direction reconstruction of air showers self-consistent to $0.1^\circ$ \cite{CorstanjeLOFAR_wavefront2014}.
However, no firm conclusion on the direction accuracy can be made from this value, since there could be a systematic offset to the true direction of the air shower. 
Such a systematic shift would indeed be expected from the small asymmetry of the radio wavefront, which is not yet investigated deeply. 
If this is understood, maybe even higher accuracies will be achievable with next-generation arrays like GRAND or SKA. 
While unimportant for cosmic-ray physics, high direction accuracy is crucial for the detection of photons or neutrinos, which are not deflected by magnetic fields on their way to Earth.

\subsubsection{Shower core}
Although practically every experiment published results on the direction reconstruction, there seem to be no publications on the reconstruction of the shower core (location where the shower axis hits the ground).
Thus, no quantitative conclusions can be made on the achievable accuracy, but several ideas and methods have been discussed in the community. 

Since all properties of the radio signal measured by a detector depend on the distance to the shower axis, in principle every quantity can be used to determine the shower core:
\begin{itemize}
 \item Wavefront: The asymptotic cone of the hyperbolic wavefront points directly to the shower axis, at least when correcting for the small asymmetry of the wavefront \cite{LOPESwavefront2014}. 
 The vertex of the cone only has a slight offset not more than a few meters to the shower plane.
 In principle this could provide an accuracy for the core position of the order of a meter, provided sufficient measurement accuracy of the arrival time in individual antenna stations. 
 \item Footprint: The Cherenkov ring visible in the footprint is centered around the shower axis \cite{LOFARcherenkovRing2014}, and the core position is one of the free parameters when fitting a lateral-distribution function to the measured amplitudes at different positions \cite{NellesLOFAR_LDF2014} - a method already used by AERA \cite{AERAenergyPRD}. 
 Moreover, simulated radio footprints can be matched with the measured one to determine the core position \cite{BuitinkLOFAR_Xmax2014}.
 \item Frequency spectrum: The slope of the frequency spectrum measured in an individual antenna depends on the distance to the shower axis, but also on other quantities like the distance to the shower maximum \cite{Grebe_ARENA2012}. 
 \item Polarization: The polarization of the radio signal originating from the Askaryan effect points directly to the shower core. 
 This can provide a method for events with high signal-to-noise ratio, for which the deviation between the measured polarization and the polarization expected from the dominant geomagnetic effect is really due to the sub-dominant Askaryan effect and not due to background.
\end{itemize}

The performance of all these methods and their combinations will have to be investigated in the future, and there is no obvious reason why the radio reconstruction of the core should not be as good as the reconstruction by particle or air-Cherenkov detectors.

\subsection{Energy}

\subsubsection{General considerations}

Energy conservation requires that the total energy contained in the air shower equals the energy of the primary particle. 
The air-shower energy is the sum of the energies of all shower components, in particular the muonic and the electromagnetic shower components, but also components practically invisible, like neutrinos. 
Thus, the most accurate reconstruction of the complete shower energy is possible when the sizes of the muonic and electromagnetic shower components are determined at a certain stage of shower development, e.g., at the shower maximum \cite{Matthews2005}, where the exact relation depends on the hadronic interaction models \cite{KASCADEGrandeHadronicModels2014}.  
This means that in addition to a muon detector at the surface or underground, a measurement of the electromagnetic component is required, which can be by electron detectors on ground \cite{KASCADEGrandeEnergySpectrum2012}, or indirectly by the measurement of electromagnetic radiation emitted by the shower. 
Moreover, the distance to the shower maximum has to be known in order to properly interpret the measured sizes of the muonic and electromagnetic components. 

Radio detection has an intrinsic advantage here because the radio signal is always created in the region around the shower maximum, with the strongest emission originating from slightly before the shower maximum \cite{GlaserShortAuthor2016}.
Thus, the radio signal provides a measure for the size of the electromagnetic component approximately at the shower maximum. 
This means that except for clipped showers whose shower maximum is very close to or under ground, the energy in the radio signal does not depend on the shower age \cite{AERAenergyPRD}. 
Therefore, the distance to the shower maximum is only required to accurately interpret the number of muons. 
Again, radio detection is of advantage: as explained in the next section, the radio technique provides several methods to measure this distance to the shower maximum.
Hence, the combination of radio and muon detectors seems to be ideal for the purpose of energy reconstruction. 
No results of such combined methods have been published yet, but they are investigated at the Pierre Auger Observatory \cite{Holt_TAUP} and at Tunka \cite{TunkaRex_NIM2015}, which both operate muon and radio detectors together. 

Current efforts concentrate on reconstructing the energy of the electromagnetic shower component from the radio measurement alone.
Then the correlation of the energy in the electromagnetic shower component with the total energy is used to estimate the energy of the primary particle, similar to the technique used for fluorescence  detection \cite{AugerFDinstrument2010}. 
The fraction of air-shower energy going into the electromagnetic component statistically depends on the mass of the primary particle.
The difference between the extreme cases of protons and iron nuclei as primary particles is about $10\,\%$ for primary energies around $10^{17}\,$eV \cite{EngelReview2011}.
This translates into a corresponding composition-dependent systematic uncertainty of the primary energy, which can be reduced, but not eliminated when estimating the average mass composition from the available measurements. 
Despite that $10\,\%$ systematic uncertainty, several radio experiments already achieve energy precisions and scale accuracies of better than $20\,\%$, both of which are comparable to other detection techniques. 

Two different methods have been used to reconstruct the energy from the signal strength of the radio emission, and both seem to provide roughly the same accuracy: 
The first method determines the total radiation energy by integration of the recorded power over the footprint and over the duration of the radio pulse. 
The second method uses the radio amplitude at a detector-specific reference distance from the shower axis.

\begin{figure}[t]
  \centering
  \includegraphics[width=0.5\linewidth]{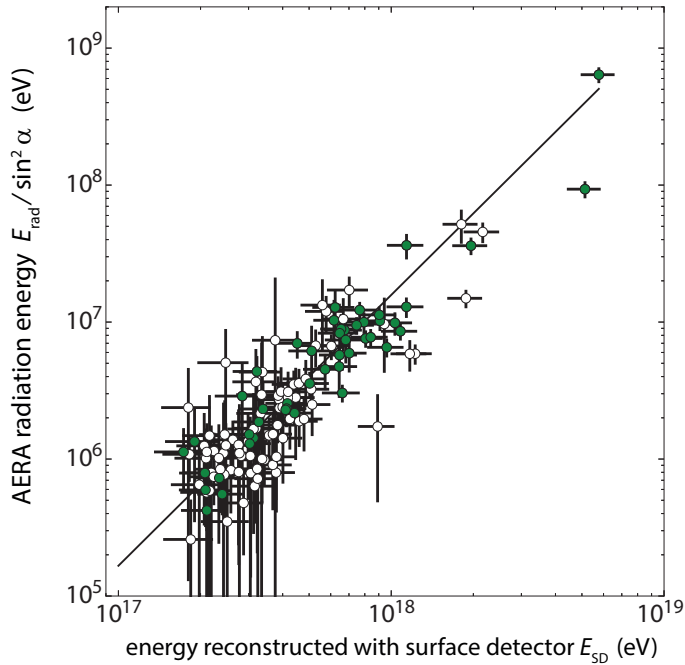}
  \caption{The reconstructed radiation energy of AERA events scales quadratically with the energy of the primary particle reconstructed by the Auger surface detector (slightly modified from Ref.~\cite{GlaserEnergyICRC2015}).}
  \label{fig_AERAenergyCorrelation}
\end{figure}

\subsubsection{Integrating over the footprint}
Because absorption is negligible for radio frequencies in the atmosphere, the radiation energy emitted by the shower equals the radiation energy passing any sufficiently large (compared to the wavelength) and closed surface around the shower. 
Since the radiation is relativistically beamed in a small cone, in fact an open surface is adequate for integration, as long as it intersects the full radiation cone. 
This surface can be the ground plane of the detector, or the shower plane perpendicular to the shower axis. 
The surface and time integral of the radiation power yields the total radiation energy of the shower, i.e., the energy in the radio signal emitted by the electromagnetic shower component \cite{AERAenergyPRL, AERAenergyPRD}. 

Due to the coherent nature of the radio emission, the field strength of the radio signal is proportional to the number of electrons and positrons contributing to the emission, which is proportional to the energy of the electromagnetic shower component.
Therefore, the radiation energy scales quadratically with the field strength and with the energy of the electromagnetic component, where small deviations from an exact quadratic scaling are indicated by recent simulations studies \cite{GlaserAERAenergyARENA2012} and accelerator measurements \cite{SLAC_T510_PRL2016}. 
The approximately quadratic scaling has been confirmed experimentally by the Pierre Auger Observatory (see figure \ref{fig_AERAenergyCorrelation}) with the result that a shower of total energy of $1\,$EeV emits on average $16\,$MeV energy in the frequency band of AERA, if perpendicular to the geomagnetic field of $24\,$\textmu T at that site.
This results in the following formula for the radiation energy, which is expected to be independent of the integration method and independent of all detector properties, except of the frequency band:

\begin{equation}
 E_\mathrm{rad}(30-80\,\mathrm{MHz}) = \big(15.8 \pm 0.7 (\mathrm{stat}) \pm 6.7 (\mathrm{syst})\big)\,\mathrm{MeV} \cdot \bigg( \sin(\alpha) \frac{E_\mathrm{shower}}{\mathrm{EeV}} \frac{B_\mathrm{geo}}{24\,\textrm{\textmu T}} \bigg)^2
\end{equation}

with $E_\mathrm{rad}(30-80\,\mathrm{MHz})$ the radiation energy in the AERA frequency band, $\alpha$ the angle between the shower axis and the geomagnetic field, $E_\mathrm{shower}$ the total energy of the shower, which is assumed equal to the energy of the primary particle, and $B_\mathrm{geo}$ the strength of the geomagnetic field. 
The geomagnetic scaling in this formula neglects that a part of the radiation energy originates from the Askaryan effect, which could be taken into account by replacing the term $\sin^2 \alpha$ by $(\epsilon + \sin \alpha)^2$ with $\epsilon$ the relative strength of the Askaryan effect (cf.~section \ref{sec_emissionMechanisms}). 
However, due to the quadratic relation, the error by neglecting the Askaryan effect is small against the measurement uncertainties for most geomagnetic angles $\alpha$.

The integration over the complete footprint brings the principle advantage that the method is independent of the observation altitude, since absorption of radio waves is negligible in the atmosphere. 
Different technical methods for the integration have not yet been compared.
They should not matter for the theoretical principle, but of course might influence the experimentally achievable energy precision, since different methods can be differently sensitive to background or systematic uncertainties. 
For the AERA events a two-dimensional lateral distribution function has been fit to the individual measurements (cf.~section \ref{sec_footprint}), and then the integral of this function has been used. 
With this approach an energy precision of $17\,\%$ has been achieved, improving with the number of stations contributing to the event \cite{AERAenergyPRD}. 
Cross-correlation beamforming provides an implicit way of integration, but the correlation of the beam amplitude on the total radiation energy is not easy to understand, since it depends in a non-trivial way not only on the amplitude, but also on the pulse shape in each antenna. 
Still a comparison of LOPES and KASCADE-Grande could show that the energy precision achievable by cross-correlation beamforming is at least as good as the precision of the KASCADE-Grande particle detector array, which is about $20\,\%$ \cite{SchroederLOPESsummaryARENA2012}.
For arrays with events featuring many antennas with detected signal an alternative could be to determine the integral of the radio footprint directly, without fitting a function.
This might be a general option for inclined air-showers with huge radio footprints, or for all air-showers when measured by very dense arrays like SKA. 

\begin{figure}[t]
  \centering
  \includegraphics[width=0.48\linewidth]{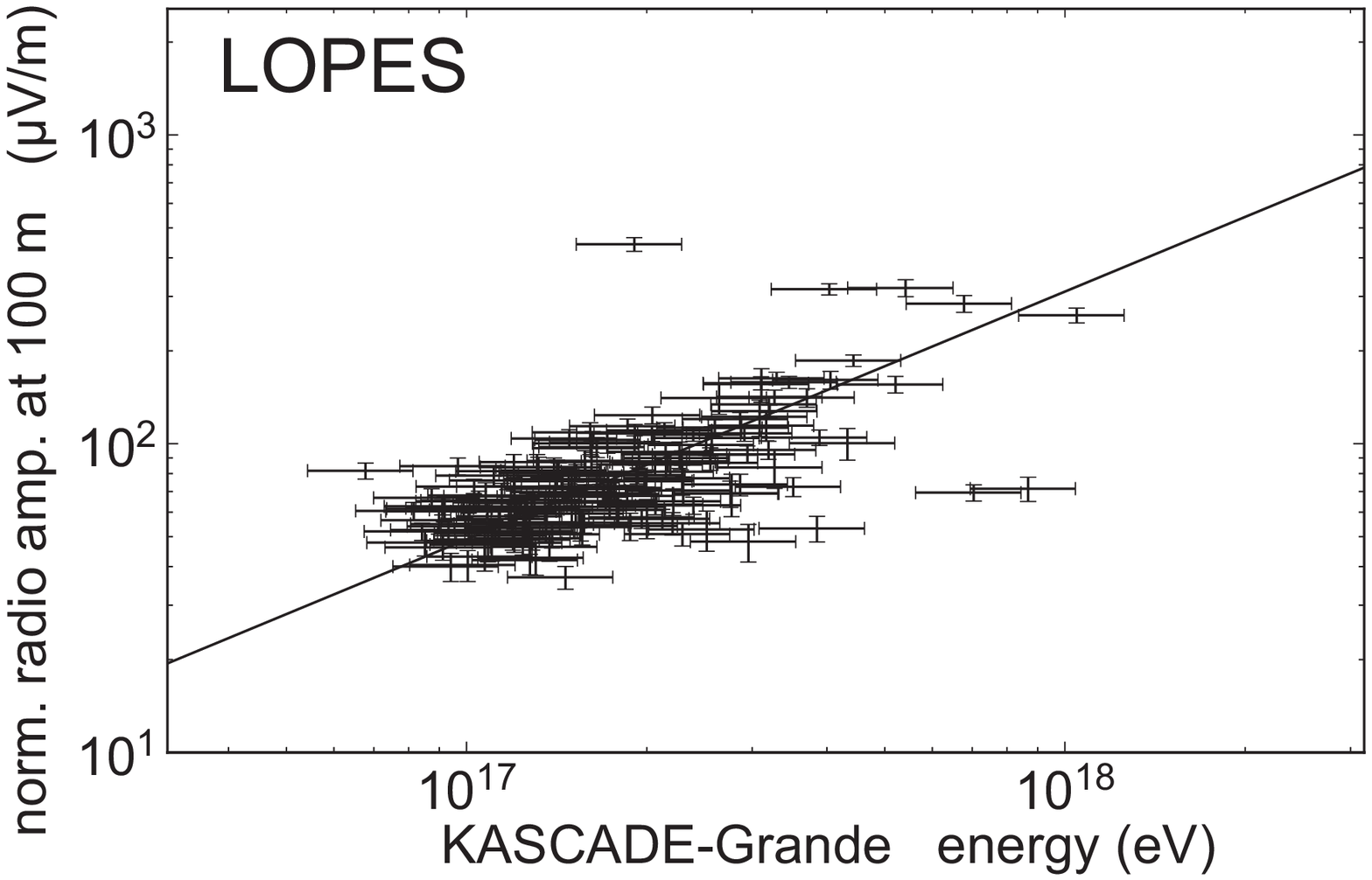}
  \hfill
  \includegraphics[width=0.48\linewidth]{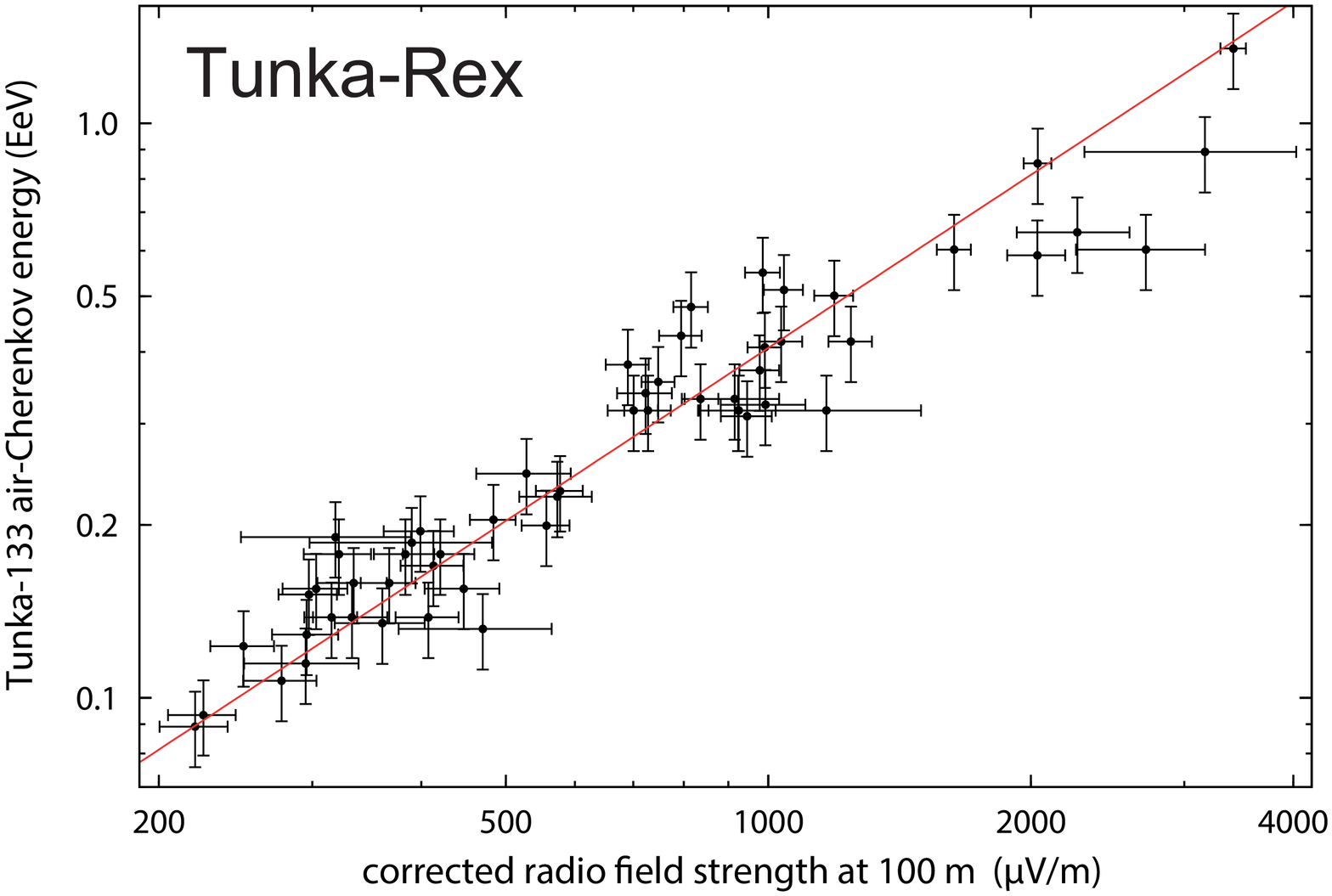}
  \caption{Correlations of the amplitude at $100\,$m normalized by the sine of the geomagnetic angle (LOPES, left) and additionally corrected for the asymmetry of the footprint (Tunka-Rex, right) with the energy of the primary particle reconstructed by the host experiments KASCADE-Grande and Tunka-133, respectively (from references \cite{TunkaLOPESenergyScale2016} and \cite{TunkaRex_UHECR2014}).}
  \label{fig_LOPESandTunkaRexEnergyCorrelation}
\end{figure}

\subsubsection{Amplitude at reference distance}
Similar to the established methods for air-shower arrays of non-imaging air-Cherenkov \cite{Tunka133_NIM2014} or particle detectors \cite{KASCADEGrandeS500}, the signal strength at a fixed reference distance $r_\mathrm{ref}$ to the shower axis can be used as an energy estimator, where the amplitude at $r_\mathrm{ref}$ is determined by fitting a lateral distribution function to the measurements of individual antenna stations. 

The reference distance $r_\mathrm{ref}$ is chosen such that maximum precision for the energy reconstruction is obtained: 
first of all, the amplitude at the reference distance should have minimal dependence on the distance to the shower maximum.
This is given close to the Cherenkov ring at ground whose radius is around $100\,$m for vertical showers at typical observation levels, and increases with increasing zenith angle. 
Thus, to first order the reference distance depends only on observation altitude and zenith angle, and indeed is around $100\,$m for all existing air-shower arrays.
With second order effects, however, the optimum distance slightly depends on the detector. 
Since the lateral distribution of the radio signal is flatter for lower frequencies, the frequency band plays a role. 
Moreover, the influence of background and other measurement uncertainties will depend on the detector, e.g., on the antenna spacing.

The reference distance has been determined with CoREAS simulations and measurements of various detectors. 
For LOPES which featured a relatively dense spacing with typical antenna distances smaller than $r_\mathrm{ref}$, the optimum distance increases from $70\,$m for small zenith angles below $20^\circ$ up to $100\,$m for zenith angles from  $32^\circ - 40^\circ$ \cite{2014ApelLOPES_MassComposition}. 
AERA and Tunka-Rex feature antenna spacings larger than $r_\mathrm{ref}$, and have determined optimum values for $r_\mathrm{ref}$ of $110\,$m for AERA \cite{GlaserAERAenergyARENA2012}, and $120\,$m for Tunka-Rex \cite{KostuninTheory2015}. 

After correction for the geomagnetic angle, the amplitude at the reference distance is directly proportional to the energy of the electromagnetic shower component and, therefore, also correlated with the energy of the primary particle (see figure \ref{fig_LOPESandTunkaRexEnergyCorrelation}). 
The precision for the energy of the electromagnetic shower component has been determined by a comparison of Tunka-Rex radio measurements and Tunka-133 air-Cherenkov measurements to about $15\,\%$ \cite{TunkaRex_XmaxJCAP2016}, with additional $5\,\%$ systematic uncertainty for the energy of the primary particle, if the mass is unknown.
For LOPES a precision of better than $20\,\%$ for the energy of the primary particle has been achieved \cite{2014ApelLOPES_MassComposition}, similar to the precision of AERA when using this method \cite{GlaserAERAenergyARENA2012}. 
However, in all cases the energy resolution cannot be determined independently, but only in comparison to the energy measurement of another detector at the same site, which itself comes with some uncertainties.
Thus, the real resolution of the radio measurements might be slightly worse or better depending on how accurately these uncertainties are known.
In any case, the achieved precision is not yet at the theoretical limit of better than $10\,\%$ predicted by CoREAS simulations made for LOPES and Tunka-Rex.
Consequently, the precision likely can be improved by further developing the reconstruction methods, by increasing the antenna density, or by selecting events with higher signal-to-noise ratios.

\begin{figure}[t]
  \centering
  \includegraphics[width=0.99\linewidth]{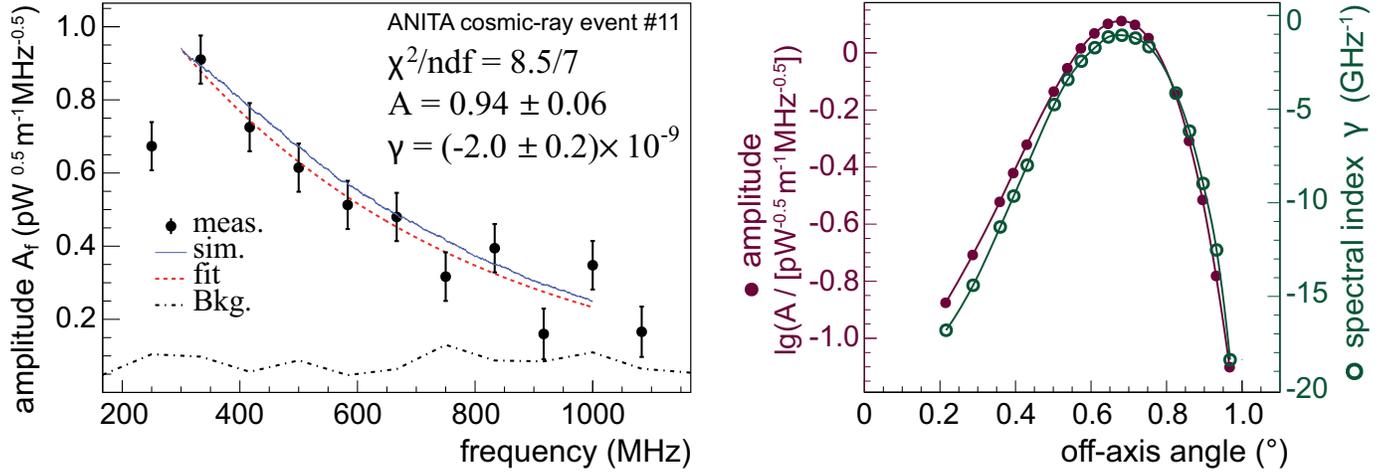}
  \caption{Left: Measurement of the frequency spectrum of a cosmic-ray air-shower event measured by ANITA.
  From a fit to the spectrum an amplitude parameter $A$ and a spectral index $\gamma$ are determined. 
  Right: According to simulations, $A$ and $\gamma$ show a correlated behavior versus the off-axis angle with a peak at the Cherenkov angle (modified from reference \cite{ANITA_CR_2016}).}
  \label{fig_ANITAenergyReconstruction}
\end{figure}

\subsubsection{Signal at a single antenna station}
\label{sec_singleStationEnergy}
Under certain conditions the shower energy can be estimated from the amplitude measured in a single antenna station. 
This is possible when either the distance from the antenna station to the characteristic reference distance $r_\mathrm{ref}$ (see above) is given by an independent measurements, or when the distance can be reconstructed from other features of the radio signal, e.g., the slope of the frequency spectrum.

The first method has recently been demonstrated by Tunka-Rex \cite{HillerARENA2016}: 
The shower axis is reconstructed by the co-located air-Cherenkov array Tunka-133, which has a lower threshold and is instrumented more densely than the radio array. 
Thus, even if only one antenna station has detected the radio signal of the air shower, its distance to $r_\mathrm{ref}$ is known from the air-Cherenkov detector, and the radio amplitude at $r_\mathrm{ref}$ can be extrapolated using an average lateral distribution. 
The energy precision in this case is about $20\,\%$, which is only slightly worse than the precision of $15\,\%$ achieved when requiring at least three antenna stations with signal. 
The advantage is a lower detection threshold when requiring only one antenna with signal, which in the case of Tunka-Rex triples the number of events. 
Consequently, the single-station method is an interesting approach for any hybrid detectors, e.g., combinations of particle and radio detectors. 

The second method is pioneered by the balloon-borne experiment ANITA, which could show that the shower energy can be reconstructed with about $30\,\%$ accuracy when using not more than the radio measurement in a single station \cite{ANITA_CR_2016}. 
However, in the case of ANITA this single station is more sophisticated. 
It consists of several closeby radio antennas hanging at the balloon, which allows for a rough measurement of the arrival direction of the radio signal when detected simultaneously by a few antennas. 
Since ANITA features a wide frequency band of up to about $1\,$GHz, not only the total amplitude, but also the slope of the frequency spectrum can be measured for individual events. 
Simulations show that this spectral slope depends on the distance from the Cherenkov angle (see figure \ref{fig_ANITAenergyReconstruction}).
This enables an estimation of the amplitude at the Cherenkov angle, corresponding to the amplitude at $r_\mathrm{ref}$ in the method described above.
Subsequently, the shower energy can be estimated, since the geomagnetic angle is known from the rough direction measurement. 
In principle, this method should be applicable also for ground-based radio detectors, if the design of a single station corresponds roughly to the ANITA design, i.e., a station would have to consist of at least three antennas with wide frequency bands.

\subsubsection{Energy scale}
The total accuracy of the energy reconstruction does not only depend on the precision, but also on the accuracy of the scale.
The scale accuracy itself depends on the absolute amplitude calibration of the radio detector and on the accuracy of the conversion coefficient between radiation energy or amplitude and the shower energy.

The absolute amplitude calibration of current radio arrays has uncertainties between $14\,\%$ and $18\,\%$ as discussed in section \ref{sec_calibration}. 
The conversion coefficient can be determined either from simulations or by comparison to a reference detector, e.g., the particle-detector array KASCADE-Grande measuring in coincidence with LOPES, or the fluorescence detectors of the Pierre Auger Observatory measuring in coincidence with AERA. 
In the first case of using simulations, the accuracy is about $20\,\%$, because CoREAS simulations have been experimentally confirmed to this level \cite{LOPESimprovedCalibration2016, TunkaRex_NIM2015}, where this $20\,\%$ already included the uncertainty of the amplitude calibration. 
In the second case of comparison with a reference experiment, there is no systematic uncertainty from simulations nor from the scale uncertainty due to the absolute calibration, because the uncalibrated radio signal can be compared directly to the absolute energy measurements of the reference experiment.
However, in this case the scale uncertainty can never be smaller than the uncertainty of the reference experiment (e.g., $14-16\,\%$ for the fluorescence detector of the Pierre Auger Observatory \cite{AugerNIM2015}).  In practice the scale uncertainty is slightly larger around $20\,\%$ \cite{2014ApelLOPES_MassComposition, AERAenergyPRD}. 

Consequently, the total accuracy of radio measurements of the shower energy is dominated by a $20\,\%$ scale uncertainty.
In addition to the scale uncertainty, the relative precision for individual events is between $15\,\%$ and $20\,\%$ depending on the experiment and reconstruction methods. 
If the absolute calibration of antennas could be improved, and if simulations of the radio emission by air-showers turn out to be trustworthy to a level of a few percent, then radio detection could become even more accurate than the currently leading fluorescence technique, whose scale accuracy is about $14\,\%$ \cite{AugerNIM2015}.
Radio detection already is accurate to $10\,\%$ for the relative comparison of the energy scale of two different air-shower arrays, when a consistent calibration of both radio arrays is used.
This has recently been shown in a comparison of the KASCADE-Grande and Tunka-133 experiments via their radio extensions LOPES and Tunka-Rex \cite{TunkaLOPESenergyScale2016}.

\begin{figure}[t]
  \centering
  \includegraphics[width=0.7\linewidth]{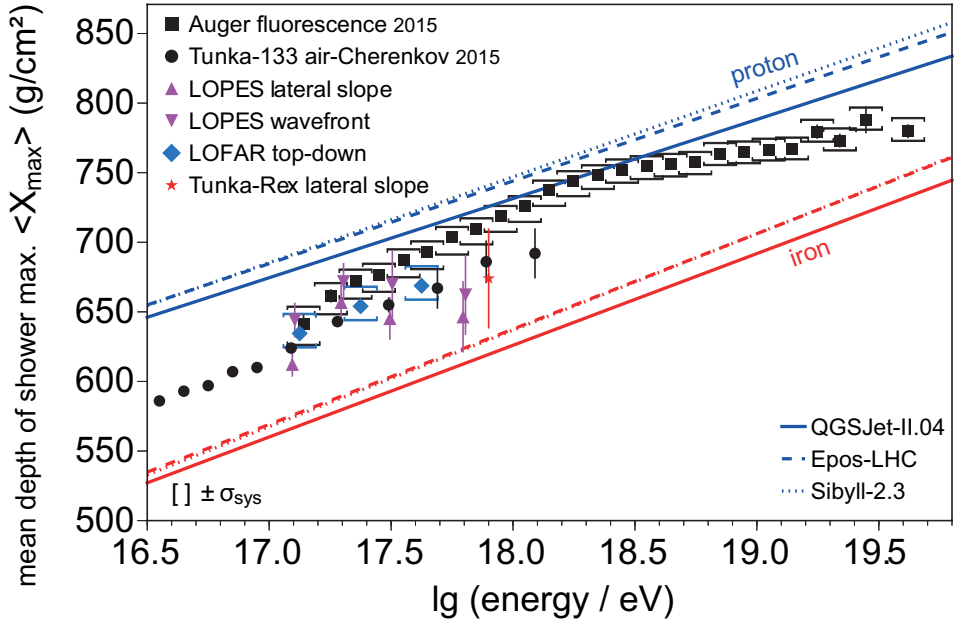}
  \caption{Overview on different measurements of the mean $X_\mathrm{max}$ by selected air-Cherenkov \cite{Tunka133_ISVHECRI2014}, air-fluorescence \cite{AugerHEATXmaxICRC2015}, and radio measurements \cite{LOPES_ECRS2014, LOFAR_Nature2016, KostuninPhDThesis2015} (Tunka and LOPES values not corrected for selection biases), and the predictions by CORSIKA simulations based on different hadronic interaction models for pure proton and iron compositions (in the simulations $X_\mathrm{max}$ has been determined as atmospheric depth of the maximum energy deposit $dE/dX$) \cite{RiehnPhDThesis2015} (figure also in Ref.~\cite{SchroederARENA2016}).}
  \label{fig_XmaxOverview}
\end{figure}

\subsection{Shower maximum: $X_\mathrm{max}$}
\label{sec_XmaxReconstruction}
The position of the shower maximum is one of the most sensitive statistical estimators for the mass-composition of cosmic rays. 
While the shower development depends mostly on the atmospheric depth, $X$, which is the traversed column density (integrated mass of the atmosphere per area) measured in g/cm$^2$, the radio signal on ground depends mostly on the geometrical distance to the shower maximum measured in km - very similar to the characteristics of the Cherenkov light emitted by air showers. 
For a given distance in km, the atmospheric depth or the shower inclination only play the role of second order effects, since the shape of the longitudinal shower profile, and thus the extension of the shower region relevant for the radio emission depends slightly on these parameters. 
However, these second order effects on the radio signal have not yet been investigated, and current efforts concentrate on the first order effect, namely the position of the shower maximum.
Several methods have been developed to reconstruct the distance from the detector to the shower maximum from various measurable characteristics of the radio signal, summarized in this section. 

In a subsequent step, the measured distances to the shower maximum have to be converted to atmospheric depths, $X_\mathrm{max}$, taking into account atmospheric models. 
Depending on the desired accuracy this requires continuous monitoring of the atmospheric conditions above the experiment \cite{AugerAtmMonitoring2012}, or the interpolation of the atmospheric conditions above the experiment from databases such as GDAS \cite{AugerGDAS2012}. 
The derived $X_\mathrm{max}$ distributions then can be converted into estimates for the composition of the primary cosmic-ray particles \cite{KampertUnger2012}. 
Showers initiated by light particles on average have their maximum closer to the detector: 
in the relevant energy range above $10^{17}\,$eV, the mean $X_\mathrm{max}$ is about $100\,$g/cm$^2$ deeper (= the $X_\mathrm{max}$ value is higher) for showers initiated by protons than for showers initiated by iron nuclei (see figure \ref{fig_XmaxOverview}). 
Moreover, the $X_\mathrm{max}$ distribution is narrower for heavier primary particles.
This relation can be used to statistically distinguish mass groups of nuclei, as well as to search for photons \cite{AugerPhotonLimit2009}, or neutrinos \cite{AugerInclinedNeutrinos2011, AugerTauNeutrinos2012}, because for inclined showers only weakly interacting neutrinos have a reasonable probability to interact deep in the atmosphere causing a shower maximum relatively close to the detector.

Nevertheless, the most precise $X_\mathrm{max}$ measurement alone cannot be used to determine the type of an individual primary particle, but only for a statistical determination of the mass composition, because shower-to-shower fluctuations give a certain probability that the same $X_\mathrm{max}$ can be caused by different primary particles. 
Consequently, further shower parameters are required to better estimate the primary particle type, like shape parameters of the shower profile \cite{DoctersThesis2015}, or complementary information of other shower components.
In particular the size ratio between the electromagnetic and muonic components provides complementary information on the mass composition \cite{KASCADEGrandeMassGroups2013}.
Hence, a combination of radio and muon detectors should be ideal not only for the most accurate measurements of the shower energy, but also to estimate the primary particle type of individual events. 
This already is under investigation, but current research still focuses on the optimization of methods for accurate $X_\mathrm{max}$ reconstruction.

Whether using the shower maximum alone as a mass estimator or also other parameters, the interpretation for the mass composition always depends on the hadronic interaction models assumed for the shower development, which is one of the largest systematic uncertainties in ultra-high-energy astroparticle physics.
Luckily, the electromagnetic component and the emitted radio signal are less dependent on these hadronic models than the muonic component, which is one of the advantages of the radio technique in comparison to particle detector arrays. 
Less dependent means that the predicted mean values of $X_\mathrm{max}$ differ between recent hadronic interaction models by about $20\,\%$ of the difference between a pure proton and a pure iron composition \cite{SybillICRC2015}. 
This systematic uncertainty is at the same level as the current record for $X_\mathrm{max}$ precision achieved by LOFAR radio measurements \cite{BuitinkLOFAR_Xmax2014}, as well as by Auger air-fluorescence measurements \cite{AugerNIM2015}, and hampers any more precise interpretation of measured $X_\mathrm{max}$ values in terms of mass composition. 
In particular, the proton-to-helium ratio currently comes with huge systematic uncertainties \cite{LOFAR_Nature2016, AugerXmaxImplications2014}.

\begin{figure}[t]
  \centering
  \includegraphics[width=0.53\linewidth]{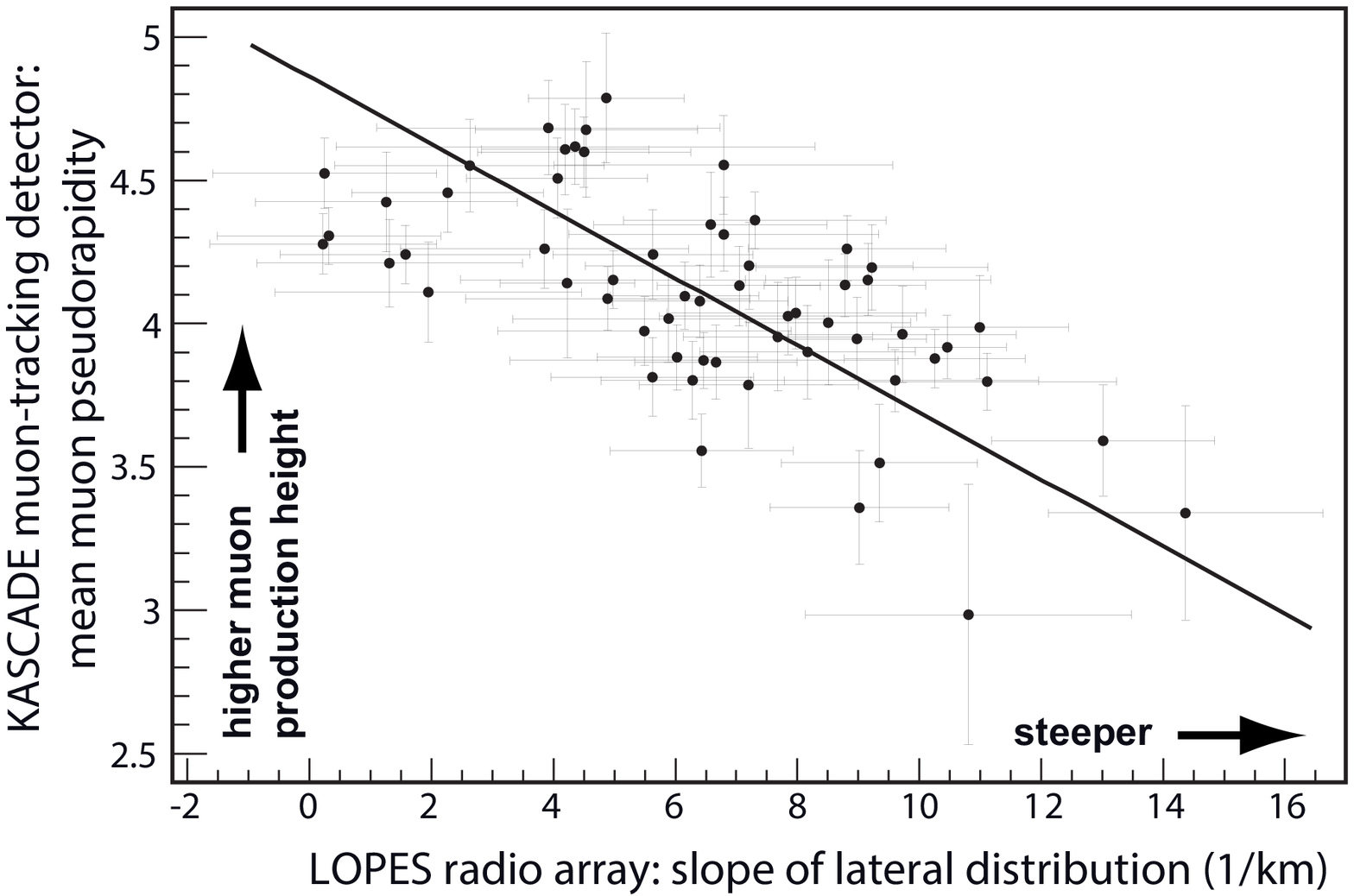}
  \hfill
  \includegraphics[width=0.45\linewidth]{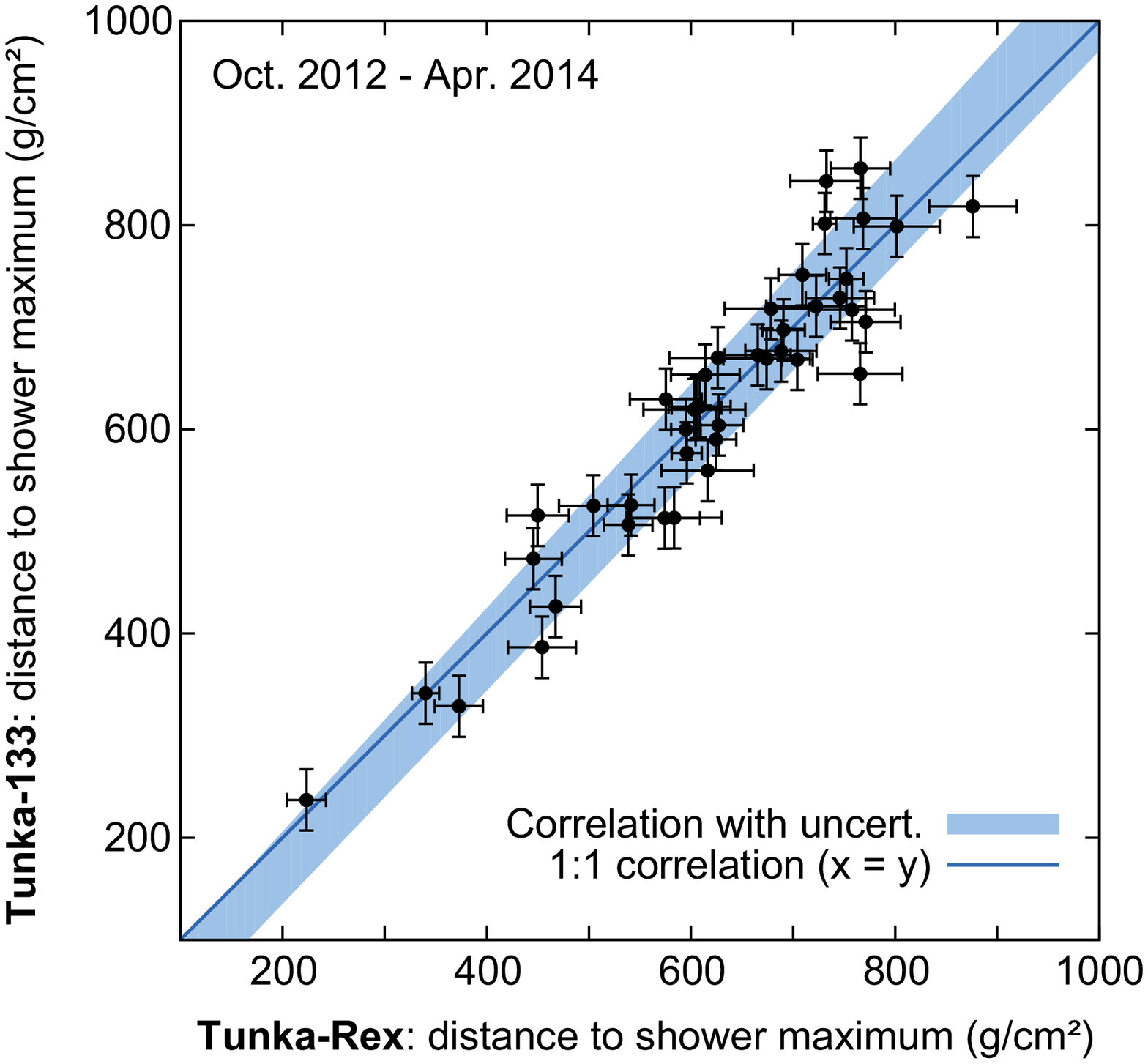}
  \caption{Left: Correlation of the slope parameter of the radio lateral distribution measured by LOPES and the mean muon pseudorapidity measured for the same air showers by the muon-tracking detector of KASCADE-Grande;
  right: Correlation of the distance to $X_\mathrm{max}$ reconstructed by coincident Tunka-Rex radio and Tunka-133 air-Cherenkov measurements (slightly modified from Refs.~\cite{2012ApelLOPES_MTD} and \cite{TunkaRex_XmaxJCAP2016}, respectively).}
  \label{fig_footprintXmaxCorrelation}
\end{figure}

\subsubsection{Shape of the footprint}
The footprint of the radio signal, which is the two-dimensional lateral distribution of the radio amplitude at the detector, has two main features sensitive to the position of the shower maximum.
First, the diameter of the Cherenkov ring, which is especially pronounced at higher frequencies above $100\,$MHz \cite{LOFARcherenkovRing2014}.
At lower frequencies the Cherenkov ring corresponds to the distance from the shower axis at which the exponential decrease of the lateral distribution sets in \cite{AllanICRC1971, NellesLOFAR_LDF2014}. 
In the lateral-distribution functions introduced in section \ref{sec_footprint} this diameter of the Cherenkov ring is described by the width of a Gaussian.
Consequently, the size of the width parameter can be used to determine the distance to the shower maximum. 
LOFAR achieved a precision of slightly better than $40\,$g/cm$^2$ using this method \cite{LOFARmeasuredLDF2015}, which demonstrates the principle feasibility of the method, but is worse than for the top-down approach described in the next section.

Second, the slope of the exponential tail depends on the distance to the shower maximum \cite{HuegeUlrichEngel2008, deVries2010, Kalmykov2012}.
This feature has been used to experimentally prove that radio measurements are sensitive to the longitudinal shower development by comparing LOPES radio measurements to KASCADE muon measurements (see left part of figure \ref{fig_footprintXmaxCorrelation}, \cite{2012ApelLOPES_MTD}). 
A more sophisticated analysis of this correlation includes a dependence on the zenith angle as a second-order effect and has been used for $X_\mathrm{max}$ reconstruction by LOPES \cite{2014ApelLOPES_MassComposition}. 
The slope of the lateral distribution is also employed for the $X_\mathrm{max}$ reconstruction of Tunka-Rex. 
For this radio measurement of $X_\mathrm{max}$ by Tunka-Rex a correlation with the $X_\mathrm{max}$ reconstruction by the Tunka-133 air-Cherenkov detector has been observed, which is another experimental evidence for the sensitivity of radio measurements to the position of the shower maximum \cite{TunkaRex_XmaxJCAP2016} (see right part of figure \ref{fig_footprintXmaxCorrelation}). 
The achieved precision of the radio reconstruction is only $90\,$g/cm$^2$ for LOPES, which was at a radio-loud site, and $40\,$g/cm$^2$ for Tunka-Rex at a radio-quiet site, which still is two times worse than the accuracy of the leading air-fluorescence method.  
Future analyses have to check whether this can be improved by using both features of the footprint simultaneously, or by even including possible further, more subtle dependences, as done implicitly  by a top-down simulation approach introduced by LOFAR.

\begin{figure}[p]
  \centering
  \includegraphics[width=0.54\linewidth]{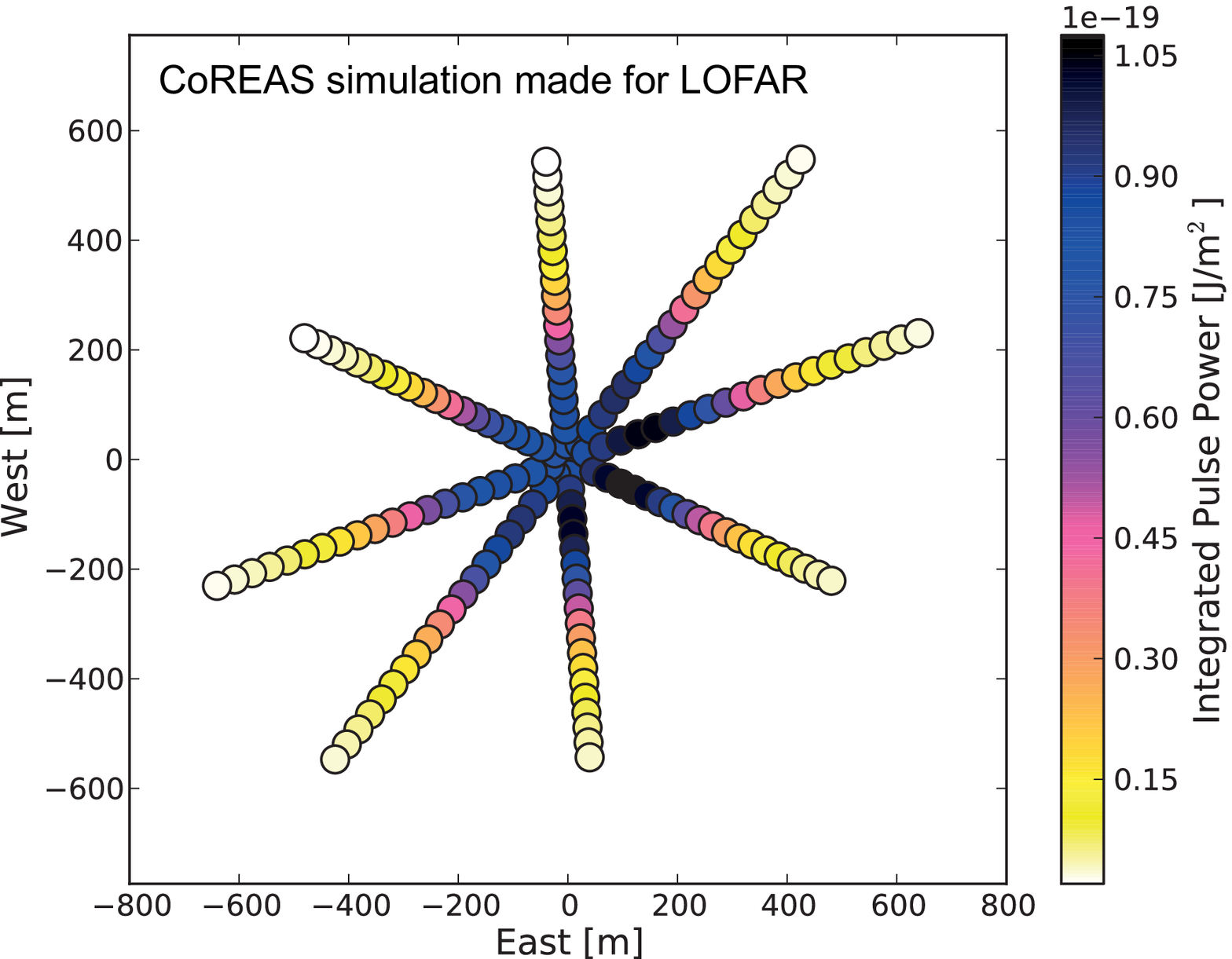}
  \hfill
  \includegraphics[width=0.44\linewidth]{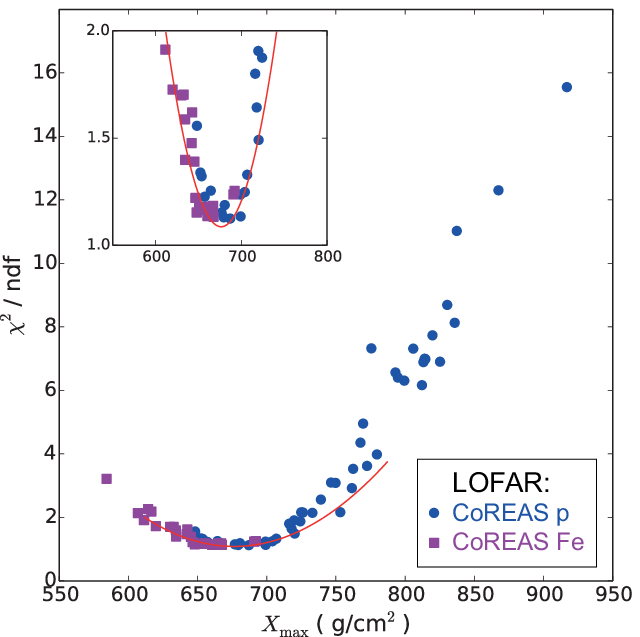}
  \caption{Left: simulated pattern of the radio amplitude of an inclined air-shower used to interpolate the full footprint on ground (the star-shape pattern is regular in the shower plane perpendicular to the shower axis).
  Right: reduced $\chi^2$ values of several CoREAS simulations made for the LOFAR event already shown in figure \ref{fig_footprints} over the true $X_\mathrm{max}$ values of the simulations (slightly modified from Refs.~\cite{NellesLOFAR_LDF2014} and \cite{LOFAR_Xmax_ICRC2015}, respectively).}
  \label{fig_LOFAR_XmaxReco}
\end{figure}

\begin{figure}[p]
  \centering
  \includegraphics[width=0.99\linewidth]{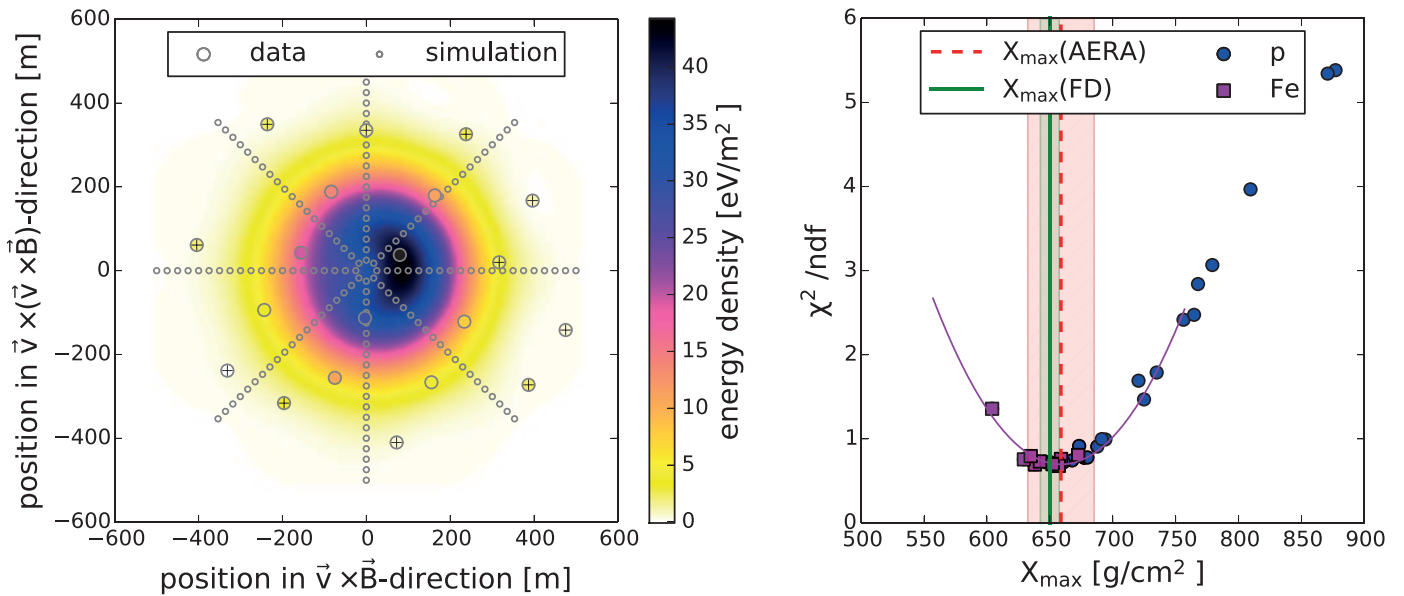}
  \caption{$X_\mathrm{max}$ reconstruction with the top-down method for an AERA event. 
  Left: the radio footprint is simulated with CoREAS on a star-shape grid, interpolated and matched with the data by scaling the amplitude and moving the core; for stations with signal the energy density is marked with the color inside the circle, sub-threshold stations are marked with a '+' (left). 
  Right: the $\chi^2$ values of many such radio simulations are used to determine $X_\mathrm{max}$ for the measurement, which for this event is in agreement with the measurement of the Auger fluorescence detector (figures from Ref.~\cite{AERAoverviewICRC2015}).}
  \label{fig_AERAtopDownXmax}
\end{figure}

\subsubsection{Top-down simulation approaches}
LOFAR introduced a more general and at the same time more precise approach for the reconstruction of $X_\mathrm{max}$, whose only disadvantage is the large demand of computing resources \cite{BuitinkLOFAR_Xmax2014, LOFAR_Nature2016}. 
For each measured event many air-shower simulations are produced for different primary particles, in particular protons and iron nuclei, where the incoming direction used for the simulations is reconstructed by arrival-time measurements.
To limit the computing time per shower the complete footprint on ground is determined by interpolation from the values simulated at positions in a star-shape grid. 
Moreover, all simulations are performed for only one single energy of the primary particle assuming linear scaling of the radio amplitude with the primary energy. 
Thus, for each simulated shower the best fitting energy and core position can be determined and quantified through a $\chi^2$ value.
These minimum $\chi^2$ values of each simulation are then plotted versus the true $X_\mathrm{max}$ values of the simulations, and the minimum of a parabolic fit to these $\chi^2$-values is considered the most likely $X_\mathrm{max}$ of the measured event (see figure \ref{fig_LOFAR_XmaxReco}). 

By this method an $X_\mathrm{max}$ uncertainty of better than $20\,$g/cm$^2$ has been achieved by LOFAR, which is approximately as good as for the leading air-fluorescence technique. 
Systematic uncertainties, e.g., due to the effect of humidity on the refractive index of air at radio frequencies or due to hadronic interaction models, have been taken into account by the LOFAR collaboration and are already contained in the $20\,$g/cm$^2$ uncertainty. 
Interestingly this uncertainty is twice as low as with the parametrization approach of the footprint \cite{LOFARmeasuredLDF2015}, which shows the clear advantage of the top-down approach for LOFAR.
AERA already applied the same method for a comparison to fluorescence measurements of the same air-showers, but so far only few selected events have been investigated (see figure \ref{fig_AERAtopDownXmax} for an example) \cite{AERAoverviewICRC2015}.
Future studies with larger statistics have to show whether this top-down approach is also advantageous for sparse arrays, which measure the amplitude only at few positions per event.

Another intrinsic advantage of the top-down method is that it can easily be extended to include additional information from the radio measurements, e.g., the arrival time or the frequency spectrum in individual antenna stations, or information from coincident measurements by other detectors. 
The LOFAR reconstruction already includes the particle numbers measured by the LORA scintillator array, but only amplitude information from the radio signal. 
Future improvements could also involve other mass-sensitive parameters in addition to $X_\mathrm{max}$, e.g., shape parameters of the shower profile, which might be accessible with sufficiently dense antenna arrays like the SKA.

\subsubsection{Steepness of the wavefront}
A different method for $X_\mathrm{max}$ reconstruction makes use of the hyperbolic radio wavefront reconstructible from arrival-time measurements (cf.~section \ref{sec_wavefront}). 
The angle between the shower plane and the cone limiting the hyperboloid depends on the shower inclination as well as on $X_\mathrm{max}$. 
However, the offset between the apex of the cone and the hyperboloid also depends on $X_\mathrm{max}$, which requires a complicated reconstruction formula for $X_\mathrm{max}$ depending on both, the offset and the cone angle.
As practical simplification alternatively either the offset or the cone angle can be fixed in the reconstruction such that the remaining free parameter carries the whole $X_\mathrm{max}$ sensitivity. 

\begin{figure}[t]
  \centering
  \includegraphics[width=0.99\linewidth]{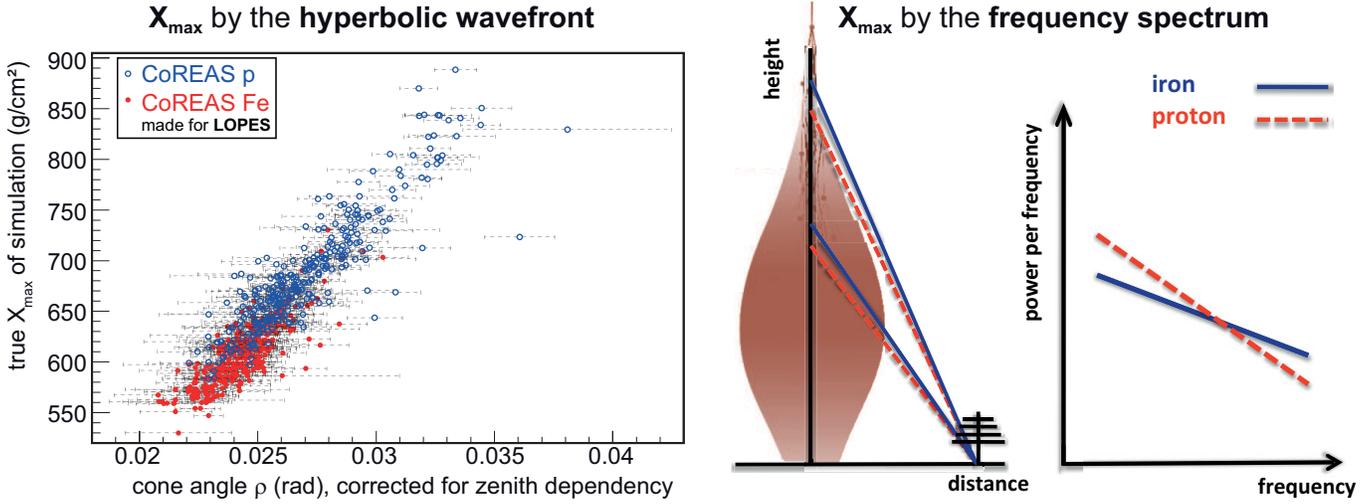}
  \caption{Left: Correlation of $X_\mathrm{max}$ and the wavefront angle $\rho$ (between the shower plane and the asymptotic cone of the hyperbolic wavefront) for CoREAS simulations made for the LOPES detector (from Ref.~\cite{LOPESwavefront2014}).
  Right: Sketch of the dependence of the spectral slope on the position of the shower maximum (sightly modified from Ref.~\cite{Grebe_ARENA2012}).}
  \label{fig_XmaxWavefront}
\end{figure}

The latter approach has been chosen by LOPES \cite{LOPESwavefront2014}, where the cone angle of the wavefront has been determined by cross-correlation beamforming fixing the offset to $3\,$ns. 
The LOPES measurements contain various hints that the cone angle really is sensitive to the shower development, e.g., weak correlations have been observed with the slope of the lateral distribution and with the shower age determined by the KASCADE-Grande particle detectors.
Although the $X_\mathrm{max}$ uncertainty achieved by LOPES is too large for practical applications, CoREAS simulations show that under ideal conditions, i.e., negligible uncertainties on the arrival time measurement and the reconstructed shower axis, the wavefront method can be as precise as the footprint methods (see left part of figure \ref{fig_XmaxWavefront}). 
This makes it worth to investigate the wavefront method further at arrays with lower uncertainties, and worth to study how arrival time measurements can contribute to multivariate approaches for $X_\mathrm{max}$ reconstruction.

\subsubsection{Slope of the frequency spectrum}
The frequency spectrum measured at an individual antenna station depends not only on the position relative to the shower axis, but also on $X_\mathrm{max}$ \cite{GrebeThesis2013, JansenThesis2016}. 
Generally the amplitude of the radio signal decreases with increasing frequency, and the spectral index of this decrease is slightly higher for deeper positions of the shower maximum: 
the closer the shower maximum to the detector is the steeper the frequency spectrum.
This means that at a given distance to the shower axis, showers initiated by iron nuclei on average have a softer frequency spectrum than proton initiated showers (see right part of figure \ref{fig_XmaxWavefront}). 
As for the wavefront and the slope of the lateral distribution, the effect of $X_\mathrm{max}$ is weak though: 
the average difference between proton and iron showers on the slope parameter, the wavefront angle, and the spectral index is of the order of $10\,\%$ of the total values. 
This problem is especially relevant for the spectral slope method, since the frequency spectrum generally is harder to measure than the simple amplitude or arrival time at an antenna station. 
While it was possible to show that the spectral slope can indeed be measured for single events with good signal-to-noise ratio \cite{Grebe_ARENA2012, NiglFrequencySpectrum2008}, the $X_\mathrm{max}$ precision available from such measurements is not yet clear.

The advantage of the spectral slope method versus other methods is its applicability to the measurement of a single antenna station. 
If the spectral measurement in one antenna is accurate enough, and if the shower axis is reconstructed accurately enough by another cosmic-ray detector, then in principle one single radio antenna would suffice to measure $X_\mathrm{max}$, while all other reconstruction methods need at least 3 or 4 antennas with signal. 
Nevertheless, first an experimental proof-of-principle is required that the frequency spectrum can be used for $X_\mathrm{max}$ measurements under realistic conditions.

\subsubsection{Dependency of the polarization}
As discussed in section \ref{sec_polarization}, the polarization of the radio signal emitted by air showers is slightly elliptical, where the direction and size of the ellipticity are determined by the interference of the geomagnetic and Askaryan effects at the observer location \cite{HuegeCoREAS_ARENA2012}.
The ratio between the strengths of the Askaryan and geomagnetic emissions depends on the shower inclination \cite{SchellartLOFARpolarization2014} and, consequently, on the distance to the shower maximum. 
For the asymmetry of the radio footprint, which is also caused by the interference of both emission mechanisms, a slight dependence on $X_\mathrm{max}$ has already been shown for CoREAS simulations \cite{KostuninTheory2015}, but not yet explicitly with measured data. 
Nevertheless, just like the asymmetry of the footprint does, the direction and ellipticity of the polarization at each given location must also depend slightly on $X_\mathrm{max}$, because its origin is the same interplay of the Askaryan and the geomagnetic effects.
Although not yet investigated, provided sufficient measurement precision, this could well be used to reconstruct $X_\mathrm{max}$ as a stand-alone method or in combination with the other methods as part of a multivariate analysis.

\begin{table}[t]
\centering
\caption{Reconstruction methods and best achieved precisions for radio measurements of the three most important shower parameters. 
Where available, theoretical predictions just give a rough idea of the potentially achievable accuracy, which is neither a strict theoretical optimum nor a value expected under realistic conditions 
(only some references given here, for further references see text).} \label{tab_showerParameters}
\vspace{0.3cm}
\small
\begin{tabular}{lcc}
\hline
parameter / method&experimental accuracy& theoretical prediction\\
\hline
\underline{Direction:}&&\\
Triangulation (plane wave) & $2^\circ$ & \\
Hyperbolic wavefront \cite{LOPESwavefront2014, CorstanjeLOFAR_wavefront2014} & $\mathbf{0.7}\boldsymbol{^\circ}$ &  $< 0.1^\circ$ \\
\underline{Energy:}&&\\
Total radiation energy \cite{AERAenergyPRD}& $17\,\%$ (precision) + $\mathbf{14\,}$\textbf{\%} (scale) & $< 10\,\%$ \\
Amplitude at reference distance \cite{TunkaRex_XmaxJCAP2016}& $\mathbf{15}\,$\textbf{\%} (precision) + $20\,\%$ (scale) & $< 10\,\%$ \\
\underline{Depth of shower maximum, $X_\mathrm{max}$:}&&\\
Footprint, exponential slope \cite{TunkaRex_XmaxJCAP2016} & $40\,$g/cm$^2$ & $30\,$g/cm$^2$ \\
Footprint, width \cite{LOFARmeasuredLDF2015} & $\lesssim 40\,$g/cm$^2$ & -\\
Footprint, top-down simulations \cite{BuitinkLOFAR_Xmax2014} & $\mathbf{\lesssim 20}\,$\textbf{g/cm}$\mathbf{^2}$ & - \\ 
Hyperbolic wavefront \cite{LOPESwavefront2014}& $\lesssim 140\,$g/cm$^2$ &  $30\,$g/cm$^2$ \\ 
Frequency spectrum \cite{JansenThesis2016}& $135\,$g/cm$^2$ &  $60\,$g/cm$^2$ \\ 
Polarization & - & - \\ 
\hline
\end{tabular}
\end{table}

\subsection{Summary on air-shower parameters}
In summary, recent experimental analyses have demonstrated in practice that the radio technique can be used to measure the most important shower parameters, which are direction, energy, and $X_\mathrm{max}$. 
For all parameters the accuracy achieved is not yet at its limit, and further improvements of the reconstruction methods likely will increase the precision. 
For the accuracy of the energy scale not the reconstruction methods, but instead the absolute calibration of the antennas seems to be the limiting factor, which might be improved by future efforts. 
Table \ref{tab_showerParameters} gives an overview on the different reconstruction methods, the experimentally demonstrated, and the theoretical predicted precisions under idealized conditions, where the latter depend on assumptions made, and is not necessarily the theoretical optimum.
On the one hand, this is very promising, since it means that radio detection has the potential to become the most precise of all techniques.
On the other hand, a lot of work still has to be done for making the radio measurement of shower parameters a reliable standard method, in particular all systematic uncertainties have to be well understood.
Still, the achieved precisions meanwhile are of the same order as for other air-shower techniques, such that observatories with multiple detector systems already could profit from including radio measurements in a hybrid reconstruction of air-shower parameters.

\clearpage

\section{Reconstruction of the primary particle in dense media}
\label{sec_neutrinoReco}

This chapter shortly discusses the reconstruction of the arrival direction, energy and type of the primary particle based on the radio signal emitted by cascades in dense media.
For lunar observations all types of primary particles are of interest, but it seems unlikely that observations by radio telescopes on Earth can be accurate enough to distinguish fine differences in the shower development related to the type of primary particle. 
Thus, for lunar observations only the arrival direction and the energy of the primary particle can be estimated.
For ice as medium, cosmic-ray nuclei and photons are shielded by the atmosphere and only neutrinos are considered as primary particles, which implies that the neutrino flavor is a quantity of interest. 
In this case of a neutrino as primary particle, an additional difficulty arises from the fact that neutrinos transfer only a part of their momentum and energy when scattering with nuclei, i.e., the cascade developing in the medium contains an unknown fraction of the initial neutrino energy (on average about $20\,\%$ \cite{ARA_2012}).
This means that the shower direction and energy are correlated with the neutrino direction and energy, but in particular the energy correlation has a large spread leading to significant uncertainties of $50\,\%$ or more for the reconstruction of the neutrino energy.

Generally, techniques for the reconstruction of shower properties in dense media are not yet as advanced as the corresponding methods for air showers. 
Since showers in dense media have been detected only by optical detectors and at accelerators, reconstruction methods for the radio signals are investigated theoretically, with simulations and calibration devices, but not yet with real natural showers \cite{ARA_2016, ARIANNA_timedomain2015}.
Although Askaryan emission in dense media has similar properties as in air, the larger Cherenkov angle makes a difference.
Due to the large opening angle and the narrow width of the Cherenkov cone in dense media, typically only a small part of the Cherenkov cone is observed by one signal station, though each station features several antennas for most detector design. 
Thus, signal properties such as the arrival time, the amplitude, the pulse shape, and the polarization have to be measured accurately in the individual antennas of one station in order to estimate the properties of the primary particle.

\subsection{Arrival direction}
Due to the large Cherenkov angle in dense media the direction of the shower axis differs significantly from the arrival direction of the radio signal given by the Poynting vector.
Furthermore, simple triangulation of the shower direction is not possible for typical detector designs because the radio signal is recorded by only a few nearby antennas on the Cherenkov cone. 
In the best case, the polarization, the arrival direction of the radio signal, and the gradients of the arrival time and amplitude over the width of the Cherenkov cone can be measured \cite{ARA_2012}.
The latter gives information on the distance to the vertex of the Cherenkov cone: the smaller the gradients the more distant the vertex. 
Since the radio signal is emitted from the vertex of the Cherenkov cone under the known Cherenkov angle of the medium, and since the polarization of the Askaryan emission points towards the shower axis, the geometry of the shower axis can be reconstructed from this information. 
Then, the arrival direction of the initial particle, e.g., a neutrino interacting in the ice or a cosmic-ray nucleus hitting the lunar regolith, is assumed to coincide approximately with the direction of the shower axis due to momentum conservation.
The accuracy of this method for direction reconstruction can be tested in practice by comparing it to the established optical Cherenkov-light detection, if and only if hybrid arrays will be built in ice.
In principle this method for direction reconstruction can also work for lunar observations when the polarization is measured accurately, since the distance to the vertex is known by the distance to the origin of the emission on the lunar surface \cite{SKAlunar_ICRC2015}. 

\subsection{Energy}
The shower energy can be reconstructed from the measurement of the pulse amplitude provided that the distance to the shower can be estimated. 
As for air showers, also for cascades in dense media the emission is coherent and the total energy in the radio signal is expected to scale quadratically with the energy of the electromagnetic shower component.
For purely electromagnetic showers this is equal to the energy of the primary particle, e.g., an electron originating from a charged-current interaction of an electron neutrino or from a tau decay. 
However, since high-energy neutrinos have much larger interaction probability with nuclei than with electrons, electromagnetic cascades will regularly occur jointly with a hadronic cascade. 
Nevertheless, also in this case the energy of the primary particle initiating the shower can be estimated, because for hadronic cascades the fraction of the total energy in the electromagnetic component increases with shower energy in a known way \cite{AlvarezMuniz_ZHAires_ice2012}.

Accelerator experiments generally have confirmed the quadratic dependence of the radio power on the shower energy, but further investigations are necessary in order to study whether the scaling is exactly or only approximately quadratically \cite{ANITA_SLAC_PRL_2007, TA_ELS_ICE_ICRC2015}. 
A major difficulty in practice will be that the amplitude depends strongly on the observation angle relative to the Cherenkov angle of the medium, and thus on the angle relative to the shower axis. 
Consequently, an accurate reconstruction of the shower direction is a prerequisite for an accurate energy reconstruction.
Another way to make the energy reconstruction more accurate might be to exploit the correlation between the slope of the frequency spectrum and the distance to the Cherenkov angle used by ANITA for the energy reconstruction of cosmic-ray air showers \cite{ANITA_CR_2016} (cf.~section \ref{sec_singleStationEnergy}). 
Dedicated simulations for showers in ice will be necessary to study how this correlation depends on the density and on the refractive index of the medium.
Then, the method should be applicable also in ice, provided that the frequency band of the detector is wide enough for an accurate measurement of the slope of the frequency spectrum. 
Finally, combinations with acoustic detectors might help to make the energy reconstruction more accurately, since the sound amplitude depends on the total energy of the hadronic shower \cite{ACoRNE2009} and features different systematic uncertainties than the radio signal.

\subsection{Neutrino flavor}
In dense media the position of the shower maximum is of limited used to distinguish different types of primary particles. 
Showers in dense media are much more compact with a typical length of few meters, which is almost point-like with respect to the extensions of antenna arrays like ARA or ARIANNA, except for ZeV energies at which shower extensions of a few $100\,$m are possible \cite{HansonLPM2016}. 
This means that for PeV to EeV energies the position of the shower as a whole can be reconstructed, but not details in the shower development.
Even if the position of the shower maximum could be reconstructed, it is not clear whether this could help to distinguish neutrino flavors. 
For neutral-current interactions the neutrino flavors should be indistinguishable, but for charged-current interactions producing a charged lepton in addition to a hadronic cascade, there might be a chance.
It seems that no studies have been performed how the shower maximum and properties of the radio signal statistically depend on the flavor of the primary neutrino.
Nevertheless, basic considerations outlined below give some ideas how the flavor ratio of neutrinos can be determined statistically, once neutrinos are measured by radio arrays in ice. 

Also in dense media the radio signal is primarily emitted by electrons and positrons, i.e., either by an electromagnetic shower or by the electromagnetic component of hadronic showers. 
The radio signal of both cases is slightly different (cf.~section \ref{sec_radioFeaturesDenseMedia}) \cite{AlvarezMuniz_ZHAires_ice2012}, but it has not yet been investigated whether these differences will be detectable under realistic conditions. 
If they are, then these difference between hadronic and electromagnetic showers might be useful to statistically determine the flavor ratio of the neutrinos. 
However, pure electromagnetic cascades are only expected with reasonable rate for anti-electron neutrinos at the Glashow resonance around $6\,$PeV, but hadronic shower originate from all types of charged and neutral-current interactions for all neutrino flavors at all relevant energies.
As additional complication the few electromagnetic cascades as consequence of charged-current interactions of electron neutrinos are not initiated separately but together with hadronic cascades.
Therefore, the real challenge is the separation of coincident electromagnetic and hadronic cascades from sole hadronic cascades, which probable requires percent precision of the radio measurements.

When combining radio with other techniques, the separation power between different neutrino flavors should be enhanced. 
Optical measurements can distinguish muon tracks from showers, which might provide a statistical measure for the ratio of muon neutrinos.
Moreover, the ratio between the acoustic and radio amplitudes could in principle be used to determine whether a shower is purely hadronic or joint with an electromagnetic shower.
More detailed studies of the differences between hadronic and electromagnetic cascades with respect to the radio and the acoustic emissions will be necessary to judge the potential of this method.
Nevertheless, this general considerations give hints that also for in-ice radio detectors hybrid combinations with other techniques might increase the accuracy for the properties of the primary particle.

%\clearpage
\section{Applications of the radio technique in high-energy astroparticle physics}
Based on the current understanding of the characteristics of the radio signal, possible science cases for the radio technique are discussed in this chapter. 
These are mostly the scientific goals of running and proposed experiments.
They give an idea what can be done with the radio technique and also what cannot.
When the radio technique was revived by digital technology more than ten years, ago, there was hope that antenna arrays could completely replace existing cosmic-ray observatories, and allow for huge observatories with highest exposure for all kinds of ultra-high-energy particles at lowest cost.

Meanwhile the knowledge of cosmic-ray physics has advanced, in particular it is now known that ultra-high-energy cosmic rays are not only protons, but instead there seems to be a significant fraction of heavier nuclei at all energies \cite{AugerXmaxImplications2014}. 
Only at the very highest energies around and above $10^{20}\,$eV it is still open whether the composition is mixed or almost purely protons or purely iron nuclei, which likely will be answered by the upgraded Pierre Auger Observatory \cite{AugerUpgrade_CRIS2015}. 
Hence, experiments focusing only on energy and arrival direction are insufficient.
Instead, cosmic-ray observatories need methods to determine the type of the primary particle with maximum accuracy, which requires a combination of different detection techniques. 
Consequently, for cosmic-ray physics stand-alone radio arrays will be of limited use, but by contributing to hybrid-detector observatories there is plenty of scientific application for radio detection. 
The radio technique can play a crucial role to measure air showers in the energy range above $10^{16}\,$eV with unprecedented accuracy, which will be necessary to understand the origin of the most energetic galactic and the ultra-high-energy extragalactic cosmic-rays.
Moreover, thunderstorms form an interesting research topic, and radio detection remains the most promising technique for the search for ultra-high-energy neutrinos.

\subsection{Enhancing accuracy of air-shower arrays}
As already demonstrated, e.g., by Tunka-Rex, antennas can be a very cost-effective extension arrays of any other type of air-shower detector. 
Since the radio signal contains complementary information in particular to muon measurements, particle-detector arrays benefit most from radio extensions.
The existing particle-detector array provides trigger, data acquisition, and infrastructure for the radio detector for minimum additional costs.
The complementary information in the radio signal then can be used to increase the total accuracy for the properties of the primary particle in several aspects:

\begin{itemize}
 \item \textbf{Energy scale:} Radio detection provides an accurate measure for the energy content of the electromagnetic shower component at the shower maximum. 
 Since the radiation energy of showers can be calculated by simulation codes like ZHAireS or CoREAS on an absolute scale, this can be used to calibrate and compare the energy scales of different experiments \cite{AERAenergyPRL, TunkaLOPESenergyScale2016}.
 \item \textbf{Energy precision:}  The energy content of the electromagnetic shower component in combination with the electron and muon numbers at ground should provide a significantly more precise reconstruction of the energy of the primary particle than the particle numbers alone.
 \item \textbf{Type of particle:} Radio detection can contribute in two ways to this science goal. First, by measuring $X_\mathrm{max}$. Second by combining the muon number measured at ground with the size of the electromagnetic shower component obtained from the radio measurements. 
 Each of the two complementary methods can be used to reconstruct the composition of the primary particles statistically.
 With both methods combined there might be a chance to reconstruct the type of the primary particle for individual events.
\end{itemize}

\begin{figure}[t]
  \centering
  \includegraphics[width=0.99\linewidth]{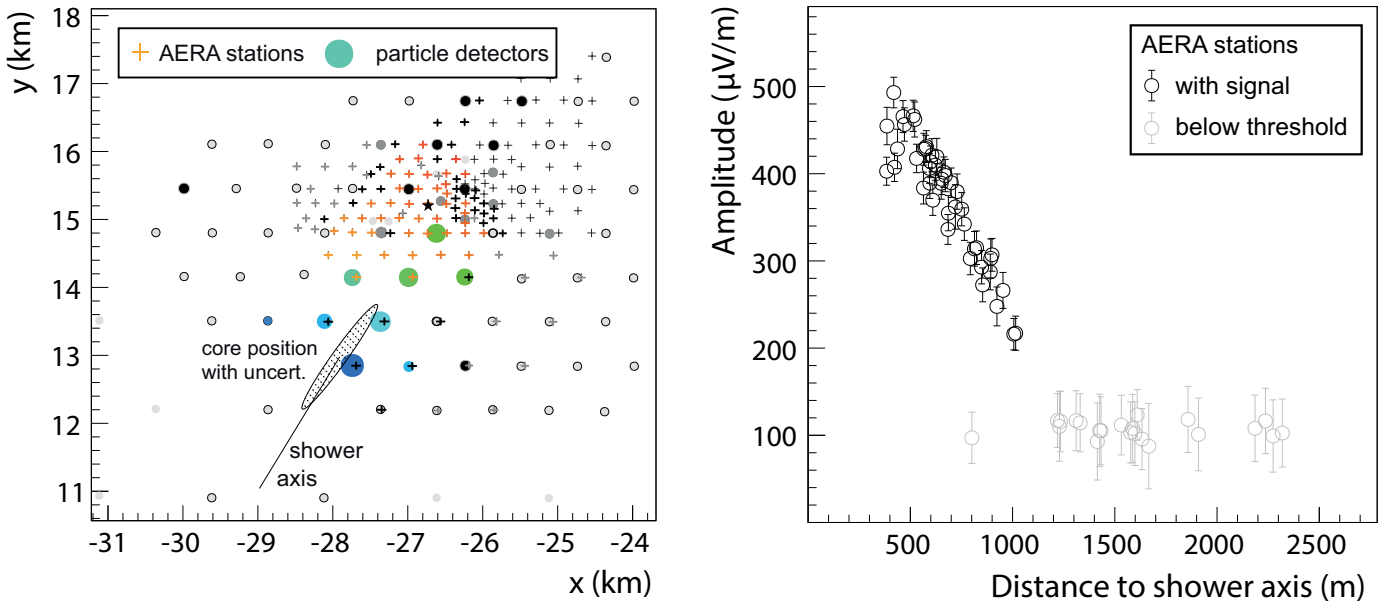}
  \caption{Footprint and one-dimensional lateral distribution of an inclined shower measured by AERA and the Auger surface-detector array of $1.3\,$EeV energy and $78^\circ$ zenith angle; in the left figure colored stations detected a signal, gray stations are below threshold, and black stations did not take part in the measurement of this event (from Ref.~\cite{AERAoverviewICRC2015}).}
  \label{fig_inclinedAERAevent}
\end{figure}

\subsection{Inclined showers}
The radio technique has a unique advantage for inclined showers, because unlike for optical methods or air-shower particles, there is almost no absorption in the atmosphere. 
Moreover, for inclined showers the radio footprint becomes large and can be seen at distances up to a kilometer or more from the shower axis at energies above $1\,$EeV (see figure \ref{fig_inclinedAERAevent}) \cite{AERAoverviewICRC2015, Gousset2004, LOPESinclined2007}. 
Due to the additional projection effect for inclined showers this means that radio antennas can be spaced with distances of a few kilometers to each other, where the optimum spacing depends drastically on the targeted zenith angle range. 
Consequently, focusing on inclined air-showers makes the radio technique scalable to large areas. 
For full sky coverage including showers until about $60^\circ$ zenith angle, the small footprint of a few $100\,$m requires a dense antenna spacing and makes radio detection expensive,
but for inclined showers radio antennas can provide large exposure for moderate costs.
Moreover, for inclined showers antennas would be the ideal complement to water- or ice-Cherenkov particle detectors, since for inclined air showers the electromagnetic shower component is mostly absorbed before reaching ground. 
Thus, the particle detectors measure almost purely muons, and the radio signal carries the pure information on the electromagnetic shower component.
The combination of both measurements might provide the only possible way to determine the mass composition for inclined cosmic rays.

Current analyses at the Pierre Auger Observatory have to show the accuracy for such hybrid methods under practical conditions. 
There is no principle reason which should limit the accuracy for the shower energy, but for $X_\mathrm{max}$ the question is completely open:
on the one hand, the larger distance to the shower maximum ought to make the reconstruction of $X_\mathrm{max}$ more difficult. 
On the other hand, the lever arm is much larger, since the radio signal can be measured at larger distances from the shower axis, which should increase the accuracy of the lateral-slope and wavefront methods for $X_\mathrm{max}$.
For stand-alone operation of a radio array, efficient self-triggering would be necessary in addition. 
This requires further investigations as currently done by the pathfinder experiment GRANDproto in China \cite{GRAND_ICRC2015}.

\subsection{Highest statistics for the highest energies}
To find out whether the cosmic-ray energy spectrum ends at a few $100\,$EeV, or whether nature features even more powerful particle accelerator, the exposure of the currently leading Pierre Auger Observatory might be insufficient. 
Huge radio arrays for inclined air-showers are one way, but not the only way for increasing the accumulated world exposure by an order of magnitude.
Since radio arrays for inclined showers would observe only a limited zenith angle range, the area has to be even larger than for particle-detector arrays, i.e., of the order of $100,000\,$km\textsuperscript{2}. 
This requires significant efforts, but is not completely unrealistic.
With GRAND such a huge array has been proposed \cite{GRAND_ICRC2015}, which might still be more economic than cosmic-ray space missions. 
Moreover, GRAND has the additional benefit that inclined showers can simultaneously be used to search for EeV neutrinos. 
Future studies on the accuracy of the arrival direction, energy and mass-composition have to show which type of cosmic-ray physics can be done with such a huge array detecting inclined showers. 
Nevertheless, if the upgraded Pierre Auger Observatory should show that the mass-composition is nearly pure at the highest energies (e.g., almost only protons or only iron nuclei as primary particles) \cite{AugerUpgrade_CRIS2015}, then mass sensitivity will be less important, and only good direction and energy resolution count, which is much easier to realize.

\begin{figure}[t]
  \centering
  \includegraphics[width=0.49\linewidth]{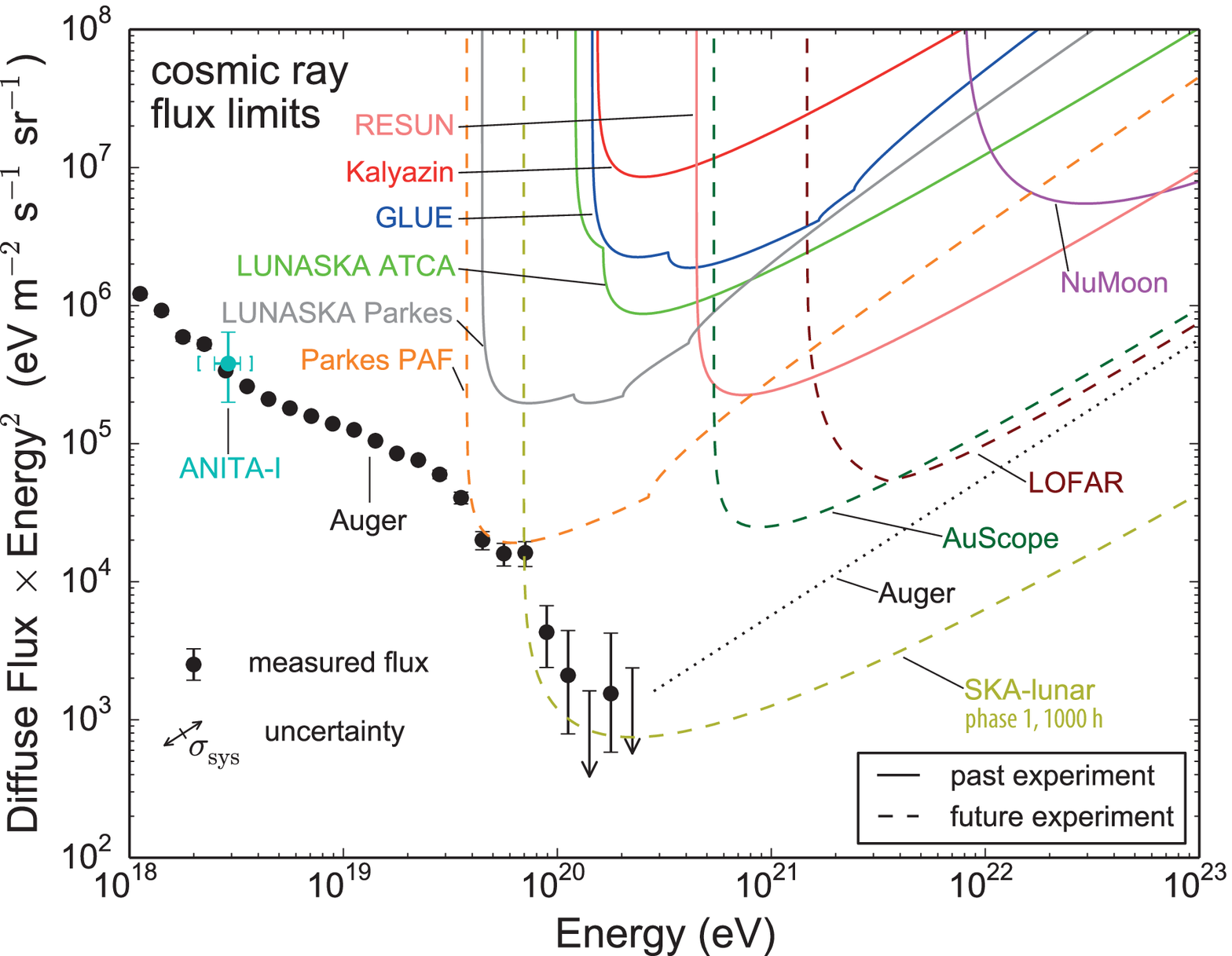}
  \hfill
  \includegraphics[width=0.49\linewidth]{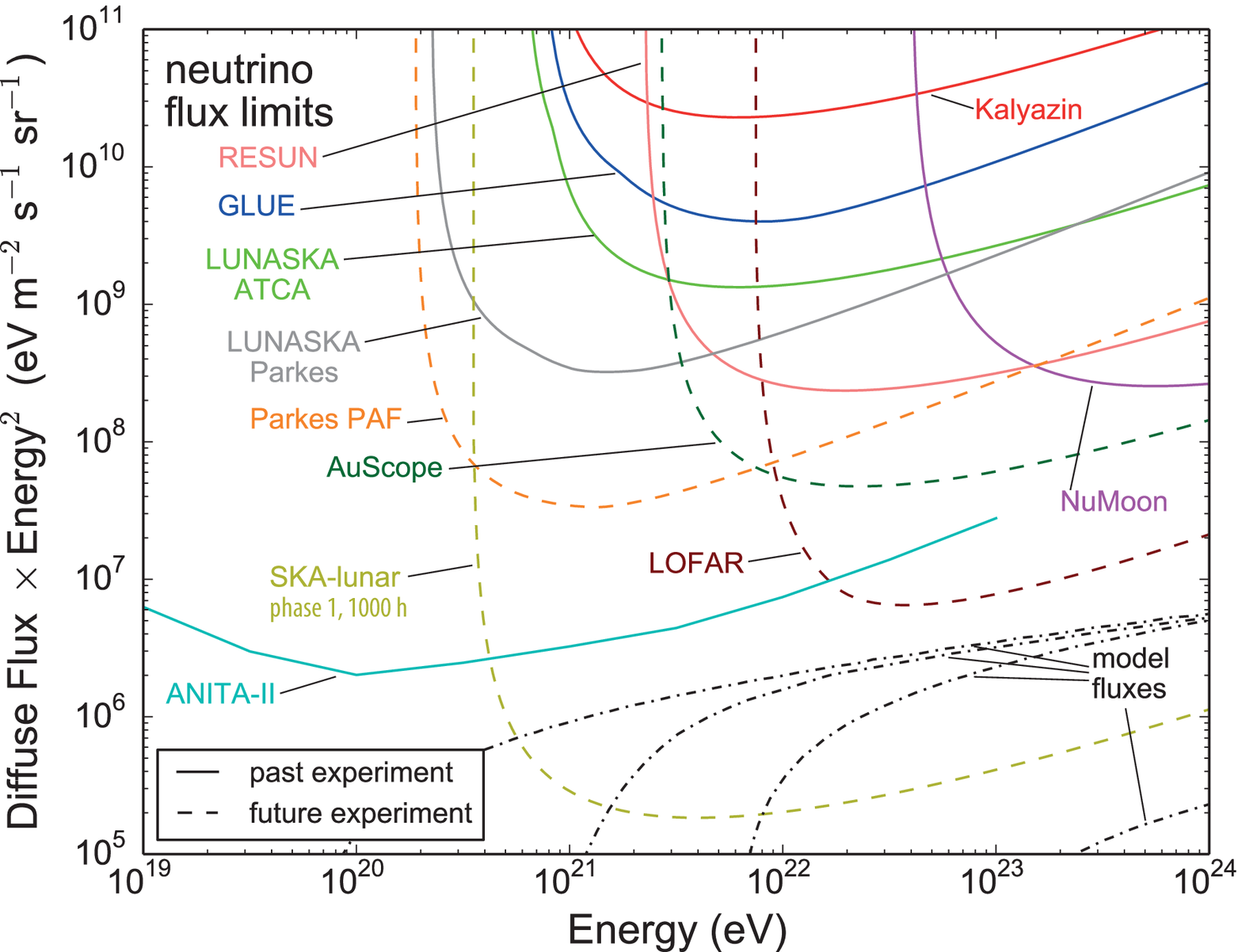}
  \caption{Sensitivity of various radio telescopes for the detection of lunar Askaryan emission of showers initiated by cosmic rays or neutrinos in the Moon (figures by J. Bray \cite{BrayReview2016}: see Refs.~\cite{SKAlunarScienceChapter2015, SKAcosmicRayScienceChapter2015} for the SKA, Ref.~\cite{AugerEnergySpectrum2010} for Auger measurements, and Ref.~\cite{ExcoticNeutrinos2012} for the neutrino models). 
  For comparison also  the cosmic-ray measurement \cite{ANITA_CR_2016}, and the neutrino limits \cite{ANITA_neutrino_PRD_2012} of the balloon-borne ANITA experiment are shown.}
  \label{fig_lunarSensitivity}
\end{figure}

There are at least two other ideas how radio detection could drastically increase the statistics for the highest-energy cosmic rays:
First, the radio signal of particle showers in the lunar regolith is expected to be observable by next-generation radio telescopes like the SKA \cite{SKAlunar_ICRC2015}. 
Due to the large area of the Moon the exposure can be large yielding competitive sensitivities for ultra-high-energy cosmic-rays and neutrinos (see figure \ref{fig_lunarSensitivity}), although with limited measurement accuracy.
In principle larger objects than the Moon could be observed, such as Jupiter, but even with next-generation radio observatories the estimated energy threshold is much higher than for lunar observations \cite{BrayNellesJupiter2016}. 
Second, detection of air-showers in the atmosphere is not only possible with antenna arrays on ground, but also from satellites or balloons as demonstrated by ANITA \cite{ANITA_CR_PRL_2010}. 
Plans for space-borne radio detection of air showers are not yet as advanced as the plans for space-borne fluorescence-light detection with the JEM-EUSO project \cite{JemEuso2013}.
Radio satellites could yield a similarly high exposure as JEM-EUSO, but ionospheric disturbances of the radio signal might cause additional measurement uncertainties compared to balloon-borne experiments \cite{MotlochSatelliteExposure2014}. 
Like for the lunar method, the achievable accuracy for the properties of the primary particles is questionable, but could be sufficient to find out whether particles with extreme energies beyond the known range do exist.

\subsection{Ultra-high precision for air showers around 100 PeV}
After the LOFAR superterp observing air showers with a few $100$ antennas, the next step in precision could be done by the LOFAR extension NenuFAR consisting of almost $2,000$ antennas \cite{NenuFAR2015}, if a larger fraction of these antennas would be used for air-shower detection in addition to astronomical observations.
Just a few years later, the low-frequency instrument of the SKA will be built.
SKA-low will be an ultra-dense, interferometric array with several $10,000$ antennas inside of an area of $1\,$km\textsuperscript{2}.
It can be used to measure the radio emission of air showers in unprecedented detail and commensally with astronomical observations \cite{SKA_ICRC2015}. 
The covered area is slightly larger than the KASCADE-Grande experiment was \cite{Apel2010KASCADEGrande}, and similar to the size of IceTop \cite{IceTopNIM2013}. 
Consequently, the SKA can be used to study cosmic rays in the same energy range around $100\,$PeV.
This will improve our understanding of the second knee in the cosmic-ray energy spectrum and of the transition from galactic to extragalactic cosmic rays.
Having in the order of $10,000$ measurement points per event, it should be possible to reconstruct the shower profile much more accurately than just determining the position of the shower maximum. 
In addition to cosmic-ray physics, this detail of information will be useful for several purposes:
first, to test hadronic interaction models for the air-shower development; 
second, to determine the type of the primary particles more accurately; 
third, for particle physics in this energy range, like the measurement of cross sections \cite{AugerProtonCrosssection2012}.

\subsection{Origin of lightning strikes and structure of thunderstorm clouds}
Thunderstorms are well visible at all radio frequencies.
At lower frequencies of a few MHz, distant lightning strikes are one of the main sources of atmospheric radio background \cite{ITUnoise2015}, and radio measurements are traditionally used to map flashes of lightning \cite{KrehbielLightningMapping2004}. 
Moreover, nearby thunderstorms influence the radio signal emitted by air-showers due to the large electric field in thunderclouds accelerating charged particles of the air showers. 
Earlier analyses at LOPES confirmed this effect \cite{LOPESthunderstorm2007, LOPESthunderstorm2011}, but saw it mainly as a disturbance for cosmic-ray measurements. 
LOFAR, however, has demonstrated that the polarization pattern of the radio emission by air-shower recorded during thunderstorms contains information on the structure of the thundercloud.
Therefore, air-showers can be used to probe thunderclouds over the antenna array \cite{SchellartLOFARthunderstorm2015}. 
This method of thunderstorm research is just at its beginning, and might be continued with higher precision at the SKA. 
A particularly interesting aspect of cosmic-ray related thunderstorm research is whether and in which way the initiation of lightning strikes is linked to air-showers \cite{DwyerThunderstormReview2014}.

\subsection{Search for PeV and EeV photons}
Like cosmic nuclei, high-energy photons also cause air-showers when colliding with the atmosphere.
But unlike hadronic showers, photon induced showers contain almost no muons, i.e., the energy of the photon is almost completely transferred in the electromagnetic shower component. 
Moreover, on average photons penetrate deeper in the atmosphere such that the shower maximum of photon-induced showers is closer to the detector, and the footprint of the radio signal is smaller. 
Nevertheless, the radiation energy of photon-induced showers at radio frequencies is roughly the same or even slightly larger than for hadronic showers. 
Thus, for the same energy of the primary particle, photons should be detectable with roughly the same or slightly lower energy threshold as protons or nuclei. 
In addition to the discrimination power of $X_\mathrm{max}$, antenna arrays could strongly discriminate photons from hadrons if operated in coincidence with muon detectors by searching for air-showers with strong radio signal, but without muons.
So far no gamma rays with PeV energies have been detected with any technique \cite{KASCADEGrandeGammaLimits_ICRC2015}, but the search for them is an important scientific goal \cite{HiSCORE2014}:
gamma rays at PeV energies are absorbed over extragalactic distances, so that their detection can indicate the most energetic cosmic-ray sources in the Milky Way. 
Hence, this science goal is especially interesting for the southern hemisphere from which the Galactic Center can be observed.
For example, a radio extension of the IceCube surface array could be beneficial when searching for inclined photon-induced showers.
Also at higher energies no gamma rays have been detected yet, and the best limits at EeV energies are set by the Pierre Auger Observatory combining particle and air-fluorescence detection \cite{AugerPhotonLimits2009, AugerPhotonPointSources2014}, which exclude some exotic models for the origin of ultra-high-energy cosmic rays. 
Assuming that photons mostly originate from the GZK effect during the propagation of protons and nuclei, the expected photon flux is of the order of $10^{-3}$ to $10^{-4}\,$km$^{-2}$sr$^{-1}$year$^{-1}$ at $1\,$EeV for proton and iron nuclei as primary particles, respectively \cite{KampertUnger2012}, which is about an order of magnitude below current limits.
Thus, a first observation of photons could be expected with a tenfold increase of the exposure the Pierre Auger Observatory currently features for air-fluorescence detection. 
The additional information on the air-showers obtained by radio measurements can help to better separate photons from other primary particles during daytime and bad weather, when no fluorescence detection is possible. 
Hence, radio technique can provide the necessary increase in statistics for a first detection of photons at EeV energies when larger parts of the Pierre Auger Observatory or other experiments will be equipped with antennas.

\begin{figure}[t]
  \centering
  \includegraphics[width=0.69\linewidth]{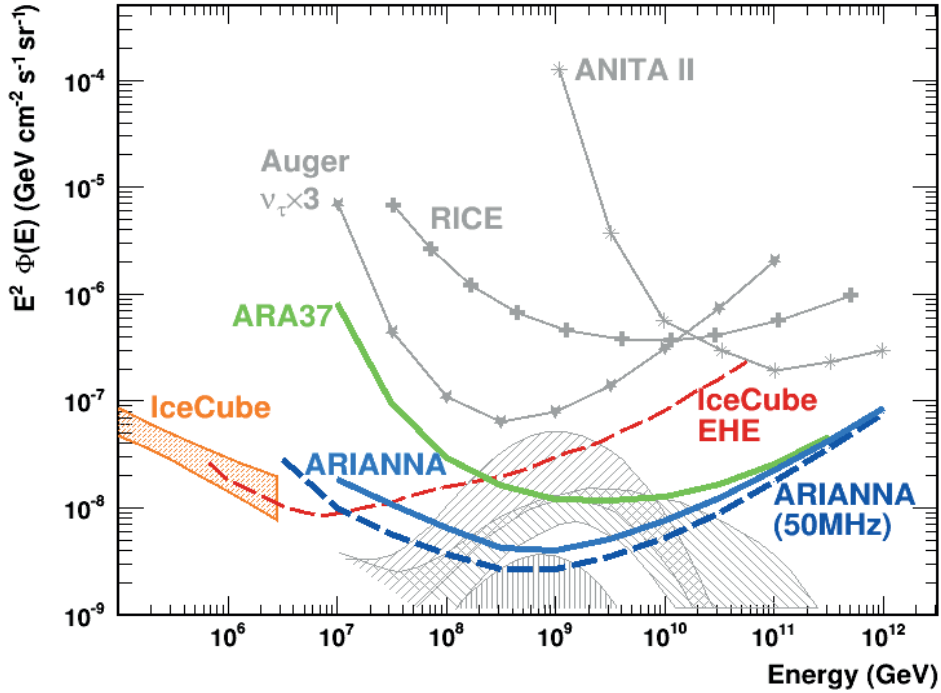}
  \caption{Sensitivity of current and planned radio detectors measuring Askaryan emission from showers in ice compared to the neutrino flux measured by IceCube and to the sensitivity of a future high-energy extension of IceCube. 
  Gray shaded areas indicate various model predictions for the flux of GZK neutrinos.
  The main reason for the different sensitivities of ARA and ARIANNA is the different size of the arrays assumed for the calculation; this figure is in different units than figure \ref{fig_lunarSensitivity}, but the ANITA neutrino limit can be used as reference for comparison (figure from Ref.~\cite{InIceNeutrinoSensitivity_ICRC2015}).}
  \label{fig_inIceSensitivity}
\end{figure}

\subsection{Search for ultra-high-energy neutrinos}
Ultra-high-energy neutrinos can be detected in several ways by particle showers initiated in solid or liquid media (in principle also by air-showers, but the cross section is too low for significant detection statistics). 
IceCube already observed neutrinos of a few PeV by the detection of Cherenkov light from particle showers in ice \cite{IceCubeNeutrinosScience2013}, but the search for neutrinos at higher energies so far has been unsuccessful with any technique, including traditional air-shower detection \cite{AugerNeutrinoPointSources2012}. 
Still, there is a good reason to believe that neutrinos at EeV energies must exist, because they should be produced by interactions of primary cosmic-ray protons or nuclei either in the source or during propagation. 
Therefore, the search is continued.
Radio detection seems to be the most promising technique for observing target volumes much larger than the few km\textsuperscript{3}, which are reasonably possible with optical techniques. 

For this purpose, two main ideas are followed by several experiments for ultra-high-energy neutrinos:
First, the instrumentation or observations of large volumes of ice, like with ANITA \cite{ANITA_neutrino_PRD_2012}, ARA \cite{ARA_2015}, or ARIANNA \cite{ARIANNA_2015} at Antarctica (see figure \ref{fig_inIceSensitivity}). 
The measurement accuracies of the neutrino energy and arrival direction of such arrays have to be investigated more deeply, once the first detection is made. 
At least in principle also the flavor ratio should be measurable when combining radio with optical detectors at lower energy around $10\,$PeV, and with acoustic detectors at EeV energies, since the different detection techniques have different signal responses to charged-current interactions of neutrinos (cf.~chapter \ref{sec_neutrinoReco}).  
Second, the observation of the Moon as target material by several radio telescopes \cite{BrayReview2016} - in the same way as planned for the detection of charged cosmic rays with the SKA \cite{SKAlunar_ICRC2015}. 
Although no neutrino has been detected yet, radio experiments already have provided competitive limits on the neutrino flux.
Next-generation experiments will achieve a sensitivity at which a detection or non-detection of neutrinos will be important to decide between different scenarios for the origin of ultra-high-energy cosmic rays. 
Once neutrinos will have been detected, these observations can be exploited for particle physics in addition to astrophysics, e.g., the angular distribution of showers induced by neutrinos in the Earth contains information on the neutrino cross section at ultra-high energies \cite{ANITAupwardEvent2016}.

%\clearpage
 
\section{Conclusions}
Significant progress has been achieved in the last years regarding the digital radio technique for high-energy cosmic rays and neutrinos:
in several regions of the world antenna arrays have been constructed and measure successfully cosmic-ray air showers.
Moreover, prototypes for large-scale radio arrays aiming at neutrinos have been deployed in Antarctica, and radio emission by particle cascades was measured under laboratory conditions in a variety of accelerator measurements.
The mechanisms of the radio emission now seem to be sufficiently well understood in order to apply the radio technique for serious measurements in astroparticle physics. 
This is a substantial advance compared to the situation in the 1970's when analog radio experiments measured air showers, but were not able to interpret the measurements with sufficient accuracy.
This is also a clear advance to the situation just a few years ago when the digital measurements started, but the interpretation was hampered by a lack of theoretical understanding.
Nonetheless, optical detection techniques are still leading in high-energy astroparticle physics:
in particular the detection of Cherenkov light in water and ice for neutrino detection and in air for photon detection, and the combination of particle detector arrays with air-fluorescence telescopes for extensive air-showers initiated by ultra-high-energy cosmic rays.
However, this might change soon: 
the current generation of antenna arrays for air showers has demonstrated that digital radio detection together with sophisticated data processing can compete with the established techniques in accuracy. 
Even though the accuracy recently achieved for radio detection is yet slightly worse than that of air-fluorescence detection when taking into account all systematic uncertainties, this is compensated by the higher duty cycle, since the radio technique is not limited to clear nights.

So can radio detection completely replace the established techniques?
The answer likely is no, but certainly there are some aspects for which the radio technique is taking over: 
for the search for neutrinos above $100\,$PeV energy, radio detection already is the most promising option.
For cosmic rays, there is a competition between different techniques when aiming at huge exposures, and radio will have a realistic chance to become the technique of choice, since there at least three ideas how radio detection can provide huge exposure.
All ideas need further investigation, but might be realizable with a sensible amount of resource: 
first, huge antenna arrays of several $10,000\,$km\textsuperscript{2} for inclined air showers \cite{GRAND_ICRC2015}; second, the observation of particle showers in the lunar regolith \cite{SKAlunar_ICRC2015}; third, the observation of the atmosphere with antennas from space \cite{SWORDarxiv2013}.
In other aspects, radio detection will not replace existing techniques.
Simply because of the high threshold it makes no sense to use radio detection for photon or neutrino detection below the PeV energy range. 
For air showers the true potential of the radio technique is not in the replacement, but in the combination with other techniques. 
This is especially interesting to enable further progress in this research fields without the need for significant new resources, because radio extensions are relatively economic compared to other air-shower detectors. 
The additional information provided by radio measurements of air showers can increase the total accuracy of the reconstructed energy and mass of the primary particle, which is of utmost importance in order to find and understand the sources of ultra-high-energy cosmic rays.
Therefore, the addition of radio antennas to air-shower observatories might bring the necessary step in accuracy to distinguish between competing scenarios for the origin of cosmic rays.

While the principle advantages of radio detection are clear now, a lot of work still has to be done on the details for further advancing the technique. 
For neutrinos, a successful proof-of-principle is needed, which likely requires further extending the size of existing experiments by at least an order of magnitude. 
For cosmic-ray air showers, the principle issues are solved, so one has to go deeper in order to improve.
This means a better study of systematic uncertainties, a better absolute calibration of antennas, and subsequently a more accurate testing of simulation codes representing our 
understanding of air-shower physics and the associated radio emission. 
If this is done, there is a reasonable chance that radio measurements can become even more accurate for the shower energy and for $X_\mathrm{max}$ than the established air-fluorescence and air-Cherenkov techniques, because current antenna arrays have already reached equal precision, but are not yet at the theoretical limit. 
Moreover, accurate radio measurements of the energy scale and of the shower development will help to improve our understanding of particle cascades at energies beyond the range of LHC, and consequently be valuable input to high-energy particle and astroparticle physics in general. 
Concluding, it seems highly advisable to equip any future cosmic-ray observatory with additional radio antennas, and to spend the minimal additional resources required to observe air-showers commensally with any astronomical radio observatories operating in the frequency range of a few MHz to a few GHz.

\vspace{3cm}

\section*{Acknowledgments}
Thanks go to all authors and collaborations providing permission to reuse their work for this review. 
Discussions with colleagues at the institute, and of the LOPES, Pierre Auger and Tunka-Rex Collaborations have been very useful. 
Especially I would like to thank Johannes Bl\"umer, Andreas Haungs, Tim Huege, and several other colleagues for proofreading and for providing ideas on how to improve this article. 
Finally I would like to thank my employer, the Karlsruhe Institute of Technology (KIT), for funding my position, in particular since this article is submitted as my habilitation thesis.

\end{document}